\documentclass[a4paper, oneside,english,12pt]{amsart}
\usepackage[T1]{fontenc}
\usepackage[utf8]{inputenc}

\usepackage{geometry}
\usepackage{enumitem}
\usepackage{amstext}
\usepackage{amsmath}
\usepackage{amsthm}
\usepackage{amssymb}
\usepackage{mathtools}
\usepackage{bm}
\usepackage{subcaption}
\usepackage{blkarray}
\usepackage{soul}
\usepackage{tablefootnote}
\usepackage{tikz-qtree}
\usepackage[toc,page,header]{appendix}
\usepackage{makecell}

\newcommand{\bH}{\mathbf{H}}      
\usepackage{chngcntr}

\usepackage[font=tiny,labelfont=bf]{caption}

\usepackage{titletoc}

\usepackage[authoryear,round]{natbib}

\allowdisplaybreaks

\usepackage[bb=boondox]{mathalfa}

\usepackage{tikz}
\usetikzlibrary{arrows}
\usetikzlibrary{shapes}
\usetikzlibrary{plotmarks}

\makeatletter

\usepackage{algorithm}
\usepackage[noend]{algpseudocode}

\algnewcommand{\lIfThenElse}[3]{%
  \State \algorithmicif\ #1\ \algorithmicthen\ #2\ \algorithmicelse\ #3}

\newcommand{\bX}{\mathbf{X}}

\algnewcommand{\lIfThen}[2]{%
  \State \algorithmicif\ #1\ \algorithmicthen\ #2}

\algdef{SE}[SUBALG]{Indent}{EndIndent}{}{\algorithmicend\ }%
\algtext*{Indent}
\algtext*{EndIndent}

\makeatother

\usepackage{babel}

\DeclareMathOperator*{\argmin}{argmin}

\newcommand{\norm}[1]{\left\lVert#1\right\rVert}
\newcommand{\abs}[1]{\left\lvert#1\right\rvert}

\usepackage{setspace}
\usepackage[colorlinks,linkcolor=darkbrown,citecolor=green,urlcolor=blue]{hyperref}
\usepackage{cleveref}
\usepackage{graphicx,xcolor}

\definecolor{orange}{rgb}{1,0.5,0}
\definecolor{blue}{rgb}{0.22, 0.58, 0.82}
\definecolor{green}{rgb}{0.2, 0.65, 0.32}
\definecolor{red}{rgb}{0.91, 0.26, 0.2}
\definecolor{purple}{rgb}{0.46, 0.21, 0.68}
\definecolor{bluegray}{rgb}{0.04,0,0.7}
\definecolor{darkbrown}{rgb}{0.40,0.2,0.05}
\definecolor{forestgreen}{RGB}{34, 139, 34}

\definecolor{green}{RGB}{46,139,87}

\newcommand{\mc}{\mathcal}

\newcommand{\E}{\mathbb{E}}
\newcommand{\loss}{\mathcal{L}}

\newcommand{\bigO}{\mathcal{O}}

\newcommand{\child}{\mathrm{child}}
\newcommand{\fix}{\mathfrak{f}}

\renewcommand{\Pr}{\mathbb{P}}
\newcommand{\y}{\mathbf{y}}
\newcommand{\D}{\mathbf{D}}

\newcommand{\Z}{\mathbf{Z}}

\newcommand{\X}{\mathbf{X}}
\newcommand{\F}{\mathbf{F}}

\newcommand{\balpha}{\boldsymbol{\alpha}}
\newcommand{\bbeta}{\boldsymbol{\beta}}
\newcommand{\bgamma}{\boldsymbol{\gamma}}
\newcommand{\bdelta}{\boldsymbol{\delta}}

\newcommand{\bmu}{\boldsymbol{\mu}}
\newcommand{\bSigma}{\boldsymbol{\Sigma}}
\newcommand{\bLambda}{\boldsymbol{\Lambda}}
\newcommand{\bP}{\boldsymbol{P}}

\newcommand{\bt}{\boldsymbol{t}}

\newcommand{\incomp}{\not\lessgtr}

\newtheorem{theorem}{Theorem}
\newtheorem{corollary}{Corollary}
\newtheorem{lemma}{Lemma}
\newtheorem{definition}{Definition}
\newtheorem{proposition}{Proposition}

\theoremstyle{definition}

\newtheorem{exampleplain}{Example}
\newtheorem*{exampleplains}{Example}
\newenvironment{example}
 {\pushQED{$\hfill\square$}\exampleplain}
 {\popQED\endexampleplain}
\newenvironment*{example*}
 {\pushQED{$\hfill\square$}\exampleplains}
 {\popQED\endexampleplain}

\newenvironment{examplecont}[1][]{
\begin{trivlist}
   \item[\hskip \labelsep {\bfseries Example #1 (continued).}]}{\hfill{}$\square$
   \end{trivlist}
}

{\hfill $\square$}

      \newenvironment{proofof}[1][]{\begin{trivlist}
   \item[\hskip \labelsep {\bfseries Proof of #1.}]}{\hfill{}$\square$\end{trivlist}}

\theoremstyle{remark}

\renewcommand{\appendixpagename}{Supplement to Robustly estimating heterogeneity in factorial data using Rashomon Partitions}

\title[Rashomon Partitions for Heterogeneity]{Robustly estimating heterogeneity in factorial data using Rashomon Partitions}
\author{Aparajithan Venkateswaran$^{\S}$}
\author{Anirudh Sankar$^{\ddagger}$}
\author{Arun G. Chandrasekhar$^{\ddagger,\star}$}
\author{Tyler H. McCormick$^{\S,\P,\dagger}$ }

\thanks{$^{\S}$Department of Statistics, University of Washington, USA}
\thanks{$^{\ddagger}$Department of Economics, Stanford University, USA}
\thanks{$^{\star}$J-PAL, NBER, USA}
\thanks{$^{\P}$Department of Sociology, University of Washington, USA}
\thanks{$^{\dagger}$Corresponding author, tylermc@uw.edu, (206)543-5396.}

\begin{document}

\begin{abstract} 
In both observational data and randomized control trials, researchers select regression models to articulate how the outcome of interest varies with combinations of observable covariates.  
Choosing a model that is too simple can obfuscate important heterogeneity in outcomes between covariate groups, while too much complexity risks identifying spurious patterns.  
In this paper, we propose a novel Bayesian framework for model uncertainty in regression models called \emph{Rashomon Partition Sets} (RPSs).  The RPS consists of \emph{all} models that have posterior density close to the \emph{maximum a posteriori (MAP)} model. We construct the RPS by enumeration, rather than sampling, which ensures that we explore all models with high evidence in the data,  even if they offer dramatically different substantive explanations. 
We use a $\ell_0$ prior, which allows us to capture complex heterogeneity without imposing strong assumptions about the associations between effects, showing that this prior is minimax optimal from an information-theoretic perspective. We characterize the approximation error of (functions of) parameters computed conditional on being in the RPS relative to the entire posterior.  We propose an algorithm to enumerate the RPS from the class of models that are interpretable and unique, then provide bounds on the size of the RPS. We give simulation evidence along with three empirical examples: price effects on charitable giving, heterogeneity in chromosomal structure, and the introduction of microfinance.

\end{abstract}

\maketitle

\newpage 

\section{Introduction}
\begin{quote}
 \small ``You didn't come here to make the choice, you've already made it. You're here to try to understand why you made it. I thought you'd have figured that out by now.''

    \hspace{2.75in} --- The Oracle, \emph{The Matrix Reloaded}
\end{quote}

We explore model uncertainty in a setting with heterogeneous effects and a discrete, partially ordered covariate space. Suppose there are $n$ units and each has $M$ features, each taking one of $R$ partially ordered values. Let $\D$ be the indicator matrix with entries $D_{ik}=1$ if observation $i$ has feature combination $k$, where $\mathcal{K}$ is the set of $K := R^M$ feature combinations (cells).
The dataset is $\Z := (\y,\X)$ and we consider the regression model
\begin{align}\label{eq:mainreg}
   \y = \D \bbeta + \boldsymbol{\zeta},
\end{align}
where $\beta_k = \E[Y_i \mid D_{ik}=1]$ is the expected outcome in cell $k$ and $\boldsymbol{\zeta}$ is unstructured noise.
To ground our discussion, we introduce the following example, which will run throughout. 
\begin{example}[Running example]\label{ex:amox-ibu}
There are two medications, Amoxicillin and Ibuprofen.  A group of patients is already taking both medications, but at different doses.  Patients take Amoxicillin at 250 mg or 500 mg; Ibuprofen is taken at 200mg or 400mg. The outcome is a pain score reported by the patient. There are four possible $\beta_k$'s in  Equation~\ref{eq:mainreg}: $\beta_{250,200}$,$\beta_{250,400}$, $\beta_{500,200}$, and $\beta_{500,400}$. Model selection amounts to choosing which $\beta_k$'s to set equal, if any.  Setting $\beta_{250,200}=\beta_{500,200}$ and $\beta_{250,400}=\beta_{500,400}$, for example, allows tests whether a higher dose of Ibuprofen is associated with a lower pain score, regardless of Amoxicillin levels.  There's likely association between these local dose effects.  If increasing Ibuprofen changes pain at one Amoxicillin dose, the corresponding effect at the other Amoxicillin dose is likely related, though the direction and magnitude of that relationship are not known \emph{a priori}.
\end{example}
We propose a paradigm to address model uncertainty in this setup: linear regression with discrete, partially ordered covariates and unknown correlation structure in the covariates.  In a Bayesian framework, we develop \emph{Rashomon Partition Sets (RPSs)}, which consist of models that are in the neighborhood of the \emph{maximum a posteriori} (MAP) model. We find the RPS by enumeration rather than sampling, ensuring we see all possible models with a high posterior density.      
Models with similar posterior support can differ dramatically in their scientific interpretation or policy implications.  Exploring the RPS, therefore, gives the scientist or policymaker a holistic view of uncertainty in the space of mechanisms that enjoy a high level of support from the data at hand.  A solid consensus within the RPS corresponds to a robust archetype of the scientific phenomenon.  However, multiple disparate explanations among the models in the RPS indicates that either (i) a small change in the observed data could dramatically impact substantive conclusions or (ii) the space of models is not sufficiently rich to distinguish among multiple scientific explanations.  
Moreover, if the goal is to choose a specific model (e.g., to enact a specific policy), the researcher can weigh additional considerations (e.g., cost of an intervention, equity, privacy), since all models in the RPS enjoy posterior support similar to the MAP. 

We prove several appealing results and provide new insights across three data examples. We propose an $\ell_0$ prior that encourages simple models by penalizing the \emph{number} of unique $\beta_k$'s in the model. This prior does not require specifying the correlation structure between variable effects \emph{a priori} and, we show, is minimax optimal in the sense of minimizing the worst-case discrepancy to the posterior obtained under a prior better aligned with the underlying correlation structure (which is, of course, unknown in practice). %
  We characterize the approximation error of (functions of) parameters computed conditional on being in the RPS relative to the entire posterior.  Finally, we propose conditions to restrict the set to candidate models only to those that are unique and interpretable, and, under these conditions, derive bounds on the size of the RPS.  We propose an algorithm to enumerate the models in the RPS and demonstrate its potential to arrive at new scientific insights in the context of three data examples from diverse applied domains.  In all three examples, we find divergent conclusions compared to those in the extant literature.

In the remainder of this section, we provide background and a road map for the rest of the paper.
The term ``Rashomon'' references \citet{breiman2001statistical}'s ``Statistical Modeling: The Two Cultures'' paper, which describes ``a multitude of different descriptions [equations $f(x)$] in a class of functions giving about the same minimum error rate'' as the \emph{Rashomon Effect}.
The term is an allusion to a movie directed by Akira Kurosawa (based on a short story, \emph{In the Grove}, by Akutagawa Ry\={u}nosuke) that recounts the same event from multiple perspectives, representing several credible interpretations of the same situation. 
Breiman presents the Rashomon effect as an observation, and subsequent work in the context of prediction demonstrates that Rashomon sets will essentially always exist in contexts with inherent noise~\citep{semenova2022existence, xin2022exploring}.

We propose leveraging the Rashomon Effect for statistical model uncertainty.  Our approach represents a balance between a preference for robustness and a desire to remain anchored by the observed data. %
In their seminal paper developing an \emph{Occam's Window} approach to Bayesian Model Averaging (BMA) for graphical models,~\cite{madigan1994model} %
articulate a similar philosophical perspective.  They say:
\begin{quote}
\small
[standard BMA] does not accurately represent model uncertainty. Science is an iterative process in which competing models of reality are compared on the basis of how well they predict what is observed; models that predict much less well than their competitors are discarded. Most of the models in [standard BMA] have been discredited [...] so they should be discarded. %
\end{quote}

To instantiate this philosophy, we use the MAP as an anchor and then enumerates a list of models with similar posterior density.  Enumerating rather than sampling allows us to find the \emph{entire} set of high posterior models, and improves efficiency by focusing \emph{only} on high posterior models. Models in the RPS can achieve the same level of fit with wildly different substantive explanations, so enumerating them is critical to ensure we explore the entire space of explanations supported by the data at hand. Recent papers use related approaches to find examples, rather than the full set, of high posterior models (e.g., \citet{muller2011ppmx,rovckova2018particle} or \citet{balocchi2023crime}).

\begin{figure}
    \centering
\includegraphics[width=.45\textwidth]{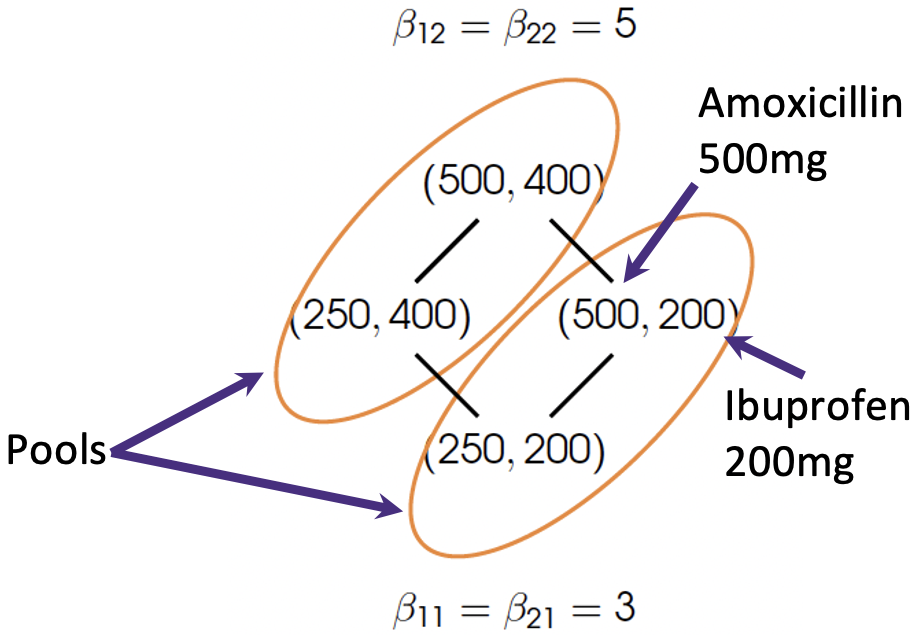}    \caption{\tiny Running example.  Each node in the Hasse diagram corresponds to a specific level of each factor.  \emph{Pools} (see Definition~\ref{def:pool}) aggregate across nodes.}
    \label{fig:hass-running}
\end{figure}

{ We find the RPS by searching over a geometry for partially ordered sets known as the Hasse diagram. ~\citet{banerjee2021selecting} introduced the Hasse diagram as a representation of the model space. However, their work aims to select a single model using a frequentist approach, whereas our Bayesian setup addresses model uncertainty. In this discrete covariate space, each statistical model corresponds to a partition of the Hasse diagram, constructed setting some of the $\beta$'s in Equation~\ref{eq:mainreg} to be equal and, thus, some feature combinations to have the same expected outcome.  %
Figure~\ref{fig:hass-running} shows a Hasse diagram, which we now interpret in the context of our running example.} 

\begin{examplecont}[\ref{ex:amox-ibu}]
  The nodes in the Hasse diagram correspond to feature combinations, each consisting of unique factor levels. The first level in each node in Figure~\ref{fig:hass-running} corresponds to the dose of Amoxicillin (250mg or 500mg) and the second corresponds to the dose of Ibuprofen (200mg or 400mg). %
  \end{examplecont}

{ %
Each statistical model corresponds to a single partition of the Hasse diagram.  This contrasts with other related geometries, such as trees. Trees split hierarchically among variables with no ordering (e.g., splitting a tree on Ibuprofen or Amoxicillin first is arbitrary). As the complexity of the tree grows, so does the number of equivalent trees.  Without additional weighting, using trees implies a preference for more complex models. Since there are more duplicates of complex trees, they are more likely to be sampled in a sampling-based approach (e.g.,~\cite{wager2018estimation}) or will be overrepresented in the Rashomon set in an enumeration approach.}   

We also impose \emph{permissibility conditions} based on scientific principles that limit our search to the space of interpretable models. We think of the world in increments: how changing a level in one variable marginally affects the outcome.  The overall effect of a feature combination is the sum of its marginal effects. We also rule out partitions that correspond to measure zero events. Since the drugs work in different ways and for different purposes, we would not expect a partition whose interpretation requires the marginal effect of increasing Amoxicillin to be exactly canceled by the marginal effect of increasing Ibuprofen, for example. Such a partition would arise only under an extremely rare coincidence
(e.g. Amoxicillin causes upset stomach that is \emph{exactly} offset by the pain reliever Ibuprofen). %
\citet{banerjee2021selecting} introduces related restrictions within a modeling framework that is bespoke to randomized trials. %

The Hasse diagram allows us to explore complex interactions between factors, making it extremely difficult to supply a prior over the covariance between all of the marginal effects.  What's more, they are rarely independent. As discussed previously, for example, the effect of taking more Ibuprofen with a low level of Amoxicillin is not independent of taking more Ibuprofen at a higher level of Amoxicillin.  %
To address this, we use a prior that controls the \emph{number} of unique $\beta_k$'s. Our prior is the $\ell_0$ prior, which does not impose any structure on the relationship between the variable effects. We show this prior is minimax optimal. Among priors with the same number of distinct ($\beta_k$)'s, it is least vulnerable, in worst-case posterior total variation distance, to misspecifying the unknown dependence structure among effects.

Together, our prior and permissibility restrictions, both motivated scientifically, substantially improve computational efficiency, making it possible to show that the RPS is bounded in size polynomially (in the number of features and levels), and enumerate the entire RPS in realistic data examples.  They also allow us to accommodate an arbitrary dependence structure among covariates, which contrasts with related work that either requires independence (e.g., ~\citet{rovckova2018particle}) or known structure (e.g.,~\citet{balocchi2023crime}). For example, \citet{banerjee2021selecting} adopt a LASSO ($\ell_1$) framework that treats marginal effects as independent \emph{a priori}. However, in practice, we expect a positive correlation (``more at higher doses'') or a negative correlation (``plateau or decline'') between marginals.  %
We show that these theoretical and operational issues with $\ell_1$ matter in finite samples in Section \ref{section:simulations}. %

The remainder of the paper is structured as follows. In Section~\ref{sec:defnRPS}, we define the RPS formally. Then, we give statistical properties for an arbitrary set of partitions and introduce our minimax optimal $\ell_0$ penalty.  We then give a formal definition of our permissible partition structure in Section~\ref{section:environment}.
We show that this combination allows us to bound the size of the RPS in Section~\ref{section:rps-size} and enumerate it entirely in Section~\ref{section:enumeration}. Section~\ref{section:simulations} provides simulation evidence and Section~\ref{section:real-data} gives three empirical examples, highlighting robust archetypes in each setting. Finally, Sections~\ref{section:related-body} and~\ref{section:conclusion} provide a discussion of related work and future directions, respectively. 
All of our code is available at \url{https://github.com/AparaV/rashomon-partition-sets}.

\section{Rashomon Partition sets}
\label{sec:defnRPS}
In this section, we define Rashomon Partition Sets.  We also explore the statistical properties, first considering the posterior over the model coefficients, then the poseterior over the space of models.
To begin, we formally introduce the notion of a pool:

\begin{definition}[Pool]
\label{def:pool}
{A \emph{pool} $\pi$ is a set of feature combinations having identical expected outcomes.}
\end{definition}
{For a given pool $\pi$, two feature combinations $k^{(1)}, k^{(2)} \in \pi$ only if $\beta_{k^{(1)}} = \beta_{k^{(2)}}$. The converse is not true. That is, we could have $k^{(1)} \in \pi_1$ and $k^{(2)} \in \pi_2$ for $\pi_1 \neq \pi_2$ even though $\beta_{k^{(1)}} = \beta_{k^{(2)}}$.}
Pools are similar to clusters, but we are not grouping observations based on similar covariates but, instead, based on similar (expected) outcomes.
Observations in the same pool have the same expected outcome. A \emph{partition}, $\Pi$, is a set of pools such that every observation is assigned to a single pool.   

\begin{definition}[Partition]
\label{def:partition}
{Given $M$ features taking on $R$ partially ordered values each, a \emph{partition} $\Pi$ is a partitioning of this feature space into pools.}
\end{definition}

\begin{examplecont}[\ref{ex:amox-ibu}]
  Figure~\ref{fig:hass-running} illustrates a partition with two pools.  The first corresponds to a high dose of Ibuprofen (400mg) and any dose of Amoxicillin, while the second corresponds to a low dose (200mg) of Ibuprofen and any dose of Amoxicillin. Testing for a difference in outcomes across the pools in Figure~\ref{fig:hass-running} corresponds to testing for the marginal effect of increasing Ibuprofen, since we have aggregated over the two possible doses of Amoxicillin.  
\end{examplecont}

The partition, $\Pi$, in the space of all partitioning models, $\mathcal{P}$, is a model of heterogeneity such that for every pool $\pi \in \Pi$, possibly a singleton, if feature combinations $k,k'\in \pi$, then $\beta_k = \beta_{k'}$. Identifying heterogeneity in outcomes, then, becomes a search across $\mathcal{P}$. The posterior given the data $\Z$ is $\Pr(\Pi \mid \Z)$. Let $\mathcal{P}^{\star} \subseteq \mathcal{P}$ be the set of permissible partitions that obey some permissibility rules (to be defined in \Cref{section:environment}).
\begin{definition}[Rashomon Partition Set (RPS)]
\label{def:RPS}
For some posterior probability threshold $\tau \in [0,1]$, the \emph{Rashomon Partition Set} relative to a reference partition $\Pi_0$, $\mathcal{P}_\tau(\Pi_0)$, is %
\begin{align}
\mathcal{P}_{\tau}(\Pi_0) = \{ \Pi \in \mathcal{P}^{\star} : \ \Pr(\Pi \mid \Z) \geq (1-\tau) \cdot \Pr(\Pi_0 \mid \Z) \} .\label{eq:RPS}   
\end{align}
\end{definition}
The RPS relative to $\Pi_0$ is the set of partitions that have a similar or higher posterior value than the reference. In our analysis, we are interested in $\Pi^{\text{MAP}}$---the \emph{maximum a posteriori (MAP)} partition, so we will focus on $\mathcal{P}_\tau(\Pi^{\text{MAP}})$. That is, in our setting the RPS is the set of partitions that are sufficiently close to the posterior of the MAP partition. We write this as $\mathcal{P}_\tau$, dropping the reference argument unless explicitly needed. We could, more generally, define an RPS based on any posterior threshold, $\theta$, such that $\mathcal{P}_{\theta} = \{ \Pi \in \mathcal{P}^{\star} : \ \Pr(\Pi \mid \Z) \geq \theta \}$.  
Defining the RPS as a Bayesian model allows us to use the infrastructure of posterior probabilities to compare multiple models and quantify uncertainty.
We begin with an initialization partition, $\Pi_0$. We then enumerate $\mathcal{P}_{t}(\Pi_0)$ with $t = 0$, which by definition includes $\Pi^{\text{MAP}}$. We construct the RPS by moving down the list of partitions, ordered by relative posterior, until we reach $(1-\tau)\Pr(\Pi^{\text{MAP}}|\Z)$.

We now elaborate on the statistical framework underlying the RPS.
Our first goal is to describe how well we can approximate the distribution over effects, $P_{\bbeta \mid \Z}$, using the RPS. We then discuss how to construct the posterior over partitions, $\Pr(\Pi \mid \Z)$, using generalized Bayesian inference. Technical details and full proofs are in \Cref{appendix:posterior-approx}.

\subsection{Posterior over effects.}

We start with the posterior over the entire set of permissible pools:
\begin{align*}
    P_{\bbeta \mid \Z}(\bbeta) &= \sum_{\Pi \in \mc{P}^\star} \Pr(\bbeta \mid \Z, \Pi)\, \Pr(\Pi \mid \Z),    
\end{align*}
where $P_{\bbeta \mid \Z}$ denotes the marginal posterior distribution of $\bbeta$ given $\Z$, and analogously for measurable functions of $\bbeta$.
Throughout our analysis, we assume that $P_{\bbeta \mid \Z}$ is a proper distribution, i.e., it satisfies the Kolmogorov axioms. We will approximate functions of $P_{\bbeta \mid \Z}$ using only the RPS. We give results for the general definition of the RPS with respect to a threshold $\theta$, though in practice we define the RPS relative to the MAP, setting $\theta = (1-\tau)\cdot\Pr(\Pi^{\text{MAP}} \mid \Z)$ so that $\mc{P}_{\tau}(\Pi^{\text{MAP}}) = \mc{P}_{\theta}$.

 As discussed above, when referencing~\citet{madigan1994model}, we expect that, in many settings, the scientific object of interest will be the posterior within the RPS.  It remains useful, however, to understand how the restricted posterior compares to the full posterior.  To do this, we characterize the uniform approximation error of the posterior distribution of $\bbeta$, and measurable functions of it, restricting to the RPS (Theorem \ref{thm:marginal_approx_rashomon}). 
For a given threshold $\theta$ and corresponding RPS $\mc{P}_{\theta}$, define the posterior over partitions restricted to this set,
\begin{align*}
\Pr(\Pi \mid \Z, \mc{P}_\theta)
  &:= \frac{\Pr(\Pi \mid \Z)}
           {\sum_{\Pi^{\prime} \in \mc{P}_{\theta}} \Pr(\Pi^{\prime} \mid \Z)} \,,
  \qquad \Pi \in \mc{P}_\theta.
\end{align*}
The posterior for $\bbeta$ restricted to $\mc{P}_\theta$ is then
\begin{align*}
P_{\bbeta \mid \Z, \mc{P}_{\theta}}(\bbeta)
  &:= \Pr(\bbeta \mid \Z, \Pi \in \mc{P}_\theta)
   = \sum_{\Pi \in \mc{P}_\theta} \Pr(\bbeta \mid \Z, \Pi)\,
      \Pr(\Pi \mid \Z, \mc{P}_\theta),
\end{align*}
and analogously for measurable functions of $\bbeta$.
This approximation only evaluates models in the RPS and normalizes the posterior over partitions within the RPS. The quality of the approximation depends on both the shape of the posterior (i.e., how concentrated the posterior is around the highest probability models) and the structure of the RPS.

\begin{theorem}[Rashomon approximation of posterior effects]
\label{thm:marginal_approx_rashomon}
Let $f: \mathbb{R}^K \to \mathbb{R}^m$ be a measurable function of the effects $\bbeta$, where $K$ is the number of unique feature combinations and $m \geq 1$. Then the posterior distribution of $f(\bbeta)$ over the Rashomon Partition Set uniformly approximates the entire posterior of $f(\bbeta)$ in the sense that
\begin{align*}
    \sup_{\bt} \abs{ F_{\bbeta \mid \Z, \mc{P}_\theta}(\bt) - F_{\bbeta \mid \Z}(\bt ) } &\leq \begin{cases}
        \min \left\{1, 2 \left( 1 - \abs{\mc{P}_{\theta}} \theta \right) \right\}, & \theta > 1 / \abs{\mc{P}^{\star}}, \\[0.5em]
        \min \left\{ 1, 2 \left(\abs{\mc{P}^{\star}} - \abs{\mc{P}_{\theta}} \right) \theta \right\}, &\text{else},
    \end{cases}
\end{align*}
where $F_{\bbeta \mid \Z}$ is the distribution function of $f(\bbeta) \mid \Z$ and $F_{\bbeta \mid \Z, \mc{P}_\theta}$ is the same but conditioned on the RPS, for all $\theta \in (\min_{\Pi} \Pr(\Pi \mid \Z), \max_{\Pi} \Pr(\Pi \mid \Z))$.
\end{theorem} 

For $\theta > \max_{\Pi} \Pr(\Pi \mid \Z)$, the RPS is empty, so the Rashomon approximation of the posterior is not defined. For $\theta < \min_{\Pi} \Pr(\Pi \mid \Z)$, the RPS encompasses the entire space, so the RPS recovers the exact posterior.
The behavior of the error in \Cref{thm:marginal_approx_rashomon} depends on the size of the RPS, $\abs{\mc{P}_{\theta}}$, relative to the full space, $\abs{\mc{P}^{\star}}$, for a given $\theta$.
The threshold for the regime change, $1 / \abs{\mc{P}^{\star}}$, can be viewed as the probability of choosing a partition $\Pi \in \mc{P}^{\star}$ from a uniform distribution.
When $\theta$ is larger than this uniform prior, the error depends only on the posterior mass accumulated by the RPS, i.e., $\abs{\mc{P}_{\theta}} \theta$. When $\theta$ is smaller than the uniform prior, the error depends on how much posterior mass has been left out of the RPS, i.e., $\left(\abs{\mc{P}^{\star}} - \abs{\mc{P}_{\theta}} \right) \theta$.
It is only the choice between the calculated bound and the trivial bound of $1$ that depends on the behavior of the posterior distribution, i.e., on $\abs{\mc{P}_{\theta}} \theta$. We visualize the full error bound in simulations in \Cref{section:simulations} %
the behavior of $\mc{O}(\theta \abs{\mc{P}_{\theta}})$ in \Cref{section:real-data}.

Setting $f(\bbeta) = \bbeta$ recovers the posterior of $\bbeta$. The function $f$ also covers other useful quantities derived from $\bbeta$. One example is $f(\bbeta) = \max_k \beta_k$, the maximum expected outcome over feature combinations $k$. Conditional on a given feature combination being estimated as the one with the maximum effect, there is a winner's curse: selecting the maximum induces positive bias, so the posterior needs to be adjusted to have a lower mean in order to correct this bias \citep{andrews2019inference}. Other examples include the variability of outcomes across feature combinations, $f(\bbeta) = \sum_{k=1}^K (\beta_k - \bar{\beta})^2$ where $\bar{\beta} = \frac{1}{K} \sum_{k=1}^K \beta_k$,
and quantiles of the expected outcome distribution.

We now focus specifically on estimating the full posterior mean using only the RPS. Let $\bbeta_{\Pi} := \E(\bbeta \mid \Z, \Pi)$ denote the posterior mean of $\bbeta$ under partition $\Pi$. The overall posterior mean can be written as the mixture
\[
\E_{\Pi \mid \Z} \bbeta
  := \E(\bbeta \mid \Z)
  = \sum_{\Pi \in \mc{P}^{\star}} \bbeta_{\Pi} \, \Pr(\Pi \mid \Z).
\]
For some priors on $\bbeta$, we could approximate $\Pr(\Pi \mid \Z)$ directly, but this requires specifying a prior on $\bbeta$ and an approximation with adequate accuracy (Appendix~\ref{appendix:gprior-marginal} gives an example using Gaussian priors). More generally, we instead work with the posterior restricted to the RPS,
and approximate this restricted posterior mean using self-normalized posterior weights over $\mc{P}_\theta$. Define the mean conditional effect estimator
\begin{align}
    \bar{\bbeta}_{\mc{P}_{\theta}}
    &:= \sum_{\Pi \in \mc{P}_{\theta}} \bbeta_{\Pi} \,
       \frac{\Pr(\Pi \mid \Z)}
            {\sum_{\Pi^{\prime} \in \mc{P}_{\theta}} \Pr(\Pi^{\prime}  \mid \Z)}.
    \label{eq:conditional-effects-RPS}
\end{align}
If the RPS captures most of the posterior mass, then $\bar{\bbeta}_{\mc{P}_\theta}$ will be a good approximation to the full posterior mean $\E(\bbeta \mid \Z)$. More generally, we view $\bar{\bbeta}_{\mc{P}_\theta}$ as summarizing the effects implied by the high-posterior partitions in the RPS.
We can then characterize the quality of this approximation for a given RPS.
\begin{corollary}    
\label{lemma:posterior-effects-error}
The mean conditional effect in \Cref{eq:conditional-effects-RPS} approximates the posterior mean effect restricted to the Rashomon set, $\E_{\Pi, \mc{P}_{\theta}} \bbeta$, as
\begin{align*}
    \frac{\norm{\bar{\bbeta}_{\mc{P}_{\theta}} - \E_{\Pi, \mc{P}_{\theta}} \bbeta}}{\norm{\E_{\Pi, \mc{P}_{\theta}} \bbeta} }
    &= \mathcal{O} \left( \frac{1}{\abs{\mc{P}_{\theta}} \theta} - 1 \right).
\end{align*}
If we further have that the effects are bounded, $\norm{\bbeta_{\Pi}} < \infty$ for all $\Pi \in \mc{P}^{\star}$, then the mean conditional effect in \Cref{eq:conditional-effects-RPS} approximates the full posterior mean effect, $\E_{\Pi \mid \Z} \bbeta$, as
\begin{align*}
    \norm{\bar{\bbeta}_{\mc{P}_{\theta}} - \E_{\Pi \mid \Z} \bbeta}
    &= \begin{cases}
         \mc{O} \left( 1 - \abs{\mc{P}_{\theta}} \theta \right), & \theta > 1 / \abs{\mc{P}^{\star}}, \\[0.25em]
        \mc{O} \left( \left(\abs{\mc{P}^{\star}} - \abs{\mc{P}_{\theta}} \right) \theta \right), &\text{else}.
    \end{cases}
\end{align*}
\end{corollary}

The behavior of the error in \Cref{lemma:posterior-effects-error} can be studied similarly to \Cref{thm:marginal_approx_rashomon}. These results extend to functions of $\bbeta$ as well. 

\subsection{Posterior over partitions.}

We now turn to constructing a posterior distribution over partitions,
$\Pr(\Pi \mid \Z) \propto \Pr(\y \mid \X,\Pi) \cdot \Pr(\Pi)$.
We follow a generalized Bayesian inference approach~\citep{bissiri2016general}, which replaces the likelihood contribution with an
exponentiated loss, while retaining a prior over the model class. The advantage in our
setting is that it lets us compare partitions directly through the empirical criterion, without requiring a full joint prior over all
possible interactions in the factorial space. Appendix~\ref{appendix:generalized-bayes}
connects this loss-based formulation back to a fully specified Gaussian Bayesian model:
under an information-scaled Gaussian prior on the pool means, integrating out the pool
means yields a marginal posterior over partitions with the same scaled-loss-and-pool-count-penalty form.  Our algorithm will find the same RPS in either specification. 
Let \(\loss(\Pi;\Z)\) be the loss incurred by partition \(\Pi\), let \(\eta>0\)
be a global loss-scale or learning-rate parameter, and let
\(\exp\{-\lambda H(\Pi)\}\) be the prior over \(\mc P^\star\), where
\(H(\Pi)\) measures partition complexity. We suppress the dependence of the posterior
on the fixed tuning parameters \((\eta,\lambda)\). Then
\begin{align}
    \Pr(\Pi \mid \Z )
      &\propto
      \exp \{ -\eta \loss(\Pi;\Z ) \}
      \cdot
      \exp \{ -\lambda H(\Pi) \}
       =: \exp\{ -Q(\Pi) \},
\label{eq:objective_fcn} \\
       Q(\Pi) &= \eta \loss(\Pi; \Z) + \lambda H(\Pi). \nonumber
\end{align}
The learning rate \(\eta\) calibrates the empirical loss to the information scale of the
data. Since the loss below is an average squared error, a likelihood-scale Gaussian
calibration corresponds to \(\eta\) growing with \(n\); Appendix~\ref{appendix:generalized-bayes}
gives the corresponding expression in terms of the common noise scale and prior
hyperparameters.
In determining the RPS, only the relative scale of the loss and penalty matters, though the scaling does impact posterior probabilities.
We use the mean-squared error for the loss function,
\begin{align}
 \loss(\Pi; \Z)
   &=  \frac{1}{n} \sum_{\pi \in \Pi} \sum_{k(i) \in \pi}
        \bigl(y_i - \widehat{\mu}_{\pi}\bigr)^2,
   \qquad
   \widehat{\mu}_{\pi}
     = \frac{\sum_{k(i) \in \pi} y_i}{ \sum_{k(i) \in \pi} 1}.
 \label{eq:mse-and-predictions}
\end{align}

We define a prior over the \emph{number of distinct pools}, i.e.,
\(H(\Pi) \propto \abs{\Pi}\), the size of the partition. The prior plays a regularizing
role, putting more weight on less granular aggregations. It corresponds to an
\(\ell_0\) penalty: conditional on the number of pools in a partition, all permissible
partitions are equally likely. 
Critically, this prior regularizes the number of pools rather than imposing a particular
correlation structure on effects or interactions across the factorial space. %
The $\ell_0$ prior allows the researcher to express a preference for more
parsimonious models without specifying a full joint distribution over all interactions.
In \Cref{cor:other-losses}, we show that our key algorithm,
yet to be described, can work with any non-negative loss and a penalty
(prior) that is increasing in \(\abs{\Pi}\).

The RPS, taken together with the $\ell_0$ penalty, is similar in spirit to the Occam's window approach used in the context of Bayesian model averaging by~\citet{madigan1994model} and~\citet{madigan96}. These papers use a stochastic search over the discrete space of models that ultimately results in a set of high-posterior models and discards more complicated models if simpler models are found to have higher posterior probability. Our approach includes a prior with an $\ell_0$ penalty as part of the model, rather than using it only to guide the search. In \Cref{thm:l0-minimax}, we show that this choice of prior is minimax optimal.

Let $\mc{Q}$ be a family of priors for the expected outcomes $\bbeta$. For any prior $Q \in \mc{Q}$, denote the posterior over $\bbeta$ given data $\Z$ as $\textrm{P}_{Q, \Z}$, i.e.,
\begin{align*}
    \textrm{P}_{Q, \Z}(\bbeta)
      &= \Pr(\bbeta \mid \Z, \bbeta \sim Q)
       = \frac{\Pr(\y \mid \X, \bbeta)\, Q(\bbeta)}{\Pr( \y \mid \X)}.
\end{align*}

Fix the sparsity at $h$, meaning there are $h$ distinct pools in the partition. Define the restricted space of partitions as
$\mc{P}_{\mid h} = \{\Pi \in \mc{P}^{\star} : H(\Pi) = h\}$.
Let $N(h) = \abs{ \mc{P}_{\mid h}}$. The $\ell_0$ penalty imposes a sparsity restriction on the number of pools. Therefore, at a fixed sparsity level $h$, the $\ell_0$ penalty corresponds to a uniform prior over $\mc{P}_{\mid h}$. Denote this $\ell_0$ prior as $P_{\ell_0}$. For any $\Pi \in \mc{P}_{\mid h}$, $P_{\ell_0}(\Pi) = {1}/{N(h)}$.

For any given $\bbeta$, there is a corresponding permissible partition $\Pi_{\bbeta} \in \mc{P}^{\star}$. We can then define $\mc{Q}_{\mid h}$ to be the family of priors for the restricted space of $\bbeta$ such that $\Pi_{\bbeta} \in \mc{P}_{\mid h}$. Let $\mc{Q}_{\mc{P} \mid h}$ denote the family of priors, derived from $\mc{Q}_{\mid h}$, over partitions in $\mc{P}_{\mid h}$. We can move from $\mc{Q}_{\mid h}$ to $\mc{Q}_{\mc{P} \mid h}$ by noting that for a given $\bbeta$ there is a corresponding permissible partition $\Pi_{\bbeta} \in \mc{P}^{\star}$. Thus, for any prior $Q \in \mc{Q}_{\mid h}$, we can define a prior over $\mc{P}_{\mid h}$ as
\begin{align*}
    Q_{\mc{P} \mid h}(\Pi)
      &= \int_{\bbeta} \mathbb{I}(\Pi_{\bbeta} = \Pi)\, Q(\bbeta)\, d\bbeta,
      \qquad \Pi \in \mc{P}_{\mid h}.
\end{align*}
For reference, we define the supports for these priors in Table \ref{tab:notation-robust-prior}.

For two priors $P, Q \in \mc{Q}_{\mc{P} \mid h}$, define the total variation distance between the corresponding posteriors over partitions as
\begin{align*}
    \delta(P, Q)
      &= \sup_{\Pi \in \mc{P}_{\mid h}}
          \abs{\Pr_{P, \Z}(\Pi) - \Pr_{Q, \Z}(\Pi)},
\end{align*}
where $\Pr_{P,\Z}(\Pi)$ denotes the posterior probability of $\Pi$ given $\Z$ under the prior $P$.

\begin{theorem}
\label{thm:l0-minimax}
For a given sparsity level $h$, the $\ell_0$ penalty is minimax optimal in the sense that
\begin{align*}
    \sup_{Q \in \mc{Q}_{\mc{P} \mid h}}
      \delta(P_{\ell_0}, Q)
    &= \inf_{P \in \mc{Q}_{\mc{P} \mid h}}
        \sup_{Q \in \mc{Q}_{\mc{P} \mid h}}
          \delta(P, Q).
\end{align*}
\end{theorem}

In other words, if one is unwilling to commit to a specific correlation structure for the model coefficients, the $\ell_0$ penalty, which puts a prior on the \textit{number} of selected features, is optimal for model selection. The $\ell_0$ prior is also agnostic to correlation on the scale of variable effects (the change in outcome when moving from one level to another). This property is essential for scientific interpretation, where we often expect some correlation between the effects of changing the level of a feature, regardless of whether or not other features are present. This feature stands in stark contrast to priors built on the $\ell_1$ penalty, which make the unrealistic assumption of independence among variable effects, such as~\citet{banerjee2021selecting}. RPS can, in principle, be built using other priors, but we advocate using this robust prior, which is practical given the complexity of the factorial space of interactions in many scientific settings (e.g., multiple arms and levels in a complex clinical trial, complex economic concepts like entrepreneurship or market context). We do not want to impose false independence or unwarranted assumptions on correlations among the $\beta_k$'s; instead, we aim to be robust in an environment with a complex and unknown correlational structure. We show how this choice lends to computational tractability in \Cref{section:rps-size}.

\subsection{Loss-space characterization of the RPS.}
\label{subsec:loss-space-rps}

It is useful to characterize the Rashomon partition set directly in the score space.
Recall from \eqref{eq:objective_fcn} that
\[
    \Pr(\Pi \mid \Z) \propto \exp\{-Q(\Pi)\},
\]
where \(Q(\Pi)\) is the penalized score for partition \(\Pi\). In the generalized
Bayesian formulation above, this score balances empirical fit and model complexity.
For example, when using the mean-squared-error loss in \eqref{eq:mse-and-predictions},
we have \(Q(\Pi)=\loss(\Pi;\Z)+\lambda H(\Pi)\), or more generally
\(Q(\Pi)=\eta\loss(\Pi;\Z)+\lambda H(\Pi)\) if an explicit loss-scale parameter
\(\eta>0\) is included.

Because posterior probabilities are proportional to \(\exp\{-Q(\Pi)\}\), posterior
ratios are equivalent to differences in the penalized score. Specifically, for any
reference partition \(\Pi_0\),
\[
    \frac{\Pr(\Pi\mid \Z)}{\Pr(\Pi_0\mid \Z)}
    =
    \exp\{-[Q(\Pi)-Q(\Pi_0)]\}.
\]

The normalizing constant cancels out, since it depends only on the data and not the partition. Thus, thresholding posterior probabilities is equivalent to thresholding the penalized
score. This gives the following loss-space characterization of the RPS.

\begin{proposition}[Loss-space characterization of the RPS]
\label{prop:equiv}
Fix a reference partition \(\Pi_0\in\mc P^\star\), and suppose \(Q(\Pi_0)>0\).
For any posterior-ratio threshold \(\tau\in[0,1)\), the RPS
\[
    \mc P_\tau(\Pi_0)
    =
    \{\Pi\in\mc P^\star:
    \Pr(\Pi\mid\Z)\ge (1-\tau)\Pr(\Pi_0\mid\Z)\}
\]
can be written as an additive score-neighborhood
\[
    \mc P_\tau(\Pi_0)
    =
    \{\Pi\in\mc P^\star:
    Q(\Pi)-Q(\Pi_0)\le -\log(1-\tau)\}.
\]
Equivalently, writing $\epsilon = {-\log(1-\tau)} / {Q(\Pi_0)}$, we obtain the relative loss-space form
\begin{align*}
    \mc P_\epsilon(\Pi_0) &= \{\Pi\in\mc P^\star: Q(\Pi)\le (1+\epsilon)Q(\Pi_0)\}
\end{align*}
Conversely, any \(\epsilon\ge0\) corresponds to the posterior-ratio threshold $ \tau = 1-\exp\{-\epsilon Q(\Pi_0)\}$.
\end{proposition}

When \(\Pi_0=\Pi_{\mathrm{MAP}}\), the set
\[
    \mc P_\epsilon(\Pi_{\mathrm{MAP}})
    =
    \{\Pi\in\mc P^\star:
    Q(\Pi)\le (1+\epsilon)Q(\Pi_{\mathrm{MAP}})\}
\]
consists of all permissible partitions whose penalized score, balancing fit and
simplicity, lies within a factor \(1+\epsilon\) of the MAP score. This relative
form is the one we use in practice. It has the useful property that multiplying the
entire score \(Q\) by a positive constant does not change the set
\(\mc P_\epsilon(\Pi_{\mathrm{MAP}})\).

The tolerance \(\epsilon\) is specified by the researcher and is interpreted on the
scale of the chosen score \(Q\). If the goal is to approximate the full posterior, then
\(\epsilon\) should be chosen based on computational constraints, since adding more
models to the RPS captures more posterior mass and improves the approximation to
the full posterior. If the goal is scientific interpretation, we choose \(\epsilon\) so
that all partitions in the RPS remain high-quality explanations of the data; adding
models with little support would dilute the scientific interpretation. We give an
example of how to choose \(\epsilon\) in practice in Section~\ref{section:real-data}.

In practice, our strategy has three steps. First, we define a reference partition using
an off-the-shelf algorithm. Second, we use this partition to get a sense of the magnitude
of the score and to enumerate an initial neighborhood of candidate partitions. We choose
this initial tolerance large enough that the enumerated neighborhood contains the MAP
over \(\mc P^\star\). Third, after identifying the MAP, we report
\[
    \mc P_\epsilon(\Pi_{\mathrm{MAP}})
    =
    \{\Pi\in\mc P^\star:
    Q(\Pi)\le (1+\epsilon)Q(\Pi_{\mathrm{MAP}})\}.
\]
We demonstrate this procedure with empirical examples in Section~\ref{section:real-data}.

\section{Permissible partitions}
\label{section:environment}
In this section, we describe our definition of a permissible partition. We limit ourselves to partitions that are interpretable and substantively meaningful, and we show that doing so also substantially reduces the computational burden. Our approach is modular, however, and we could build the RPS with a different set of permissibility restrictions if warranted by the scientific context.
To begin, recall that we have $M$ features taking on $R$ discrete values and $\mathcal{K}$ is the set of all $K = R^M$ unique feature combinations. 
We equip the feature combinations with a partial order. For a feature combination $k \in \mc{K}$, let  $k_m$ denote the value that the $m$-th feature takes. We say $k \geq k^{\prime}$ if and only if $k_m \geq k^{\prime}_m$ for all $m = 1, \dots, M$. We say $k > k^{\prime}$ if $k \geq k^{\prime}$ but $k \neq k^{\prime}$, and say that $k$ and $k^{\prime}$ are incomparable if there are two features $m_1$ and $m_2$ such that $k_{m_1} > k^{\prime}_{m_1}$ and $k_{m_2} < k^{\prime}_{m_2}$. We denote this $k \incomp k^{\prime}$
We denote the expected outcome of feature combination $k$ by $\beta_k$. We will return to our running example from the introduction (see Figure~\ref{fig:hass-running}).
\subsection{Hasse diagrams.}
We now formalize permissibility in terms of the geometry of the Hasse diagram. \citet{banerjee2021selecting} introduced Hasse diagrams to identify heterogeneity in outcomes in the context of selecting a single model in the frequentist paradigm with an $\ell_1$ penalty.  We leverage this geometry to construct the RPS while also generalizing their implementation to address the strong assumptions the~\citet{banerjee2021selecting} method requires.  
We begin by formally defining the Hasse diagram.
\begin{definition}[Hasse diagram]
\label{def:hasse}
The Hasse diagram, $\mc{H} = (\mc{K}, \mc{E})$, is a graph with nodes $\mc{K}$ and edges $\mc{E}$ relating the feature combinations through the partial ordering. Specifically, for two feature combinations $k, k^{\prime} \in \mc{K}$, the edge $\langle k, k^{\prime} \rangle \in \mc{E}$ if and only if $k > k^{\prime}$ and there does not exist a $k^{\prime\prime} \in \mc{K}$ such that $k > k^{\prime\prime} > k^{\prime}$, i.e., $k$ and $k^{\prime}$ are adjacent in the partial order. We will also denote the edge $\langle k, k^{\prime} \rangle$ as $e_{k, k^{\prime}}$.
\end{definition}

\begin{figure}
\centering
\begin{subfigure}[t]{0.3\textwidth}
    \begin{centering}
     \begin{center}
     \begin{tikzpicture}[scale =0.38]
         \node[] at (0,-3) (v1) {\footnotesize  $[250, 200]$};
         \node[] at (-3,0) (v2) {\footnotesize $[250, 400]$};;
         \node[] at (3,0) (v4) {\footnotesize $[500, 200]$};
         \node[] at (0,3) (v6) {\footnotesize $[500, 400]$};

        \draw[rotate around={135:(-1.5,1.5)}, orange, line width=0.3mm] (-1.5,1.5) ellipse (2cm and 5cm);
        \draw[rotate around={135:(1.5,-1.5)}, orange, line width=0.3mm] (1.5,-1.5) ellipse (2cm and 5cm);
         
        \draw[line width = 0.3mm, >=latex] (v1) to (v2);
        \draw[line width = 0.3mm,  >=latex] (v1) to (v4);
        \draw[line width = 0.3mm,  >=latex] (v2) to (v6);
        \draw[line width = 0.3mm,  >=latex] (v4) to (v6);
        \draw[line width = 0.3mm, >=latex] (v2) to (v6);
    \end{tikzpicture}
     \end{center} 
    \par
    \end{centering}
    \caption{Permissible}
    \label{fig:hasse-amoxicillin-ibuprofen-panel-a}
\end{subfigure}%
~
\begin{subfigure}[t]{0.3\textwidth}
    \begin{centering}
     \begin{center}
     \begin{tikzpicture}[scale =0.38]
         \node[] at (0,-3) (v1) {\footnotesize  $[250, 200]$};
         \node[] at (-3,0) (v2) {\footnotesize $[250, 400]$};;
         \node[] at (3,0) (v4) {\footnotesize $[500, 200]$};
         \node[] at (0,3) (v6) {\footnotesize $[500, 400]$};

        \draw[rotate around={135:(-1.5,1.5)}, blue, line width=0.3mm] (-1.5,1.5) ellipse (2cm and 5cm);
        \draw[blue, line width=0.3mm] (0,-3) ellipse (2cm and 2cm);
        \draw[blue, line width=0.3mm] (3,0) ellipse (2cm and 2cm);
         
        \draw[line width = 0.3mm, >=latex] (v1) to (v2);
        \draw[line width = 0.3mm,  >=latex] (v1) to (v4);
        \draw[line width = 0.3mm,  >=latex] (v2) to (v6);
        \draw[line width = 0.3mm,  >=latex] (v4) to (v6);
        \draw[line width = 0.3mm, >=latex] (v2) to (v6);
    \end{tikzpicture}
     \end{center} 
    \par
    \end{centering}
    \caption{Not permissible}
    \label{fig:hasse-amoxicillin-ibuprofen-panel-b}
\end{subfigure}%
~
\begin{subfigure}[t]{0.3\textwidth}
    \begin{centering}
     \begin{center}
     \begin{tikzpicture}[scale =0.38]
         \node[] at (0,-3) (v1) {\footnotesize  $[250, 200]$};
         \node[] at (-3,0) (v2) {\footnotesize $[250, 400]$};;
         \node[] at (3,0) (v4) {\footnotesize $[500, 200]$};
         \node[] at (0,3) (v6) {\footnotesize $[500, 400]$};

        \draw[orange, line width=0.3mm] (0,-3) ellipse (2cm and 2cm);
        \draw[orange, line width=0.3mm] (3,0) ellipse (2cm and 2cm);
        \draw[orange, line width=0.3mm] (-3,0) ellipse (2cm and 2cm);
        \draw[orange, line width=0.3mm] (0,3) ellipse (2cm and 2cm);
         
        \draw[line width = 0.3mm, >=latex] (v1) to (v2);
        \draw[line width = 0.3mm,  >=latex] (v1) to (v4);
        \draw[line width = 0.3mm,  >=latex] (v2) to (v6);
        \draw[line width = 0.3mm,  >=latex] (v4) to (v6);
        \draw[line width = 0.3mm, >=latex] (v2) to (v6);
    \end{tikzpicture}
     \end{center} 
    \par
    \end{centering}
    \caption{Permissible}
    \label{fig:hasse-amoxicillin-ibuprofen-panel-c}
\end{subfigure}%

\caption{\tiny Hasse diagrams for Amoxicillin and Ibuprofen example.}
\label{fig:hasse-amoxicillin-ibuprofen}
\end{figure}

\begin{examplecont}[\ref{ex:amox-ibu}]
{\Cref{fig:hasse-amoxicillin-ibuprofen} illustrates three possible partitions.  In \Cref{fig:hasse-amoxicillin-ibuprofen-panel-a} we see two pools, corresponding to the effect of increasing from $200$~mg to $400$~mg of Ibuprofen.  The pools effectively marginalize over the levels of Amoxicillin dosage to identify the effect of increasing Ibuprofen. \Cref{fig:hasse-amoxicillin-ibuprofen-panel-b} is difficult to interpret.  We have heterogeneity between a high and low dose of Amoxicillin with a low dose of Ibuprofen, but no heterogeneity at a high dose of Ibuprofen.  For this to happen, we would need, for example, the increased dose of Amoxicillin to cause stomach irritation that is exactly offset by the higher (but not lower) dose of Ibuprofen, which corresponds to a measure zero configuration. \Cref{fig:hasse-amoxicillin-ibuprofen-panel-c} represents the full set of interactions.}
\end{examplecont}

{Not all pools are scientifically meaningful. Ibuprofen and Amoxicillin are different drugs with different mechanisms, so it would not be scientifically coherent, for example, to combine an Ibuprofen-only treatment with an Amoxicillin-only treatment, even if both have identical expected outcomes. To address this, we propose permissibility rules to ensure that we only consider interpretable and unique models.  We represent these restrictions geometrically using possible splits of the Hasse diagram}. In Appendix \ref{appendix:environemnt}, we discuss the implications of permissibility restrictions in the context of several other commonly used models and demonstrate how they can be represented as restrictions on the Hasse diagram.

\subsection{The geometry of permissibility.}
We now demonstrate how we can use the geometry of the Hasse diagram to articulate permissible partitions.  The key insight is that we can define permissibility in terms of the \emph{edges} of the Hasse, rather than the nodes. We define \emph{splitting} as the process of creating partitions by severing edges of the Hasse diagram (see \Cref{fig:hasse-amoxicillin-ibuprofen}).  In plain language, we define permissible partitions as those that result from splitting, in either direction, across the entire Hasse diagram.  More formally, we define permissible partitions as follows:

\begin{definition}[Permissible partition]
\label{def:permissibility-profile-measure-zero}
A partition $\Pi_0$ is \emph{permissible} if and only if
\begin{enumerate}[label=(\arabic*)]
    \item \label{profile-permissibility:1} every $\pi \in \Pi_0$ is a \emph{pool} (cf. \Cref{def:pool}), %
    \item \label{profile-permissibility:2} Every $\pi \in \Pi_0$ is a \emph{closed interval} in the partial order, i.e. there are well defined unique endpoints $\min \pi$ and $\max \pi $ (possibly equal to each other) such that $\pi = \{k \mid \min \pi \leq k\leq \max\pi \}$, and
    \item \label{profile-permissibility:3} $\Pi_0$ respects \emph{parallel splits}, i.e., for every pair of distinct pools $\pi_i, \pi_j \in \Pi_0$
    \begin{enumerate}[label=(\alph*), ref=(\theenumi \alph*),leftmargin=2\parindent]
        \item \label{profile-permissibility:3a} if $\min \pi_i \incomp \min \pi_j$, then there exists a $\pi^\prime \in \Pi_0$ such that $\min \pi^{\prime} = p^{\prime}$, where for each feature $m$, $p^{\prime}_m := \max\{ p^{(i)}_m, p^{(j)}_m\}$, with $p^{(i)} = \min \pi_i$ and $p^{(j)}= \min \pi_j$, and  
        \item \label{profile-permissibility:3b} if $\max \pi_i \incomp \max \pi_j$, then there exists a $\pi^{\prime \prime} \in \Pi_0$ such that $\max \pi^{\prime \prime} = p^{\prime \prime}$, where for each feature $m$, $p^{\prime \prime}_m := \min \{ \tilde{p}^{(i)}_m , \tilde{p}^{(j)}_m \}$, with $\tilde{p}^{(i)} = \max \pi_i$ and $\tilde{p}^{(j)} = \max \pi_j$.
    \end{enumerate}
\end{enumerate}
We denote the set of all permissible partitions by $\mc{P}^{\star}$.
\end{definition}

Operationally, a partition is permissible if and only if it can be obtained by cutting a set of parallel edge-families in the Hasse diagram. Condition~(1) ensures that we only consider valid partitions, in the
sense that each element of $\Pi_0$ is a pool. Condition~(2) means that each pool must form a contiguous block in the Hasse, with a single ``bottom'' and ``top'' corner and no ``holes'' along any monotone path. This rules out L-shaped or N-shaped pools. In a partial (as opposed to total) order, an interval need not resemble a line segment and can instead have ``thickness”; for example, the pool consisting of all elements in the Hasse diagram \Cref{fig:hasse-amoxicillin-ibuprofen} is a closed interval.

Condition~(3) then requires that
these intervals line up in parallel on the Hasse: whenever two pools
start or end at incomparable corners, there must be another pool whose
corner sits at the coordinate-wise max or min of those points. This
rules out partial splits that stop in the middle of the grid and
ensures that permissible partitions correspond to clean%
splits that run all the way through the Hasse diagram.

This definition of permissibility produces partitions that are interpretable in terms of marginal effects (how changing a dosage in one drug marginally affects the outcome), while disregarding partitions that are ``measure zero.''  These measure zero partitions are not robust since the only rationalization for these splits relies on exact marginal effects that offset in a specific way.
Such partitions require tremendous coincidence.

 We could also represent permissibility in terms of the difference in outcome between adjacent levels of factors in our enumeration algorithm.  We present this in \Cref{def:permissible-profile}, along with a formal proof of equivalence. We also provide additional technical details and a direct comparison with~\citet{banerjee2021selecting} in Appendix \ref{appendix:environemnt}. We now describe these conditions in the context of our running example.

\begin{examplecont}
\Cref{fig:hasse-amoxicillin-ibuprofen-panel-a} illustrates a simple permissible partition
with two pools. One pool contains the low dose of Ibuprofen,
$\{(250 \text{ mg}, 200 \text{ mg}), (500 \text{ mg}, 200 \text{ mg})\}$, and the other
contains the high dose of Ibuprofen,
$\{(250 \text{ mg}, 400 \text{ mg}), (500 \text{ mg}, 400 \text{ mg})\}$.
Each pool is a valid pool of the profile (Condition~(1)) and is a closed interval: moving left or right along the Amoxicillin axis keeps us inside a pool,
and each pool has a unique minimum and maximum (Condition~(2)).
Moreover, the two pools are obtained by a single horizontal split of the
Hasse diagram that runs all the way through the grid. The minima
$(250,200)$ and $(250,400)$ and the maxima $(500,200)$ and $(500,400)$
are all comparable in the partial order, so Condition~(3) is satisfied.

\Cref{fig:hasse-amoxicillin-ibuprofen-panel-b}  is \emph{not} permissible. Geometrically, \Cref{fig:hasse-amoxicillin-ibuprofen-panel-b} 
has a split that starts at the edge between $(250 \text{ mg}, 200 \text{ mg})$ and $(500 \text{ mg}, 200 \text{ mg})$ but does not run all the way through the Hasse diagram. This ``partial''
split violates Condition~(3).

To interpret \Cref{fig:hasse-amoxicillin-ibuprofen-panel-b} in terms of treatment effects, we would
have to say that increasing Ibuprofen from $200$ mg to $400$ mg has an
effect when Amoxicillin is at $250$ mg, but \emph{no} effect when
Amoxicillin is at $500$ mg, because $(500 \text{ mg}, 200 \text{ mg})$
is isolated in its own pool while $(500 \text{ mg}, 400 \text{ mg})$
shares a pool with $(250 \text{ mg}, 400 \text{ mg})$. At the same
time, moving from $(250 \text{ mg}, 200 \text{ mg})$ to
$(500 \text{ mg}, 400 \text{ mg})$ must produce exactly the same
average effect as moving from $(250 \text{ mg}, 200 \text{ mg})$ to
$(250 \text{ mg}, 400 \text{ mg})$. As discussed, this requires a measure zero interaction between the drugs.

In our permissibility rules, Condition~(3a) rules out these measure zero splits. Take the pools
$\pi_1 = \{(250 \text{ mg}, 400 \text{ mg}), (500 \text{ mg}, 400
\text{ mg})\}$ and $\pi_2 = \{(500 \text{ mg}, 200 \text{ mg})\}$. The
minima of these pools, $\min \pi_1 = (250 \text{ mg}, 400 \text{ mg})$ and
$\min \pi_2 = (500 \text{ mg}, 200 \text{ mg})$, are incomparable: one has more
Amoxicillin but less Ibuprofen. Condition~(3a) then requires that there
exist another pool whose minimum is the coordinate-wise maximum
of $\min \pi_1$ and $\min \pi_2$, $(500 \text{ mg}, 400 \text{ mg})$. But no such pool exists. In fact, the coordinate-wise maximum $(500 \text{ mg}, 400 \text{ mg}) \in \pi_1$.

\Cref{fig:hasse-amoxicillin-ibuprofen-panel-c} shows another permissible
partition, this time with four pools, each containing a single treatment
combination. Each pool is trivially a pool and a closed interval, so
Conditions~(1)--(2) hold automatically. The only non-trivial check is
Condition~(3). Consider the pools containing $\pi_1 = \{(250 \text{ mg}, 400 \text{
mg})\}$ and $\pi_2 = \{(500 \text{ mg}, 200 \text{ mg})\}$. Their minima (and maxima)
are the points themselves, which are incomparable in the partial order:
one has more Amoxicillin but less Ibuprofen. Condition~(3a) requires that
whenever this happens there exists another pool whose minimum is the
coordinate-wise maximum of $\min \pi_1, \min \pi_2$. Here the coordinate-wise
maximum is $(500 \text{ mg}, 400 \text{ mg})$, and there is indeed a pool
whose minimum is exactly this point: the singleton pool
$\pi_3 = \{(500 \text{ mg}, 400 \text{ mg})\}$.
Condition~(3b) requires that there is another pool whose maximum is the coordinate-wise minimum of $\max \pi_1, \max \pi_2$. Here, the coordinate-wise minimum is $(250 \text{ mg}, 200 \text{ mg})$, and the pool $\pi_4 = \{ (250 \text{ mg}, 200 \text{ mg})\}$ satisfies this.

Geometrically, we can view
\Cref{fig:hasse-amoxicillin-ibuprofen-panel-c} as the result of splitting
all the way across in both directions (once along the Amoxicillin axis
and once along the Ibuprofen axis) so that every cell is its own block
but all splits still run cleanly through the Hasse diagram.
\end{examplecont}

Our permissibility definition leads to substantial improvements in memory and computation requirements.  
We can define partitions on the Hasse diagram by removing (splitting on) edges in $\mc{E}$ to form disjoint components in $\mc{K}$. The removed edges correspond to non-zero marginal changes. This guarantees that all sets in a partition consist of parallel, closed intervals.

At a high level, permissibility restrictions generate equivalence classes among the edges, $\mc{E}$. The equivalence classes are those edges that can only be removed together to generate a partition. Suppose we decompose $\mc{E}$ into $n$ mutually disjoint and exhaustive sets of edges $E_1, \dots, E_n$, where each set $E_j$ is an equivalence class. A partition $\Pi$ induced by these equivalence classes satisfies: $\Pi$ removes edge $e$ if and only if $\Pi$ removes $e^{\prime}$ for every $e^{\prime}$ such that $e, e^{\prime} \in E_i$ for some $i = 1, \dots, n$. The equivalence classes correspond to partitions where we pool along one of these edges if and only if we pool along the other.
Let $\mc{E}^{\prime}$ represent the set of edges that remain after the partition. Then the pruned graph $(\mc{K}, \mc{E}^{\prime})$ specifies the corresponding partition.

Specifically, permissible partitions can be generated by identifying a unique decomposition of $\mc{E}$ into equivalent edges. The decomposition is given by
\begin{align*}
    \mc{E} &= \bigcup_{m=1}^M \bigcup_{r=1}^{R-1} E_{m, r},
\end{align*}
where for some feature $m$ taking on value $r$, the equivalence class is
\[
E_{m,r} = \{e_{k, k^{\prime}} \in \mc{E} \mid k_m = r,\; k^{\prime}_m = r+1\}.
\]
In other words, $E_{m,r}$ contains all edges between pairs of feature combinations $k$ and $k^{\prime}$ that differ only in the $m$-th feature, with the $m$-th feature increasing from level $r$ to level $r+1$. This decomposition of $\mc{E}$ into equivalent edges corresponds to all ``parallel'' edges on the Hasse diagram (see \Cref{fig:hasse-amoxicillin-ibuprofen} for an example).

This equivalence allows us to store partitions efficiently. If there are $n$ equivalence classes, then there are $2^n$ possible partitions -- we either split or pool across the edges in the $i$-th equivalence class.
Rather than storing the partition as a set of pools or through a tree data structure, we can reduce the storage and calculation by a logarithmic factor by just keeping track of the hyperplanes induced by splits. That is, we can simply store a binary vector of length $n$. For interpretability purposes, we can reshape this vector into the $\bSigma \in \{0, 1\}^{M \times (R-1)}$ matrix. Here $\Sigma_{mr} = 1$ if and only if feature combinations with level $r$ are pooled with feature combinations with level $r+1$ in feature $m$. We walk through detailed examples in \Cref{appendix:environemnt}. Without loss of generality, we can extend to the case where feature $i$ takes on $R_i$ values such that not all $R_i$ are equal (some entries in the partition matrix will not be defined). Henceforth, we consider only permissible partitions and drop the ``permissible'' quantifier unless specifically needed.

\subsection{Profiles.}
{In some scientific settings, we prioritize heterogeneity among a subset of variables while holding others fixed. In an experiment, for example, we are likely interested in particular in differences in outcomes between the treatment and control groups for individuals with certain characteristics (i.e., the conditional average treatment effect, which we illustrate in Section~\ref{section:real-data}).  Alternatively, we may be interested in heterogeneity in outcomes among people with a certain demographic, income, or geographic feature.  In general, the permissibility restrictions define a scientifically coherent set of possible interactions among a set of variables, described through marginal changes in outcomes when moving among levels on the Hasse diagram. If we know that there is an interaction term that, when introduced, dramatically changes the interaction pattern, then using that term allows us to have different profiles for different levels of that interaction.  If we do not have that information, it amounts to assuming that we have a homogeneous or regular pattern subject to the permissibility rules. Each profile implies conditioning on the levels of one or more variables and corresponds to a Hasse diagram.  %
Specifically, we define:}

\begin{figure}
\begin{subfigure}[t]{0.5\textwidth}
    \begin{centering}
     \begin{center}    %
     \begin{tikzpicture}[scale = 0.5] \def \n {5} \def \radius {2cm} \def \margin {8}
     \node[ minimum size=19pt] at (0,-4) (v1){\footnotesize $[0,0]$};
     \node[ minimum size=19pt] at (-2,-2) (v2){\footnotesize $[0,200]$};
     \node[ minimum size=19pt] at (-4,0) (v3){\footnotesize $[0,400]$};
     \node[ minimum size=19pt] at (2,-2) (v4){\footnotesize $[250,0]$};
     \node[ minimum size=19pt] at (4,0) (v5){\footnotesize $[500,0]$};
     \node[ minimum size=19pt] at (0,0) (v6){\footnotesize $[250,200]$};
     \node[ minimum size=19pt] at (-2,2) (v7){\footnotesize $[250,400]$};
      \node[ minimum size=19pt] at (2,2) (v8){\footnotesize $[500,200]$};
       \node[ minimum size=19pt] at (0,4) (v9){\footnotesize $[500,400]$};
      \draw[line width = 0.3mm, rotate around={132:(-2,2)}, orange] (-2,2) ellipse (1.5cm and 5cm);
       \draw[line width = 0.3mm, rotate around={132:(0,0)}, orange] (0,0) ellipse (1.5cm and 5cm);
        \draw[line width = 0.3mm, rotate around={135:(2,-2)}, orange] (2,-2) ellipse (1.5cm and 5cm);
     
     \draw[line width = 0.3mm, >=latex] (v1) to (v2);
     \draw[line width = 0.3mm,  >=latex] (v1) to (v4);
     \draw[line width = 0.3mm,  >=latex] (v2) to (v3);
     \draw[line width = 0.3mm,  >=latex] (v2) to (v6);
     \draw[line width = 0.3mm,  >=latex] (v4) to (v6);
     \draw[line width = 0.3mm, >=latex] (v2) to (v6);
     \draw[line width = 0.3mm, >=latex] (v3) to (v7);
     \draw[line width = 0.3mm, >=latex] (v4) to (v5);
       \draw[line width = 0.3mm, >=latex] (v6) to (v7);
      \draw[line width = 0.3mm, >=latex] (v6) to (v8);
      \draw[line width = 0.3mm, >=latex] (v5) to (v8);
     \draw[line width = 0.3mm, >=latex] (v7) to (v9); 
      \draw[line width = 0.3mm, >=latex] (v8) to (v9);
    \end{tikzpicture}
     \end{center} 
    \par
    \end{centering}
    \caption{\tiny Permissible; increase in Ibuprofen}
    \label{fig:hasse-profile-a}
\end{subfigure}%
\begin{subfigure}[t]{0.5\textwidth}
    \begin{centering}
     \begin{center}    %
     \begin{tikzpicture}[scale = 0.5] \def \n {5} \def \radius {2cm} \def \margin {8}
     \node[ minimum size=19pt] at (0,-4) (v1){\footnotesize $[0,0]$};
     \node[ minimum size=19pt] at (-2,-2) (v2){\footnotesize $[0,200]$};
     \node[ minimum size=19pt] at (-4,0) (v3){\footnotesize $[0,400]$};
     \node[ minimum size=19pt] at (2,-2) (v4){\footnotesize $[250,0]$};
     \node[ minimum size=19pt] at (4,0) (v5){\footnotesize $[500,0]$};
     \node[ minimum size=19pt] at (0,0) (v6){\footnotesize $[250,200]$};
     \node[ minimum size=19pt] at (-2,2) (v7){\footnotesize $[250,400]$};
      \node[ minimum size=19pt] at (2,2) (v8){\footnotesize $[500,200]$};
       \node[ minimum size=19pt] at (0,4) (v9){\footnotesize $[500,400]$};
       \draw[line width = 0.3mm, rotate around={48:(-3,-1)}, blue] (-3,-1) ellipse (1cm and 3cm);
        \draw[line width = 0.3mm, rotate around={130:(3,-1)}, blue] 
        (3,-1) ellipse (1.2cm and 3cm);
          \draw[line width = 0.3mm, rotate around={130:(1,1)}, blue] 
        (1,1) ellipse (1.2cm and 3cm);
        \draw[line width = 0.3mm, rotate around={130:(-1,3)}, blue] 
        (-1,3) ellipse (1.2cm and 3cm);
       \draw[fill=none, line width = 0.3mm, blue] (0,-4) circle (1cm);

     \draw[line width = 0.3mm, >=latex] (v1) to (v2);
     \draw[line width = 0.3mm,  >=latex] (v1) to (v4);
     \draw[line width = 0.3mm,  >=latex] (v2) to (v3);
     \draw[line width = 0.3mm,  >=latex] (v2) to (v6);
     \draw[line width = 0.3mm,  >=latex] (v4) to (v6);
     \draw[line width = 0.3mm, >=latex] (v2) to (v6);
     \draw[line width = 0.3mm, >=latex] (v3) to (v7);
     \draw[line width = 0.3mm, >=latex] (v4) to (v5);
       \draw[line width = 0.3mm, >=latex] (v6) to (v7);
      \draw[line width = 0.3mm, >=latex] (v6) to (v8);
      \draw[line width = 0.3mm, >=latex] (v5) to (v8);
     \draw[line width = 0.3mm, >=latex] (v7) to (v9); 
      \draw[line width = 0.3mm, >=latex] (v8) to (v9);
    \end{tikzpicture}
    \end{center} 
    \par
    \end{centering}
     \caption{\tiny Not permissible; increase in Ibuprofen while taking Amoxicillin}
    \label{fig:hasse-profile-b}
\end{subfigure}%
\label{fig:hasse-profiles}
\caption{\tiny Extended Hasse diagram including the control (0mg) level.  Panel A shows a permissible partition corresponding to increasing Ibuprofen while averaging over all levels of Amoxicillin (including control).  Panel B shows increasing Ibuprofen while already taking Amoxicillin (e.g., comparing $\{(250,200),(500,200)\}$ to $\{(250,400),(500,400)\}$) but is not permissible.}
\end{figure}

\begin{definition}[Profile]
\label{def:generalized-profile}
A profile, $\rho(k)$, is a binary vector indicating, for each of the $M$ features,
whether the feature is ``turned on'' in the Hasse diagram.
In the special case of experiments, a profile is a binary vector indicating which features have a level above the control.
\end{definition}

To be explicit, the hierarchy between \textit{profiles}, \textit{partitions}, and \textit{pools} is as follows: a \textit{profile} fixes which features are ``active'' or are being considered; within a profile, a \textit{partition} divides the active feature combinations, and; each element of a partition is a \textit{pool} with constant expected outcome.

In any factorial analysis, the choice of features and their levels (milligrams of dosages, socioeconomic categories, etc.) implicitly defines a coordinate system. The Hasse diagram is determined by this scientific or design-based representation. %
 Importantly, choosing coordinates does not mean assuming monotonicity.
Like any estimator of heterogeneity, RPS characterizes model uncertainty within the factorial representation chosen by the researcher. If the chosen coordinates do not align with the simplest possible parametrization of the underlying phenomenon, the RPS will naturally introduce additional splits to represent the complexity in those coordinates. This is informative: a fragmented or alternating block structure alerts the researcher that an alternative representation (for example, a cyclic ordering of time) might yield a simpler or more interpretable pattern.

Profiles do not define or alter the underlying coordinate system. Rather, they are conditional Hasse diagrams that encode scientifically meaningful regimes -- contexts in which different subsets of features, or different mechanisms, are expected to operate. Profiles are therefore downstream of the coordinate choice, not replacements for it. When such distinctions are scientifically justified, profiles offer a principled way to focus attention on interpretable partitions. When such distinctions are not present, the method always admits a single-profile specification.

Since profiles operate within the chosen factorial representation, a natural sensitivity check is to run the RPS procedure under multiple plausible profile specifications and compare the conclusions. %
As discussed above, if the researcher’s representation forces the RPS to use many splits along a dimension, this fragmentation itself is evidence that the effect is structurally richer along that axis than the chosen coordinates capture.

We present algorithms that search for profiles, but in general this is an NP hard problem with no specific scientific hypothesis in mind or policy-based reason to condition on particular levels of a feature.  Discovering profiles represents a level of flexibility beyond the scope of our current work, as it essentially corresponds to learning a Hasse diagram for each possible combination of levels of conditioning among all features. We apply Definition \ref{def:permissibility-profile-measure-zero} to define permissibility within a single profile, but we may also want to consider pooling across profiles, which we describe in Appendix~\ref{appendix:environemnt}.

\begin{examplecont}
{Imagine we create a control where patients take $0$~mg of Amoxicillin and/or Ibuprofen. We could then expand that Hasse diagram by adding $0$~mg for both Ibuprofen and Amoxicillin.  Figure~\ref{fig:hasse-profile-a} shows the pooling structure to measure increasing Ibuprofen dosage.  The key difference is that now we are also averaging $0$~mg of Amoxicillin.  However, we say that we want to measure the impact of increasing Ibuprofen specifically on individuals who already take Amoxicillin.  In Figure~\ref{fig:hasse-profile-b} we show a partition where we measure this effect by comparing $\{(250,200),(500,200)\}$ to $\{(250,400),(500,400)\}$.  This partition, however, is not permissible, precisely because it does not identify a marginal effect.  Notice, however, that comparing $\{(250,200),(500,200)\}$ with $\{(250,400),(500,400)\}$ is permissible in Figure~\ref{fig:hasse-amoxicillin-ibuprofen-panel-a}, which we can view as a profile conditional on Ibuprofen and Amoxicillin dosages $>0$.}
\end{examplecont}

\section{Size of the Rashomon Partition Set}
\label{section:rps-size}

Given that we would like to enumerate $\mathcal{P}_{\theta}$ it is useful to calculate bounds on both its size and also  $\mathcal{P}^{\star}$. Since any permissible partition requires each profile to respect \Cref{def:permissibility-profile-measure-zero}, we can consider each profile independently. We will use $m$ to denote the number of features with non-zero values in the profile we are focusing on, so $m \in \{1,\dots,M\}$. Without loss of generality, we assume that every feature $i$ takes on $R_i = R$ ordered values. Our proofs naturally extend to the general case. All technical details are deferred to Appendix \ref{appendix:rps-size}. 
First, $\mathcal{P}^{\star}$ is small relative to the total number of potential partitions.

\begin{proposition}
\label{prop:num-permissible-poolings}
In each profile, the total number of
    all possible partitions is $\bigO \left(2^{2(R-1)^m} \right)$, and
    permissible partitions is $\bigO(2^{m(R-2)})$.
\end{proposition}

Next, we show that the size of the RPS is only polynomial in $M$ and $R$. In Lemma \ref{lemma:num-leaves-threshold}, we observe that the $\ell_0$ prior bounds the number of pools in any Rashomon partition.

\begin{lemma}
\label{lemma:num-leaves-threshold}
For a given Rashomon threshold $\theta$ and regularization parameter $\lambda$, any partition in the RPS, $\mathcal{P}_{\theta}$, can have at most $H_{\theta}(\lambda)$ pools,
\begin{align*}
    H_{\theta}(\lambda) = \left\lfloor - \frac{\ln (c \theta)}{\lambda}  \right\rfloor,
\end{align*}
where $c := c(\Z)$ is a normalization constant depending only on $\Z$ and $\lfloor \cdot \rfloor$ is the floor function.
\end{lemma}

Lemma \ref{lemma:num-leaves-threshold} allows us to further reduce the number of the partitions in Proposition \ref{prop:num-permissible-poolings} by considering only partitions that meet this requirement. Even when the regimes of scientific action i.e., profiles, are unknown, we show that the size of the RPS is bounded polynomially in \Cref{thm:rashomon-set-polynomial-size}. In \Cref{lemma:k-profiles-num-partitions} we bound the size of the RPS when the profiles are known apriori. Such a relationship between regularization and size of the model class was previously hypothesized and shown for empirical data by \citet{semenova2022existence}.

\begin{theorem}\label{thm:rashomon-set-polynomial-size}
Let $\theta$ be the Rashomon threshold and let $H := H_\theta(\lambda)$ be an upper bound on the number of pools among
Rashomon-admissible partitions.
Then the size of the Rashomon Partition Set satisfies the following bound:
\[
\abs{\mathcal{P}_{\theta}}
\in
\begin{cases}
\bigO\!\left(M^{2H-1}R^{H-1}\right), & \text{if } R > M^{c_{\mathrm{crit}}},\\[4pt]
\bigO\!\left((MR)^{\gamma H - 1}\,(\log_2(MR))^{-1}\right), & \text{if } R \le M^{c_{\mathrm{crit}}},
\end{cases}
\]
where $c_{\mathrm{crit}} := \dfrac{\log_2 3 - 1}{\,2 - \log_2 3\,}$ and
$\gamma := \dfrac{2}{e\ln 2} \approx 1.061$.
In particular, for fixed $H$, $\abs{\mathcal{P}_\theta}$ grows polynomially in $M$ and $R$ in either regime.
\end{theorem}

Observe that $c_{\mathrm{crit}} \approx 1.41$. In many settings, the number of factor levels $R$ is fixed (or grows sufficiently slowly) while the number of features
$M$ increases. In particular, if $R = o(M^{c_{\mathrm{crit}}})$, then for all sufficiently large $M$ we are in the second
regime $R \le M^{c_{\mathrm{crit}}}$.
Moreover, regardless of which regime applies, for any fixed $H$ the theorem implies that $\abs{\mathcal{P}_\theta}$ grows
at most polynomially in $M$ and $R$. Because the Rashomon set contains models of varying granularity (ranging from $k=1$ to $H$ profiles), the bound is derived by maximizing over the number of profiles $k$ to identify the worst-case complexity. In our empirical examples, the realized Rashomon sets are in fact far smaller than these worst–case bounds: with only a few hundred partitions in the RPS we already obtain a close approximation to the full posterior.

\section{Enumerating Rashomon Partitions}
\label{section:enumeration}

 We will first develop intuition to present an algorithm to enumerate the RPS for a single profile. Since we do not pool across profiles, we can enumerate the Rashomon Partition Set for each profile independently and then finally combine them in the end.
The intuition behind our enumeration is that any split we make introduces a new set of pools. If for some reason this split is very bad, then no matter what other split we make,  we can never recover. Theorems \ref{thm:rashomon-fixed-bound} and \ref{thm:rashomon-equivalent-bound} help us identify those poor splits. They rely on the fact that equivalent points having the exact same feature values will always belong to the same pool. However, equivalent units may not have the same outcome. Therefore, we will always incur some loss from these equivalent units (see \citet{angelino2017learning} or \citet{xin2022exploring} for implementations of related strategies). We defer technical details of the results to Appendix \ref{appendix:enumeration-proofs}.

Consider some partition matrix $\bSigma$, where the partition is given by $\Pi := \Pi(\bSigma)$. Given some data $\Z = (\X, \y)$, we will use the mean squared error and the average outcome in pool $\pi \in \Pi$, $\widehat{\mu}_{\pi}$, as defined in \Cref{eq:mse-and-predictions}.
However, the results generalize to any non-negative loss as we will see in \Cref{section:generalization} with weighted mean-squared error.

Suppose we fix some indices $\mathcal{M}$ in $\bSigma$. Define a new matrix $\bSigma_{\fix}$,
\begin{align*}
    \Sigma_{\fix, (i, j)} &= \begin{cases}
        \Sigma_{(i, j)}, & (i, j) \in \mathcal{M} \\
        0, & \text{else}
    \end{cases}.
\end{align*}
In other words, $\bSigma_{\fix}$ is a partition where all heterogeneity splits made by $\bSigma$ corresponding to indices in $\mathcal{M}$ are obeyed and we maximally split at all other places. Let $\Pi_{\fix} := \Pi(\bSigma_{\fix})$ correspond to this maximal partition respecting $\bSigma$ at indices $\mathcal{M}$. Next, define
\begin{align*}
    \pi_{\fix} &= \{ k \in \mathcal{K} \mid  k_i \leq j+1 \iff (i, j) \in \mathcal{M} \}
\end{align*}
to be the set of all feature combinations covered by indices in $\mathcal{M}$. And we define the complement $\pi_{\fix}^c = \mathcal{K} \setminus \pi_{\fix}$. Finally, define $H(\Pi, \mathcal{M}) = \sum_{\pi \in \Pi} \mathbb{I}\{\pi \cap \pi_{\fix} \neq \varnothing\}$ to be the number of pools in $\Pi$ consisting of feature combinations corresponding to indices $\mathcal{M}$.

Consider a procedure where we keep $\bSigma$ constant at $\mathcal{M}$ and make further splits (not already implied by $\mathcal{M}$) at other indices only. Define $\child(\bSigma, \mathcal{M})$ to be all such $\bSigma^{\prime}$.
Our search algorithm in Algorithm \ref{alg:r-aggregate-profile} starts at some partition and fixes some heterogeneity splits. Theorem \ref{thm:rashomon-fixed-bound} says that if the loss incurred by these fixed heterogeneity splits is already too high, then we should discard this partition and its children.

\begin{theorem}
\label{thm:rashomon-fixed-bound}
Let $\theta_{\epsilon}$ be the Rashomon threshold in the score space i.e., $\Pi \in \mathcal{P}_{q, \epsilon}$ if and only if $Q(\Pi) < \theta_{\epsilon}$. Given a partition $\Pi := \Pi(\bSigma)$ for partition matrix $\bSigma$, a set of fixed indices $\mathcal{M}$, and data $\Z$ consisting of $n$ observations, define
\begin{align}
    b(\Sigma,M;\Z)
=
\eta\frac{1}{n}
\sum_{\pi\in\Pi_f}
\sum_{k(i)\in\pi}
\mathbf 1\{k(i)\in\pi_f\}
(y_i-\widehat\mu_\pi)^2
+
\lambda H(\Pi,M).\label{eq:fixed-point-loss}
\end{align}
If $b(\bSigma, \mathcal{M}; \Z) > \theta_{\epsilon}$, then $\bSigma$ and all $\bSigma^{\prime} \in \child(\bSigma, \mathcal{M})$ are not in the Rashomon set $\mathcal{P}_{q, \epsilon}$.
\end{theorem}

Theorem \ref{thm:rashomon-equivalent-bound} ``looks ahead'' to see if this partition is of poor quality. If the loss incurred by feature combinations yet to be split is too high, then we abandon this partition.

\begin{theorem}
\label{thm:rashomon-equivalent-bound}
Consider the same setting as Theorem \ref{thm:rashomon-fixed-bound}. Define
\begin{align}
b_{\mathrm{eq}}(\Sigma,M;\Z)
&=& b(\bSigma, \mathcal{M}; \Z) + b_{eq}(\bSigma, \mathcal{M}; \Z). \label{eq:equivalent-point-loss}
\end{align}
If $B(\bSigma, \mathcal{M} ; \Z) > \theta_{\epsilon}$, then $\bSigma$ and all $\bSigma^{\prime} \in \child(\bSigma, \mathcal{M})$ are not in the Rashomon set $\mathcal{P}_{q, \epsilon}$.
\end{theorem}

Theorems \ref{thm:rashomon-fixed-bound} and \ref{thm:rashomon-equivalent-bound} help aggressively cut down the search space by combining the lowest penalty on the splits already made and the lowest mean-squared error on the splits yet to be made. If this is already too high, then we abandon our search. The $\ell_0$ prior is critical to this enumeration strategy, since it means extending a branch corresponds to a discrete increase in the number of pools, which yields sharp and easily computable bounds. We illustrate this in Algorithm \ref{alg:r-aggregate-profile}. Here, we start with all feature combinations pooled together. We begin our search at the first feature trying to split the two feature combinations with the lowest dosages into separate pools. We keep a queue of possible splits to consider. Whenever we remove a possible split from the queue, we check its viability using Lemma \ref{lemma:num-leaves-threshold}, and Theorems \ref{thm:rashomon-fixed-bound} and \ref{thm:rashomon-equivalent-bound}. If this is a bad split, we go to the next split in the queue. And if this is a good split (so far), we check if it already meets the Rashomon threshold and recursively add other further possible splits to this queue. We also maintain a cache of splits that have been added to the queue at some point to avoid doubling back on old splits. {We walk through a detailed example in Appendix \ref{appendix:enumeration-proofs}, Example \ref{ex:algorithm-raggregate-profile}.} As noted before, we can solve each profile  independently. In Algorithm \ref{alg:r-aggregate}, we explicitly show how to do this. Note that in line \ref{line:eq-bound}, we once again leverage Theorem \ref{thm:rashomon-equivalent-bound} by noting that each profile will always incur some loss. Once we solve each profile independently, Algorithm \ref{alg:r-aggregate-across-profiles} describes how to pool across profiles as defined in \Cref{def:permissible-partition}. {Appendix \ref{appendix:enumeration-proofs} describes this in more detail.}  Together, we have:

\begin{theorem}
\label{thm:enumeration-algorithm}
Algorithm \ref{alg:r-aggregate} correctly enumerates the Rashomon partition set.
\end{theorem}

{Theorems \ref{thm:rashomon-fixed-bound} and \ref{thm:rashomon-equivalent-bound} look very similar to Theorems 3.1 and 3.2 of \cite{xin2022exploring}. Indeed, they are inspired by and follow a very similar proof strategy to \cite{xin2022exploring}. However, there are three key differences (in addition to using a Bayesian perspective). First, we generalize their idea to allow for discrete ordered values, which represents a nontrivial extension from their setting  with binary variables. Second, while they only consider classification, we allow for any regression problem with any non-negative error (likelihood) function and a penalty that is increasing in the size of the partition (see \Cref{cor:other-losses}). Third, and perhaps the most important difference is that we work with a geometry for partitioning the feature space that is very different from trees, which we briefly alluded to in the Introduction and will discuss in \Cref{section:related-body} and \Cref{appendix:crf}.}

 {Our algorithm falls under a class of algorithms referred to as branch-and-bound. The worst case time complexity is $\mc{O}(2^{M(R-2)})$ for \Cref{alg:r-aggregate-profile} as we may need to check every partition. The average case time complexity of branch-and-bound algorithms is harder to analyze. In \Cref{appendix:additional-sims}, we perform some simulations to show the runtime for various $(M, R, \epsilon)$ parameters.  The runtime depends on the complexity of the space (determined by $M,R$) and the threshold, with larger Rashomon sets taking longer to enumerate.  All of our empirical examples ran in less than three hours using a standard personal computer.
Given our setup, it is easy to see how Algorithm \ref{alg:r-aggregate} can be parallelized by delegating calls to \Cref{alg:r-aggregate-profile} and profile to separate threads.}

\begin{algorithm}[!tb]
\caption{\texttt{EnumerateRPS}($M, R, H, \Z, q, \epsilon)$}
\label{alg:r-aggregate}
\begin{algorithmic}[1]
\Require $M$ features, $R$ factors per feature, max pools $H$, data $\Z$, reference score $q$, threshold $\epsilon$
\Ensure Rashomon set $\mathcal{P}_{q, \epsilon}$

\State $\mc{P}_{q, \epsilon} = \varnothing$
\State $\mc{R}$ all sets of candidate profiles
\For {set of profiles $\rho \in \mc{R}$}
    \State $H^{\prime} = H - \abs{\rho} + 1$
    \State \label{line:eq-bound} $\mathcal{E} = [ b_{eq} \text{ of profile $\rho_i$ for $\rho_i \in \rho$} ]$ \Comment{$b_{eq}$ in \Cref{thm:rashomon-equivalent-bound} with zero matrix}
    
    \State $\mathcal{P} = \texttt{dict()}$
    \For {$\rho_i \in \rho$}
        \State $q_i = q(1+\epsilon) - \texttt{sum}(\mathcal{E}) + \mathcal{E}_{\rho_i}$
        \State $M_i = $ active features in $\rho_i$
        \State $R_i = R[M_i]$
        \State $\mathcal{P}[\rho_i] = \texttt{EnumerateRPS\_genprofile}(M_i, R_i, H^{\prime}, \Z, q_i)$ \Comment{See \Cref{alg:r-aggregate-profile}}
        \State Sort partition matrices in $\mathcal{P}[\rho_i]$ on score \(Q\)
    \EndFor
    
    \State $\mc{P}^{\prime} = \bigtimes_{\rho_i \in \rho} \mc{P}[\rho_i]$    \Comment{Obtain candidate partitions with Cartesian product}
    
    \State $\mc{P}_{q, \epsilon} = \mc{P}_{q, \epsilon} \cup \texttt{PoolProfiles}(\mc{P}^{\prime}, \rho_0, \Z, q(1+\epsilon))$ \Comment{See \Cref{alg:r-aggregate-across-profiles}}

\EndFor

\State \Return $\mc{P}_{q, \epsilon}$

\end{algorithmic}
\end{algorithm}

\section{Simulations}
\label{section:simulations}
We present two simulation studies to illustrate the performance of our method. %
In Appendix~\ref{appendix:additional-sims}, we provide additional details of our simulation studies, including runtime, as well as an additional simulation using posterior densities.

\begin{figure}

\begin{centering}
\begin{center}

\begin{tikzpicture}[scale = 0.4] \def \n {5} \def \radius {2cm} \def \margin {8}

    \node[] at (0,-4) (v1){\footnotesize$(1,1)$};
    \node[] at (-2,-2) (v2){$(1,2)$};
    \node[] at (-4,0) (v3){$(1,3)$};
    \node[] at (2,-2) (v4){$(2,1)$};
    \node[] at (4,0) (v5){$(3,1)$};
    \node[] at (0,0) (v6){$(2,2)$};
    \node[] at (-2,2) (v7){$(2,3)$};
    \node[] at (2,2) (v8){$(3,2)$};
    \node[] at (0,4) (v9){$(3,3)$};
    \node[] at (-6,2) (v10){$(1,4)$};
    \node[] at (-4,4) (v11){$(2,4)$};
    \node[] at (-2,6) (v12){$(3,4)$};
    \node[] at (0,8) (v13){$(4,4)$};
    \node[] at (6,2) (v14){$(4,1)$};
    \node[] at (4,4) (v15){$(4,2)$};
    \node[] at (2,6) (v16){$(4,3)$};

    \draw[line width = 0.3mm, >=latex] (v1) to (v2);
    \draw[line width = 0.3mm,  >=latex] (v1) to (v4);
    \draw[line width = 0.3mm,  >=latex] (v2) to (v3);
    \draw[line width = 0.3mm,  >=latex] (v2) to (v6);
    \draw[line width = 0.3mm,  >=latex] (v4) to (v6);
    \draw[line width = 0.3mm, >=latex] (v2) to (v6);
    \draw[line width = 0.3mm, >=latex] (v3) to (v7);
    \draw[line width = 0.3mm, >=latex] (v4) to (v5);
    \draw[line width = 0.3mm, >=latex] (v6) to (v7);
    \draw[line width = 0.3mm, >=latex] (v6) to (v8);
    \draw[line width = 0.3mm, >=latex] (v5) to (v8);
    \draw[line width = 0.3mm, >=latex] (v7) to (v9); 
    \draw[line width = 0.3mm, >=latex] (v8) to (v9);
    \draw[line width = 0.3mm, >=latex] (v3) to (v10);
    \draw[line width = 0.3mm, >=latex] (v11) to (v10);
    \draw[line width = 0.3mm, >=latex] (v7) to (v11); 
    \draw[line width = 0.3mm, >=latex] (v12) to (v11);
    \draw[line width = 0.3mm, >=latex] (v9) to (v12); 
    \draw[line width = 0.3mm, >=latex] (v13) to (v12);
    \draw[line width = 0.3mm, >=latex] (v9) to (v16); 
    \draw[line width = 0.3mm, >=latex] (v13) to (v16);
    \draw[line width = 0.3mm, >=latex] (v8) to (v15); 
    \draw[line width = 0.3mm, >=latex] (v16) to (v15);
    \draw[line width = 0.3mm, >=latex] (v5) to (v14); 
    \draw[line width = 0.3mm, >=latex] (v15) to (v14);

    \draw[rotate around={45:(-3,-1)}, orange, line width=0.3mm] (-3,-1) ellipse (1.5cm and 6cm);
    \draw[rotate around={45:(-1,1)}, orange, line width=0.3mm] (-1,1) ellipse (1.5cm and 6cm);
    \draw[rotate around={45:(1,3)}, orange, line width=0.3mm] (1,3) ellipse (1.5cm and 6cm);
    \draw[rotate around={45:(3,5)}, orange, line width=0.3mm] (3,5) ellipse (1.5cm and 6cm);

    \node[] at (0, -6) (pi1) {$\pi_1, \beta_{1} = 0$};
    \node[] at (-6, 6) (pi2) {$\pi_2, \beta_{2} = 1.5$};
    \node[] at (6, -2) (pi3) {$\pi_3, \beta_{3} = 3$};
    \node[] at (0, 10) (pi4) {$\pi_4, \beta_{4} = 4.5$};
    
\end{tikzpicture}

 \end{center} 

\end{centering}

\caption{\small Hasse diagram illustrating partition used in the two drugs experiment.  The outcome is total bacterial load, so the treatment effect increases in the level of Amoxicillin but the level of Ibuprofen does not alter the outcome.}
\label{fig:hasse-sim-1}
\end{figure}

Our first simulation extends the running example with Ibuprofen and Amoxicillin from
two dosage levels per drug to four. We also use a different out outcome of interest to simplify our exposition.  In the simulation, the outcome is the inverse of the total Gram-positive bacteria present after 10 days, so larger values
correspond to lower bacterial burden and therefore more effective treatment. The true mean outcome increases with the dose of
Amoxicillin, while Ibuprofen has no effect on bacterial load. We depict the Hasse diagram of this setup in Figure~\ref{fig:hasse-sim-1}. The true top pool consists of all dosage combinations with the highest
dose of Amoxicillin, regardless of the dose of Ibuprofen. Since overall MSE averages error across all $16$ dosage combinations, a method can have
low overall MSE while still failing to recover the full top pool.  This happens because small errors among
the tied best combinations, or among nearby combinations, may barely affect average
squared error while still changing which treatments are identified as most effective. This setting is challenging for $\ell_1$-based selection approaches that decide whether increasing a dosage matters enough to cleave two dosage levels apart or instead leave them pooled. The $\ell_1$ prior treats these marginal effects as independent draws from a Laplace distribution. However, the true marginal effects of Amoxicillin dosage increases are perfectly correlated: each step up has the same marginal effect of 1.5. Treating these dosage increments as independent makes each one “fight for survival” separately, degrading finite-sample recovery of the top pool.

In each dataset, we fixed the number of samples
per dosage combination to $n_k$ and drew outcomes for each combination from a
$\mathcal{N}(\beta_k,1)$ distribution. We varied
$n_k \in \{10,20,50,100,500,1000\}$ and, for each $n_k$, simulated $r=100$ datasets.
\begin{figure*}[!tb]
    \centering
    \includegraphics[height=1.75in]{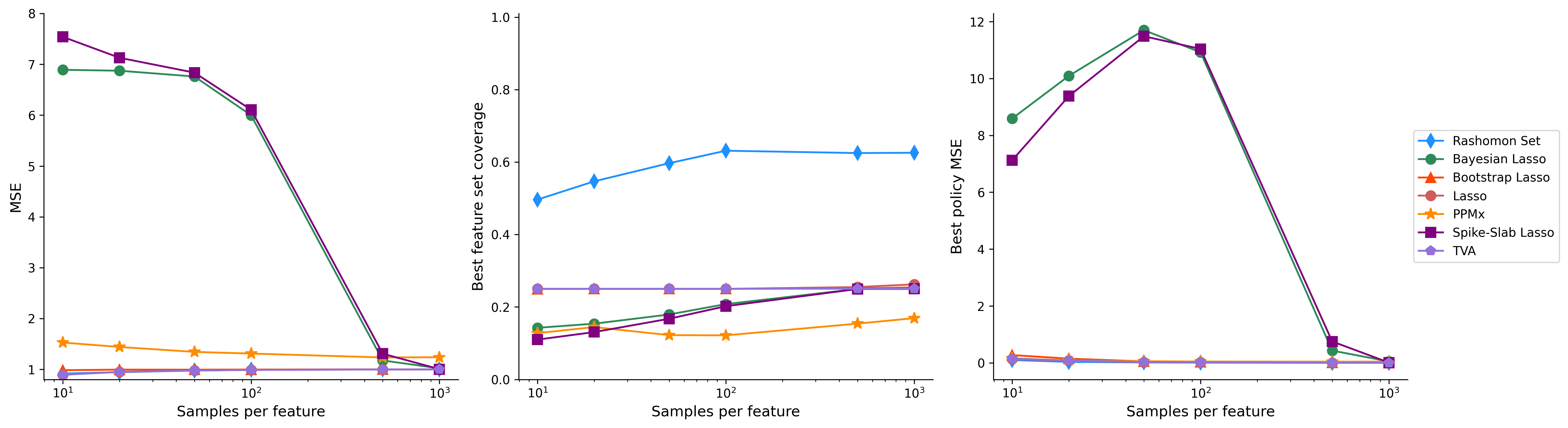}
    \caption{\small Results for the drug combination simulation. From left to right: mean squared error, best policy set coverage and best policy mean squared error. Best-policy-set coverage is the posterior-mean overlap with the true top pool, not the probability that the exact true partition is contained in the posterior support.}
    \label{fig:sim-1-results}
\end{figure*}

For the baseline comparisons, we implemented a Gibbs sampler for Bayesian Lasso of \citet{park2008bayesian} and Spike-and-Slab Lasso of \citet{rovckova2018spike} in Python. We also implemented PPMx of \citet{muller2011ppmx} using used the \texttt{ppmSuite} package in R \citep{ppmSuite}.  We also implemented Lasso and TVA~\citet{banerjee2021selecting} to give a comparison to single-model-selection methods. %
We evaluate performance using three complementary metrics. Overall MSE measures how well a method recovers the simulated outcome surface across all dosage combinations, including combinations that are not optimal in terms of treatment effectiveness. We then separately evaluate whether the method identifies the best-performing (most effective) dosage combinations, meaning the combinations in the top pool of the true partition with the highest true efficacy. This measure reflects how concentrated the near-optimal set is around the truth.  Finally, we measure how accurately the method estimates the outcome attained by that highest efficacy treatment combination. We provide the specific formulas and details for each metric in Appendix~\ref{appendix:additional-sims}. For the Bayesian methods, including RPS, we reported the performance metric averaged across all partitions in the posterior i.e., the posterior mean (or the approximate mean for the RPS).

The results from the simulations are presented in \Cref{fig:sim-1-results}. Overall, most methods perform similarly in terms of overall MSE and the MSE for the most effective treatment combination. Bayesian Lasso and Spike-and-Slab Lasso initially suffer from small-sample issues, but converge toward the other methods as sample size increases. RPS outperforms the other methods in recovering the most effective treatment combination. Single-point-estimate methods such as TVA and Lasso often achieve competitive MSE yet miss part of the true best set, even with larger sample sizes. By contrast, RPS more reliably recovers the full set of best-performing treatments (all treatments with a high dose of Amoxicillin) because it searches over all near-optimal partitions rather than committing to a single selected model.  These results highlight the distinction between model fit, a metric that all the models excel at, and identifying the true most effective treatment combination. It helps explain why some baselines achieve competitive MSE while performing poorly
on best-policy-set coverage: they can approximate the average outcome surface without
recovering the top-pool archetype. RPS is designed for the latter task, since enumeration
makes the full set of high-quality permissible partitions visible.
We present additional implementation details for the baseline methods, as well as a more detailed discussion of the results in Appendix~\ref{appendix:additional-sims}.  In the same Appendix, we also show heatmaps of the RPS and provide a comparison of the RPS to the treatment combinations identified by the other methods that incorporate model uncertainty.

Here we compare with~\citet{banerjee2021selecting}, which imposes an $\ell_1$ penalty on marginal effects. But the issue is not specific to $\ell_1$: it applies to any approach that indirectly recovers pools by sparsely estimating marginals. Such approaches must take a stance on correlations between marginals, and those assumptions can conflict with the underlying science and hurt finite-sample performance. RPS instead penalizes the number of pools directly through an $\ell_0$ prior. Permissibility preserves an interpretation in terms of active marginals, but because RPS never evaluates the marginal effects, it never has to impose a correlation structure on them. %

In our second simulation study, we study the treatment effect in a binary treatment setting.  We look at~\citet{wager2018estimation}'s Causal Random Forests (CRFs), which we discuss in \Cref{section:related-body} and \Cref{section:related}.  CRFs sample over the space of decision trees, so we ask how many causal trees are permissible and are in the RPS? Causal trees partition the features to find heterogeneity in the treatment effect directly. This is in contrast to partitions in the RPS that find heterogeneity in the outcome. To make the comparison fair, we set the outcome of the control group to a constant, 0, without any noise.

We simulate data with four features, the first feature being a binary treatment variable. The second feature takes on 3 ordered levels and the last two features take on 4 ordered levels. The following are the outcomes for the treatment group:
\begin{align*}
    \beta_{(1, 1, 1:2, 1:3)} &= 2, \ \beta_{(1, 1, 1:2, 4)} = 4, \ 
    \beta_{(1, 1, 3:4, 1:3)} = 2, \ \beta_{(1, 1, 3:4, 4)} = 0, \\
    \beta_{(1, 2, 1:2, 1:3)} &= 3, \ \beta_{(1, 2, 1:2, 4)} = 5, \
    \beta_{(1, 2, 3:4, 1:3)} = 7, \ \beta_{(1, 2, 3:4, 4)} = 1, \\
    \beta_{(1, 3, 1:2, 1:3)} &= 1, \ \beta_{(1, 3, 1:2, 4)} = -1, \
    \beta_{(1, 3, 3:4, 1:3)} = -1, \ \beta_{(1, 3, 3:4, 4)} = -2.
\end{align*}
\begin{table}[!tb]
    \centering
    \caption{\tiny Results for second simulation study. We compare how often CRFs find permissible partitions and how often they are present in the RPS. We vary both the number of trees in the CRF and the Rashomon threshold. Each cell shows the fraction of CRF trees inside the RPS (within parentheses are absolute counts). The numbers are averaged over 100 simulations.}
    \label{tab:crf-rps-sims}
    \begin{tabular}{c|c|c|c}
          & \makecell{ \# trees = 20 \\ {\footnotesize (\# permissible = 1.29)} } & \makecell{ \# trees = 50 \\ {\footnotesize (\# permissible = 2.67)} } & \makecell{ \# trees = 100 \\ {\footnotesize (\# permissible = 10.32)} } \\
         \hline
         \hline
         $\epsilon = 0.1$ {\footnotesize ($\abs{\mc{P}_{\theta}} = 7.46$)} & 0\% {\footnotesize (0)} & 0\% {\footnotesize (0)} & 0\% {\footnotesize (0)} \\
         $\epsilon = 0.2$ {\footnotesize ($\abs{\mc{P}_{\theta}} = 46.6$)} & 0\% {\footnotesize (0)} & 0\% {\footnotesize (0)} & 0\% {\footnotesize (0)} \\
         $\epsilon = 0.3$ {\footnotesize ($\abs{\mc{P}_{\theta}} = 126.54$)} & 0.41\% {\footnotesize (0.52)} & 0.91\% {\footnotesize (1.15)} & 3.35\% {\footnotesize (4.24)} \\
         $\epsilon = 0.5$ {\footnotesize ($\abs{\mc{P}_{\theta}} = 823.81$)} & 0.16\% {\footnotesize (1.29)} & 0.32\% {\footnotesize (2.67)} & 1.25\% {\footnotesize (10.32)} \\
    \end{tabular}
\end{table}
We generate $n_a = 10$ data points per feature combination. In the treatment group, we drew outcomes from a $\mc{N}(\beta_a, 1)$ distribution. We averaged simulations over 100 iterations.
The results are presented in Table \ref{tab:crf-rps-sims}. The vast majority of partitions sampled by CRFs are not scientifically coherent (permissible) partitions and thus cannot be interpreted as plausible explanations. This result is not specific to CRFs and would hold for any algorithm using unrestricted trees. The number of trees that are in the RPS is also very small, meaning that, although averaging over trees has appealing asymptotic properties, the trees included in particular sample are unlikely to be high-quality explanations.  %

\section{Empirical data examples}
\label{section:real-data}

For each model in the RPS, $\Pi$, we calculate a treatment effect or outcome $t: \mathcal{P} \to \mathbb{R}$. Then, for some interval $I \subseteq \mathbb{R}$, we measure the confidence of the RPS in $t$ falling in the interval $I$, weighted by the posterior as 
\begin{align}
    c(t, I) = \frac{1}{\sum_{\Pi \in \text{RPS}}\Pr(\Pi \mid \Z)} \sum_{\Pi \in \text{RPS}} \mathbb{I} (t(\Pi) \in I) \Pr(\Pi \mid \Z). \label{eq:RPS-confidence-counter}
\end{align}
We choose five intervals based on the spread of $t$ within the RPS. We use a fixed learning rate to estimate tempered posterior probabilities, following our generalized Bayes setup (though we could also recover the posterior density under the fully specified Gaussian setup in the Appendix). Specifically, $I_1 = (-\infty, -2 \sigma)$, $I_2 = [-2 \sigma, 0)$, $I_3 = \{0\}$, $I_4 = (0, 2 \sigma]$, and $I_5 = (2 \sigma, \infty)$ where $\sigma$ is the standard deviation of $\{t( \Pi) \mid \Pi \in \text{RPS}\}$.  The category \(I_3=\{0\}\) records structural zeros: cases in which the relevant
contrast is exactly zero because, under partition \(\Pi\), the two feature combinations
being compared are placed in the same pool. (Thus the point mass at zero is induced by
the discrete posterior over partitions, not by a continuous posterior over effect sizes.) Together, these indicate how often there is a large negative, small negative, zero, small positive, and large positive effect, weighted by the posterior density over all partitions in the RPS. When there is homogeneity, we would expect $c(t, I_j) \to 1$ for some $I_j$ and $c(t, I_k) \to 0$ for $k \neq j$. When there is heterogeneity, we see disagreement in $c$ such as $c(t, I_j) \to p_j$ and $c(t, I_k) \to p_k$ where $k \neq j$ and both $p_j > 0, p_k > 0$. We plot $c(t, I_i)$ for $i = 1, \dots, 5$ in each empirical dataset as a heatmap to easily visualize which intervals ``light up'' indicating heterogeneity or homogeneity of $t$ within the RPS. For instance, along each column in the heatmaps in Figures \ref{fig:karlan-rset}, \ref{fig:nhanes-results}, and \ref{fig:microcredit-mosiac-fx}, a single dark red cell indicates homogeneity while several orange or peach cells indicate heterogeneity, especially when they are far from each other. In Appendix \ref{appendix:real-data}, we visualize the posteriors over the RPS for the functions, $t$, used below.

\subsection{Does price matter in charitable giving?}
\label{section:real-data-karlan}

\citet{karlan2007does} used mail solicitations to prior donors of a non-profit political organization to study the effect of price on charitable donations. The data contains 50,083 individuals in the United States who had previously donated to the organization. All individuals received a letter soliciting donations. Those in the treatment group (33,396 people) included an additional paragraph describing that their donation will be matched in some way. %
 The letters were identical otherwise. Three treatment arms were cross-randomized: (i) the maximum size of the matching gift across all donations (\$25k, \$50k, \$100k, unspecified), (ii) the ratio of price match (1:1, 2:1, 3:1), and (iii) an example suggested donation amount ($1\times$, $1.25\times$, $1.5\times$ the person's highest previous contribution).  %
 Additionally, %
they classify states as red or blue depending on whether they voted for George Bush or John Kerry in the 2004 U.S. presidential election. We restrict our analysis to individuals who received the treatment, with the goal of discovering robust patterns of treatment heterogeneity. {Our feature space is a $ 4 \times 3 \times 3 \times 2$ Hasse, corresponding to maximum gift match, match ratio, suggested donation, and Democrat/Republican state. The outcome is the amount, in hundreds of dollars, that were donated.} %

\begin{figure}[!tb]
    \centering
    \begin{subfigure}[t]{0.35\textwidth}
        \centering
        \includegraphics[height=1.75in]{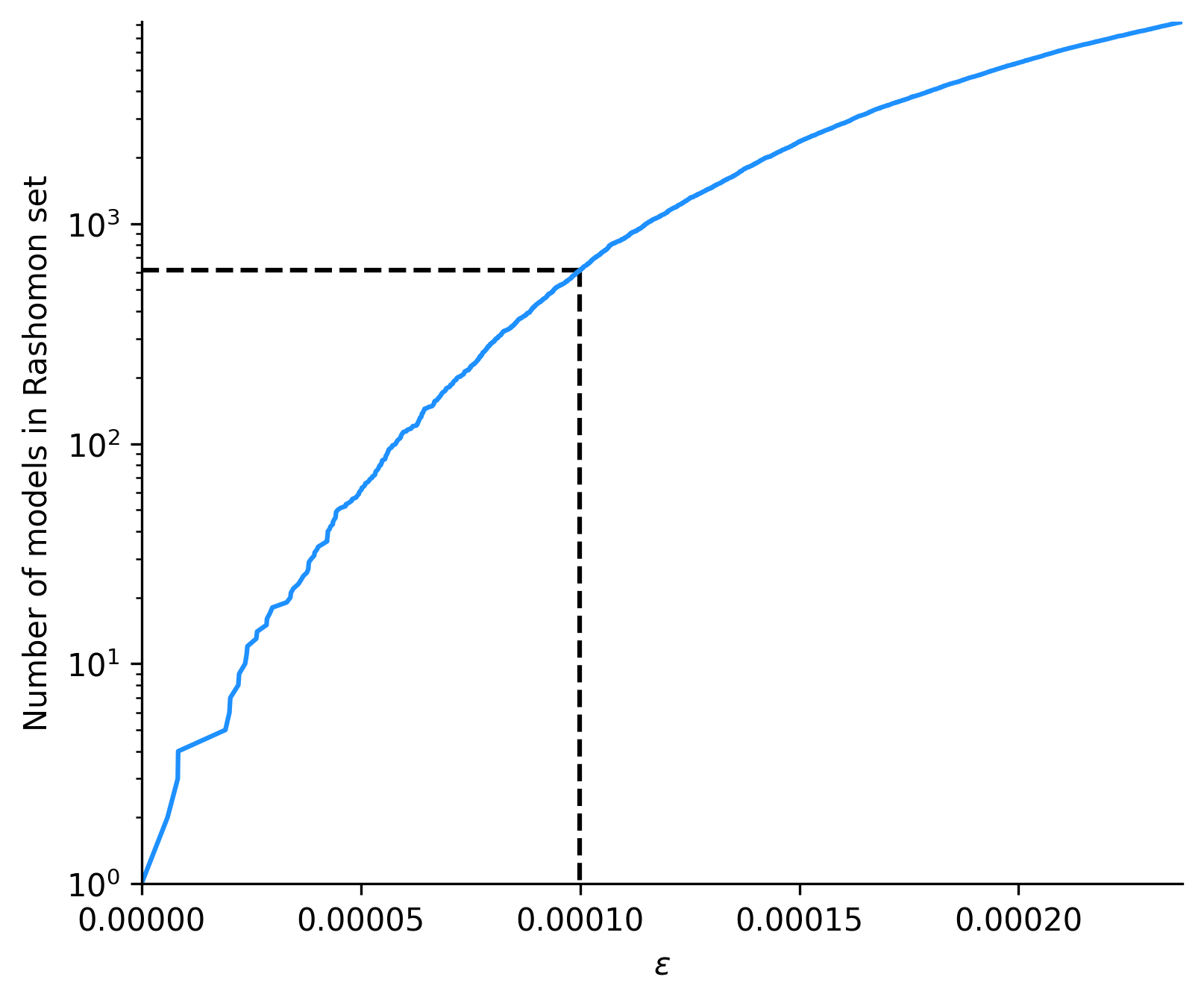}
    \end{subfigure}%
    ~ 
    \begin{subfigure}[t]{0.35\textwidth}
        \centering
        \includegraphics[height=1.75in]{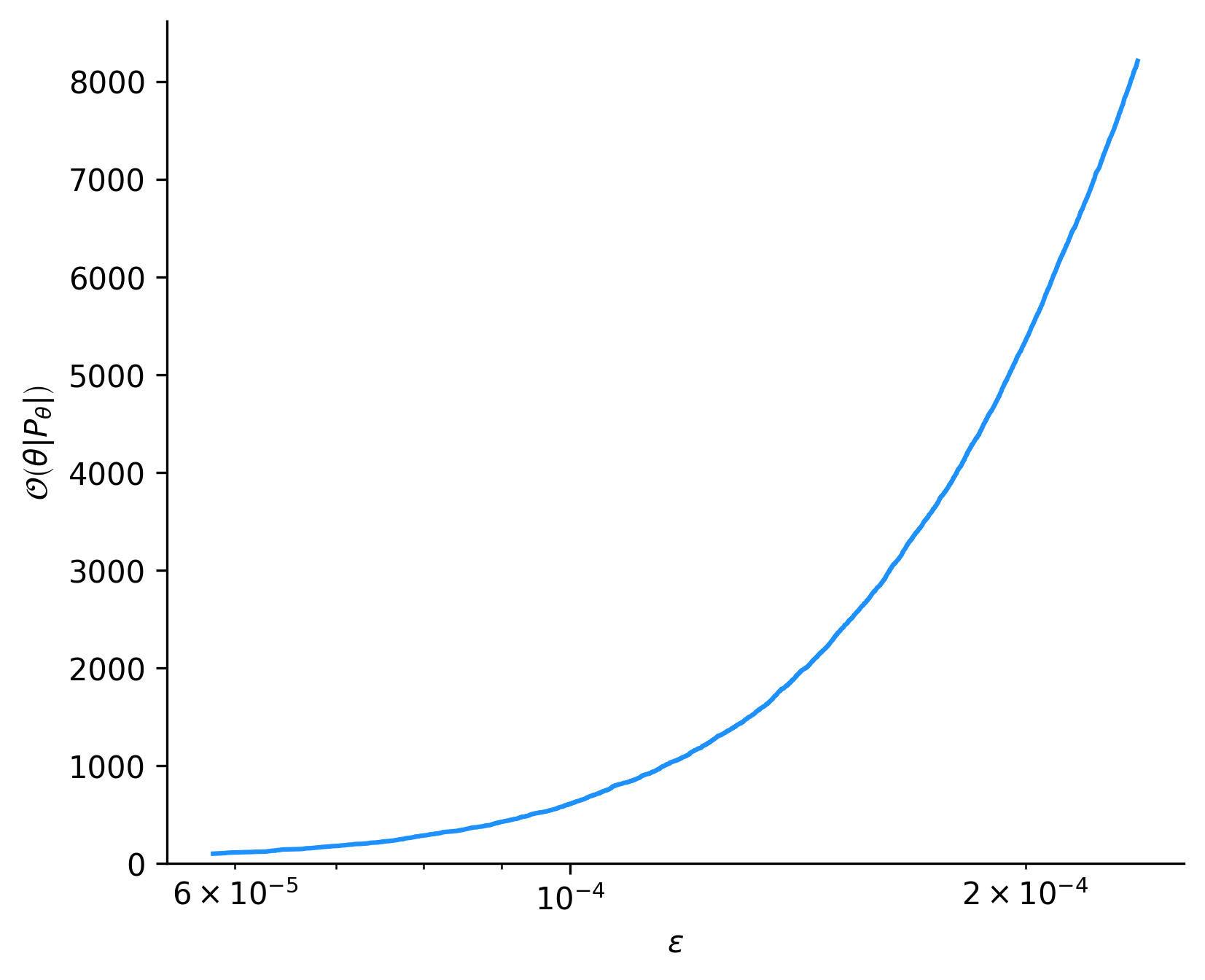}
    \end{subfigure}%
    \\
    \begin{subfigure}[t]{1\textwidth}
        \centering
        \includegraphics[height=2.25in]{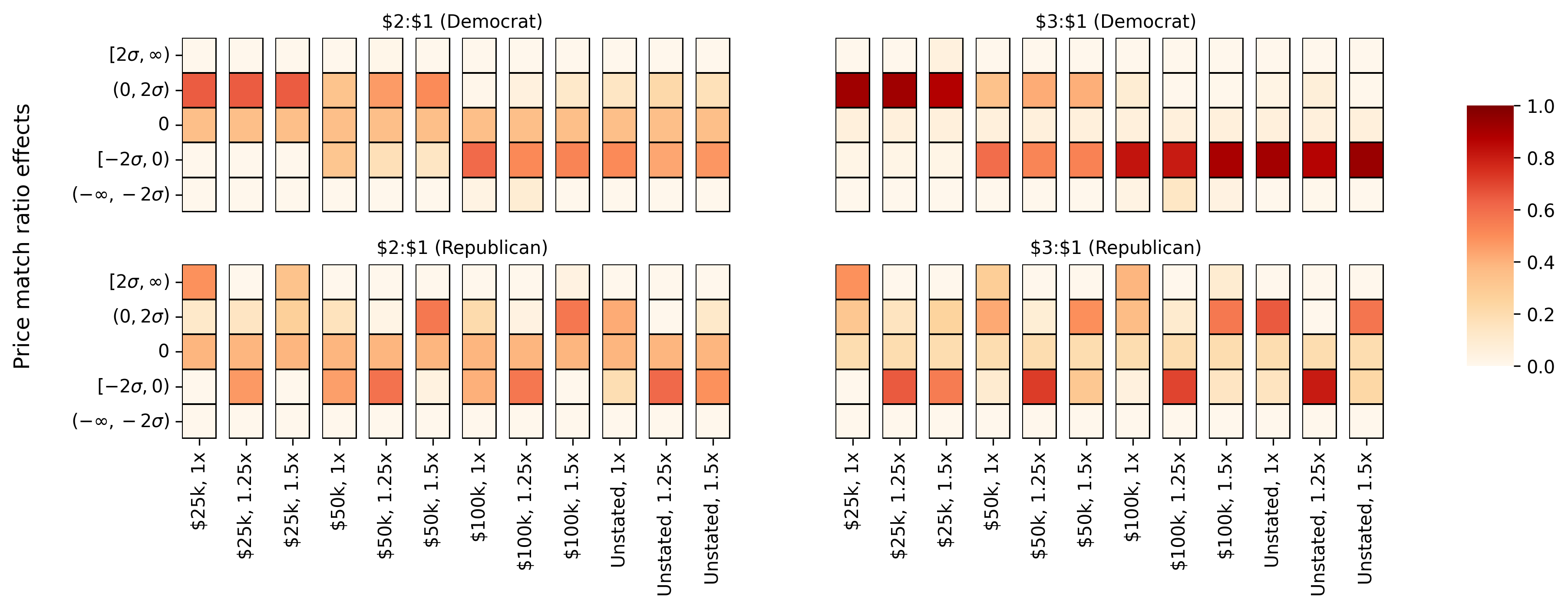}
    \end{subfigure}
    \caption{{\tiny Results for the \citet{karlan2007does} dataset. The top two panels show the size of the RPS and a partial error term from \Cref{thm:marginal_approx_rashomon} as a function of $\epsilon$. Our choice of $\epsilon = 10^{-4}$ is highlighted by the black dashed line. The bottom panel shows the RPS approximation of the effects of price match for \$2:\$1 and \$3:\$1 stratified by political leaning. Each column represents  $c(t_{1, \mathbf{x}}, I)$ and  $c(t_{2, \mathbf{x}}, I)$ for various $\mathbf{x}$ (gift size and suggested contributions). For example, (\$25k, 1x, \$3:\$1, Democrat) shows a strong positive effect (higher donations) while (Unstated 1.5x, \$3:\$1:, Democrat) shows a strong negative effect (lower donations) compared to \$1:\$1 (all else equal).}}
    \label{fig:karlan-rset}
\end{figure}
Figure \ref{fig:karlan-rset} shows how the set size and error bound change with $\epsilon$. Using Figure~\ref{fig:karlan-rset} as a guide, we chose $\epsilon$ so that adding additional models to the RPS does not dramatically increase the approximation quality (akin to choosing the number of components in principal components using a scree plot).  We choose $\epsilon = 10^{-4}$. A larger $\epsilon$ would dilute the RPS by adding more models that have little support in the data.  

The bottom panel Figure~\ref{fig:karlan-rset} provides a visual summary of the heterogeneity in charitable giving (compared to a reference of a 1:1 match) across models in the RPS. We computed $c(t_{1, \mathbf{x}}, I)$ and $c(t_{2, \mathbf{x}}, I)$ in Equation \ref{eq:RPS-confidence-counter} for $t_{1, \mathbf{x}}( \Pi) = \widehat{\E}[y(2\mathbin{:}1, x, z, p) \mid \Pi] - \widehat{\E}[y(1\mathbin{:}1, x, z, p) \mid \Pi]$ and $t_{2, \mathbf{x}}( \Pi) = \widehat{\E}[y(3\mathbin{:}1, x, z, p) \mid \Pi] - \widehat{\E}[y(1\mathbin{:}1, x, z, p) \mid \Pi]$. We estimated $t_1$ and $t_2$ for each $\mathbf{x}$: maximum limit $x$, suggested donation $z$, and state political affiliation $p$, using the $\bbeta(\Pi)$ vector for each model and combination of features.
We categorized each effect into five intervals defined by the standard deviation of the effects across all models in the RPS including a separate bin for ``0'' indicating no difference in effect compared to a 1:1 match. Figure \ref{fig:karlan-het-dist-appendix}, we shows tempered posterior densities in the RPS of $t_{1, \mathbf{x}}$ and $t_{2, \mathbf{x}}$.

{\cite{karlan2007does} makes three claims, all of which are strongly rejected by the RPS.  First, \cite{karlan2007does} find that match ratio does not matter.  For \$3:\$1 in Democrat states, however, we see a robust increase in donations compared to a \$1:\$1 match for a maximum limit of \$25,000.  We also see a robust \emph{decrease} in donations compared to a \$1:\$1 match for \$100,000 or unspecified limits. This robust discouragement effect is particularly interesting and may warrant further research. The result on match ratios mattering is of great policy relevance for the same reasons as argued in \cite{karlan2007does}: if they did not matter, low ratio matches could be used to save money, but if they robustly do matter, when they are positive high ratio matches ought to be leveraged and may have excess returns and when they are negative the costs are even more damaging.
Second, \cite{karlan2007does} find that the matching gift maximum does not matter. In contrast, we see that Democrats are robustly encouraged by lower gift maxima and deterred by the very high/unrestricted ones.  Further, in Republican states, we see a consistent increase in donations for a suggested donation of 1 $\cdot$ HPC and a robust negative effect for 1.25 $\cdot$ HPC.
Third,~\cite{karlan2007does} find that having a match motivates Republicans and not Democrats, though again the ratio does not matter. Our analysis confirms that political leaning matters, with over 99\% of the models in the RPS splitting on political leaning.  Our conclusion, however, is much more subtle.  Democrats did give more under some matching conditions, specifically when the match ratio was high but the suggested donation and maximum were low.  As previously noted, however, they also gave less under other conditions, indicating that, while there is heterogeneity in behavior based on political leaning, there is substantial, policy relevant, heterogeneity that was omitted from the previous analysis.}

\subsection{Heterogeneity in telomere length.}
\label{section:real-data-nhanes}

Telomeres are regions of repeated nucleotide sequences near the end of the chromosome that protect the chromosome from damage. They  reduce in length every time a cell divides, eventually becoming so short that the cell can no longer divide. %
A recent literature examines features are associated with (or possibly cause) changes in telomere length. Telomere shortening has been associated with cellular senescence and may hold target biomarkers for genetic predispositions and anti-cancer therapies \citep{rossiello2022telomere, srinivas2020telomeres}. %
Recent research suggests that there may only be a narrow range of healthy telomere lengths; anything extremal is at increased risk of immune system problems or cancer \citep{alder2018diagnostic, protsenko2020long}.
 Research has found heterogeneity by race, ethnicity, age, and even stress  \citep{chae2014discrimination, geronimus2015race, hamad2016racial, vyas2021telomere}. %

 We use the National Health and Nutrition Examination Survey (NHANES) collected in 1999 and 2002. The survey included blood draws, and DNA analyzes were performed from the samples and the length of the telomere was estimated. Specifically, we consider the mean T/S ratio (telomere length relative to standard reference DNA).\footnote{See \url{https://wwwn.cdc.gov/nchs/nhanes/1999-2000/TELO_A.htm}  for details. Website last accessed on 2024-01-29.}  The dataset also contains socio-economic variables. To speak to the emerging literature on telomere heterogeneity, we focus on hours worked (a proxy for stress), age, gender, race, and education. Our goal is to study the RPS of this heterogeneity on T/S.\footnote{We removed all individuals who were missing data for relevant covariates. We binned the number of hours worked: $\leq 20$ hours, $21-40$ hours, and $\geq 40$ hours. Age was categorized into five ordered discrete factors -- $\leq 18$ years, $19-30$ years, $31-50$ years, $51-70$ years, and $>70$ years. Education was categorized into 3 ordered discrete factors (no GED, GED no college, completed college).}

\begin{figure}[!tb]
    \centering
    \begin{subfigure}[t]{0.4\textwidth}
        \centering
        \includegraphics[height=1.75in]{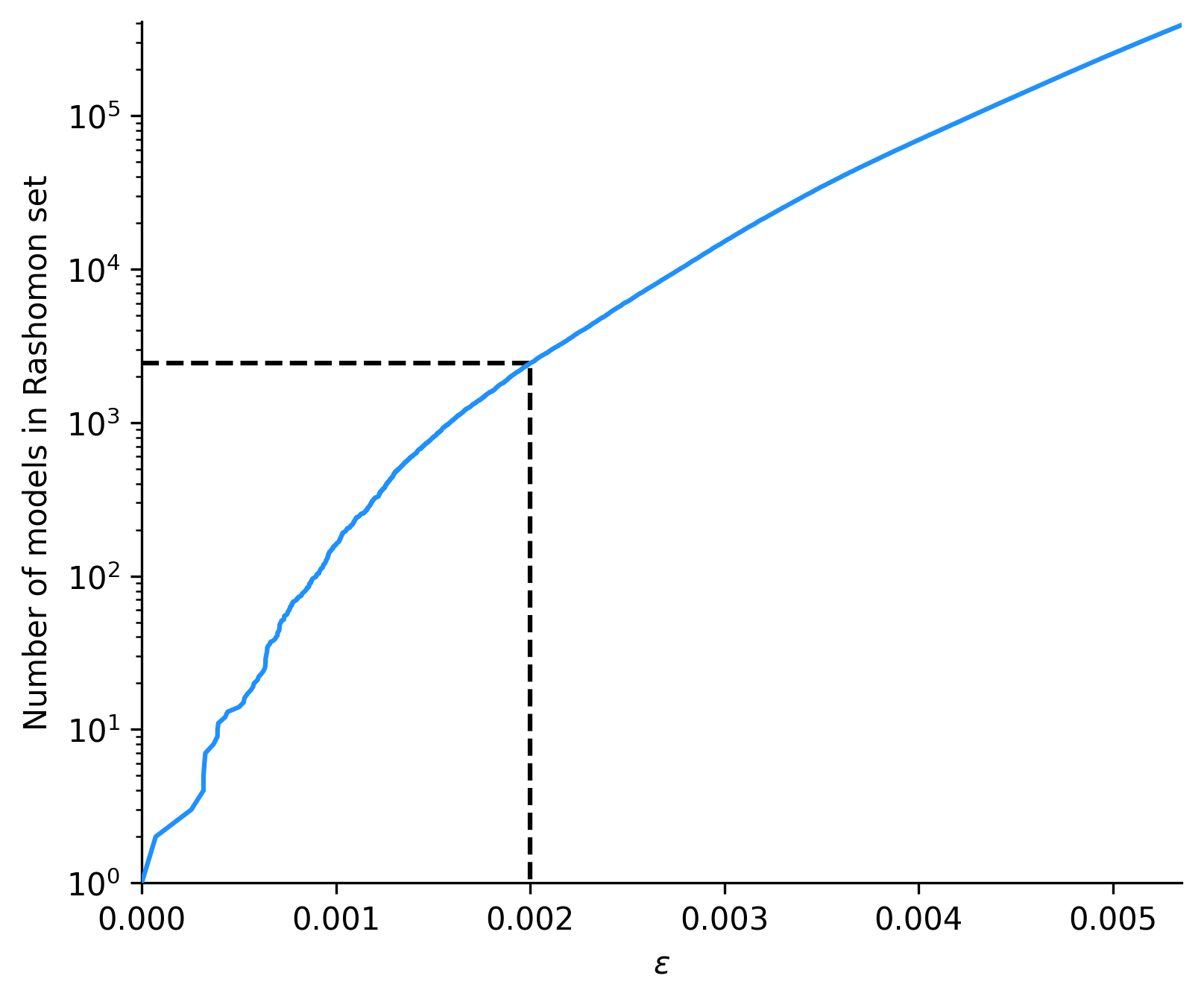}
    \end{subfigure}%
    ~ 
    \begin{subfigure}[t]{0.4\textwidth}
        \centering
        \includegraphics[height=1.75in]{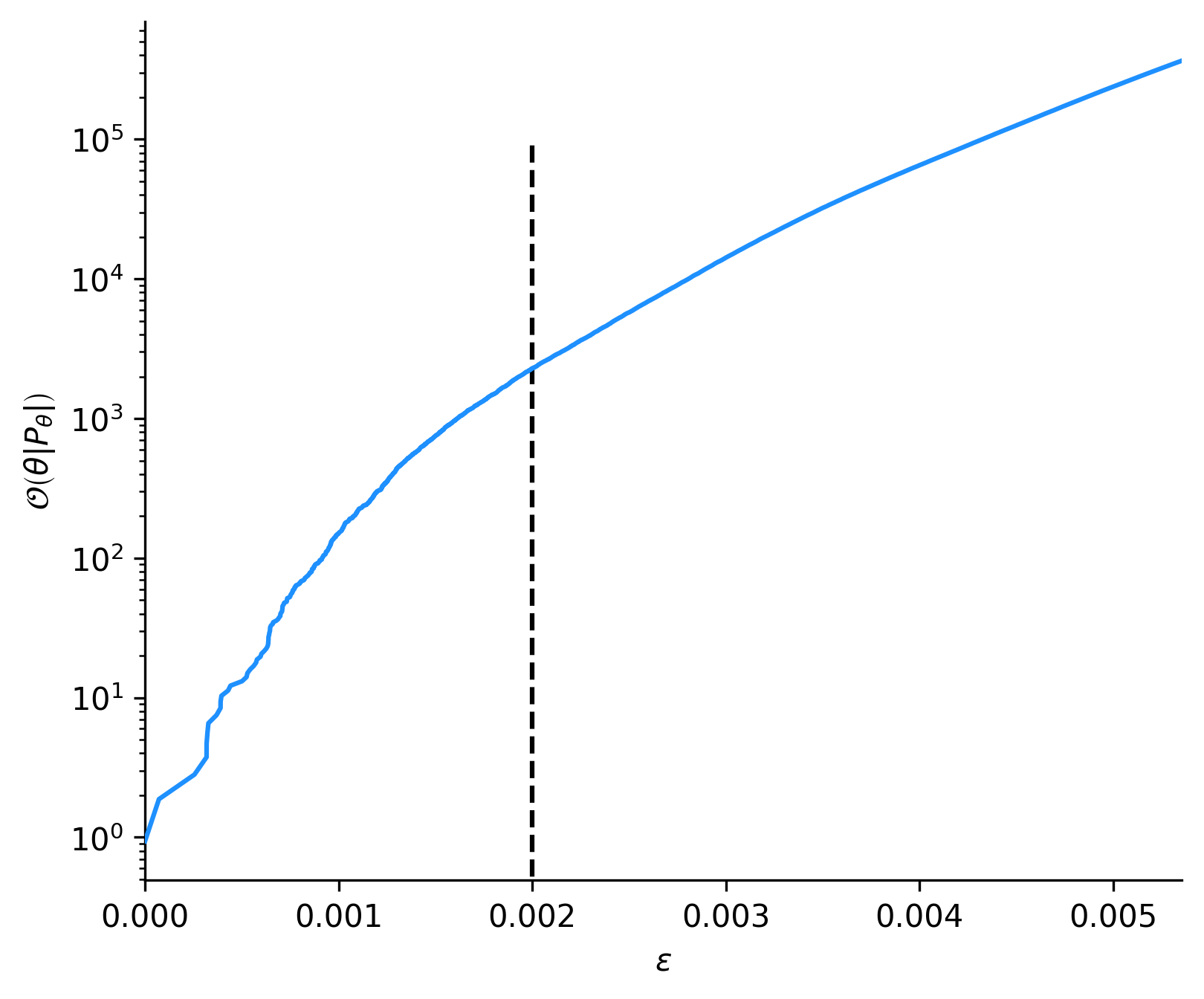}
    \end{subfigure}%
    \\
    \begin{subfigure}[t]{0.9\textwidth}
        \centering
        \includegraphics[height=2.25in]{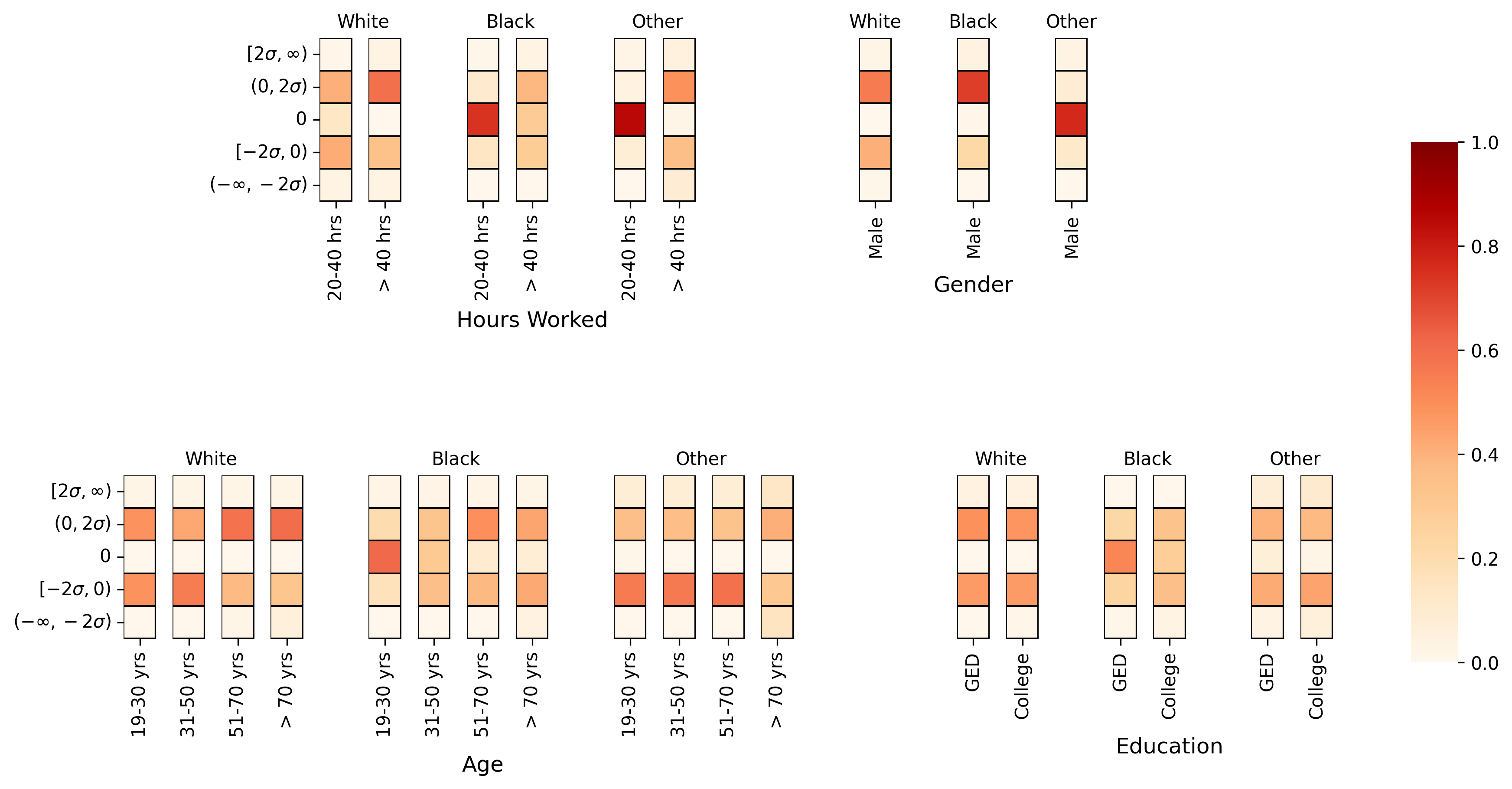}
    \end{subfigure}
    \caption{\tiny The top two panels show what happens as we increase $\epsilon$ in the NHANES dataset to the size of the RPS and the partial error term from \Cref{thm:marginal_approx_rashomon}, highlighting our choice of $\epsilon$.
    {In the bottom panel, we highlight heterogeneity in telomere length across the four features (hours worked, gender, age, and education) within the RPS. Each panel marginalizes one feature and stratifies it by race. Each column corresponds to a feature level. Each cell represents the difference in telomere length relative to the lowest level taken by that feature, representing $c(t_{r, \mathbf{x}}, I)$ for race $r$ and features $\mathbf{x}$.}
    }
    \label{fig:nhanes-results}
\end{figure}
We show our choice of $\epsilon$ for the Rashomon threshold in the top two panels of Figure \ref{fig:nhanes-results}, and a visual summary of the conclusions of the RPS in the bottom panel, similar to Figure~\ref{fig:karlan-rset}. In the RPS, we found robust heterogeneity in race -- specifically, we found no partition that pools features across races. 
We found robust evidence of male/female heterogeneity in only in White and Black races. All partitions for these races split males and females into separate pools. This was absent in Other races -- only 23\% of the partitions in the RPS split on male/female.

For each race $r$, we find the length of telomeres stratified by each feature $m \in $ \{Hours worked, Gender, Age, Education\} relative to the lowest level of that feature, $t_{r, \mathbf{x}}( \Pi) = \widehat{\E}[Y_i(x_m, \mathbf{x}_{-m}, r) \mid \Pi] - \widehat{\E}[Y_i(1, \mathbf{x}_{-m}, r) \mid \Pi]$ using Equation~\ref{eq:mainreg}. We compare telomere lengths of \textit{Male} ($x_m = 2)$ relative to \textit{Female} ($x_m = 1$).
 As previously, we compute $c(t_{r, \mathbf{x}}, I)$ in Equation \ref{eq:RPS-confidence-counter} across five intervals based on the overall standard deviation of $t_r$ in the RPS. We report this as a heatmap in the bottom panel of Figure \ref{fig:nhanes-results} and visualize posteriors over the RPS in Figure \ref{fig:nhanes-het-dist-appendix}. Again, the interval ``0'' corresponds to the case where the features were pooled together thereby having no difference in telomere lengths from the lowest feature level.
We find very few robust patterns. As discussed earlier, we find robust differences in telomere lengths across males and females in the Black population and a robust non-difference in Other races. Similarly, we find a robust non-difference in Black and Other races in telomere lengths for people who work fewer than 40 hours.

Our findings reveal an absence of robust evidence supporting the patterns highlighted in existing literature. Moreover, of the few robust patterns we do identify, several findings contradict prior research. We find Black males have longer telomeres than females. Among White people, we find older people have longer telomeres, which also contradicts existing research. This underscores the necessity for further exploration in this field using comprehensive data and appropriate statistical methods.  This contradiction reveals the fragility of empirical research that relies on a single model.

\subsection{Heterogeneity in the impact of microcredit access.}
\label{section:real-data-microcredit}

A large literature has looked at the impact of microfinance on several outcomes, ranging from private consumption to business outcomes to social outcomes (e.g., female empowerment). Mostly, the literature has found little beyond basic consumption effects \citep{angelucci2015microcredit,attanasio2015impacts,augsburg2015impacts,banerjee2015miracle,crepon2015estimating,tarozzi2015impacts,meager2019understanding}, though there is suggestive evidence of some potential heterogeneity. One specific heterogeneity of interest concerns entrepreneurs: %
 those with pre-existing businesses may be particularly benefited by the access to microfinance loans \citep{banerjee2019can}. Another concerns family size \citep{baland2008microfinance}: the returns to credit access may vary by whether the household has more children.

We analyze data from \cite{banerjee2015miracle} generated from a randomized controlled trial in which 102 neighborhoods in Hyderabad, India were randomly assigned to treatment or control, each with equal probability,  where treatment meant that a partner microfinance organization, Spandana, entered. At baseline a number of characteristics of sampled individuals were collected, including the gender of the head of the household, the education status of the head of the household, the number of businesses previously owned by the household, and the number of children in the household. Additionally, at the neighborhood level, information about the share of households with debt, the share of households with businesses,  total expenditure per capita in the region, and average literacy rates in the region were also collected at baseline. Motivated by the literature, we look at the regional debt and business variables.
We consider outcomes from the second (longer term) endline, focusing on four spheres: (i) loans, (ii) household response (total expenditure, durables, temptation goods, labor supply), (iii) business (revenue, size, assets, profits), (iv) female empowerment (female business participation, education of daughters). We discretized the regional characteristics and the number of businesses previously owned into four levels based using quartiles. We set the first quartile as the ``base control,'' so the feature is active in the profile if it is a higher level.

\begin{figure}[!tb]
    \centering
    \includegraphics[height=2.25in]{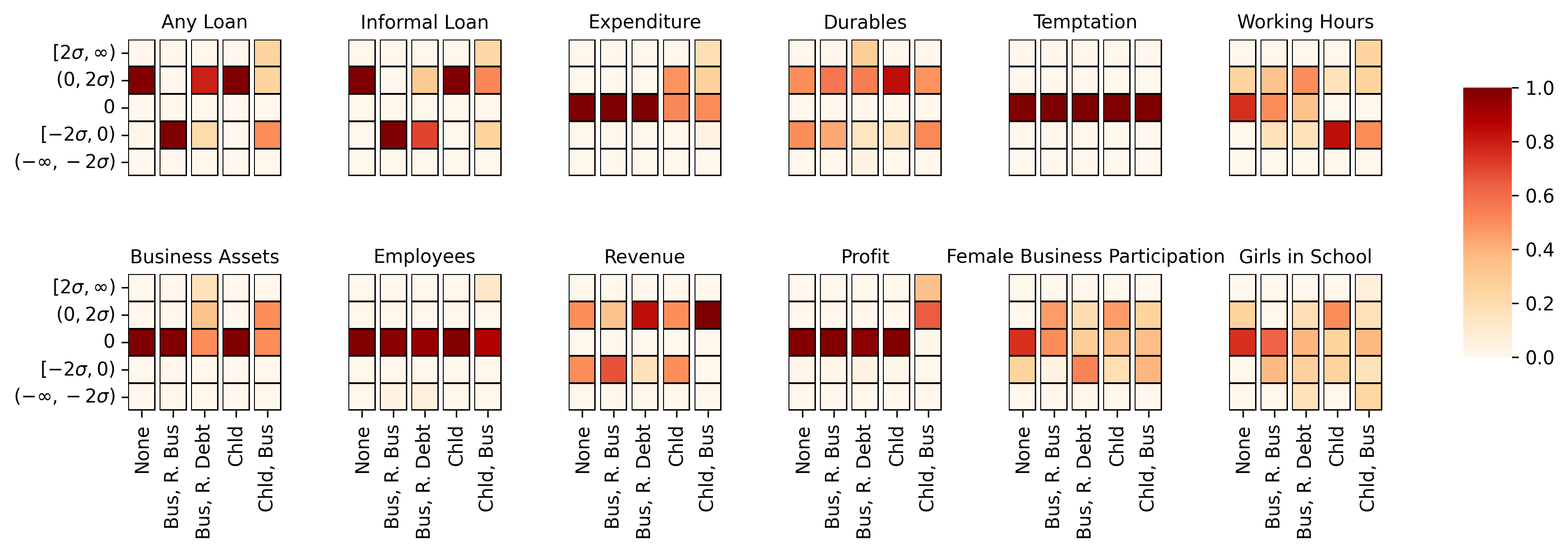}
    \caption{
    \tiny {This plot visualizes the heterogeneity of treatment effects across partitions in the RPS, for each feature combination. Each column corresponds to one of the five robust feature profiles described where the label denotes which features are active (i.e., do not take the lowest level). ``None'' means that all features are taking these lowest values, and `R.' is shorthand for regional characteristics. Each cell plots $c(\textrm{CATE}_{\mathbf{x}}, I)$.}}
    \label{fig:microcredit-mosiac-fx}
\end{figure}

To study the impact of access to microcredit, we allow features across treatment and control profiles to be pooled together (see \Cref{def:permissible-partition}).  This allows for differences in the differences in the structure of heterogeneity (i.e. different Hasse diagrams) between treatment and control.
Then, we measure the heterogeneous impact as the conditional average treatment effect, $\textrm{CATE}_{\mathbf{x}}( \Pi) = \E[Y_i(1, \mathbf{x}) \mid \Pi] - \E[ Y_i(0, \mathbf{x}) \mid \Pi]$. We estimate $\widehat{\textrm{CATE}}_{\mathbf{x}}( \Pi) = \widehat{\E
}[y(1, \mathbf{x}) \mid \Pi] -  \widehat{\E}[y(0, \mathbf{x}) \mid \Pi]$ where $\widehat{\E}[y(1, \mathbf{x}) \mid \Pi]$ is the estimated potential outcome for a household assigned to treatment with feature combination $\mathbf{x}$, and $\widehat{y}(0, \mathbf{z})$ is the estimated potential outcome for a household assigned to control with feature combination $\mathbf{z}$ in model $\Pi$, computed using Equation~\ref{eq:mainreg}.
If feature $\mathbf{x}$ is pooled across the treatment and control profiles, then $\widehat{\textrm{CATE}}_{\mathbf{x}}( \Pi) = 0$ indicating no treatment effect heterogeneity in feature $\mathbf{x}$.Otherwise, $\widehat{\textrm{CATE}}_{\mathbf{x}}( \Pi) \neq 0$ indicating treatment effect heterogeneity.
We present $c(\textrm{CATE}_{\mathbf{x}}, I)$ in Equation \ref{eq:RPS-confidence-counter} categorized into five intervals based on effect sizes across the RPS, which captures robust (or non-robust) qualitative patterns in Figure \ref{fig:microcredit-mosiac-fx}. In Appendix \ref{appendix:real-data}, Figure \ref{fig:microcredit-mosiac-fx-full} shows the full set of profiles and outcomes. We show the posterior densities of these effects, restricted to the RPS, in Figures \ref{fig:microcredit-het-dist-appendix-loan}, \ref{fig:microcredit-het-dist-appendix-temp}, and \ref{fig:microcredit-het-dist-appendix-profit}.%

{We identify a few key archetypes, highlighted in Figure~\ref{fig:microcredit-mosiac-fx}.  First, consider the archetype of large households, with no previous businesses, in a region with low baseline debt and business presence. (labeled ``Chld'' in Figure~\ref{fig:microcredit-mosiac-fx} to indicate the lowest level on all variables except number of children/household size).  These households tended to take more loans (including informal), consume more (including durables but not temptation goods), and supply less labor (work fewer hours) in response to the intervention.  We see uncertain effects for profits and revenue for this group. Large households in similar economic environment, but with previous business experience (labeled ``Chld, Bus'') had robustly increased profits and revenue in response to the intervention, despite unclear results on loans and expenditures.  For small households with no previous businesses, in a region with low baseline debt and business presence (labeled ``None'') we see increases in loans (including informal), but no other robust outcomes.  We also see no conclusive evidence of the impact of microcredit on female empowerment indicators across any archetype. In Appendix \ref{appendix:real-data}, we also look at the treatment effect heterogeneity by gender. For the most part, we find no robust heterogeneity. We also see a robust finding that introducing microcredit did not change spending on temptation goods across all the archetypes. }

The RPS provides an avenue for the researcher to identify patterns of robust treatment effects across outcomes of interest. It also clearly demonstrates when, for numerous profiles, there is little robustness to be had. Without strong priors, data cannot robustly speak to the impacts of microcredit in most cases.  Research, therefore, that makes conclusions in such settings based on a very specific treatment of heterogeneity is relying heavily on their (highly consequential) priors over the structure of such heterogeneity.  %

The RPS also gives policymakers guidance on robust interventions. If the policymaker considered regions with high baseline debt, since robustly there are no positive profits and half the RPS suggests negative profits, they may not wish for the microcredit firm to enter this market.  But in contrast, in other markets, e.g., low debt and business presence, for large non-entrepreneurial families since there are robustly no effects on profits and robustly positive effects on consumption and leisure, they can proceed with confidence.

\section{Related work}\label{section:related-body}

We provide an overview of related work in this section; more details are in Appendix~\ref{section:related}. Our work is related to a flourishing literature on the Rashomon effect~\citep{chatfield1995model, breiman2001statistical, mcallister2007model,tulabandhula2014robust,pawelczyk2020counterfactual,black2022model,d2022underspecification,kobylinska2023exploration,zhong2023exploring}.
One line identifies sets of estimands that generate similar objective function values~\citep{marx2020predictive,coker2021theory,watson2023predictive} and has been explored in the context of variable importance~\citep{fisher2019all, dong2020exploring}. %
The most related is~\citet{xin2022exploring}, who identify $\epsilon$-Rashomon sets and a decision tree algorithm to enumerate the set of estimands (trees)  that have squared loss smaller than a threshold slightly higher than that of a reference model.  Our work focuses on (Bayesian) inference with discrete (and continuous, as we outline in the Appendix) variables in the general regression setting, while they address prediction with binary variables only for classification.  %
Our work opts for a geometric representation based on Hasse diagrams, encoding scientific plausibility using permissibility rules.%
Additionally, \citet{semenova2022existence} hypothesized and showed using simulations that regularization changes the size of the RPS. We establish and prove this relationship for the $\ell_0$ penalty in Theorem \ref{thm:rashomon-set-polynomial-size}.

Next, we relate our work to literature on Bayesian model uncertainty. %
Our setup is reminiscent of other work %
that leverage priors over trees (e.g.,~\citet{chipman1998bayesian}, ~\citet{denison1998bayesian}, ~\citet{wu2007bayesian} or Bayesian Additive Regression Trees (BART) \citep{chipman2010bart}).  %
We use Hasse diagrams that obey permissibility criteria and our computational approach does not involve sampling from the posterior, which allows researchers to focus on a set of the highest posterior explanations for heterogeneity while avoiding the computational issues associated with sampling the massive space of trees.  We demonstrate in Appendix~\ref{section:generalization} how to extend our framework to functions across pools (e.g.~\citet{chipman2002bayesian}).

Our approach is also related to Bayesian Model Averaging (BMA)~\citep{raftery1997bayesian,clyde2003model}.  %
Unlike BMA, the dimension of $\bbeta$ stays fixed throughout, though there are restrictions on $\bbeta$ given a particular partition.  We thereby avoid searching the extremely large space of highly correlated models of different dimensions~\citep{raftery1997bayesian,hans2007shotgun,onorante2016dynamic} while preserving a unified Bayesian inference framework. \citet{tian2009computing} and \citet{chen2014finding} use a related strategy for causal discovery by finding high posterior equivalence classes of Bayesian networks.

Finally, our work speaks to literature that leverages ideas from machine learning to estimate treatment effect heterogeneity. We previously discussed~\cite{banerjee2021selecting}. %
Our work is also related to existing tree-based methods for identifying treatment heterogeneity. ~\cite{wager2018estimation}, which we also discuss in detail in Appendix~\ref{section:related}, construct regression trees (every tree corresponding to some partition $\Pi$ in our language) to describe heterogeneity in the space of covariates and then sample from the distribution over trees to (honestly) estimate conditional average treatment effects.  Similar to the comparison with Bayesian tree models, our work differs in moving from trees to Hasse diagrams, which do not impose false hierarchies on a partially ordered set, considering only plausible heterogeneity through permissibility rules, and emphasizing enumeration of high quality explanations rather than sampling across all possible trees. 
Finally, we contrast with recent work that uses machine learning proxies, or predictions of the outcome that use features flexibly, to study heterogeneity in treatment effects.  %
\citet{chernozhukov2018generic}, conduct inference while using an arbitrary machine learning algorithm to construct proxies and then cluster respondents into groups with the highest and lowest treatment efficacy using treatment outcomes predicted based on the proxies.
These clusters, which are derived from amalgamating covariates through ``black box'' machine learning algorithms, can then be related back to observables.  %
 Our work, in contrast, enumerates a set of plausible explanations in the domain of observed covariates directly.

\vspace{-10pt}
\section{Discussion}
\label{section:conclusion}

In this paper, %
we derive a Bayesian framework and an algorithm to enumerate \emph{all} possible pooling across feature combinations with the highest posterior density: the Rashomon Partition Set.  %
By only considering scientifically plausible pools in a geometry that allows for partial ordering (Hasse diagrams), we substantially reduce the number of possible explanations for heterogeneity without sacrificing flexibility. These choices mean that the resulting high posterior partitions are \emph{interpretable} and useful for researchers and policymakers when designing future interventions or generalizations.
Since we are in a factorial space, for a fixed Rashomon threshold, the enumeration runtime increases exponentially with the size of the space. On the other hand, we also expect the size of the Rashomon Partition Set to increase as it covers roughly the same fraction of the entire model space. We recommend choosing a smaller threshold while still having an acceptable error tolerance based on Theorem \ref{thm:marginal_approx_rashomon} (as illustrated in the empirical examples).

We now highlight two additional philosophical points about our approach.  First, our approach is fundamentally generative in the sense that it produces insights that are directly interpretable. We expect that Rashomon partitions themselves will be of interest for researchers or policymakers. They allow for the identification of the most robust conclusions, settings where policymakers can intervene without worrying about likely negative consequences, and defining ``archetypes'' for theory-building. 
In this way, our work contributes to a growing literature in artificial intelligence (AI) and machine learning that pushes back on the use of black box algorithms to make high-stakes decisions (see e.g.,~\citet{rudin2019stop}).  We show that, by leveraging the correct geometry and permissibility structure, we eliminate the need to use complex, under-identified data mining tools that require post-hoc explanation for interpretation.%

Second, our work highlights the aperture that exists between statistical and practical decision-making.  %
Our work posits that the quest for the ``best'' statistical model is Sisyphean in essentially any scientifically interesting setting.  While this may seem dire, it actually presents an opportunity to involve additional factors beyond model performance that are often critical in practice for making decisions. Amongst models in the RPS, a policymaker could choose based on, for example, implementation cost, equity considerations, or preserving privacy without sacrificing statistical performance.

There are many promising areas for future work in extending the framework we present here.  First, we present results in terms of a posterior in a Bayesian framework.  We could, however, also construct a similar structure under a frequentist paradigm.  In such a setup, we would need to explore a re-splitting strategy (see~\cite{wager2018estimation}, for example) to construct an ``honest'' set of Hasse diagrams.  Furthermore, we could use our approach to identify groups that are systematically underrepresented in randomized trials (see~\citet{parikh2024we}, for example) and, as a further generalization, to compare results across trials (see for example~\cite{meager2019understanding}). We could also blend PPMx’s covariate‐driven cohesion with RPS’s guaranteed enumeration.  By adopting a prior that upweights pools of treatment arms whose covariate profiles are more similar, RPS could focus on the most plausible partitions in applications where side‐information is available.  We could likely adapt our branch‐and‐bound enumeration and associated uniform‐error and size bounds, ensuring that exhaustive uncertainty quantification is maintained even under this richer, covariate‐informed prior. Finally, our computational approach could be more generally valuable in a wide range of settings, in model selection for graphical models or in for discrete model averaging more generally.

\clearpage
\newpage
\bibliographystyle{apalike}
\bibliography{tva.bib}

\newpage

\appendix

\counterwithin{figure}{section}
\counterwithin{equation}{section}
\counterwithin{theorem}{section}
\counterwithin{lemma}{section}
\counterwithin{corollary}{section}
\counterwithin{definition}{section}
\counterwithin{proposition}{section}
\counterwithin{algorithm}{section}
\counterwithin{exampleplain}{section}
\newtheorem{observation}{Observation}[section]

\section{Permissibility and Hasse diagrams.}
\label{appendix:environemnt}
The geometric definition of permissibility is presented in \Cref{def:permissibility-profile-measure-zero}.  For computation, it is beneficial to use an equivalent definition in terms of differences in outcomes between adjacent feature combinations.  Recall that the geometric interpretation relied on edges in the Hasse diagram.  In the Hasse diagram, edges join adjacent feature combinations, so, intuitively, defining permissibility in terms of edges should match a definition in terms of the outcomes for adjacent feature combinations.  In this section, we formalize this setup and the connection to the representation presented in the main text.

To simplify this discussion, we define \emph{variants}, or feature combinations that differ by only one level, as finest grain building blocks for partitions.

\begin{definition}[Variants]
\label{def:variant}
Two feature combinations $k < k^{\prime}$ are \emph{variants} if they have the same value of features for all but one and they vary by exactly one intensity value i.e., $k^{\prime}_{-m} = k_{-m}$ and $k^{\prime}_m = k_m + 1$ for some feature $m$.
\end{definition}

We can think of the overall effect of some feature combination, then, as summing up through these marginal effects. This amounts to considering variants, or taking one step up the Hasse.
We then arrange feature combination assignments into a \emph{feature variant aggregation design matrix}, $\F \in\left\{ 0,1\right\} ^{n,K}$.   The entries of the matrix are as follows. If  $k\left(i\right)$ is the feature combination that $i$ is assigned to, we set
\begin{align*}
F_{i\ell} &:= \mathbb{I} \left\{ k\left(i\right)\geq\ell\ \cap\ \rho\left(k\left(i\right)\right) = \rho\left(\ell\right)\right\} .
\end{align*}
So the variant design matrix switches on a dummy variable for all variants that are subordinate to $k(i)$.  \cite{banerjee2021selecting} identifies active marginal effects explicitly as the first step of a two-step procedure. They first reparameterize equation \eqref{eq:mainreg} so that the estimand is the vector of marginal effects, impose sparsity, and estimate its support via LASSO ($\ell_1$) regularization. In a second step, the implied pools on the Hasse are deduced, and the pooled specification is then estimated. This first step is effectively unavoidable within their frequentist framework: it is the device that enables pooling without resorting to even more problematic multiple hypothesis testing.
By contrast, defining permissibility directly over partitions of the Hasse diagram, thereby abstracting from marginals, is the key conceptual move. Combined with an explicit Bayesian formulation, this allows us to penalize the object of interest, the number of permissible pools, rather than an auxiliary parameterization. This yields an $\ell_0$ prior that is robust to marginal effect correlation structure, extends naturally to general factorial settings beyond randomized trials, and enables Rashomon partition set enumeration.

This setup allows us to understand the marginal value of climbing the ordering up from $k(i)$, as in \cite{banerjee2021selecting}. Here, we restrict ourselves to a single profile, anticipating an extension to multiple profiles later.
To see this, it is useful to rewrite \Cref{eq:mainreg} in its variant form, 
\begin{align}
    \y  &= \F  \balpha + \epsilon, \label{eq:tva-regression}
\end{align}
which is just a linear transformation of \Cref{eq:mainreg}, with $\bbeta$ described below,
\begin{align}
    \beta_k &= \sum_{k^{\prime} \leq k; \rho(k) = \rho(k^{\prime})} \alpha_{k^{\prime}}.
    \label{eq:tva-climbing-up-hasse}
\end{align}

We imagine that moving \textit{up} from one node to its adjacent node in Hasse inherits a value that corresponds to the marginal change in the outcome moving from an immediate subordinate variant to the present variant. This says that an expected outcome of a feature combination is the sum of expected marginal values leading up to it. For instance, the treatment $(500~\textrm{mg},400~\textrm{mg})$ has value $\alpha_{500, 400} = (\beta_{500,400} - \beta_{500,200}) - (\beta_{250,400} - \beta_{250, 200})$. These will either capture a main effect of increasing a dosage (as on the sides of the Hasse diagram) or an interaction effect between multiple dosage increases (as in the interior of the Hasse diagram). 

Of course, in this particular parameterization of $\bbeta$, we chose to climb \textit{up} the Hasse. We could have alternatively chosen to climb \textit{down} the Hasse as
\begin{align}
    \y &= \mathbf{G} \bgamma + \epsilon, \\
    \beta_k &= \sum_{k^{\prime} \geq k; \rho(k) = \rho(k^{\prime})} \gamma_{k^{\prime}}, \label{eq:tva-climbing-down-hasse}
\end{align}
where $G_{i\ell} := \mathbb{I} \left\{ k\left(i\right)\leq \ell \ \cap\ \rho \left(k\left(i\right)\right) = \rho\left(\ell\right)\right\}$.

When the goal of the problem is to identify heterogeneity in $\bbeta$, there is no reason to prefer one parameterization of climbing the Hasse diagram over the other.\citet{banerjee2021selecting} used the climbing up parameterization with $\balpha$, which is a consequential choice. Here we take the view that this choice should not matter. Allowing for both parameterizations is coherent and not a contradiction -- if the researcher is interested in the marginals, they can fix their interpretation to a single parameterization. Being agnostic to the direction of Hasse diagram traversal has very important practical implications for computational feasibility, which we explore in Section \ref{section:rps-size}.

\begin{definition}[Generating a partition]
A parameter vector $\bbeta$ is said to be able to generate a partition $\Pi_0$ if it can induce exactly the same grouping of feature combinations into pools, $\pi_1, ..., \pi_n$, as $\Pi_0$.
\end{definition}

From above definition, the same $\bbeta$ can generate multiple partitions. To avoid ambiguity, we generally fix the partition $\Pi_0$ and look for compatible $\bbeta$. We now give the formal definition of permissibility using $\bbeta$ and a proof of equivalence.

\begin{definition}[Permissible partitions]
\label{def:permissible-profile}
A partition $\Pi_0$ is \emph{permissible} if
\begin{enumerate}[label=(\arabic*)]
    \item every $\pi \in \Pi_0$ is a \emph{pool} (cf. \Cref{def:pool}), and
    \item for every $\bbeta$ that generates $\Pi_0$, with respect to the Lebesgue measure:
    \begin{enumerate}
        \item the support of $\balpha(\bbeta)$, $S_{\balpha(\bbeta)} = \{\alpha_k \neq 0 \mid \alpha_k \in \balpha(\bbeta)\}$, is measurable, and
        \item the support of $\bgamma(\bbeta)$, $S_{\bgamma(\bbeta)} = \{\gamma_k \neq 0 \mid \gamma_k \in \bgamma(\bbeta)\}$, is measurable.
    \end{enumerate}

\end{enumerate}
We denote the set of all permissible partitions by $\mc{P}^{\star}$.
\end{definition}

\begin{lemma}
\label{lemma:topological-measurable-equivalent}
Definitions \ref{def:permissible-profile} and \ref{def:permissibility-profile-measure-zero} are equivalent to each other.
\end{lemma}

\begin{proofof}[Lemma \ref{lemma:topological-measurable-equivalent}]

Condition (1) of Definition \ref{def:permissible-profile} and Definition \ref{def:permissibility-profile-measure-zero} are identical. So we will focus only on equivalence between Condition (2) of \Cref{def:permissible-profile} and Conditions (2), (3) of \Cref{def:permissibility-profile-measure-zero}.

First, let's introduce some working terminology. For any two features $k \leq k^{\prime}$, define aggregated upward distance in outcome as, $u(k, k^{\prime}) = \sum_{k < k^{\prime \prime} \leq k^{\prime}} \alpha_{k^{\prime \prime}}$: the sum of marginals along all monotone paths on the Hasse from $k$ to $k^{\prime}$. Similarly, define aggregated downward distance in outcome as, $d(k, k^{\prime}) = \sum_{k \leq k^{\prime \prime} < k^{\prime}} \gamma_{k^{\prime \prime}}$: the sum of marginals along all monotone paths on the Hasse from $k^{\prime}$ to $k$.

For any pool $\pi \in \Pi_0$, $k < k^{\prime} \in \pi$ if and only if the aggregated monotone distance, $\abs{u(k, k^{\prime})} + \abs{d(k, k^{\prime})} = 0$ by construction.

\textbf{Necessity:} Assume that every $\bbeta$ that generates partition $\Pi_0$ satisfies \Cref{def:permissible-profile}.

\begin{enumerate}
\item \textbf{Parallel splits:} Let feature $u$ lie on the boundary of $\pi$.
For every $k \in \pi$ with $k \leq u$ we must have $\abs{u(k, u)} + \abs{d(k, u)} = 0$ (and symmetrically for $k \geq u$).
Suppose there is some $v \in \pi_2$ that is adjacent to $u$. (If there is no such $u, v$, then $\pi$ is the only pool in $\Pi$ and we are trivially done.) Crossing from $u$ to $v$ corresponds to changing exactly one feature by one level.

Suppose, for contradiction, that Condition (3) of \Cref{def:permissibility-profile-measure-zero} is not satisfied at this boundary. Without loss of generality, assume the contradiction is from (3a). That is, $\min \pi \incomp \min \pi_2$ and there is no distinct $\pi^{\prime}$ with $\min \pi^{\prime}$ being the coordinate-wise maximum of $\min \pi$ and $\min \pi_2$ (if they are not unique, we just pick the minimum of all of them). This means that the coordinate-wise maximum should be either in $\pi$ or $\pi_2$. Without loss of generality, assume that this coordinate-wise maximum is $m^{(2)} \in \pi$.

Let us also define the coordinate-wise minimum to be $m^{(1)} \in \pi_3$ where $\pi_3$ is distinct. By definition, $\abs{u(m^{(1)}, \min \pi)} + \abs{d(m^{(1)}, \min \pi)} \neq 0$ and $\abs{u(m^{(1)}, \min \pi_1)} + \abs{d(m^{(1)}, \min \pi_1)} \neq 0$. Since $m^{(2)} \in \pi$, $\abs{u(\min \pi, m^{(2)})} + \abs{d(\min \pi, m^{(2)})} = 0$. Further, $\abs{u(\min \pi_1, \max \pi_1)} + \abs{d(\min \pi_1, \max \pi_1)} = 0$. But $\abs{u(\max \pi_1, m^{(2)})} + \abs{d(\max \pi_1, m^{(2)})} \neq 0$

However, $\abs{u(m^{(1)}, m^{(2)})} + \abs{d(m^{(1)}, m^{(2)})} \neq 0$. And the monotone paths from $m^{(1)}$ to $m^{(2)}$ comprise of all monotone paths from $m^{(1)} \leftrightarrow \min \pi \leftrightarrow m^{(2)}$ and $m^{(1)} \leftrightarrow \min \pi_1 \leftrightarrow \max \pi_1 \leftrightarrow m^{(2)}$. Therefore, the aggregated monotone distance between $m^{(1)}$ to $\min \pi_1$ and the aggregated monotone distance from $\max \pi_1$ to $m^{(2)}$ should \textit{exactly} equal the aggregated monotone distance from $m^{(1)}$ to $\min \pi$ (all other distances are zero). And these correspond to non-zero $\balpha$ or $\bgamma$ i.e., $S_{\balpha(\bbeta)}$ and $S_{\bgamma(\bbeta)}$. The is a measure-zero event i.e., any small perturbation $\balpha + \epsilon$ (or $\bgamma + \epsilon$) can upset the exact balance and change the underlying partition. This leads to a contradiction the $\bbeta$ that generates partition $\Pi_0$ satisfies \Cref{def:permissible-profile}.

\item \textbf{Closed intervals:} Let $\pi$ be a pool. To show that $\pi$ is a closed interval, we need to show connectivity i.e., that all shortest paths between $u, v \in \pi$ is contained entirely in $\pi$, and that the endpoints of $\pi$ are unique.

First, we will show connectivity. Let $u, v \in \pi$. Consider any shortest path between $u$ and $v$ in the Hasse diagram. For contradiction, assume that this path exits $\pi$. If no such path exists, then connectivity is already satisfied. So suppose not. Then the path must cross the boundary of $\pi$ at least twice: once to leave and once to re-enter. Both boundary crossings correspond to changes in the same feature and level. Any path that crosses such a boundary twice must traverse that feature in opposite directions. This strictly increases the length of the path, contradicting the assumption that the path is shortest.

Next, let us turn to unique extrema. For sake of contraction, assume more than one minima exists. Call two of them $m, m^{\prime}$. Then, it must be that they are incomparable i.e., there are two features $i$ and $j$ such that $m_i > m^{\prime}_i$ and $m_j < m^{\prime}_j$. This leads to non-parallel splits contradicting our earlier conclusion. The proof extends without loss of generality to maxima as well.

\end{enumerate}

\textbf{Sufficiency:} Assume that $\Pi_0$ satisfies \Cref{def:permissibility-profile-measure-zero}.

Since all splits are parallel, boundaries between pools correspond to consistent changes in a single feature and direction. Since each pool is a closed interval, every feature combination within a pool can be connected to any other via shortest paths that remain entirely within the pool.

We construct marginal effects feature by feature so that crossing any split changes the aggregated monotone distance, while movements within a pool do not. Under this construction, aggregation over all monotone paths yields zero distance within pools and nonzero differences across pools.

Thus, the constructed marginal effects induce exactly the given partition. And any perturbation to the non-zero marginals will still robustly retain the partition structure since it does not rely on exact cancellations. Therefore the support of the non-zero marginals is measurable.

\end{proofof}

When we wish to learn heterogeneity in $\bbeta$, there is no reason to prefer one parameterization of climbing the Hasse over the other. Seeing that the parallel splitting criteria is linked to robustly estimating the pools of heterogeneity, we want to obey both of them together at the same time. This does run the risk of generating more granular partitions as a result of stronger restrictions, but this is a small price to pay for robustly estimating heterogeneity when one wishes to be agnostic about the system. Hence the full criterion for permissibility \Cref{def:permissibility-profile-measure-zero}, respecting parallel splits from both above (condition \ref{profile-permissibility:3a}) and below (condition \ref{profile-permissibility:3b}).
A by-product of being agnostic to the direction of Hasse traversal is that there is a bijective mapping between the $\bSigma$ partition matrices and permissible partitions. We show in \Cref{prop:num-permissible-poolings} that this significantly reduces the size of the model class, and later show in \Cref{thm:rashomon-set-polynomial-size} that the size of the RPS, which is our primary estimation goal, is only polynomial. %

We discuss specific examples of using the $\bSigma$ matrix to represent partitions in Examples \ref{example:hasse-1}, \ref{example:hasse-2}, \ref{example:hasse-3} below. %

\begin{figure}
\centering
\begin{subfigure}[t]{0.4\textwidth}
    \begin{centering}
     \begin{center}
     \begin{tikzpicture}[scale = 0.5] \def \n {5} \def \radius {2cm} \def \margin {8}
     \node[ minimum size=19pt] at (0,-4) (v1){\footnotesize $[1,1]$};
     \node[ minimum size=19pt] at (-2,-2) (v2){\footnotesize $[1,2]$};
     \node[  minimum size=19pt] at (-4,0) (v3){\footnotesize $[1,3]$};
     \node[  minimum size=19pt] at (2,-2) (v4){\footnotesize $[2,1]$};
     \node[  minimum size=19pt] at (4,0) (v5){\footnotesize $[3,1]$};
     \node[  minimum size=19pt] at (0,0) (v6){\footnotesize $[2,2]$};
     \node[  minimum size=19pt] at (-2,2) (v7){\footnotesize $[2,3]$};
      \node[  minimum size=19pt] at (2,2) (v8){\footnotesize $[3,2]$};
       \node[ minimum size=19pt] at (0,4) (v9){\footnotesize $[3,3]$};

      \draw[line width = 0.3mm, rotate around={45:(2,0)}, orange] (2,0) ellipse (3cm and 3cm);
      \draw[line width = 0.3mm, rotate around={45:(-1,-3)}, orange] (-1,-3) ellipse (1.5cm and 3cm);
      \draw[line width = 0.3mm, rotate =0, orange] (-4,0) ellipse (1.2cm and 1.2cm);
      \draw[line width = 0.3mm, rotate around={135:(-1,3)}, orange] (-1,3) ellipse (1.5cm and 3cm);
     
     \draw[line width = 0.3mm, >=latex] (v1) to (v2);
     \draw[line width = 0.3mm,  >=latex] (v1) to (v4);
     \draw[line width = 0.3mm,  >=latex] (v2) to (v3);
     \draw[line width = 0.3mm,  >=latex] (v2) to (v6);
     \draw[line width = 0.3mm,  >=latex] (v4) to (v6);
     \draw[line width = 0.3mm, >=latex] (v2) to (v6);
     \draw[line width = 0.3mm, >=latex] (v3) to (v7);
     \draw[line width = 0.3mm, >=latex] (v4) to (v5);
       \draw[line width = 0.3mm, >=latex] (v6) to (v7);
      \draw[line width = 0.3mm, >=latex] (v6) to (v8);
      \draw[line width = 0.3mm, >=latex] (v5) to (v8);
     \draw[line width = 0.3mm, >=latex] (v7) to (v9); 
      \draw[line width = 0.3mm, >=latex] (v8) to (v9);
    \end{tikzpicture}
     \end{center} 
    \par
    \end{centering}
\end{subfigure}%
~
\begin{subfigure}[t]{0.4\textwidth}
    \begin{centering}
     \begin{center}
     \begin{tikzpicture}[scale = 0.5] \def \n {5} \def \radius {2cm} \def \margin {8}
     \node[ minimum size=19pt] at (0,-4) (v1){\footnotesize $[1,1]$};
     \node[ minimum size=19pt] at (-2,-2) (v2){\footnotesize $[1,2]$};
     \node[ minimum size=19pt] at (-4,0) (v3){\footnotesize $[1,3]$};
     \node[ minimum size=19pt] at (2,-2) (v4){\footnotesize $[2,1]$};
     \node[ minimum size=19pt] at (4,0) (v5){\footnotesize $[3,1]$};
     \node[ minimum size=19pt] at (0,0) (v6){\footnotesize $[2,2]$};
     \node[ minimum size=19pt] at (-2,2) (v7){\footnotesize $[2,3]$};
      \node[ minimum size=19pt] at (2,2) (v8){\footnotesize $[3,2]$};
       \node[ minimum size=19pt] at (0,4) (v9){\footnotesize $[3,3]$};

      \draw[line width = 0.3mm, rotate around={45:(-1,-1)}, orange] (-1,-1) ellipse (3cm and 5cm);
      \draw[line width = 0.3mm, rotate around={45:(2,2)}, orange] (2,2) ellipse (1.5cm and 4.5cm);
     
     \draw[line width = 0.3mm, >=latex] (v1) to (v2);
     \draw[line width = 0.3mm,  >=latex] (v1) to (v4);
     \draw[line width = 0.3mm,  >=latex] (v2) to (v3);
     \draw[line width = 0.3mm,  >=latex] (v2) to (v6);
     \draw[line width = 0.3mm,  >=latex] (v4) to (v6);
     \draw[line width = 0.3mm, >=latex] (v2) to (v6);
     \draw[line width = 0.3mm, >=latex] (v3) to (v7);
     \draw[line width = 0.3mm, >=latex] (v4) to (v5);
       \draw[line width = 0.3mm, >=latex] (v6) to (v7);
      \draw[line width = 0.3mm, >=latex] (v6) to (v8);
      \draw[line width = 0.3mm, >=latex] (v5) to (v8);
     \draw[line width = 0.3mm, >=latex] (v7) to (v9); 
      \draw[line width = 0.3mm, >=latex] (v8) to (v9);
    \end{tikzpicture}
     \end{center} 
    \par
    \end{centering}
\end{subfigure}%

\caption{Hasse diagram for Examples \ref{example:hasse-1} and \ref{example:hasse-2}. The partition described in Example \ref{example:hasse-1} is shown in orange ellipses on the left panel. The right panel describes a different permissible partition in the right panel as in Example \ref{example:hasse-2}}
\label{fig:hasse-example}
\end{figure}

\begin{example}
\label{example:hasse-1}
Consider an example with $M=2$ features, each with $R = 3$ discrete values, $\{1, 2, 3\}$. Then there are $K=R^M = 9$ different feature combinations. The Hasse diagram is shown in Figure \ref{fig:hasse-example}. So, we end up pooling $(2, 2)$ with $(3, 2)$ and $(2, 3)$ with $(3, 3)$. The corresponding $\bSigma \in \{0, 1\}^{2 \times 2}$ matrix for this profile is
\begin{align*}
    \bSigma &= \begin{bmatrix}
       0 & 1 \\
       1 & 0
    \end{bmatrix}.
\end{align*}
This indicates that we split variants with value 1 from value 2 in the first feature (by $\Sigma_{11} = 0$) and pool variants of value 2 with value 3 in the first feature (by $\Sigma_{12} = 1$). Further, we pool variants with value 1 and value 2 in the second feature (by $\Sigma_{21} = 1$) and split variants with value 2 from value 3 in the second features (by $\Sigma_{22} = 0$).

\end{example}

\begin{example}
\label{example:hasse-2}
Consider the same setup in Example \ref{example:hasse-1} with $M=2$ features, each with $R = 3$ discrete values, $\{1, 2, 3\}$. Another permissible partition can be defined by the matrix
\begin{align*}
    \bSigma &= \begin{bmatrix}
       1 & 0 \\
       1 & 1
    \end{bmatrix}.
\end{align*}
The pools are $\pi_1 = \{(a, b) \mid a = \{1, 2\}, b = \{1, 2, 3\}\}$ and $\pi_2 = \{(a, b) \mid a = \{3\}, b = \{1, 2, 3\}\}$. This is illustrated in the right panel of Figure \ref{fig:hasse-example}.
\end{example}

\begin{figure}[!tb]
\begin{centering}
 \begin{center}
 \vfill
    \begin{tikzpicture}[scale = 0.65] \def \n {5} \def \radius {2cm} \def \margin {8}
        \node[  minimum size=19pt] at (0,-4) (v1){\footnotesize $[1,1]$};
        \node[ minimum size=19pt] at (-2,-2) (v2){\footnotesize $[1,2]$};
        \node[  minimum size=19pt] at (-4,0) (v3){\footnotesize $[1,3]$};
        \node[  minimum size=19pt] at (2,-2) (v4){\footnotesize $[2,1]$};
        \node[  minimum size=19pt] at (4,0) (v5){\footnotesize $[3,1]$};
        \node[  minimum size=19pt] at (0,0) (v6){\footnotesize $[2,2]$};
        \node[  minimum size=19pt] at (-2,2) (v7){\footnotesize $[2,3]$};
        \node[  minimum size=19pt] at (2,2) (v8){\footnotesize $[3,2]$};
        \node[  minimum size=19pt] at (0,4) (v9){\footnotesize $[3,3]$};
        \node[  minimum size=19pt] at (6,2) (v10){\footnotesize $[4,1]$};
        \node[  minimum size=19pt] at (4,4) (v11){\footnotesize $[4,2]$};
        \node[  minimum size=19pt] at (2,6) (v12){\footnotesize $[4,3]$};
        \node[  minimum size=19pt] at (8,4) (v13){\footnotesize $[5,1]$};
        \node[  minimum size=19pt] at (6,6) (v14){\footnotesize $[5,2]$};
        \node[  minimum size=19pt] at (4,8) (v15){\footnotesize $[5,3]$};

        \draw[line width = 0.3mm, rotate around={45:(4,2)}, orange] (4,2) ellipse (3cm and 3cm);
        \draw[line width = 0.3mm, rotate around={45:(0,-2)}, orange] (0,-2) ellipse (3cm and 3cm);
        \draw[line width = 0.3mm, rotate around={45:(7,5)}, orange] (7,5) ellipse (1.5cm and 3cm);
        \draw[line width = 0.3mm, rotate around={135:(-3,1)}, orange] (-3,1) ellipse (1.5cm and 3cm);
        \draw[line width = 0.3mm, rotate around={135:(1,5)}, orange] (1,5) ellipse (1.5cm and 3cm);
        \draw[line width = 0.3mm, rotate =0, orange] (4,8) ellipse (1.2cm and 1.2cm);
        
        \draw[line width = 0.3mm, >=latex] (v1) to (v2);
        \draw[line width = 0.3mm,  >=latex] (v1) to (v4);
        \draw[line width = 0.3mm,  >=latex] (v2) to (v3);
        \draw[line width = 0.3mm,  >=latex] (v2) to (v6);
        \draw[line width = 0.3mm,  >=latex] (v4) to (v6);
        \draw[line width = 0.3mm, >=latex] (v2) to (v6);
        \draw[line width = 0.3mm, >=latex] (v3) to (v7);
        \draw[line width = 0.3mm, >=latex] (v4) to (v5);
        \draw[line width = 0.3mm, >=latex] (v6) to (v7);
        \draw[line width = 0.3mm, >=latex] (v6) to (v8);
        \draw[line width = 0.3mm, >=latex] (v5) to (v8);
        \draw[line width = 0.3mm, >=latex] (v7) to (v9); 
        \draw[line width = 0.3mm, >=latex] (v8) to (v9);
        \draw[line width = 0.3mm, >=latex] (v9) to (v12);
        \draw[line width = 0.3mm, >=latex] (v8) to (v11); 
        \draw[line width = 0.3mm, >=latex] (v5) to (v10);
        \draw[line width = 0.3mm, >=latex] (v12) to (v15);
        \draw[line width = 0.3mm, >=latex] (v11) to (v14); 
        \draw[line width = 0.3mm, >=latex] (v10) to (v13);
        \draw[line width = 0.3mm, >=latex] (v10) to (v11); 
        \draw[line width = 0.3mm, >=latex] (v11) to (v12);
        \draw[line width = 0.3mm, >=latex] (v13) to (v14); 
        \draw[line width = 0.3mm, >=latex] (v14) to (v15);
    \end{tikzpicture}
 \vfill
 \end{center} 
\par\end{centering}
\caption{Hasse diagram for Example \ref{example:hasse-3}. The admissible partition is shown in orange ellipses.}
\label{fig:hasse-example-2}
\end{figure}

\begin{example}
\label{example:hasse-3}
Consider a different setup with $M=2$ features, The first feature takes on $R_1 = 5$ discrete values $\{1, 2, 3, 4, 5\}$ and the second feature takes on $R_2 = 3$ discrete values, $\{1, 2, 3\}$. An permissible partition can be defined by the matrix

\begin{align*}
    \bSigma &= \begin{bmatrix}
       1 & 0 & 1 & 0 \\
       1 & 0 & - & -
    \end{bmatrix},
\end{align*}
where we use ``$-$'' to denote that the second feature does not have dosages corresponding to those entries in the $\bSigma$ matrix.
The pools are $\pi_1 = \{(a, b) \mid a, b \leq 2\}$, $\pi_2 = \{(a, b) \mid a \leq 2, b = 3\}$, $\pi_3 = \{(a, b) \mid a = 3, 4, b \leq 2\}$, $\pi_4 = \{(a, b) \mid a = 3, 4, b = 3\}$, $\pi_5 = \{(a, b) \mid a = 5, b \leq 2\}$, and $\pi_6 = \{(5, 3)\}$. This is illustrated in Figure \ref{fig:hasse-example-2}.
\end{example}

One can quickly verify that Examples \ref{example:hasse-1} - \ref{example:hasse-3} satisfy permissibility as defined in \Cref{def:permissibility-profile-measure-zero} by visual inspection and identifying the corresponding $\bSigma$ matrices. In Example \ref{example:hasse-not-permissible} below, we show an example of a partition that is not permissible. Interestingly, there is a valid decision tree that arrives at this partition.

\begin{figure}[!tb]
\centering

\begin{subfigure}[t]{0.4\textwidth}
    \begin{centering}
     \begin{center}
     \begin{tikzpicture}[scale = 0.65] \def \n {5} \def \radius {2cm} \def \margin {8}
     \node[  minimum size=19pt] at (0,-4) (v1){\footnotesize $[1,1]$};
     \node[ minimum size=19pt] at (-2,-2) (v2){\footnotesize $[1,2]$};
     \node[  minimum size=19pt] at (-4,0) (v3){\footnotesize $[1,3]$};
     \node[  minimum size=19pt] at (2,-2) (v4){\footnotesize $[2,1]$};
     \node[  minimum size=19pt] at (4,0) (v5){\footnotesize $[3,1]$};
     \node[  minimum size=19pt] at (0,0) (v6){\footnotesize $[2,2]$};
     \node[  minimum size=19pt] at (-2,2) (v7){\footnotesize $[2,3]$};
      \node[  minimum size=19pt] at (2,2) (v8){\footnotesize $[3,2]$};
       \node[  minimum size=19pt] at (0,4) (v9){\footnotesize $[3,3]$};

      \draw[line width = 0.3mm, rotate around={45:(1,-1)}, blue] (1,-1) ellipse (1.5cm and 3cm);
      \draw[line width = 0.3mm, rotate around={45:(3,1)}, blue] (3,1) ellipse (1.5cm and 3cm);
      \draw[line width = 0.3mm, rotate around={45:(-2,-2)}, blue] (-2,-2) ellipse (1.5cm and 4.5cm);
      \draw[line width = 0.3mm, rotate around={135:(-1,3)}, blue] (-1,3) ellipse (1.5cm and 3cm);
     
     \draw[line width = 0.3mm, >=latex] (v1) to (v2);
     \draw[line width = 0.3mm,  >=latex] (v1) to (v4);
     \draw[line width = 0.3mm,  >=latex] (v2) to (v3);
     \draw[line width = 0.3mm,  >=latex] (v2) to (v6);
     \draw[line width = 0.3mm,  >=latex] (v4) to (v6);
     \draw[line width = 0.3mm, >=latex] (v2) to (v6);
     \draw[line width = 0.3mm, >=latex] (v3) to (v7);
     \draw[line width = 0.3mm, >=latex] (v4) to (v5);
       \draw[line width = 0.3mm, >=latex] (v6) to (v7);
      \draw[line width = 0.3mm, >=latex] (v6) to (v8);
      \draw[line width = 0.3mm, >=latex] (v5) to (v8);
     \draw[line width = 0.3mm, >=latex] (v7) to (v9); 
      \draw[line width = 0.3mm, >=latex] (v8) to (v9);
    \end{tikzpicture}
     \end{center} 
    \par
    \end{centering}
\end{subfigure}
 ~ 
\begin{subfigure}[t]{0.4\textwidth}
    \begin{centering}
     \begin{center}
     \begin{tikzpicture}[->,>=latex,shorten >=1pt,auto,node distance=0.8cm,scale=.8,transform shape,font = {\large\sffamily}]
          \tikzstyle{state}=[inner sep=2pt, minimum size=12pt]

          \node[state] (L0) at (-3,0) {};        
          \node[state] (L1) at (0,4) {$X_1 < 2$};
          \node[state] (L2) at (-2,2) {$\pi_1$};
          \node[state] (L3) at (2,2) {$X_2 > 2$};
          \node[state] (L4) at (0,0) {$\pi_4$};
          \node[state] (L5) at (4,0) {$X_1 > 2$};
          \node[state] (L6) at (2,-2) {$\pi_3$};
          \node[state] (L7) at (6,-2) {$\pi_2$};

        \draw[->] (L1) edge (L2);
        \draw[->] (L1) edge (L3);
        \draw[->] (L3) edge (L4);
        \draw[->] (L3) edge (L5);
        \draw[->] (L5) edge (L6);
        \draw[->] (L5) edge (L7);

    \end{tikzpicture}
     \end{center} 
    \par
    \end{centering}
\end{subfigure}

\caption{Hasse diagram with the partition that is not permissible described in Example \ref{example:hasse-not-permissible}. The pools are $\pi_1 = \{(1, 1), (1, 2), (1, 3)\}$, $\pi_2 = \{(2, 1), (2, 2) \}$, $\pi_3 = \{(3, 1), (3, 2) \}$, and $\pi_4 = \{(2, 3), (3, 3) \}$. The decision tree illustrates how to generate this partition.}
\label{fig:hasse-example-not-permissible}
\end{figure}

\begin{example}
\label{example:hasse-not-permissible}
Consider the same setup in Example \ref{example:hasse-1} with $M=2$ features, each with $R = 3$ discrete values, $\{1, 2, 3\}$. In Figure \ref{fig:hasse-example-not-permissible}, we illustrate a partition that is not permissible. This is not permissible because we have pools $\pi_1 = \{(1, 1), (1, 2), (1, 3)\}$, $\pi_2 = \{(2, 1), (2, 2) \}$, $\pi_3 = \{(3, 1), (3, 2) \}$, and $\pi_4 = \{(2, 3), (3, 3) \}$. Permissibility (see condition \ref{profile-permissibility:3} of Definition \ref{def:permissibility-profile-measure-zero}) says that if $\pi_1$ is in the partition, then feature combinations $(\cdot, 2)$ and $(\cdot, 3)$ should always be pooled together. This contradicts what we observe in $\pi_2$, $\pi_3$, and $\pi_4$. Similarly, if $\pi_4$ is in the partition, permissibility would require that feature combinations $(2, \cdot)$ and $(3, \cdot)$ need to be pooled together which contradicts $\pi_2$ and $\pi_4$. Since this partition is not permissible, we cannot represent it using the $\bSigma$ matrix.

To see why this is not permissible from the marginal perspective, let us look at $\pi_3$ and $\pi_4$. From these pools, it is evident that $\alpha_{2,3} \neq 0$, $\alpha_{3,1} \neq 0$, and $\alpha_{3,3} \neq 0$. And we know that,
\begin{align*}
    \beta_{3,3} &= \alpha_{3,3} + \alpha_{3,2} + \alpha_{3,1} + \alpha_{2,3} + C \\
    \beta_{2,3} &= \alpha_{2,3} + C,
\end{align*}
where the term $C$ is common to both equations. In the pooling currently, it so happens that $\beta_{3,3} = \beta_{2,3}$ -- the terms $\alpha_{3,3}, \alpha_{3,2}, \alpha_{3,1}$ jointly make this true by $\alpha_{3,3} + \alpha_{3,2} + \alpha_{3,1} = 0$. However, we know that $\alpha_{3,1} \neq 0$. So if we add some noise $\varepsilon > 0$ to $\alpha_{3,1}$ to get $\alpha^{\prime}_{3,1} := \alpha_{3,1} + \varepsilon$. Then, $\beta_{3,3} \neq \beta_{2,3}$ anymore. In other words, the pool $\pi_4$ is not robust to noise in the non-zero marginals as any noise will almost surely break $\pi_4$ into $\{(2,3)\}$ and $\{(3, 3)\}$ as separate pools. Hence, this partition is not permissible.

Decision trees are not robust in this sense as they may generate partitions that are not permissible. The right panel of Figure \ref{fig:hasse-example-not-permissible} illustrates a decision tree that generates the partition that is not permissible discussed in this example.
\end{example}

\subsection{Pooling across profiles.}

So far, we have been describing how to pool different feature combinations if they belong to the same profile. Now, we turn our attention to pooling \textit{across} profiles. Definition \ref{def:permissibility-profile-measure-zero} captures permissibility within a single profile, but we also want to consider pooling across profiles. For example, Definition \ref{def:permissibility-profile-measure-zero} does not speak to the question of pooling decisions for adding Ibuprofen, as a temporary pain reliever, to a prescription of Amoxicillin against a bacterial infection. Does introducing Ibuprofen make an appreciable difference (offering the patient relief while waiting for the bacterial infection to work) or not (because the antibiotic itself offers pain relief by attacking the root cause)?

In order to reason about this, we consider partially ordering of the profiles themselves using their binary representation. This also allows us to embed the profiles in an $M$-d unit hypercube with profiles as the vertices. By the same intuition behind closed intervals, we can pool two profiles if they are reachable on this hypercube. We can generalize the marginal re-parameterization to allow for marginal gains when moving between profiles through a $\bdelta$ vector,
\begin{align}
    \y &= \mathbf{F} \balpha + \mathbf{A} \bdelta + \epsilon, \label{eq:tva-regression-general} \\
    A_{i, (\ell, \rho)} &= \mathbb{I} \left\{ k (i) > \ell \ \cap\ \rho (k(i)) > \rho(\ell) = \rho \right\}, \nonumber \\
    \beta_k &= \sum_{k^{\prime} \leq k; \rho(k) = \rho(k^{\prime})} \alpha_{k^{\prime}} + \sum_{k^{\prime} < k} \sum_{\rho; \rho(k^{\prime}) < \rho(k)} \delta_{k^{\prime}, \rho}.
\end{align}
Here, observe that $\bdelta$ is indexed by the profile $\rho$ as well as the feature $k^{\prime}$. Being feature-specific gives the freedom to pool across profiles without imposing strong cross-profile restrictions that prevent measure zero events.

By setting $\bdelta = \mathbf{0}$, we can immediately see that Equation \eqref{eq:tva-regression-general} is a generalization of Equation \eqref{eq:tva-regression}. In fact, if depending on the context, we do not want to pool profile $\rho_1$ with $\rho_2$, then this corresponds to setting the appropriate entries in $\bdelta$ to 0. This is exactly what \cite{banerjee2021selecting} do in their analysis of cross-randomized behavioral nudges for improving immunization.

We formalize this in Definition \ref{def:permissibility-partition-measure-zero}.

\begin{definition}[Permissible partition]
\label{def:permissibility-partition-measure-zero}
A \emph{partition $\Pi$ of the entire feature space $\mathcal{K}$ is permissible} if:
\begin{enumerate}[label=(\arabic*)]
\item for every profile $\rho_0$, the partition induced by $\Pi$ on $\rho_0$, $\Pi_0 = \{ \pi \setminus \{k \mid \rho(k) \neq \rho_0 \} \mid \pi \in \Pi \}$ is permissible by \Cref{def:permissible-profile} (or \Cref{def:permissibility-profile-measure-zero}), and

\item for every $\bbeta$ that generates $\Pi$, with respect to the Lebesgue measure, the support of $\bdelta(\bbeta)$, $S_{\bdelta(\bbeta)} = \{\delta_{k, \rho} \neq 0 \mid \delta_{k, \rho} \in \bdelta(\bbeta)\}$, is measurable.

\end{enumerate}
\end{definition}

Specifically, by allowing to pool across different profiles, \Cref{def:permissibility-partition-measure-zero} naturally allows us to explore heterogeneity in treatment effects where treatment and control are two distinct profiles. We illustrate this in the empirical data analysis of microcredit access in \Cref{section:real-data}.

\Cref{def:permissible-partition} gives an equivalent geometric interpretation of \Cref{def:permissibility-partition-measure-zero} through the Hasse.

\begin{definition}[Permissible partition]
\label{def:permissible-partition}
A \emph{partition $\Pi$ of the entire feature space $\mathcal{K}$ is permissible} if and only if the following hold true:
\begin{enumerate}[label=(\arabic*)]
\item \label{permissibility:case1} for every profile $\rho_0$, the partition induced by $\Pi$ on $\rho_0$, $\Pi_0 = \{ \pi \setminus \{k \mid \rho(k) \neq \rho_0 \} \mid \pi \in \Pi \}$ is permissible by \Cref{def:permissibility-profile-measure-zero} (or \Cref{def:permissible-profile}),

\item \label{permissibility:case2} every $\pi \in \Pi$ is connected in feature levels across profiles i.e., if $k^{(1)}, k^{(2)} \in \pi$ such that $\rho_1 = \rho(k^{(1)})$ and $\rho_2 = \rho(k^{(2)})$ are adjacent on the hypercube, then there are feature combinations $k^{(3)}, k^{{(4)}} \in \pi$ such that $\rho(k^{(3)}) = \rho_1$, $\rho(k^{(4)}) = \rho_2$ and $\norm{k^{(3)} - k^{(4)}}_1 = 1$,\footnote{Along with \ref{permissibility:case1}, this means that we can reach $k^{(1)}$ from $k^{(2)}$ by traversing the Hasse for $\rho_1$ to $k^{(3)}$, then jumping to $k^{(4)}$ along an edge on the $M$-d hypercube, and then moving from $k^{(4)}$ to $k^{(2)}$ while respecting the Hasse for $\rho_2$.} and

\item \label{permissibility:case3} every $\pi \in \Pi$ is connected in profiles i.e., if $\pi$ contains feature combinations from profiles $\rho_0$ and $\rho_k$ where $\rho_0 < \rho_k$, then $\pi$ also contains features in profiles $\rho_1, \dots, \rho_{k-1}$ such that $\norm{\rho_i - \rho_{i+1}}_0 = 1$ for $i = 0, \dots, k-1$.\footnote{Along with \ref{permissibility:case1} and \ref{permissibility:case2}, this means that we can reach $\rho_k$ from $\rho_0$ by traversing the $M$-d hypercube while staying within $\pi$ and respecting the Hasse at each vertex of the hypercube.}

\end{enumerate}
\end{definition}

This representation agrees with our permissibility in \Cref{def:permissible-partition}. Case \ref{permissibility:case1} follows from the fact that this is a generalization of Equation \ref{eq:tva-regression}. Cases \ref{permissibility:case2} and \ref{permissibility:case3} follow from the definition of the $\mathbf{A}$ matrix. For example, consider two features $k^{(1)}, k^{(2)}$ that belong to two different profiles. We can only pool variants i.e., $||k^{(1)} - k^{(2)}||_1 = 1$. If they are variants, then the two profiles must be adjacent on the $M$-d hypercube.

At this point, it is important to note that there are no restrictions such as the parallel splitting criteria across different profiles. This is because the marginal $\delta_k$ only contributes to the outcome across profiles i.e., from the perspective within a profile, the sphere of influence of $\delta_k$ is indistinguishable from the sphere of influence of $\alpha_{k^{\prime}}$ where $k^{\prime}$ is at the lower boundary of the Hasse adjacent to the Hasse of $k$. Since each pair of feature variants from different profiles have different across-profile marginals $\delta_k$, they are not coupled together like the $\balpha$ marginals are.

Just like the $\bSigma$ matrix within each profile, we can also construct the intersection matrix $\bSigma^{\cap}$ to denote how features are pooled across two adjacent profiles. Consider partitions induced by $\Pi$ on two profiles $\rho_1$ and $\rho_2$. Let us call these $\Pi_1, \Pi_2$ respectively. $\bSigma^{\cap} = \{0, 1, \infty\}^{\abs{\Pi_1} \times \abs{\Pi_2}}$ where $\bSigma_{i, j}^{\cap} = 0$ means that pools $\pi_i \in \Pi_1$ and $\pi_j \in \Pi_2$ are poolable according to \ref{permissibility:case2} of \Cref{def:permissible-partition} but are not pooled together in $\Pi$. $\bSigma_{i, j}^{\cap} = 1$ means that these pools are poolable and are indeed pooled in $\Pi$. Finally, $\bSigma_{i, j}^{\cap} = \infty$ means that these pools are not poolable by \Cref{def:permissible-partition}. Observe that if $\bSigma^{\cap}_{i, j} = 1$, then $\bSigma^{\cap}_{i, -j} = \infty$ and $\bSigma^{\cap}_{-i, j} = \infty$ in order to respect \ref{permissibility:case1} of \Cref{def:permissible-partition}. This object is useful in our enumeration step in Algorithm \ref{alg:r-aggregate}.

\subsection{Examples of other permissibility restrictions}

Estimation strategies, in general, implicitly take some stand on partition structure. They impose permissibility restrictions, though they are not often presented formally as such.
These restrictions are generated by the choice of technique, rather than through any specific scientific consideration. In the following examples, we show how these techniques can be framed as permissibility restrictions by defining them as partitions on the Hasse. This involves identifying sets of equivalent edges for each technique that, when removed together, generate corresponding partitions learned by that technique.

\begin{example}[Long, Short, and Lasso regression]
First, take a saturated or ``long'' regression. Here, every possible feature combination is its own pool i.e., the partition is the most granular partition possible. For example, consider the treatment outcome as a function of dosages of two drugs, $A$, $B$, and weight of the treated individual, $W$. Suppose that each variable takes on three discrete levels. Then, $\Pi^{\text{long}} = \{(a,b,w): \ \text{ for every } a,b,w \in \{0,1,2\}^2\times\{\text{low, med, high}\}\}$ with 27 elements.

Second, consider a ``short'' regression where the researcher does not include all relevant variables in the regression. Then, the partition generated is identical to long regression for variables included in the model i.e., it pools across all excluded variables. For example, if the researcher ignores weight, then $\Pi^{\text{short}} = \{(a,b,:): \ \text{ for every } a,b \in \{0,1,2\}^2\}$ which pools across weight, with 9 elements.

Third, say the researcher uses Lasso to regularize the data to set marginal dosage or weight increase effects to 0, generating pools. Then the $\Pi^{\text{Lasso}}$ is bijectively determined by the support of the Lasso: the zero'ed elements generate the pooling structure.\footnote{In a frequentist perspective, under ``beta-min'' classical assumptions and irrepresentability, this will correctly identify the true generative partition with probability tending to one (as seen in~\citet{banerjee2021selecting}).}
\end{example}

\begin{example}[Decision Trees]
This is perhaps the most common approach used beyond imposed short and long regressions. In a decision tree, at every node, the statistician chooses whether or not to split based on the value of a given arm (e.g., Amoxicillin greater than 250mg). Conditional on this split, following a given decision, another variable is selected, and the process repeats recursively until termination (maybe defined by some maximum number of splits or until no more splits are possible). It is useful to note that this procedure generates parallel splits in the Hasse, conditional on previously made splits. Therefore, the leaves generated by decision trees are closed intervals.

It is useful to note that the decision trees are captured by the equivalent edge framing in a conditional setting. Initially, the set of equivalent edges for a decision tree is identical to those of the robust partitions we consider in this paper. However, the edges to split upon are chosen sequentially (rather than jointly, as in the case of robust partitions). Thus, after each split, the set of equivalent edges changes. Specifically, each set of equivalent edges could get decomposed into two smaller sets of equivalent edges upon splitting a different equivalence class (this means that there are more than $2^n$ possible partitions where there were $n$ equivalence classes originally). In other words, edges are equivalent conditional on previously made splits. The binary $\sigma$ vector we used to represent splitting of jointly equivalent edges can be generalized to a \emph{list} of indices that alternatingly indicate splitting and pooling along the edges. For example, we can order the edges in equivalence class $E_i$. Then, the pooling decisions for $E_i$ is represented by a list $\sigma_i$ such that all edges until $e_{\sigma_{i,1}}$ are pooled, all edges from $e_{\sigma_{i,1}}$ to $e_{\sigma_{i,2}}$ are split, all edges from $e_{\sigma_{i,2}}$ to $e_{\sigma_{i,3}}$ are pooled, and so on. This is essentially a tree data structure, i.e., the $\sigma$ data structure is a tree in the limit where we only have conditionally equivalent edges (but not jointly). The key point here is the perspective of encoding the partition by keeping track of splits (made in equivalence classes) is general, takes us to scientifically relevant settings beyond robust partitions and decision trees, and generates vast improvements in refinements as well.
However, the pooling restrictions imposed by trees suffer from a coherency issue, which we explore in our discussion of causal trees. We walk through a detailed example in \Cref{example:hasse-not-permissible}.
\end{example}

\begin{example}[Causal Trees and Causal Random Forests: conditionally convex splits]
Decision trees cannot natively estimate heterogeneities in treatment effects. Causal trees \citep{athey2016recursive} and causal random forests \citep{wager2018estimation} can do this natively by modifying the fit criteria used to make splits. The fit criteria for causal trees is the MSE of the treatment effect. So each leaf of the tree needs to contain both treatment and control observations to estimate the treatment effect. The partitions generated by causal trees are identical to decision trees if we ignore the treatment indicator as $(1, x)$ and $(0, x)$ are always in the same leaf or pool. Here, $x$ is the feature combination, and 1/0 is the binary treatment indicator variable.

This differs from our robust partitions because we allow for pools with $\{ (1, x_1), (1, x_2), (0, x_1)\}$ and $\{ (0, x_2) \}$ by the cross-Hasse pooling rules. On the other hand, causal trees will split them as $\{(1, x_1), (0, x_1)\}$ and $\{(1, x_2), (0, x_2)\}$ (or pool them together) even though $(0, x_1)$ and $(1, x_2)$ have the same outcome. This raises a conceptual issue. If $(1, x_2)$ and $(1, x_1)$ are equivalent to each other, and $(1, x_1)$ and $(0, x_1)$ are equivalent to each other, then by transitivity, $(1, x_2)$ and $(0, x_1)$ should be equivalent to each other as well. However, causal trees force them apart because $(0, x_1)$ and $(0, x_2)$ are not equivalent to each other. Such a pooling restriction appears incoherent. Besides this, since we allow for flexibility through the transitivity of equivalences, we enjoy statistical properties such as lower bias and variance.

Causal random forests are an aggregation over sampled partitions, $\{\Pi_b: \ b = 1,\ldots, B\}$, each of which is bijective with a partition created by causal trees.
\end{example}

\begin{figure}

    \begin{centering}
     \begin{center}
     \begin{tikzpicture}[scale =0.4]
         \node[] at (0,-3) (v1) {\small  $(1, 1)$};
         \node[] at (-3,0) (v2) {\small $(1, 2)$};;
         \node[] at (3,0) (v4) {\small $(2, 1)$};
         \node[] at (0,3) (v6) {\small $(2, 2)$};

        \draw [rounded corners=8mm, red, line width=0.3mm] (0, -5)--(-5.5, 1)--(5.5,1)--cycle;
        \draw[red, line width=0.3mm] (0,3) ellipse (1.5cm and 1.5cm);
         
        \draw[line width = 0.3mm, >=latex] (v1) to (v2);
        \draw[line width = 0.3mm,  >=latex] (v1) to (v4);
        \draw[line width = 0.3mm,  >=latex] (v2) to (v6);
        \draw[line width = 0.3mm,  >=latex] (v4) to (v6);
        \draw[line width = 0.3mm, >=latex] (v2) to (v6);
    \end{tikzpicture}
     \end{center} 
    \par
    \end{centering}

\caption{Hasse diagram admissible by TVA that is not composed of closed intervals (it contains the open interval $[(1,1),(2,2))$. Robust to measure zero cancellations in the upward direction. This cannot be represented by a decision tree.}
\label{fig:hasse-tva-not-tree}
\end{figure}

\begin{example}[Treatment Variant Aggregation of \cite{banerjee2021selecting}]
{Here, the authors consider the Hasse diagram and impose a structure, via marginal stepping through marginal effects, that is closely related to permissibility, with the difference that they require only an interpretation based on climbing \emph{up} the Hasse (as in equation \eqref{eq:tva-climbing-up-hasse}), but not the symmetric interpretation obtained by climbing \emph{down} (as in equation \eqref{eq:tva-climbing-down-hasse}). This weaker restriction may be scientifically adequate---or even desirable---in settings where effects are naturally understood as accumulating with intensity.
As a result, the admissible pools extend beyond closed intervals to a broader class of \emph{one-sided convex} sets: sets that are convex with respect to the partial order when viewed from below, but need not admit a symmetric description from above. These sets arise from equality of active lower marginals and therefore reflect an \emph{upward} construction of treatment effects. We show below, for example, that the half-open interval $[(1,1),(2,2))$ can arise as a TVA pool.
A parallel-splits condition is still imposed, but only in one direction (Condition~\ref{profile-permissibility:3a}, but not \ref{profile-permissibility:3b}), ensuring that splits propagate consistently when moving upward through the Hasse. This still retains robustness to measure-zero marginal cancellations in that direction (for example, ruling out the running example of Amoxicillin inducing a stomach upset that is exactly offset by the increment in dose of Ibuprofen).
 In this case, the equivalent-edges representation no longer applies directly.  Additionally, the $\ell_1$ prior over marginal effects used in \cite{banerjee2021selecting} requires that the penalty along a branch depends on the magnitudes of the coefficients and the global reweighting induced by the regularization path. As a result, it is difficult—if not impossible—to construct useful bounds on the minimum penalty associated with candidate splits. This effectively prevents the type of computational speed-ups that we enjoy in our branch-and-bound algorithm.}

Consider the $2 \times 2$ Hasse with $\{11, 12, 21, 22\}$ as the four feature combinations. TVA allows for the following partitions of the Hasse:
\begin{itemize}
    \item $\Pi_1 = \left\{ \{11, 12, 21\}, \{22\} \right\}$
    \item $\Pi_2 = \left\{ \{11\}, \{21\}, \{12, 22\} \right\}$
    \item $\Pi_3 = \left\{ \{11\}, \{12\}, \{21, 22\} \right\}$
    \item $\Pi_4 = \left\{ \{11, 12, 21, 22\} \right\}$
    \item $\Pi_5 = \left\{ \{11\}, \{12\}, \{21\}, \{22\} \right\}$
    \item $\Pi_6 = \left\{ \{11, 12\}, \{21, 22\} \right\}$
    \item $\Pi_7 = \left\{ \{11, 21\}, \{12, 22\} \right\}$
\end{itemize}

In this $2 \times 2$ Hasse, there are four edges $\{\langle 11, 12 \rangle$, $\langle 11, 21 \rangle$, $\langle 12, 22 \rangle$, $\langle 21, 22\rangle \}$. The partitions listed above illustrate that no edge is truly equivalent to another edge in the Hasse despite TVA imposing convexity restrictions. In other words, the set of equivalent edges appears to be degenerate i.e., each each equivalence class is a singleton. However, there is a well-defined structure. We are not free to arbitrarily split on edges. For example, $\Pi_8 = \left\{ \{11, 12, 22\}, \{21\} \right\}$ is not permissible as it violates {one-sided convexity}. This is why we say that the equivalent edges framing does not hold here. %

However, we can still generalize the $\sigma$ data structure to efficiently store this partition. This is because these partitions are ``parallel from below,'' i.e., if a feature combination $k^{(1)} = [r_1, \dots, r_i, \dots, r_m]$ and $k^{(2)} = [r_1, \dots, r_i + 1, \dots, r_m]$ are split, then all pairs of feature combinations $k^{(3)} = [s_1, \dots, r_i, \dots, s_m]$ and $k^{(4)} = [s_1, \dots, r_i + 1, \dots, s_m]$ where $k^{(3)} > k^{(1)}$ and $k^{(4)} > k^{(2)}$ are also split. Therefore, using the same set of equivalent edge decomposition as our robust partitions (described below), we allow $\sigma_i$ to denote a vector of largest levels in all features $j$, $\ell_j$, besides the $i$-th one such that $k^{(3)} = [\dots, \ell_j, \dots, r_i, \dots]$ and $k^{(4)} = [\dots, \ell_j, \dots, r_i + 1, \dots]$ are pooled.
\end{example}

\section{Exact marginal likelihood and generalized Bayesian inference}

\subsection{Generative model and exact marginal likelihood.}
\label{appendix:gprior-marginal}

Here, we briefly outline how to formally construct the full posterior using the Rashomon set. {Our goal is to estimate the marginalized posterior
\begin{align*}
    P_{\bbeta \mid \Z}(\bbeta)
    &= \sum_{\Pi \in \mc{P}^{\star}} P(\bbeta \mid \Z, \Pi)\, \Pr(\Pi \mid \Z),
\end{align*}
and then approximate this quantity using only partitions in the Rashomon set $\mc{P}_\theta$. To do this, we must derive both the conditional posterior of the effects $P(\bbeta \mid \Z, \Pi)$ and the marginal weights $\Pr(\Pi \mid \Z)$ under a fully specified generative model.}

We do this by constructing a specific data-generating process. Fix a partition $\Pi$. Define a transformation matrix $\bP \in \{0, 1\}^{K \times \abs{\Pi}}$, where $K$ is the number of possible feature combinations, that assigns each pool mean $\gamma_j$ to the feature combinations in pool $\pi_j$:
\begin{align*}
    P_{ij} &=
    \begin{cases}
        1, & \text{feature combination } i \in \pi_j, \\
        0, & \text{else.}
    \end{cases}
\end{align*}
The mean vector for the feature combinations is given by $\bbeta = \bP \bgamma$, where $\bgamma \in \mathbb{R}^{\abs{\Pi}}$. In particular, all feature combinations in a given pool share the same mean. 

{Given feature combinations $\D$, define the design matrix for the partition as $\bX_\Pi = \D \bP$. We assume a homoskedastic Gaussian working model, with $\sigma^2$ treated as fixed or estimated separately (e.g., plugged in from the residuals of a saturated regression) and held common across partitions:\footnote{{Marginalizing over $\sigma^2$ under a conjugate inverse-gamma hyperprior yields an analogous closed form featuring a multivariate-$t$ marginal, without altering the underlying geometric structure of the result.}}
\begin{align*}
    \y \mid \bgamma, \Pi, \sigma^2 \sim \mc{N}(\bX_\Pi \bgamma, \sigma^2 \mathbf{I}_n).
\end{align*}
Let $n_{\pi_j}$ denote the number of observed units whose feature combination belongs to pool $\pi_j$. Throughout this derivation, we assume $n_{\pi_j} > 0$ for every $\pi_j \in \Pi$, ensuring that $\bX_\Pi^{\intercal} \bX_\Pi$ is invertible.}

{To obtain an exact marginal likelihood with a transparent complexity penalty, we use an information-scaled $g$-prior on the pool means:
\begin{align*}
    \bgamma \mid \Pi, \sigma^2 \sim \mc{N}\bigl(\mathbf{0}, g \sigma^2 (\bX_\Pi^{\intercal} \bX_\Pi)^{-1}\bigr),
\end{align*}
where $g > 0$ is fixed across partitions.\footnote{{We assume $\y$ has been centered, so the prior shrinks pool means toward zero (equivalently, toward the grand mean of the data). A non-zero prior mean $\bgamma_0$ leaves the form of all expressions unchanged with $\mathrm{SSE}_\Pi$ replaced by $(\y - \bX_\Pi\bgamma_0)^{\intercal}(\mathbf{I}_n - \bH_\Pi)(\y - \bX_\Pi\bgamma_0)$.}}}

{Crucially, because each observation belongs to exactly one pool, the columns of $\bX_\Pi$ are disjoint indicators and $\bX_\Pi^{\intercal}\bX_\Pi = \mathrm{diag}(n_{\pi_1}, \ldots, n_{\pi_{|\Pi|}})$. The $g$-prior is therefore proper and pool-wise independent, with prior variance $g\sigma^2/n_{\pi_j}$ for each pool mean $\gamma_j$---i.e., prior precision scales with pool size, encoding the natural information-scaled belief that we should be more concentrated about pools we have more data for.}

{For a fixed partition, define the projection matrix and the ordinary least-squares estimates as:
\begin{align*}
    \bH_\Pi = \bX_\Pi(\bX_\Pi^{\intercal} \bX_\Pi)^{-1}\bX_\Pi^{\intercal}, \qquad \widehat{\bgamma}_\Pi = (\bX_\Pi^{\intercal} \bX_\Pi)^{-1}\bX_\Pi^{\intercal} \y.
\end{align*}
Standard Gaussian conjugacy gives the conditional posterior for the parameters:
\begin{align*}
    \bgamma \mid \Z, \Pi, \sigma^2 \sim \mc{N}\!\left( \frac{g}{1+g}\widehat{\bgamma}_\Pi, \, \frac{g}{1+g}\sigma^2(\bX_\Pi^{\intercal} \bX_\Pi)^{-1} \right),
\end{align*}
and $\bbeta \mid \Z, \Pi, \sigma^2$ follows immediately as the induced distribution of $\bbeta = \bP\bgamma$.}

{Integrating out $\bgamma$, the exact marginal likelihood is:
\begin{align*}
    p(\y \mid \D, \Pi, \sigma^2) &= (2\pi\sigma^2)^{-n/2} (1+g)^{-\abs{\Pi}/2} \exp\!\left\{ -\frac{1}{2\sigma^2} \y^{\intercal} \left(\mathbf{I}_n - \frac{g}{1+g}\bH_\Pi\right) \y \right\}.
\end{align*}
Equivalently, since the within-pool sum of squared errors is $\mathrm{SSE}_\Pi = \y^{\intercal}(\mathbf{I}_n - \bH_\Pi)\y$, we have, up to constants $C$ that do not depend on the partition structure $\Pi$:
\begin{align*}
    -\log p(\y \mid \D, \Pi, \sigma^2) = \frac{g}{2\sigma^2(1+g)} \mathrm{SSE}_\Pi + \frac{\abs{\Pi}}{2}\log(1+g) + C.
\end{align*}}

Exactly computing the normalizing constant over the full partition space is NP-hard and is reminiscent of the partition functions used in graphical models. See \citet{agrawal2021partition} for a survey of methods used to estimate or approximate such normalizing constants.

\subsection{Generalized Bayesian inference and algorithmic equivalence.}
\label{appendix:generalized-bayes}

{To search this space computationally, our algorithm utilizes a penalized empirical loss function. We take as our loss for a given partition $\Pi$ the mean squared error (MSE):}
\begin{align}
    \mathcal{L}(\Pi ; \Z)
    &= \frac{1}{n} \left( \y - \widehat{\y} \right)^{\intercal}
       \left( \y - \widehat{\y} \right), \nonumber \\
    \widehat{y}_i
    &= \frac{\displaystyle \sum_{\pi \in \Pi}  \mathbb{I}\{k(i) \in \pi\}
           \sum_{j} \mathbb{I}\{k(j) \in \pi\} y_j}
           {\displaystyle \sum_{\pi \in \Pi}  \mathbb{I}\{k(i) \in \pi\}
           \sum_{j} \mathbb{I}\{k(j) \in \pi\}}, \label{eq:loss-outcome-prediction}
\end{align}
where $\widehat{y}_i$ is the mean outcome in the pool $\pi \in \Pi$ containing the feature combination of unit $i$, $k(i)$. By definition, this empirical MSE is exactly $\mathrm{SSE}_\Pi / n$.

{Our goal here is to show that the algorithmic objective used in our enumeration is, up to a $\Pi$-independent constant, the negative log of the exact posterior $\Pr(\Pi \mid \Z)$ derived in \Cref{appendix:gprior-marginal}. As a consequence, the Rashomon set $\mc{P}_\theta$ is a level set of the exact finite-sample Bayesian posterior, not an approximation to one.}

{Combining the exact marginal likelihood above with an explicit $\ell_0$ prior on partition complexity, $\Pr(\Pi) \propto \exp\{-\lambda_0 \abs{\Pi}\}$, the negative log posterior score for the partition becomes, up to constants independent of $\Pi$:}
\begin{align*}
    {Q_B(\Pi) = \frac{g}{2\sigma^2(1+g)} \mathrm{SSE}_\Pi + \left\{\lambda_0 + \frac{1}{2}\log(1+g)\right\} \abs{\Pi}.}
\end{align*}

{Substituting our algorithmic loss $\mathcal{L}(\Pi; \Z) = \mathrm{SSE}_\Pi / n$, this trivially rearranges to:}
\begin{align*}
    {Q_B(\Pi) = \eta \mathcal{L}(\Pi; \Z) + \lambda \abs{\Pi},}
\end{align*}
{where the learning rate is $\eta = n g / \{2\sigma^2(1+g)\}$ and the complexity penalty is $\lambda = \lambda_0 + \frac{1}{2}\log(1+g)$.}

{Because minimization is invariant to positive rescaling, the equivalent algorithmic objective $Q'(\Pi) = \mathcal{L}(\Pi; \Z) + (\lambda/\eta)\abs{\Pi}$ used in our enumeration has an effective $\ell_0$ penalty of $\lambda_{\mathrm{eff}} = \lambda_B/\eta$, and the Rashomon threshold $\theta$ rescales accordingly. Thus, the minimum of the loss-plus-$\ell_0$-penalty objective shares the exact same ordering as the negative log posterior. However, for computing correct posterior weights over the set, the learning rate $\eta$ must be retained:}
\begin{align*}
    {\Pr(\Pi \mid \Z) \propto \exp\{-\eta\mathcal{L}(\Pi;\Z) - \lambda\abs{\Pi}\}.}
\end{align*}

{Setting $g=n$ gives the standard unit-information prior calibration, for which:}
\begin{align*}
    {\frac{\abs{\Pi}}{2}\log(1+g) = \frac{\abs{\Pi}}{2}\log(n+1) \sim \frac{\abs{\Pi}}{2}\log n.}
\end{align*}
{Hence the marginal likelihood contains the usual BIC-style Occam factor asymptotically, with any additional preference for parsimony encoded through the explicit prior $\lambda_0\abs{\Pi}$.}

{Furthermore, this connects our computational construction directly to loss-based generalized Bayesian inference \citep{bissiri2016general}. The pseudo-posterior $P(\Pi \mid \Z) \propto \exp\{-\eta \mathcal{L}(\Pi; \Z) - \lambda \abs{\Pi}\}$ is therefore not an ad-hoc computational heuristic: for the calibration of $\eta$ and $\lambda$ given above, it coincides exactly with the marginal posterior under the conjugate Gaussian model with a $g$-prior on pool means.} See Section~4 of \citet{chipman2010bart} for examples of such priors used for BART.

\section{Approximating the posterior}
\label{appendix:posterior-approx}

\begin{proofof}[Theorem \ref{thm:marginal_approx_rashomon}]
By the triangle inequality, we can write
\begin{align*}
     \sup_{\bt} \abs{ F_{\bbeta \mid \Z, \mc{P}_\theta}(\bt) - F_{\bbeta \mid \Z}(\bt ) }
     &= \sup_{\bt} \abs{ \sum_{\Pi \in \mc{P}_{\theta}} F_{\bbeta \mid \Z}(\bt \mid \Pi)
        \frac{\Pr(\Pi \mid \Z )}{\sum_{\Pi^{\prime} \in \mc{P}_{\theta}} \Pr(\Pi^{\prime} \mid \Z )}
        - \sum_{\Pi \in \mc{P}^{\star}} F_{\bbeta \mid \Z}(\bt \mid \Pi) \Pr(\Pi \mid \Z )} \\
     &\leq (\textrm{I}) + (\textrm{II}), \\
     \textrm{I}
     &= \sup_{\bt} \abs{ \sum_{\Pi \in \mc{P}_{\theta}} F_{\bbeta \mid \Z}(\bt \mid \Pi)
        \frac{\Pr(\Pi \mid \Z )}{\sum_{\Pi^{\prime} \in \mc{P}_{\theta}} \Pr(\Pi^{\prime} \mid \Z )}
        - \sum_{\Pi \in \mc{P}_{\theta}} F_{\bbeta \mid \Z}(\bt \mid \Pi) \Pr(\Pi \mid \Z )}, \\
     \textrm{II}
     &= \sup_{\bt} \abs{ \sum_{\Pi \in \mc{P}_{\theta}} F_{\bbeta \mid \Z}(\bt \mid \Pi) \Pr(\Pi \mid \Z )
        - \sum_{\Pi \in \mc{P}^{\star}} F_{\bbeta \mid \Z}(\bt \mid \Pi) \Pr(\Pi \mid \Z )}.
\end{align*}

Let us denote $K = \sum_{\Pi^{\prime} \in \mc{P}_{\theta}}\Pr(\Pi^{\prime} \mid \Z)$. Then the first term is
\begin{align*}
    (\textrm{I})
    &= \sup_{\bt} \abs{ \sum_{\Pi \in \mc{P}_{\theta}} F_{\bbeta \mid \Z}(\bt \mid \Pi) \frac{\Pr(\Pi \mid \Z )}{K}
        - \sum_{\Pi \in \mc{P}_{\theta}} F_{\bbeta \mid \Z}(\bt \mid \Pi) \Pr(\Pi \mid \Z )} \\
    &\leq \abs{\frac{1}{K} - 1}
      \sup_{\bt} \abs{ \sum_{\Pi \in \mc{P}_{\theta}} F_{\bbeta \mid \Z}(\bt \mid \Pi) \Pr(\Pi \mid \Z )} \\
    &\leq \abs{\frac{1}{K} - 1}
      \sup_{\bt} \abs{ \sum_{\Pi \in \mc{P}_{\theta}} \Pr(\Pi \mid \Z )}
      = \abs{\frac{1}{K} - 1} \, K \\
    &= 1 - \sum_{\Pi^{\prime} \in \mc{P}_{\theta}}\Pr(\Pi^{\prime} \mid \Z),
\end{align*}
where in the third line we bound $F_{\bbeta \mid \Z}(\bt \mid \Pi) \leq 1$, and in the last equality we use $0 < K \leq 1$, so that $\abs{1/K - 1} = 1/K - 1$.

Moving on to the second term,
\begin{align*}
    (\textrm{II})
    &= \sup_{\bt} \abs{ \sum_{\Pi \in \mc{P}_{\theta}} F_{\bbeta \mid \Z}(\bt \mid \Pi) \Pr(\Pi \mid \Z )
        - \sum_{\Pi \in \mc{P}^{\star}} F_{\bbeta \mid \Z}(\bt \mid \Pi) \Pr(\Pi \mid \Z )} \\
    &= \sup_{\bt} \abs{ \sum_{\Pi \in \mc{P}^{\star}\setminus \mc{P}_{\theta}} F_{\bbeta \mid \Z}(\bt \mid \Pi) \Pr(\Pi \mid  \Z )} \\
    &\leq \sup_{\bt} \abs{ \sum_{\Pi \in \mc{P}^{\star}\setminus \mc{P}_{\theta}} 1 \cdot \Pr(\Pi \mid  \Z )}
      = \sum_{\Pi \in \mc{P}^{\star}\setminus \mc{P}_{\theta}} \Pr(\Pi \mid  \Z ) \\
    &= 1 - \sum_{\Pi^{\prime} \in \mc{P}_{\theta}}\Pr(\Pi^{\prime} \mid \Z),
\end{align*}
where in the third line we again bound $F_{\bbeta \mid \Z}(\bt \mid \Pi) \leq 1$.

Therefore,
\begin{align*}
    \sup_{\bt} \abs{ F_{\bbeta \mid \Z, \mc{P}_\theta}(\bt) - F_{\bbeta \mid \Z}(\bt ) }
    &\leq (\textrm{I}) + (\textrm{II}) \\
    &\leq 2\Bigl(1 - \sum_{\Pi^{\prime} \in \mc{P}_{\theta}}\Pr(\Pi^{\prime} \mid \Z)\Bigr).
\end{align*}

There are two ways to bound the term
$1 - \sum_{\Pi^{\prime} \in \mc{P}_{\theta}}\Pr(\Pi^{\prime} \mid \Z)$.
First,
\begin{align*}
    1 - \sum_{\Pi^{\prime} \in \mc{P}_{\theta}}\Pr(\Pi^{\prime} \mid \Z)
    &\leq 1 - \sum_{\Pi^{\prime} \in \mc{P}_{\theta}} \theta
     = 1 - \abs{\mc{P}_{\theta}} \theta.
\end{align*}
Second,
\begin{align*}
    1 - \sum_{\Pi^{\prime} \in \mc{P}_{\theta}}\Pr(\Pi^{\prime} \mid \Z)
    &= \sum_{\Pi \in \mc{P}^{\star}\setminus \mc{P}_{\theta}} \Pr(\Pi \mid  \Z )
      \leq \sum_{\Pi \in \mc{P}^{\star}\setminus \mc{P}_{\theta}} \theta
      = \bigl(\abs{\mc{P}^{\star}} - \abs{\mc{P}_{\theta}}\bigr) \theta.
\end{align*}
The first bound is smaller whenever $\theta > 1 / \abs{\mc{P}^{\star}}$. Therefore,
\begin{align*}
    \sup_{\bt} \abs{ F_{\bbeta \mid \Z, \mc{P}_\theta}(\bt) - F_{\bbeta \mid \Z}(\bt ) }
    &\leq
    \begin{cases}
        2 \bigl( 1 - \abs{\mc{P}_{\theta}} \theta \bigr), & \theta > 1 / \abs{\mc{P}^{\star}}, \\[4pt]
        2 \bigl(\abs{\mc{P}^{\star}} - \abs{\mc{P}_{\theta}} \bigr) \theta, & \text{else}.
    \end{cases}
\end{align*}
Finally, we add a $\min$ operator because
$\sup_{\bt} \abs{ F_{\bbeta \mid \Z, \mc{P}_\theta}(\bt) - F_{\bbeta \mid \Z}(\bt ) }$
is trivially bounded by 1.
\end{proofof}

\begin{proofof}[\Cref{lemma:posterior-effects-error}]
This argument is similar to \Cref{thm:marginal_approx_rashomon} except for how we bound the expectations. We have
\begin{align*}
    \norm{\bar{\bbeta}_{\mc{P}_{\theta}} - \E_{\Pi, \mc{P}_{\theta}} \bbeta}
    &= \norm{\sum_{\Pi \in \mc{P}_{\theta}} \bbeta_{\Pi}
      \frac{\Pr(\Pi \mid \Z)}{\sum_{\Pi^{\prime} \in \mc{P}_{\theta}} \Pr(\Pi^{\prime}  \mid \Z)}
      - \sum_{\Pi \in \mc{P}_{\theta}} \bbeta_{\Pi} \Pr(\Pi \mid \Z)} \\
    &= \abs{\frac{1}{\sum_{\Pi^{\prime} \in \mc{P}_{\theta}}\Pr(\Pi^{\prime} \mid \Z)} - 1}
       \, \norm{\sum_{\Pi \in \mc{P}_{\theta}} \bbeta_{\Pi} \Pr(\Pi \mid \Z)} \\
    &= \abs{\frac{1}{K} - 1} \, \norm{\E_{\Pi, \mc{P}_{\theta}} \bbeta},
\end{align*}
where $K = \sum_{\Pi^{\prime} \in \mc{P}_{\theta}}\Pr(\Pi^{\prime} \mid \Z)$. Note that, by definition, $K \geq \abs{\mc{P}_{\theta}} \theta$. Further, $K \leq 1$ implies $1 / K - 1 > 0$. Therefore,
\begin{align*}
    \norm{\bar{\bbeta}_{\mc{P}_{\theta}} - \E_{\Pi, \mc{P}_{\theta}} \bbeta}
    &\leq \left(\frac{1}{\abs{\mc{P}_{\theta}} \theta} - 1 \right)
           \norm{\E_{\Pi, \mc{P}_{\theta}} \bbeta} \\
    \implies
    \frac{\norm{\bar{\bbeta}_{\mc{P}_{\theta}} - \E_{\Pi, \mc{P}_{\theta}} \bbeta}}
         {\norm{\E_{\Pi, \mc{P}_{\theta}} \bbeta} }
    &= \mathcal{O} \left( \frac{1}{\abs{\mc{P}_{\theta}} \theta} - 1 \right).
\end{align*}

If we further assume that $\norm{\bbeta_{\Pi}} < \infty$, then define
$C = \max_{\Pi \in \mc{P}^{\star}} \norm{\bbeta_{\Pi}} < \infty$.
The remainder of the proof is then identical to
\Cref{thm:marginal_approx_rashomon}, except for carrying the multiplicative constant $C$ from the expectations.
\end{proofof}

\begin{table}[!tb]
    \centering
    \caption{\small Notation used in Theorem \ref{thm:l0-minimax}.}
    \begin{tabular}{r|l}
         \textbf{Notation} & \textbf{Definition} \\
         \hline \hline
         $\mc{P}_{\mid h}$ & Set of permissible partitions with $h$ pools \\
         $Q \in \mathcal{Q}$ & Prior over all $\bbeta$ \\
         $Q \in \mathcal{Q}_{\mid h}$ & Prior over $\bbeta$ such that there is some partition $\Pi_{\bbeta} \in \mc{P}_{\mid h}$ \\
         $Q \in \mathcal{Q}_{\mc{P} \mid h}$ & Prior for partitions $\Pi \in \mc{P}_{\mid h}$ \\
         $P_{\ell_0}$ & Uniform prior over $\mc{P}_{\mid h}$ (induced by $\ell_0$ over $\mc{P}^{\star}$) \\
         $\textrm{P}_{Q, \Z}$ & Posterior distribution (over partitions or $\bbeta$) with prior $Q$ \\
         $\delta(P, Q)$ & Total variation distance between $P$ and $Q$
    \end{tabular}
    \label{tab:notation-robust-prior}
\end{table}

\begin{proofof}[Theorem \ref{thm:l0-minimax}]
For any prior $P \in \mathcal{Q}_{\mc{P} \mid h}$, we have
\begin{align*}
    \sup_{Q \in \mathcal{Q}_{\mc{P} \mid h}} \delta(\textrm{P}_{P, \Z}, \textrm{P}_{Q, \Z})
    &= \sup_{Q \in \mathcal{Q}_{\mc{P} \mid h}} \sup_{\Pi \in \mc{P}_{\mid h}}
       \abs{\textrm{P}_{P, \Z}(\Pi) - \textrm{P}_{Q, \Z}(\Pi)} \\
    &= \sup_{\Pi \in \mc{P}_{\mid h}} \sup_{Q \in \mathcal{Q}_{\mc{P} \mid h}}
       \abs{\textrm{P}_{P, \Z}(\Pi) - \textrm{P}_{Q, \Z}(\Pi)} \\
    &= \frac{1}{\Pr(\y \mid \X)} \sup_{\Pi \in \mc{P}_{\mid h}} \Pr(\y \mid \X, \Pi)
       \sup_{Q \in \mathcal{Q}_{\mc{P} \mid h}}  \abs{P(\Pi) - Q(\Pi)}.
\end{align*}

First, consider the $\ell_0$ prior:
\begin{align*}
    \sup_{Q \in \mathcal{Q}_{\mc{P} \mid h}} \delta(\textrm{P}_{P_{\ell_0}, \Z}, \textrm{P}_{Q, \Z})
    &= \frac{1}{\Pr(\y \mid \X)} \sup_{\Pi \in \mc{P}_{\mid h}} \Pr(\y \mid \X, \Pi)
       \sup_{Q \in \mathcal{Q}_{\mc{P} \mid h}}  \abs{\frac{1}{N(h)} - Q(\Pi)}.
\end{align*}
Choose an adversarial prior $Q^{\star}$ such that
$Q^{\star}(\Pi^{\star}) = 1$ for some arbitrary $\Pi^{\star} \in \mc{P}_{\mid h}$.
Then
\begin{align*}
    \sup_{Q \in \mathcal{Q}_{\mc{P} \mid h}}  \abs{\frac{1}{N(h)} - Q(\Pi)}
    &= \abs{\frac{1}{N(h)} - Q^{\star}(\Pi^{\star})}
      = 1 - \frac{1}{N(h)} \\
    \implies
    \sup_{Q \in \mathcal{Q}_{\mc{P} \mid h}} \delta(\textrm{P}_{P_{\ell_0}, \Z}, \textrm{P}_{Q, \Z})
    &= \left(1 - \frac{1}{N(h)} \right)
       \frac{\sup_{\Pi \in \mc{P}_{\mid h}}\Pr(\y \mid \X, \Pi)}{\Pr(\y \mid \X)}.
\end{align*}

Next, consider any other prior $P \in \mathcal{Q}_{\mc{P} \mid h}$ with $P \neq P_{\ell_0}$.
Let $\Pi_m = \argmin_{\Pi \in \mc{P}_{\mid h}} P(\Pi)$ and denote $P(\Pi_m) = p$.
Observe that $p < 1 / N(h)$ because $P \neq P_{\ell_0}$.
Construct an adversarial prior $Q^{\star}$ such that $Q^{\star}(\Pi_m) = 1$.
Therefore,
\begin{align*}
    \sup_{Q \in \mathcal{Q}_{\mc{P} \mid h}}  \abs{P(\Pi) - Q(\Pi)}
    &= \abs{P(\Pi_m) - Q^{\star}(\Pi_m)} = 1 - p \\
    \implies
    \sup_{Q \in \mathcal{Q}_{\mc{P} \mid h}} \delta(\textrm{P}_{P, \Z}, \textrm{P}_{Q, \Z})
    &= \frac{1}{\Pr(\y \mid \X)} \sup_{\Pi \in \mc{P}_{\mid h}} \Pr(\y \mid \X, \Pi) (1 - p) \\
    &= (1 - p)\,
       \frac{\sup_{\Pi \in \mc{P}_{\mid h}}\Pr(\y \mid \X, \Pi)}{\Pr(\y \mid \X)} \\
    &> \sup_{Q \in \mathcal{Q}_{\mc{P} \mid h}} \delta(\textrm{P}_{P_{\ell_0}, \Z}, \textrm{P}_{Q, \Z}).
\end{align*}

Thus, the $\ell_0$ prior is minimax optimal:
\begin{align*}
    \sup_{Q \in \mathcal{Q}_{\mc{P} \mid h}} \delta(\textrm{P}_{P_{\ell_0}, \Z}, \textrm{P}_{Q, \Z})
    &= \inf_{P \in \mathcal{Q}_{\mc{P} \mid h}}
       \sup_{Q \in \mathcal{Q}_{\mc{P} \mid h}} \delta(\textrm{P}_{P, \Z}, \textrm{P}_{Q, \Z}).
\end{align*}
\end{proofof}

\begin{proofof}[Proposition \ref{prop:equiv}]
\begin{proof}
Since \(\Pr(\Pi\mid\Z)\propto \exp\{-Q(\Pi)\}\), posterior ratios satisfy
\[
    \frac{\Pr(\Pi\mid\Z)}{\Pr(\Pi_0\mid\Z)}
    =
    \exp\{-[Q(\Pi)-Q(\Pi_0)]\}.
\]
Therefore,
\[
    \Pr(\Pi\mid\Z)\ge (1-\tau)\Pr(\Pi_0\mid\Z)
\]
if and only if
\[
    Q(\Pi)-Q(\Pi_0)\le -\log(1-\tau).
\]
Writing \(\delta=-\log(1-\tau)\) gives the additive score-neighborhood form.
If \(Q(\Pi_0)>0\), set \(\epsilon=\delta/Q(\Pi_0)\). Then
\[
    Q(\Pi)\le Q(\Pi_0)+\epsilon Q(\Pi_0)
    =
    (1+\epsilon)Q(\Pi_0).
\]
Conversely, given any \(\epsilon\ge0\), taking
\[
    \tau=1-\exp\{-\epsilon Q(\Pi_0)\}
\]
recovers the corresponding posterior-ratio threshold.
\end{proof}

\end{proofof}

\section{Appendix to Size of the Rashomon Set}
\label{appendix:rps-size}

\begin{proofof}[Proposition \ref{prop:num-permissible-poolings}] \
To count the number of all possible partitions, we cast this as a decision tree problem. There are $(R-1)^M$ possible feature combinations in the profile with all arms turned on. These constitute possible nodes in a binary decision tree. The leaves in the decision tree are the pools. The number of binary trees with $n$ nodes is given by
\begin{align*}
    C_n &= \frac{1}{n+1} \binom{2n}{n},
\end{align*}
where $C_n$ is the $n$-th Catalan number; see, for example, \citet{flajolet2009analytic}. Therefore, the number of trees we can construct (that may or may not be admissible) is
\begin{align*}
    T &= \sum_{n=1}^{(R-1)^M} C_n
      = \bigO \left(2^{2(R-1)^M} \right),
\end{align*}
where the big-O bound is given by \citet{topley2016computationally}.

To count the number of permissible partitions, conceptualize the binary matrix $\bSigma \in \{0, 1\}^{M \times (R-2)}$ again. Each element of $\bSigma$ tells us whether a particular pair of adjacent levels in a feature is pooled. In particular, we define $\Sigma_{ij} = 1$ if and only if feature combinations with level $j$ are pooled with feature combinations with level $j+1$ in feature $i$. Therefore, $\bSigma$ enumerates all admissible partitions. This gives us the desired result.
\end{proofof}

\begin{proofof}[Lemma \ref{lemma:num-leaves-threshold}]
From the definition of the Rashomon set, if $\Pi \in \mathcal{P}_{\theta}$, then
\begin{equation*}
    \Pr(\Pi \mid \Z) \geq \theta \\
    \implies \frac{\exp \left\{-\eta\loss(\Pi; \Z) - \lambda H(\Pi)\right\}}{c} \geq \theta.
\end{equation*}
Then, since \(\eta\loss(\Pi;\Z)\ge 0\), we have
\[
\exp\{-\lambda H(\Pi)\}
\ge
\exp\{-\eta\loss(\Pi;\Z)-\lambda H(\Pi)\}
\ge c\theta.
\]
Thus
\[
H(\Pi)\le -\frac{\log(c\theta)}{\lambda}.
\]
\end{proofof}

\begin{proofof}[Theorem~\ref{thm:rashomon-set-polynomial-size}]
We upper bound $\abs{\mathcal{P}_{\theta}}$ by bounding the number of permissible partitions with at most $H$ pools.

Fix $k$ with $1 \le k \le H$ and suppose the partition induces $k$ profiles. Profiles arise by choosing $k-1$
split locations among the $M(R-1)$ possible split positions, hence the number of possible profile-sets is at most
\[
\binom{M(R-1)}{k-1} \le C\,(MR)^{k-1}
\]
for a universal constant $C>0$.

For a fixed set of $k$ profiles, Lemma~\ref{lemma:k-profiles-num-partitions} gives two bounds, depending on the
relationship between $R$ and $M$.

\textbf{Dense regime ($R > M^{c_{\mathrm{crit}}}$).}
Lemma~\ref{lemma:k-profiles-num-partitions} yields
$\abs{\mathcal{P}^{(k)}} \le C_1\,M^k R^{H-k}$.
Therefore, summing over possible profile counts $k$:
\[
\abs{\mathcal{P}_{\theta}}
\le
\sum_{k=1}^H \binom{M(R-1)}{k-1}\,\abs{\mathcal{P}^{(k)}}
\le
C_2 \sum_{k=1}^H (MR)^{k-1}\,M^k R^{H-k}
=
C_2 \sum_{k=1}^H M^{2k-1}R^{H-1}.
\]
The summand is increasing in $k$, so the maximum occurs at $k=H$ and
$\abs{\mathcal{P}_{\theta}} \in \bigO\!\left(M^{2H-1}R^{H-1}\right)$.

\textbf{Sparse regime ($R \le M^{c_{\mathrm{crit}}}$).}
Lemma~\ref{lemma:k-profiles-num-partitions} yields
$\abs{\mathcal{P}^{(k)}} \le C_3\,(MR)^{k\log_2(H/k)}\,(\log_2(MR))^{-1}$.
Thus
\[
\abs{\mathcal{P}_{\theta}}
\le
C_4(\log_2(MR))^{-1}\sum_{k=1}^H (MR)^{\,k(1+\log_2(H/k)) - 1}.
\]
Let $f(k):=k(1+\log_2(H/k))$ and note that (under a continuous relaxation) $f$ is concave on $(0,H]$ with
\[
f'(k)=1+\log_2(H/k)-\frac{1}{\ln 2}.
\]
Setting $f'(k)=0$ gives $H/k^\star = e/2$ and hence $k^\star = 2H/e$. Evaluating at this maximum:
\[
f(k^\star)
=
\frac{2H}{e}\Bigl(1+\log_2(e/2)\Bigr)
=
\frac{2H}{e\ln 2}.
\]
Therefore $\max_{1\le k\le H} f(k) \le \gamma H$ with $\gamma := 2/(e\ln 2)$, and consequently
\[
\sum_{k=1}^H (MR)^{\,f(k)-1}
\le
H\,(MR)^{\gamma H - 1}.
\]
Absorbing the prefactor $H$ into the $\bigO(\cdot)$ (and retaining the $(\log_2(MR))^{-1}$ factor) yields
$\abs{\mathcal{P}_{\theta}} \in \bigO\!\left((MR)^{\gamma H - 1}\,(\log_2(MR))^{-1}\right)$.
\end{proofof}

\subsection{Helpful results}

We state a useful result that helps us count the number of pools generated by a partition matrix $\bSigma$.

\begin{lemma}
\label{lemma:sigma-ones-pools}
Let $\bSigma$ be the partition matrix for a profile with $M$ active features. Suppose there are $z_i$ ones in the $i$-th row of $\bSigma$. Then the number of pools created by $\bSigma$ is
\begin{align*}
    H(\bSigma)
    &= (R-1)^M
       - (R-1)^{M-1} \sum_i z_i
       + (R-1)^{M-2} \sum_{i_1 < i_2} z_{i_1} z_{i_2} \\
    &\quad - (R-1)^{M-3} \sum_{i_1 < i_2 < i_3} z_{i_1} z_{i_2} z_{i_3}
       + \dots
       + (-1)^M z_1 \dots z_M .
\end{align*}
\end{lemma}

\begin{proofof}[Lemma \ref{lemma:sigma-ones-pools}]
Observe that there are $(R-1)^M$ feature combinations in total ($R-1$ because we are assuming the $R$ discrete values include the control). Suppose we set $\Sigma_{ij} = 1$, then we are pooling features of type
\[
[r_1, \dots, r_{i-1}, j, r_{i+1}, \dots, r_M]
\]
with
\[
[r_1, \dots, r_{i-1}, j-1, r_{i+1}, \dots, r_M],
\]
where each $r_{i'}$ can take on $R-1$ values. Therefore, $(R-1)^{M-1}$ feature combinations are pooled. So, if there are $\mathrm{nnz}(\bSigma) = \sum_{i} z_i$ ones in $\bSigma$, then $(R-1)^{M-1} \sum_i z_i$ features are pooled.

However, if some of those ones are in different features, then we end up double counting some feature combinations. For example, if $\Sigma_{i_1, j} = 1$ and $\Sigma_{i_2, j'} = 1$, then we remove feature combinations of type
\[
[r_1, \dots, j, \dots, j', \dots, r_M]
\]
twice, where $j$ and $j'$ are at indices $i_1$ and $i_2$. So we need to add them back once. Similarly, the remaining higher-order terms account for this double counting, which gives the inclusion--exclusion expression in the statement.
\end{proofof}

\begin{lemma}
\label{lemma:sigma-max-ones}
Let $\bSigma$ be the matrix defined in Proposition \ref{prop:num-permissible-poolings} for a profile with $M$ active features. Suppose there are $H$ pools. Then
\begin{align*}
    \sum_i z_i &\leq \frac{(2R-3)^M + 1 - 2H}{2(R-1)^{M-1}}.
\end{align*}
\end{lemma}

\begin{proofof}[Lemma \ref{lemma:sigma-max-ones}]
Rearranging Lemma \ref{lemma:sigma-ones-pools} and dropping negative terms,
\begin{align*}
    (R-1)^{M-1} \sum_i z_i
    &\leq -H + (R-1)^M
       + (R-1)^{M-2} \sum_{i_1 < i_2} z_{i_1} z_{i_2} \\
    &\quad + (R-1)^{M-4} \sum_{i_1 < \dots < i_4} z_{i_1} z_{i_2} z_{i_3} z_{i_4}
       + \dots \\
    &\leq -H + (R-1)^M
       + (R-1)^{M-2} (R-2)^2 \sum_{i_1 < i_2} 1 \\
    &\quad + (R-1)^{M-4} (R-2)^4 \sum_{i_1 < \dots < i_4} 1 + \dots \\
    &= -H + \sum_{\substack{n=0\\ n\ \mathrm{even}}}^M \binom{M}{n}(R-1)^{M-n}(R-2)^n \\
    &= -H + \frac{(2R-3)^M + 1}{2},
\end{align*}
where the second inequality uses $z_j \leq R-2$ and the last equality uses the well-known identity
\[
    \sum_{\substack{k=0 \\ k\ \mathrm{even}}}^n \binom{n}{k} a^{\,n-k} b^k
    = \frac{1}{2} \left((a+b)^n + (a-b)^n \right).
\]
Rearranging completes the proof.
\end{proofof}

Lemma \ref{lemma:sigma-ones-pools} tells us how to count the number of pools given a partition matrix. We now state another result that bounds the sparsity of the partition matrix given some number of pools in \Cref{lemma:sigma-max-ones}.

\begin{lemma}
\label{lemma:num-sigma-matrix-pools}
Let $\bSigma$ be the matrix defined in Proposition \ref{prop:num-permissible-poolings} for a profile with $M$ active features. Then the number of $\bSigma$ matrices that generate $h$ pools is
\begin{align*}
    N(h)
    &= \sum_{k=0}^M \binom{M}{k}
       \sum_{\prod_{i=1}^k (z_i + 1) = h}
       \prod_{i=1}^k \binom{R-2}{z_i},
\end{align*}
where we define $N(1) = 1$ and $\binom{n}{k} = 0$ for $k > n$.

As $M, R \to \infty$, we have
\begin{align*}
    N(h) &=
    \begin{cases}
        \bigO \left(M R^{h-1}\right), & R > M^{c_{\mathrm{crit}}} \\
        \bigO \left((MR)^{\log_2 h}\right), & \text{else},
    \end{cases}
\end{align*}
where $c_{\mathrm{crit}} = (\log_2 3 - 1) / (2 - \log_2 3)$.
\end{lemma}

\begin{proofof}[Lemma \ref{lemma:num-sigma-matrix-pools}]
This is an exercise in counting. When we make $z$ splits in one feature, we generate $z+1$ pools. When we make $z_i$ splits in feature $i$ and $z_j$ splits in feature $j$, we generate $(z_i + 1)(z_j + 1)$ pools.

When we want to generate $h$ pools, we first choose the features where we want the splits to occur. This is what the outer summation is doing. Suppose that we have chosen $k$ features in which we will perform splits. Next, we need to identify how many splits can be made in each feature. This is what the inner summation is doing with the condition $\prod_{i=1}^k (z_i + 1) = h$. Finally, we need to identify where those splits are made, which is where the binomial coefficients appear.

To get the asymptotic bound, we first consider the term where the exponent on $R$ is the largest. This is when we choose all splits in the same feature. Next, we consider the term where the exponent on $M$ is the largest. For this to happen, we need to choose as many features as possible, that is, make the smallest number of non-zero splits in each feature. This corresponds to making one split in each of $\log_2 h$ distinct features. Hence, we obtain the asymptotic bound $\bigO\bigl(\max\{ M R^{h-1}, (MR)^{\log_2 h}\}\bigr)$.

Observe that
\begin{align*}
    M R^{h-1} > (MR)^{\log_2 h}
    \quad \iff \quad
    R > M^{\frac{\log_2 h - 1}{h - \log_2 h - 1}}.
\end{align*}
The exponent on $M$ is a decreasing function of $h$. When $h = 2$, $M R^{h-1} = (MR)^{\log_2 h}$. When $h=3$, the exponent is
\[
c_{\mathrm{crit}} = \frac{\log_2 3 - 1}{2 - \log_2 3}.
\]
Therefore $M R^{h-1} > (MR)^{\log_2 h}$ whenever $R > M^{c_{\mathrm{crit}}}$, which gives the desired result.
\end{proofof}

Lemma \ref{lemma:num-sigma-matrix-pools} has a useful implication. When $h$ is a prime number, we expect $N(h)$ to be small because all of the splits need to be made in the same feature. On the other hand, when $h = 2^k$ is a power of two, we expect $N(h)$ to be very large since we can make splits in multiple features at the same time.

\begin{lemma}
\label{lemma:integral-a^logx}
For $a > 1$,
\begin{align*}
    \int a ^{\log_2 x} \, dx
    &= \frac{x a^{\log_2 x}}{1 + \log_2 a} + C.
\end{align*}
\end{lemma}

\begin{proofof}[Lemma \ref{lemma:integral-a^logx}]
We use integration by parts:
\begin{align*}
    \int a ^{\log_2 x} \, dx
    &= a^{\log_2 x} \int dx
       - \int x \cdot \frac{a^{\log_2 x} \log_2 a}{x} \, dx \\
    &= x a^{\log_2 x}
       - \log_2 a \int a^{\log_2 x} \, dx \\
    \implies
    \int a^{\log_2 x} \, dx
    &= \frac{x a^{\log_2 x}}{1 + \log_2 a} + C.
\end{align*}
\end{proofof}

\begin{lemma}
\label{lemma:k-profiles-num-partitions}
Suppose there are $k \geq 1$ fixed profiles across $M$ features, each taking on $R$ discrete ordered values. Suppose that the maximum number of pools in any partition is $H$. Then the number of permissible partitions is bounded by
\begin{align*}
    \abs{\mc{P}^{(k)}}  &=
    \begin{cases}
        \bigO \left( M^k R^{H-k} \right), & R > M^{c_{\mathrm{crit}}} \\
        \bigO \left( (MR)^{k \log_2(H/k)} (\log_2 (MR))^{-1} \right), & \text{else},
    \end{cases}
\end{align*}
where $c_{\mathrm{crit}} = (\log_2 3 - 1) / (2 - \log_2 3)$.
\end{lemma}

\begin{proofof}[Lemma \ref{lemma:k-profiles-num-partitions}]
Let $h_i$ denote the number of pools within profile $i$. Then we know that $k \leq \sum_{i=1}^k h_i \leq h$, where $1 \leq h_i \leq h - k + 1$ for every profile $i$ and $h \leq H$. Observe that partitions within each profile are strongly convex. By Lemma \ref{lemma:num-sigma-matrix-pools}, we have a bound on the number of partitions of size $h_i$, $N_i(h_i)$,
\begin{align*}
    N_i(h_i)
    &= \max \left\{ \bigO\left(MR^{h_i-1}\right),
                    \bigO\left((MR)^{\log_2 h_i}\right) \right\}.
\end{align*}
We also know from Lemma \ref{lemma:num-sigma-matrix-pools} that $MR^{h_i-1}$ dominates $(MR)^{\log_2 h_i}$ whenever $R > M^{c_{\mathrm{crit}}}$, where $c_{\mathrm{crit}} = (\log_2 3 - 1) / (2 - \log_2 3)$. For now, we suppress this condition for readability and re-introduce it at the end.

For a given set of $k$ profiles, the number of partitions is
\begin{align*}
    \abs{\mc{P}^{(k)}}
    &= \sum_{h=k}^H \prod_{\sum_{i=1}^k h_i = h} N_i(h_i).
\end{align*}
There are $\binom{h-1}{k-1}$ positive integer solutions to the equation $\sum_{i=1}^k h_i = h$. Thus,
\begin{align*}
    \abs{\mc{P}^{(k)}}
    &= \sum_{h=k}^H \binom{h-1}{k-1}
       \max \left\{
          \bigO \left( M^kR^{\sum_{i=1}^k h_i-k} \right),
          \bigO \left((MR)^{\sum_{i=1}^k \log_2 h_i} \right)
       \right\} \\
    &= \sum_{h=k}^H \binom{h-1}{k-1}
       \max \left\{
          \bigO \left( M^kR^{h-k} \right),
          \bigO \left((MR)^{\sum_{i=1}^k \log_2 h_i} \right)
       \right\}.
\end{align*}
To bound $\sum_{i=1}^k \log_2 h_i$ when $\sum_{i=1}^k h_i = h$, we use the arithmetic--geometric mean inequality:
\begin{align*}
    \left( \prod_{i=1}^k h_i \right)^{1/k}
    &\leq \frac{\sum_{i=1}^k h_i}{k}
      = \frac{h}{k}
      \quad\implies\quad
    \prod_{i=1}^k h_i \leq \left(\frac{h}{k}\right)^k \\
    \implies \sum_{i=1}^k \log_2 h_i
    &\leq k (\log_2 h - \log_2 k).
\end{align*}
Therefore,
\begin{align*}
    \abs{\mc{P}^{(k)}}
    &= \sum_{h=k}^H \binom{h-1}{k-1}
       \max \left\{
          \bigO \left( M^kR^{h-k} \right),
          \bigO \left((MR)^{ k (\log_2 (h/k))} \right)
       \right\} \\
    &= \max \left\{
          \bigO \left( M^k \sum_{h=k}^H R^{h-k} \right),
          \bigO \left( \sum_{h=k}^H (MR)^{ k \log_2 (h/k)} \right)
       \right\}.
\end{align*}
The first term simplifies as
\begin{align*}
    M^k \sum_{h=k}^H R^{h-k}
    &= \bigO(M^k R^{H-k}).
\end{align*}
The second term can be bounded by the integral in Lemma \ref{lemma:integral-a^logx}, applied to $a = MR$ and the change of variables $x = h/k$:
\begin{align*}
    \sum_{h=k}^H (MR)^{ k \log_2 (h/k)}
    &\leq \int_{h=k}^H (MR)^{ k \log_2 (h/k)} \, dh \\
    &= \frac{H (MR)^{k \log_2(H/k)} - k}{1 + k \log_2 (MR)} \\
    &= \bigO \left( (MR)^{k \log_2(H/k)} (\log_2 (MR))^{-1} \right).
\end{align*}
Thus, for a given set of $k$ profiles, the number of partitions is bounded by
\begin{align*}
    \abs{\mc{P}^{(k)}}
    &= \max \left\{
        \bigO \left( M^k R^{H-k} \right),
        \bigO \left( (MR)^{k \log_2(H/k)} (\log_2(MR))^{-1} \right)
       \right\}.
\end{align*}
Re-introducing the condition for when the first term dominates the second gives the stated cases.
\end{proofof}

\section{Appendix to Rashomon set enumeration and generalizations}
\label{appendix:enumeration-proofs}

We organize this appendix into proofs for results in Section~\ref{section:enumeration}, additional algorithms used in Section~\ref{section:enumeration}, and proofs for results in Section~\ref{section:generalization}.

\subsection{Proofs in Section \ref{section:enumeration}.}

\begin{proofof}[Theorem \ref{thm:rashomon-fixed-bound}]
Intuitively, $b(\bSigma, \mathcal{M}; \Z)$ is the error we are \emph{forced} to pay from the part of the partition that has already been fixed by $\mathcal{M}$. If that forced error is already above the Rashomon threshold, no refinement of $\bSigma$ can ever achieve a Rashomon–admissible risk.

By definition,
\begin{align*}
b(\Sigma,M;\Z)
=
\eta\frac{1}{n}
\sum_{\pi\in\Pi_f}
\sum_{k(i)\in\pi}
\mathbf 1\{k(i)\in\pi_f\}
(y_i-\widehat\mu_\pi)^2
+
\lambda H(\Pi,M).
\end{align*}
Notice that $\abs{\Pi} \geq  H(\Pi, \mathcal{M})$. Further, by making more splits, we can only reduce the total mean-squared error incurred. Therefore,
\begin{align*}
    Q(\Pi ; \Z)
    &= \eta\loss(\Pi ; \Z) + \lambda \abs{\Pi} \\
    &= \eta \frac{1}{n} \sum_{\pi \in \Pi} \sum_{k(i) \in \pi} (y_i - \widehat{\mu}_{\pi})^2 + \lambda \abs{\Pi} \\
    &\geq \eta \frac{1}{n} \sum_{\pi \in \Pi} \sum_{k(i) \in \pi}
       \mathbb{I} \left\{ k(i) \cap \pi_{\fix} \neq \varnothing \right\} (y_i - \widehat{\mu}_{\pi})^2
       + \lambda \abs{\Pi} \\
    &\geq \eta\frac{1}{n} \sum_{\pi \in \Pi_{\fix}} \sum_{k(i) \in \pi}
       \mathbb{I} \left\{ k(i) \cap \pi_{\fix} \neq \varnothing \right\} (y_i - \widehat{\mu}_{\pi})^2
       + \lambda \abs{\Pi} \\
    &\geq \eta\frac{1}{n} \sum_{\pi \in \Pi_{\fix}} \sum_{k(i) \in \pi}
       \mathbb{I} \left\{ k(i) \cap \pi_{\fix} \neq \varnothing \right\} (y_i - \widehat{\mu}_{\pi})^2
       + \lambda H(\Pi, \mathcal{M}) \\
    &= b(\bSigma_{\fix} ; \Z).
\end{align*}
So if $b(\bSigma, \mathcal{M} ; \Z) > \theta_{\epsilon}$, then $\bSigma$ is not in the Rashomon set.

Now consider $\bSigma^{\prime} \in \child(\bSigma, \mathcal{M})$. Notice that the size of the fixed set of indices $\mathcal{M}^{\prime}$ in any child of $\bSigma$ increases (because there are fewer places to make further splits). With any further split we make in $\mathcal{M}$, the number of pools increases. Finally, the loss is non-negative. These together imply
\begin{align*}
    b(\bSigma^{\prime}, \mathcal{M}^{\prime}; \Z) &\geq b(\bSigma, \mathcal{M} ; \Z) \\
    \implies Q(\Pi(\bSigma^{\prime}); \Z) &\geq b(\bSigma^{\prime}, \mathcal{M}^{\prime} ; \Z)
      \geq b(\bSigma, \mathcal{M}; \Z).
\end{align*}
Therefore, if $b(\bSigma, \mathcal{M}; \Z) > \theta_{\epsilon}$, then $\bSigma$ and all $\bSigma^{\prime} \in \child(\bSigma, \mathcal{M})$ are not in the Rashomon set.
\end{proofof}

\begin{proofof}[Theorem \ref{thm:rashomon-equivalent-bound}]
At a high level, $b(\bSigma, \mathcal{M}; \Z)$ captures the loss from units whose pools are already fully determined by $\mathcal{M}$, while $b_{eq}(\bSigma, \mathcal{M}; \Z)$ captures the \emph{best-case} loss we could ever hope to achieve from units that still look equivalent under $\mathcal{M}$. The bound
\[
B(\bSigma, \mathcal{M} ; \Z) = b(\bSigma, \mathcal{M}; \Z) + b_{eq}(\bSigma, \mathcal{M}; \Z)
\]
is therefore an optimistic lower bound on the total loss for \emph{any} refinement of $\Pi$ that respects the current fixed structure. If even this optimistic bound exceeds $\theta_{\epsilon}$, every descendant partition of $\bSigma$ on this branch is excluded from the Rashomon set, so the entire subtree can be pruned.

By definition of $b_{eq}$,
\begin{align*}
    b_{eq}(\bSigma, \mathcal{M} ; \Z)
    &\leq \eta\frac{1}{n} \sum_{\pi \in \Pi} \sum_{k(i) \in \pi}
      \mathbb{I} \left\{ k(i) \cap \pi_{\fix}^c \neq \varnothing \right\}
      (y_i - \widehat{\mu}_{\pi})^2 .
\end{align*}
The idea in the inequality above is that any further split we make must obey the splits made at $\mathcal{M}$, so the terms with $k(i) \cap \pi_{\fix}^c \neq \varnothing$ are the ones that can still change as we refine $\Pi$.

Now write
\begin{align*}
    Q(\Pi; \Z)
    &= \eta\loss(\Pi ; \Z) + \lambda \abs{\Pi} \\
    &= \eta\frac{1}{n} \sum_{\pi \in \Pi} \sum_{k(i) \in \pi}
       \mathbb{I} \left\{ k(i) \cap \pi_{\fix} \neq \varnothing \right\}
       (y_i - \widehat{\mu}_{\pi})^2 + \lambda \abs{\Pi} \\
    &\qquad\quad
       + \frac{1}{n} \sum_{\pi \in \Pi} \sum_{k(i) \in \pi}
       \mathbb{I} \left\{ k(i) \cap \pi_{\fix}^c \neq \varnothing \right\}
       (y_i - \widehat{\mu}_{\pi})^2 \\
    &\geq b(\bSigma, \mathcal{M}; \Z) + b_{eq}(\bSigma, \mathcal{M}; \Z) \\
    &= B(\bSigma, \mathcal{M} ; \Z).
\end{align*}
Therefore, if $B(\bSigma, \mathcal{M} ; \Z) > \theta_{\epsilon}$, then $Q(\Pi ; \Z) > \theta_{\epsilon}$ and $\bSigma^{\prime} \in \child(\bSigma, \mathcal{M})$.

Let $\Pi^{\prime} := \Pi(\bSigma^{\prime})$. Then
\begin{align*}
    Q(\Pi^{\prime} ; \Z)
    &= \eta\loss(\Pi^{\prime}; \Z) + \lambda \abs{\Pi^{\prime}} \\
    &= \eta\frac{1}{n} \sum_{\pi \in \Pi^{\prime}} \sum_{k(i) \in \pi}
       \mathbb{I} \left\{ k(i) \cap \pi_{\fix} \neq \varnothing \right\}
       (y_i - \widehat{\mu}_{\pi})^2 + \lambda \abs{\Pi^{\prime}} \\
    &\qquad\quad
       + \frac{1}{n} \sum_{\pi \in \Pi^{\prime}} \sum_{k(i) \in \pi}
       \mathbb{I} \left\{ k(i) \cap \pi_{\fix}^c \neq \varnothing \right\}
       (y_i - \widehat{\mu}_{\pi})^2 \\
    &\geq b(\bSigma, \mathcal{M}; \Z)
       + \frac{1}{n} \sum_{\pi \in \Pi^{\prime}} \sum_{k(i) \in \pi}
       \mathbb{I} \left\{ k(i) \cap \pi_{\fix}^c \neq \varnothing \right\}
       (y_i - \widehat{\mu}_{\pi})^2 \\
    &\geq b(\bSigma, \mathcal{M}; \Z) + b_{eq}(\bSigma, \mathcal{M}; \Z) \\
    &= B(\bSigma, \mathcal{M} ; \Z).
\end{align*}
In the steps above, we used the fact that making any split will increase the number of pools to say that $\abs{\Pi^{\prime}} \geq \abs{\Pi}$. We also used the definition of $b_{eq}$ and the idea of a minimum loss incurred by equivalent units in the final step.

Therefore, if $B(\bSigma, \mathcal{M} ; \Z) > \theta_{\epsilon}$, then $Q(\Pi^{\prime}; \Z) > \theta_{\epsilon}$ for any $\bSigma^{\prime} \in \child(\bSigma, \mathcal{M})$. So $\bSigma$ and all such $\bSigma^{\prime}$ are not in the Rashomon set.
\end{proofof}

By inspecting the proofs of Theorems \ref{thm:rashomon-fixed-bound} and \ref{thm:rashomon-equivalent-bound}, we have the following corollary that generalizes to other non-negative loss functions and penalties that are increasing in $\abs{\Pi}$.

\begin{corollary}
\label{cor:other-losses}
Appropriate versions of Theorems \ref{thm:rashomon-fixed-bound} and \ref{thm:rashomon-equivalent-bound} hold for non-negative error functions $\mathcal{L}^{\prime}(\Pi; \Z) \geq 0$ for all $\Pi, \Z$ and penalties $H^{\prime}(\Pi)$ that are an increasing function of $\abs{\Pi}$.
\end{corollary}

\begin{proofof}[\Cref{cor:other-losses}]
This follows from the proofs of Theorems \ref{thm:rashomon-fixed-bound} and \ref{thm:rashomon-equivalent-bound}. Both of these proofs rely on two key ideas: (1) adding more (mean squared) error terms will never decrease the overall loss (keeping the scaling constant $1/n$ fixed), and (2) adding more pools to $\Pi$ will only increase the overall loss. Both of these things are true for any non-negative error function and any penalty that is an increasing function of $\abs{\Pi}$.
\end{proofof}

\begin{proofof}[Theorem \ref{thm:enumeration-algorithm}]
First note that Algorithm \ref{alg:r-aggregate-profile} correctly enumerates the Rashomon set for any given profile. This follows directly from Lemma \ref{lemma:num-leaves-threshold} and Theorems \ref{thm:rashomon-fixed-bound} and \ref{thm:rashomon-equivalent-bound}.

Next, \Cref{alg:r-aggregate-across-profiles} performs a breadth-first search starting at the control profile. Since the $M$-dimensional hypercube has a unique source (the control profile) and sink (the profile with all features active), the breadth-first search will terminate after a finite time and traverse every possible path in the hypercube. When traversing an edge in the hypercube, \Cref{alg:pool-adjacent-profiles} attempts to pool adjacent profiles using the intersection matrix $\bSigma^{\cap}$ while obeying \ref{permissibility:case1} and \ref{permissibility:case2} of \Cref{def:permissible-partition}. This pooling attempt is done recursively, guaranteeing that all permissible partitions are considered for the Rashomon set.

The choice of Rashomon thresholds for each profile, described in line~\ref{line:eq-bound}, is justified by the usage of Theorem \ref{thm:rashomon-equivalent-bound}.

Correctness of Algorithm \ref{alg:r-aggregate} immediately follows.
\end{proofof}

\subsection{Additional algorithms.}

\Cref{alg:r-aggregate} calls upon two important algorithms and uses a specific caching object that we describe here. First, \Cref{alg:r-aggregate-profile} describes how to enumerate the Rashomon partitions for a single profile. The choice of starting position in Algorithm \ref{alg:r-aggregate-profile} is arbitrary. No matter with which feature we start our search, Algorithm \ref{alg:r-aggregate-profile} will eventually explore the feature space sufficiently to identify partitions outside the Rashomon set, at which point we abandon that search. Theorems \ref{thm:rashomon-fixed-bound} and \ref{thm:rashomon-equivalent-bound} guarantee this.

\begin{algorithm}[!tb]
\caption{\texttt{EnumerateRPS\_profile}($M, R, H_{\max}, \Z, \theta_{\epsilon}$)}
\label{alg:r-aggregate-profile}
\begin{algorithmic}[1]
\Require $M$ features, $R$ factors per feature, max pools $H_{\max}$, data $\Z$, Rashomon threshold $\theta_{\epsilon}$
\Ensure Rashomon set $\mathcal{P}_{q, \epsilon}$

\State $\mathcal{P}_{q, \epsilon} = \varnothing$
\State $\mathcal{S} = \texttt{cache()}$ \Comment{See Algorithm \ref{alg:r-aggregate-cache}}
\State $\mathcal{Q} = \texttt{queue()}$
\State $\bSigma = \{1\}^{M \times (R-2)}$
\State \texttt{$\mathcal{Q}$.push$(\bSigma, 1, 1)$} \Comment{Can start at any arbitrary arm}
\While{$\mathcal{Q}$ is not empty}
    \State $(\bSigma, i, j) = \mathcal{Q}\texttt{.dequeue()}$
    
    \lIfThen {$\mathcal{S}\texttt{.seen}(\bSigma, i, j)$} {continue}

    \State $\mathcal{S}\texttt{.insert}((\bSigma, i, j))$
    
    \lIfThen {$H(\bSigma) > H_{\max}$} {continue}

    \State $\bSigma^{(1)} = \bSigma$; $\bSigma^{(1)}_{i,j} = 1$ \Comment{Copy of $\bSigma$ with no split at $(i,j)$}
    \State $\bSigma^{(0)} = \bSigma$; $\bSigma^{(0)}_{i,j} = 0$ \Comment{Copy of $\bSigma$ with split at $(i,j)$}

    \For {$m = 1$ \textbf{to} $M$}    \Comment{Branch and search across features}
        \State $j_1 = \min \{j \leq R-2 \mid \text{ not } \mathcal{S}\texttt{.seen}(\bSigma^{(1)}, m, j) \}$
        \lIfThen {$j_1 \neq \varnothing$} {$\mathcal{Q}\texttt{.enqueue}(\bSigma^{(1)}, m, j_1)$}
        \State $j_0 = \min \{j \leq R-2 \mid \text{ not } \mathcal{S}\texttt{.seen}(\bSigma^{(0)}, m, j) \}$
        \lIfThen {$j_0 \neq \varnothing$} {$\mathcal{Q}\texttt{.enqueue}(\bSigma^{(0)}, m, j_0)$}
    \EndFor
    
    \lIfThen {$B(\bSigma, i, j; \Z) > \theta_{\epsilon}$} {continue}
        \Comment{Prune branch if the equivalence bound already exceeds the Rashomon threshold}
    
    \lIfThen {$Q(\bSigma^{(1)}) \leq \theta_{\epsilon}$ and $H(\bSigma^{(1)}) \leq H_{\max}$}
        {$\mathcal{P}_{q, \epsilon}.\texttt{add}(\bSigma^{(1)})$}
    \lIfThen {$Q(\bSigma^{(0)}) \leq \theta_{\epsilon}$ and $H(\bSigma^{(0)}) \leq H_{\max}$}
        {$\mathcal{P}_{q, \epsilon}.\texttt{add}(\bSigma^{(0)})$}

    \If {$j < R - 2$}    \Comment{Search deeper within the same feature}
        \lIfThen {not $\mathcal{S}\texttt{.seen}(\bSigma^{(1)}, i, j+1)$}
            {\texttt{$\mathcal{Q}$.enqueue$(\bSigma^{(1)}, i, j+1)$}}
        \lIfThen {not $\mathcal{S}\texttt{.seen}(\bSigma^{(0)}, i, j+1)$}
            {\texttt{$\mathcal{Q}$.enqueue$(\bSigma^{(0)}, i, j+1)$}}
    \EndIf
    
\EndWhile

\State \Return $\mathcal{P}_{q, \epsilon}$

\end{algorithmic}
\end{algorithm}

\begin{figure}[!tb]
    \centering
    \includegraphics[width=1\linewidth]{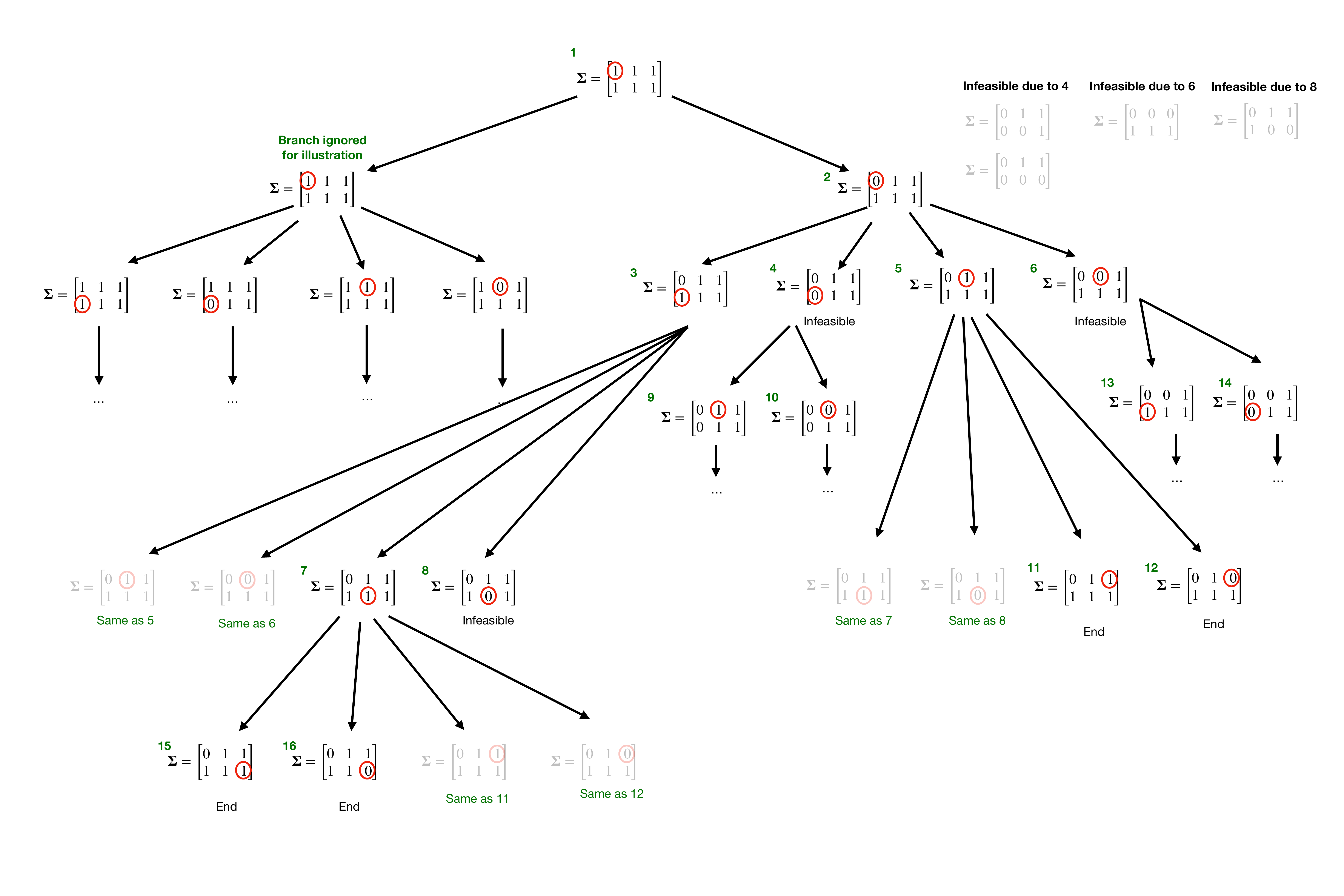}
    \caption{\small Partial search tree traced by Algorithm \ref{alg:r-aggregate-profile} as seen in Example \ref{ex:algorithm-raggregate-profile}.}
    \label{fig:raggregate-profile-example}
\end{figure}

\begin{example}[Example of Algorithm \ref{alg:r-aggregate-profile}]
\label{ex:algorithm-raggregate-profile}
Consider enumerating the RPS in a setup with $M = 2$ features and $R = 5$ levels in each feature. Let us take the profile where both features are active, i.e., $\rho = (1, 1)$. Algorithm \ref{alg:r-aggregate-profile} starts with the singleton partition, i.e., there is only one pool containing all feature combinations (see line~4). This is given by the partition matrix
\begin{align*}
    \bSigma &= \begin{bmatrix}
        1 & 1 & 1 \\
        1 & 1 & 1
    \end{bmatrix}.
\end{align*}
We start our search at $(1, 1)$ in the matrix, i.e., in the first feature at the first level, and insert $(\bSigma, 1, 1)$ into the queue $\mathcal{Q}$. Since the queue is initially not empty, we enter the loop in line~6 and remove this partition matrix. As this is the first time we encounter this partition matrix, we mark it as seen using $\mathcal{S}$.

Then, in lines~11 and 12, we create two copies of this partition matrix: $\bSigma^{(1)}$ with $\bSigma_{1,1} = 1$ (we do not split at the first level of the first feature) and $\bSigma^{(0)}$ with $\bSigma_{1,1} = 0$ (we split at the first level of the first feature).

For each copy, we scan the other features to identify the lowest feature level that we have not already seen in $\mathcal{S}$. Since we have not tested the feasibility of these partitions yet, we add them to the queue $\mathcal{Q}$. This operation happens in lines~13–17.

In line~18, we test the feasibility of the partition $\bSigma$ we dequeued being present in the RPS. If it is infeasible, we move on to the next element in the queue $\mathcal{Q}$. If it is indeed feasible, then we check whether $\bSigma^{(1)}$ and $\bSigma^{(0)}$ meet the Rashomon threshold in lines~19 and 20, respectively. If they do, then we attempt to split at the next feature level in that feature by adding those to the queue $\mathcal{Q}$ in lines~21–23.

Then, we dequeue the next item in $\mathcal{Q}$ and repeat the process until the queue is empty. Any partition or split that we did not check explicitly is a child of one of the partitions we checked and deemed infeasible.

\textbf{Walkthrough:} Figure \ref{fig:raggregate-profile-example} shows the search tree in this scenario. For the sake of illustration, we follow only certain branches of $\bSigma^{(0)}$ and show what happens when certain partitions are found to be infeasible. The green numbers show the order in which the matrices are added to the queue (and therefore inspected by the algorithm). The red circled index shows which feature and level is being considered next.

After dequeuing $\bSigma^{(0)}$, we add matrices 3 and 4 to the queue as per lines~13–17. Then, we add matrices 5 and 6 once we find that $\bSigma^{(0)}$ satisfies \Cref{thm:rashomon-equivalent-bound}.

Following the order of the queue, we dequeue matrix~3. Since the matrices that we would have added in lines~13–17 are already in the queue, we skip over them.\footnote{Strictly speaking, the algorithm would add them to the queue anyway and would later skip them when they get dequeued. For simplicity of illustration, we skip over them immediately.} Once we find that matrix~3 is feasible, we add matrices 7 and 8 to the queue as in lines~21–23.

Next, we dequeue matrix~4. We add matrices 9 and 10 as per lines~13–17. We find that matrix~4 does not satisfy Theorem \ref{thm:rashomon-equivalent-bound}, so we move on. In the top right corner of Figure \ref{fig:raggregate-profile-example}, we list matrices that never get checked because we found that 4 is infeasible.

When we dequeue matrix~5, we observe a similar behavior to matrix~3. When we dequeue matrix~6, we observe a similar behavior to matrix~4. This process continues until the queue is empty.
\end{example}

It is worth noting that Algorithm \ref{alg:r-aggregate-profile} will correctly enumerate the RPS independently of our starting search position. Obviously, some starting positions may be computationally favorable, i.e., we do not need to search for too long before we encounter low-posterior partitions. We believe domain experts will have a better understanding of the context and may be able to choose a starting location that can reduce computation costs. For instance, in Example \ref{ex:algorithm-raggregate-profile}, if we know for sure that there is heterogeneity in the lowest level in the second feature, we may wish to start at position $(2, 1)$ instead of $(1, 1)$, discarding several infeasible partitions when we dequeue the first element itself!

Next, we describe how to pool across profiles as defined by \Cref{def:permissible-partition}. The key insight here is the construction of the intersection matrix $\bSigma^{\cap}$ we discussed earlier. \Cref{alg:r-aggregate-across-profiles} describes a breadth-first search to enumerate partitions across different profiles by traversing the $M$-dimensional hypercube. This algorithm in turn relies on \Cref{alg:intersection-matrix} to obtain the intersection matrix between partitions of adjacent profiles and \Cref{alg:pool-adjacent-profiles} to pool adjacent profiles recursively. Since \Cref{alg:r-aggregate-across-profiles} is a breadth-first search, it can also be parallelized.

Finally, \Cref{alg:r-aggregate-cache} describes the implementation of the caching object used in \Cref{alg:r-aggregate}.

\begin{algorithm}[!tb]
\caption{\texttt{IntersectionMatrix}($\Pi, \rho_i, \rho_j$)}
\label{alg:intersection-matrix}
\begin{algorithmic}[1]
\Require Partition $\Pi$, adjacent profiles $\rho_i, \rho_j$ such that $\rho_i < \rho_j$
\Ensure Intersection matrix $\bSigma^{\cap}$

\State $\mathbf{m} = \rho_i \land \rho_j$    \Comment{Indices of features active in both profiles}
\State $m^{\prime} = \rho_i \oplus \rho_j$    \Comment{Index where $\rho_i, \rho_j$ differ}

\State $\Pi_{\rho_i} = \{ \pi \setminus \{k \mid \rho(k) \neq \rho_i \} \mid \pi \in \Pi \}$
\State $\Pi_{\rho_j} = \{ \pi \setminus \{k \mid \rho(k) \neq \rho_j \} \mid \pi \in \Pi \}$
\State $\bSigma^{\cap} = [\infty]^{\abs{\Pi_{\rho_i}} \times \abs{\Pi_{\rho_j}}}$
\For {$\pi_k \in \Pi_{\rho_i}$}
    \For {$\pi_{k^{\prime}} \in \Pi_{\rho_j}$}
        \State $\mathcal{A}
          = \sum_{a_1 \in \pi_{k}} \sum_{a_2 \in \pi_{k^{\prime}}}
            \mathbb{I}\{ \norm{\mathbf{x}(a_1)  - \mathbf{x}(a_2)}_1 = 1\}$
            \Comment{Count of adjacent feature combinations across the two pools}
        \If {$\mathcal{A} > 0$}
            \State $\bSigma^{\cap}_{k, k^{\prime}} = 0$
        \EndIf
    \EndFor    
\EndFor

\State \Return $\bSigma^{\cap}$

\end{algorithmic}
\end{algorithm}

\begin{algorithm}[!tb]
\caption{\texttt{PoolAdjacentProfiles}($\mathcal{P}_{q, \epsilon}, \Pi, \mathbf{z}, \bSigma^{\cap}, \Z, \theta$)}
\label{alg:pool-adjacent-profiles}
\begin{algorithmic}[1]
\Require Rashomon set $\mathcal{P}_{q, \epsilon}$, partition $\Pi$, list of pools that can be pooled across profiles $\mathbf{z}$, data $\Z$, Rashomon threshold $\theta$, intersection matrices already seen $\mathcal{S}$
\Ensure Rashomon set $\mathcal{P}_{q, \epsilon}$

\While {$\mathbf{z} \neq \varnothing$}
    \State $(k, k^{\prime}) = \mathbf{z}.\texttt{pop}()$
    \State $\bSigma^{\cap, \prime} = \bSigma^{\cap}$
    \State $\bSigma^{\cap, \prime}_{k, k^{\prime}} = 1$
    \State $\bSigma^{\cap, \prime}_{k, -k^{\prime}} = \infty$, $\bSigma^{\cap, \prime}_{-k, k^{\prime}} = \infty$   \Comment{Cannot pool $\pi_k$ or $\pi_{k^{\prime}}$ with any other pool}
    \State $\Pi^{\prime} = \left(\Pi \setminus \{\pi_k, \pi_{k^{\prime}} \}\right) \cup (\pi_k \cup \pi_{k^{\prime}})$   \Comment{Update $\Pi$ by merging $\pi_k$ and $\pi_{k^{\prime}}$}
    \If {$Q(\Pi^{\prime}; \Z) \leq \theta$}
        \State $\mathcal{P}_{q, \epsilon}
          = \Pi^{\prime} \cup \texttt{PoolAdjacentProfiles}(\mathcal{P}_{q, \epsilon}, \Pi^{\prime}, \mathbf{z}, \bSigma^{\cap, \prime}, \Z, \theta)$
    \EndIf
\EndWhile

\State \Return $\mathcal{P}_{q, \epsilon}$

\end{algorithmic}
\end{algorithm}

\begin{algorithm}[!tb]
\caption{\texttt{PoolProfiles}($\mathcal{P}, \rho_0, \Z, \theta)$}
\label{alg:r-aggregate-across-profiles}
\begin{algorithmic}[1]
\Require Candidates for Rashomon set $\mathcal{P}$, control profile $\rho_0$, data $\Z$, Rashomon threshold $\theta$
\Ensure Rashomon set $\mathcal{P}_{q, \epsilon}$

\State $\mathcal{P}_{q, \epsilon} = \varnothing$
\State $\mathcal{Q} = \texttt{queue}()$
\State $\mathcal{Q}.\texttt{enqueue}(\rho_0)$
\While {$\mathcal{Q} \neq \varnothing$}
    \State $\rho_i = \mathcal{Q}.\texttt{dequeue()}$
    \State $\mathcal{N}(\rho_i)
      = \{ \rho_j \mid \norm{\rho_i - \rho_j}_0 = 1, \rho_j > \rho_i \}$
      \Comment{Neighbors of $\rho_i$ with one additional active feature}
    \For {$\rho_j \in \mathcal{N}(\rho_i)$}
        \State $\mathcal{Q}.\texttt{enqueue}(\rho_j)$
        \For {$\Pi \in \mathcal{P}$}
            \State $\bSigma^{\cap} =
              \texttt{IntersectionMatrix}(\Pi, \rho_i, \rho_j)$
              \Comment{See Algorithm \ref{alg:intersection-matrix}}
            \State $\mathbf{z} = \{ (k, k^{\prime}) \mid \bSigma^{\cap}_{k, k^{\prime}} = 0 \}$
            \State $\mathcal{P}_{q, \epsilon}
              = \texttt{PoolAdjacentProfiles}(\mathcal{P}_{q, \epsilon}, \Pi, \mathbf{z}, \bSigma^{\cap}, \Z, \theta)$
              \Comment{See Algorithm \ref{alg:pool-adjacent-profiles}}
        \EndFor
    \EndFor
\EndWhile

\State \Return $\mathcal{P}_{q, \epsilon}$

\end{algorithmic}
\end{algorithm}

\begin{algorithm}[!tbh]
\caption{Implementation of caching object used in Algorithm \ref{alg:r-aggregate-profile}}
\label{alg:r-aggregate-cache}
\begin{algorithmic}

\State $\mathcal{S} = \texttt{cache()}$ \Comment{Initialize caching object}
\Indent
    \State $C = \{ \}$ 
\EndIndent

\State $\mathcal{S}\texttt{.insert}(\bSigma, i, j)$ \Comment{Extract and insert $\bSigma_{\fix}$}
\Indent
    \State $\bSigma[i, j:(R-2)] = \texttt{NA}$ 
    \State $C = C \cup \{ \bSigma \}$
\EndIndent

\State $\mathcal{S}\texttt{.seen}(\bSigma, i, j)$ \Comment{Extract and check presence of $\bSigma_{\fix}$}
\Indent
    \State $\bSigma[i, j:(R-2)] = \texttt{NA}$
    \State \textbf{return} $\bSigma \in C$
\EndIndent

\end{algorithmic}
\end{algorithm}

\begin{algorithm}[!tbh]
\caption{\texttt{select\_feasible\_combinations}$(K, \theta)$}
\label{alg:find-feasible-combs}
\begin{algorithmic}[1]
\Require $K$ list of $n$ sorted lists containing a numerical score, $\theta$ threshold
\Ensure $F$, list of lists of length $n$ with indices of elements from each of $K_i$ such that their sum is less than $\theta$

\State $F = \{ \}$
\State $n = \texttt{len}(K)$

\lIfThen {$n = 0$} {\textbf{return} $\{\}$}

\State $K_{1, \text{feasible indices}} = \{ i \mid K_1[i] \leq \theta \}$

\If {$n = 1$}
    \State $F = \{K_{1, \text{feasible indices}}\}$
    \State \Return $F$
\EndIf

\State $x = \sum_{j=2}^n K_j[1]$ \Comment{Smallest possible sum from remaining lists}

\For {$i \in K_{1, \text{feasible indices}}$}
    \State $\theta_i = \theta - K_1[i]$
    \lIfThen {$\theta_i < x$} {\textbf{break}} \Comment{Early stopping if even the best case exceeds $\theta$}
    \State $F_i = \texttt{select\_feasible\_combinations}(K[2:], \theta_i)$
    \For {$f \in F_i$}
        \State $F.\texttt{insert}([i]\texttt{.append}(f))$
    \EndFor
\EndFor

\Return $F$

\end{algorithmic}
\end{algorithm}

\section{Appendix to simulations}
\label{appendix:additional-sims}

\subsection{Supplementary information for \Cref{section:simulations}}

We begin by presenting additional information about the two-drug experiment (the first simulation) in the main paper.  We present additional details on the comparison methods%
 and additional results on the composition of the RPS, with reference to the Hasse diagram of Figure~\ref{fig:hasse-sim-1} in the main text.

We begin with %
the details of the comparator methods. 
For Bayesian Lasso, we set $\lambda = 5^{-1}$ and the Inverse-Gamma parameters to be $\tau_a = 10^{-1}, \tau_b = 10^{-1}$ where $\tau_a$ and $\tau_b$ are shape and scale respectively. We ran 3 chains for $5000$ iterations discarding the first $2000$ as burn-in and thinning every 2 samples.
For Spike-and-Slab Lasso, we set $\lambda_0 = 1$ and $\lambda_1 = 2$. We initialized $\theta = 0.1$ and used the hyperprior, $\text{Beta}(1, 1)$ to update $\theta$. We ran 3 chains for $5000$ iterations discarding the first $2000$ as burn-in and thinning every 2 samples.
For PPMx, we used the cohesion function from the Dirichlet Process with $M = 1$. We used the auxiliary similarity function with Gaussian density with parameters $m_0 = 0, s_0^2 = 4, v = 0.5$.
For frequentist Lasso, we used $\lambda = 5 \times 10^{-3}$ for regularization. When boostrapping, we sampled the data with replacement 500 times. For bootstrap, we used the penalized loss, $L$, as a proxy for the posterior through $\Pr(\bbeta \mid \X, \y) \propto \exp\left\{-L\right\}$, using the same score scale as in the empirical RPS implementation.

The results from the simulations are presented in \Cref{fig:sim-1-results} in the main text. The metrics reported here are as follows: 

\begin{enumerate}
\item Overall mean-squared error (MSE): Suppose $\widehat{y}_i$ and $y_i$ are the estimated and true outcomes for unit $i$, then the overall MSE is defined as
\begin{align*}
    \text{MSE} &= \frac{1}{n} \sum_{i=1}^n (\widehat{y}_i - y_i)^2.
\end{align*}
\item Best feature set coverage: Let $\pi^{\star}$ and $\widehat{\pi}^{\star}$ be the true and estimated set of features with the highest effect (the highest level of Amoxicillin with any level of Ibuprofen). Then, we define the best feature set coverage as the intersection-over-union of these two sets
\begin{align*}
    \text{IOU} &= \frac{\abs{\pi^{\star} \cap \widehat{\pi}^{\star}}}{\abs{\pi^{\star} \cup \widehat{\pi}^{\star}}}.
\end{align*}
\item MSE for feature outcome: Let $y_{\max}$ be the true best policy effect and $\widehat{y}_{max}$ be the estimated best treatment effect. Then the best feature outcome MSE is
\begin{align*}
    \text{MSE}_{\text{best}} &= (\widehat{y}_{\max} - y_{\max})^2.
\end{align*}
\end{enumerate}
For the Bayesian methods, including RPS, we reported the performance metric averaged across all partitions in the posterior i.e., the posterior mean (or the posterior mean approximation for the RPS).

By construction, the outcome depends only on the dosage of drug A and not on drug B.
Therefore, the nonzero upward marginal effects from increasing drug A recur in parallel
across all levels of drug B. In the TVA marginal-effects representation, this produces
nested, highly overlapping regressors. As a result, $\ell_1$-based support recovery is
delicate in this setting, even with preconditioning, because small perturbations can change
which marginal increments are selected as zero. Our $\ell_0$-based approach instead
regularizes only the number of pools and does not require specifying a particular
dependence structure among these marginal effects. By design, all non-zero marginal increases in outcomes as we increase dosage levels are correlated.

Additionally, there are three key observations here:
\begin{itemize}
    \item Single point estimates such as TVA and Lasso consistently miss out on coverage, even at high sample sizes. RPS, by looking at all near-optimal models is able to recover the true set of treatments with highest effect at a higher rate. In fact, the true partition in \Cref{fig:hasse-sim-1} is present in our RPS 100\% of the time. Note that the middle panel of \Cref{fig:sim-1-results} uses a metric that averages the overlap between the true top pool and the top pool induced by partitions in the posterior/RPS, so it reflects how concentrated the near-optimal set is around the truth, not whether the exact true partition appears anywhere in the RPS.
    \item Bayesian Lasso and Spike-and-Slab Lasso can also be used to estimate the set of near-optimal models (albeit without theoretical guarantees). Even these methods perform poorly compared to RPS. This is because the experiment was designed to break down the independence assumption of Lasso while RPS puts no such assumption.
    \item PPMx is particularly striking because it does not impose any independence assumptions. However, it also has very little structure since it is primarily a random clustering model. It does really well on MSE but poorly on coverage indicating that it finds \textit{some} very good representation but not one that is very useful. The interpretability of PPMx clusters is not unique to this setup \citep{argiento2024clustering}.
\end{itemize}

We visualize the RPS through a heat map. An example heatmap with instructions on how to read it is shown in \Cref{fig:heatmap-example}. We also use these heatmaps in our empirical data examples in \Cref{appendix:real-data}.

\begin{figure*}[!tb]
    \centering
    \includegraphics[height=2in]{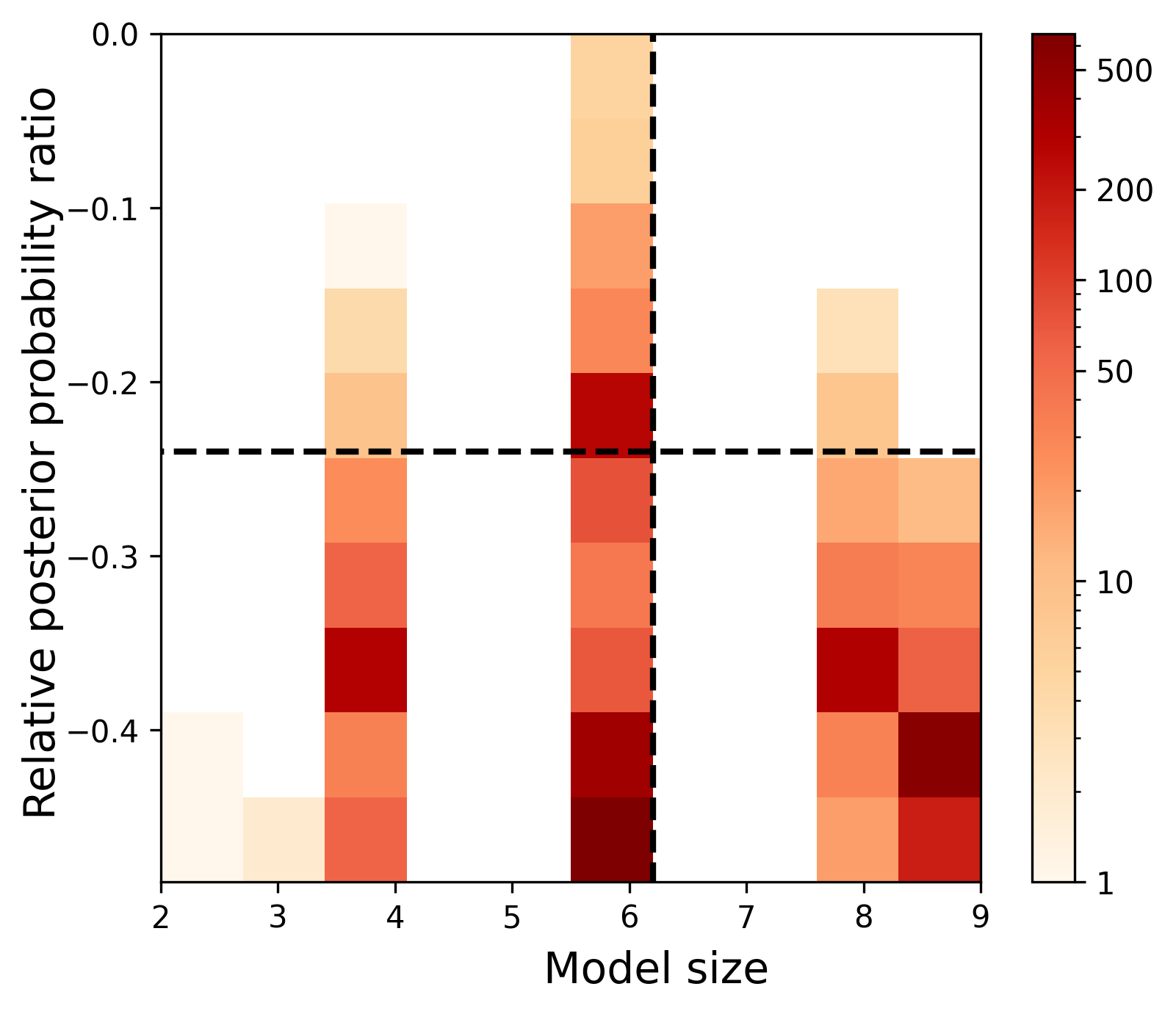}
    \caption{Visualizing the Rashomon set through a heat map. This heatmap actually reflects a 2D histogram binned by the model size (number of pools in a partition) and the relative posterior probability ratio i.e., $(\Pr(\Pi \mid \Z) - \max \Pr(\Pi \mid \Z)) / \max \Pr(\Pi \mid \Z)$. The color of the bin reflects the number of times, averaged per simulation, a model at that sparsity and probability (distinct models may be in the same bin) appear in some Rashomon set. One might refine the set of partitions further by the probability and the sparsity. For example, if we want models with a relative probability of at least -0.25, then we look only at models that are above the dashed black horizontal line. If we want models with fewer than 6 pools, then we look only at models to the left of the dashed black vertical line. If we want both criteria to be satisfied, we look at the top left box.}
    \label{fig:heatmap-example}
\end{figure*}

\begin{figure*}[!tb]
    \centering
    \includegraphics[height=3.5in]{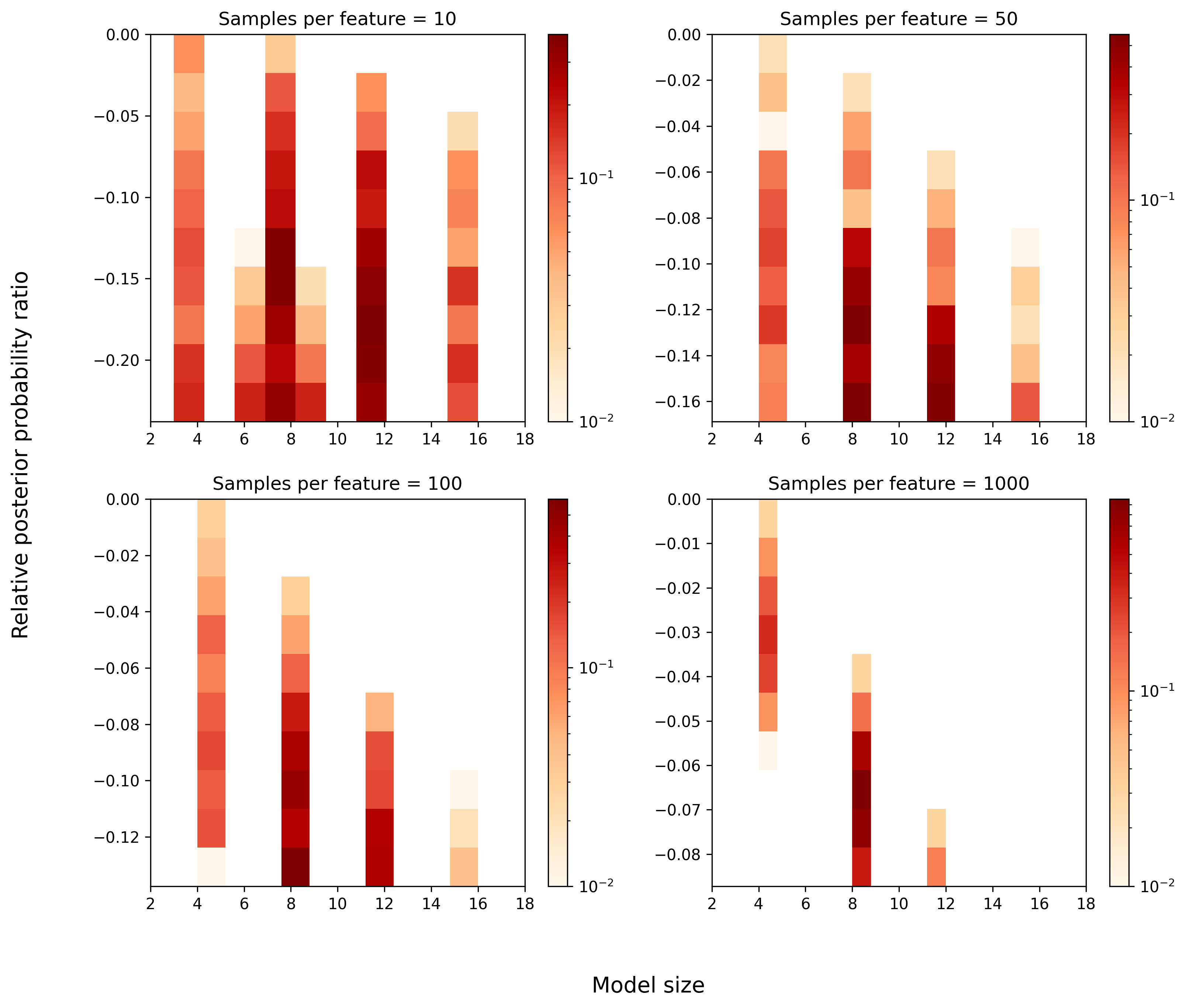}
    \caption{Visualizing the Rashomon set in Simulation 2. Notice how as the size of the data set grows, the Rashomon set concentrates around a few very good models, one of which corresponds to the data generating process.}
    \label{fig:heatmap-sim-1}
\end{figure*}

For this simulation, we visualize the RPS in a heatmap in Figure \ref{fig:heatmap-sim-1}. As the number of samples increases, the number of diverse partitions in the Rashomon set becomes smaller since we are more confident in our estimates.

\subsection{Additional simulation using posterior density}
In this simulation, we further explore the contrast between RPS and several Bayesian alternatives.  In making this comparison, we emphasize that these methods differ both in terms of their approach to model uncertainty and in terms of their prior structure.  For that reason, the relative posterior density compared to the MAP is not comparable.  The implications of looking at models within a small window of the MAP will be very different, for example, if the posterior is highly concentrated around the MAP compared to a more diffuse posterior.  We explore this distinction further in this simulation by also considering high posterior regions.
In terms of setup, consider a setting with four features. Each feature takes on four ordered factors, $\{0, 1, 2, 3\}$. There are $2^4 = 16$ possible combinations of active ($>0$) and inactive (level 0) features. We refer to ``control'' as the setting where all features are inactive, corresponding in to an experiment with four distinct interventions, each of which has four levels (control, plus low-medium-high).
{As an example, say that the data represent an information experiment.  Each feature is a technology (phonecall, social media post, letter, etc) and the levels of the feature are intensity (none, hourly, daily, weekly, etc.).  Our simulation setup corresponds to a setting where exposure to the information matters, but the outcome is the same for all contact frequencies.  This setup also corresponds to our first data example, where previous analysis found that, while matching donations increases charitable giving, the amount of the match is inconsequential~\citep{karlan2007does}.  Our analysis contradicts this conclusion. 

We define $\beta_{a}$ and $\sigma^2_a$ and the mean and variance of the normal distribution we use to simulate outcomes, where $a$ corresponds to a vector of indicators for whether or not a particular feature is at a level above control (so $\beta_{1,0,0,0}$, for example means that the first feature is at at level above control and the others are not). We set the following to have a non-zero outcome:}
\begin{align*}
    \beta_{(0, 0, 0, 1)} &= 4.4, \ \sigma^2_{(0, 0, 0, 1)} = 1, \ 
    \beta_{(0, 1, 0, 0)} = 4.3, \  \sigma^2_{(0, 1, 0, 0)} = 1, \
    \beta_{(0, 1, 0, 1)} = 4.45, \ \sigma^2_{(0, 1, 0, 1)} = 1, \\ 
    \beta_{(1, 0, 1, 0)} &= 4.5, \ \sigma^2_{(1, 0, 1, 0)} = 1.5, \
    \beta_{(1, 1, 1, 1)} = 4.35, \  \sigma^2_{(1, 1, 1, 1)} = 1. 
\end{align*}
All other settings have outcome $\beta = 0$ and variance $\sigma^2 = 1$. %
We generated data with $n_a = 30$ data points per feature combination (so $30\times 16=480$ total data points).  Each vector $a$ has a different number of feature combinations depending on which features are active.\footnote{For example,  $a=(1, 0, 0, 0)$ will have three feature combinations, $[(1, 0, 0, 0), (2, 0, 0, 0), (3, 0, 0, 0)]$ and $a=(1, 1, 0, 0)$ will have $3^2 = 9$ feature combinations, $[(1, 1:3, 0, 0), (2, 1:3, 0, 0), (3, 1:3, 0, 0)]$.} The outcomes were drawn from $\mc{N}(\beta_a, \sigma_a^2)$. The setting where $a=(1, 0, 1, 0)$ should have the highest outcomes, so it is the true ``best'' setting or profile. %
 We averaged the results over $r = 100$ simulations.

 We also contrast RPS with Bayesian Lasso \citep{park2008bayesian}, Spike-and-Slab Lasso \citep{rovckova2018spike}, and PPMx \citep{muller2011ppmx}. Lasso is shown, both, as a point estimate and through bootstrapping, where the bootstrap ensemble is thought of as a ``posterior'' (even though it shows sampling variability only). We discuss the implementation details in Appendix \ref{appendix:additional-sims}. It is important to note that the full posteriors for the Bayesian baselines were approximated based on MCMC sampling.

\begin{figure}[!tb]
    \centering
    \begin{subfigure}[t]{0.45\textwidth}
        \centering
        \includegraphics[height=2in]{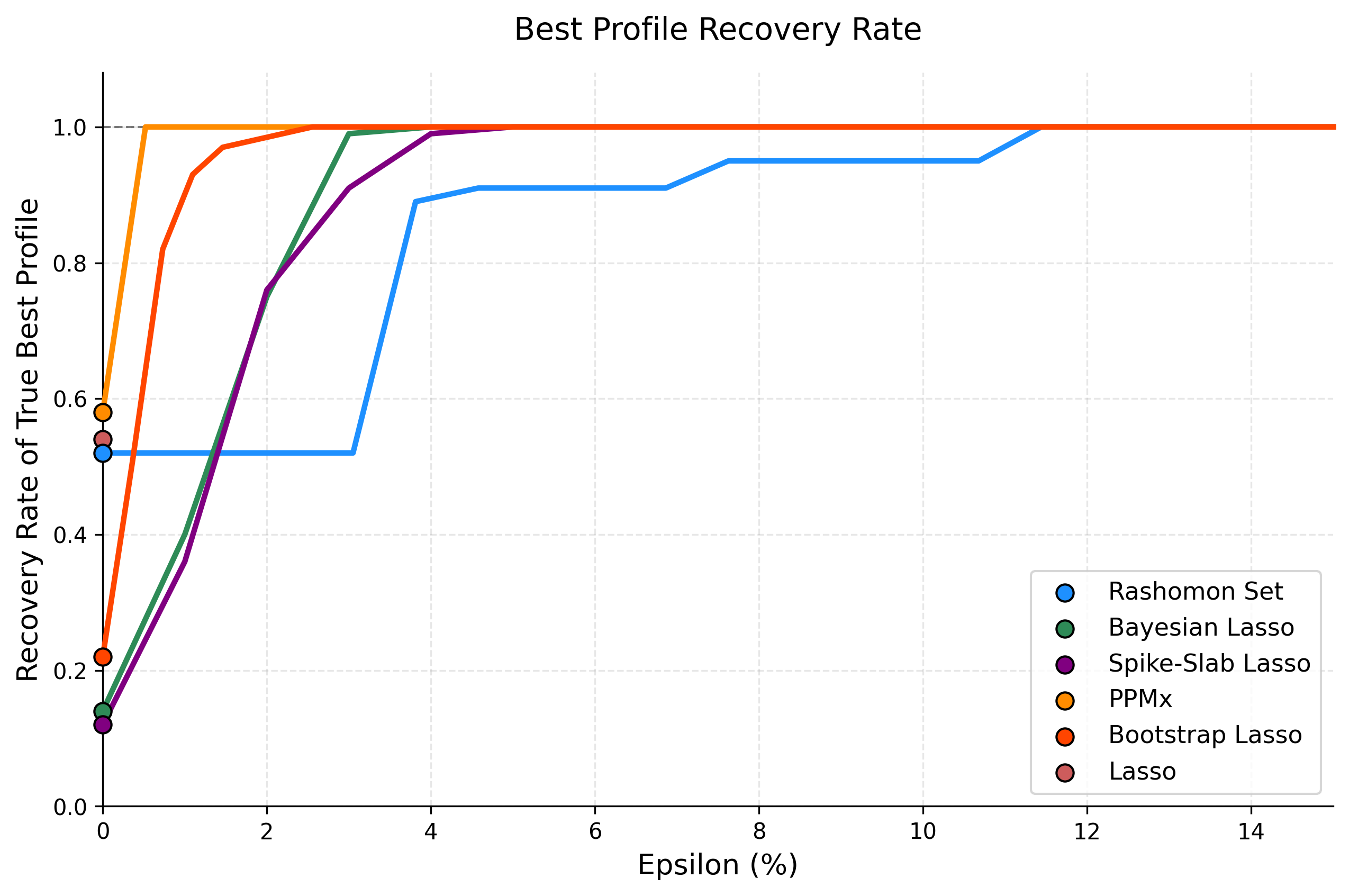}
    \end{subfigure}%
    ~ 
    \begin{subfigure}[t]{0.45\textwidth}
        \centering
        \includegraphics[height=2in]{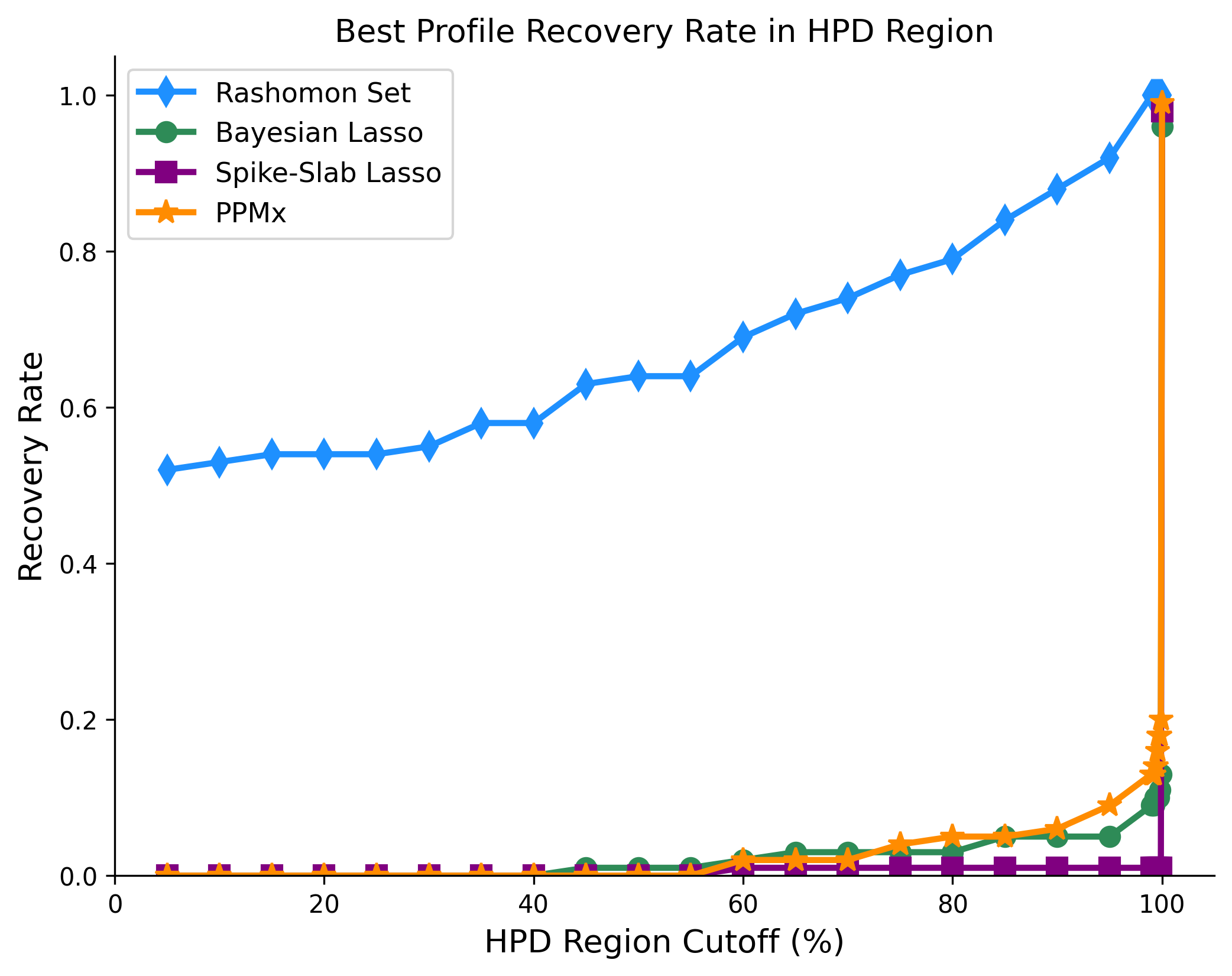}
    \end{subfigure}
    \caption{\tiny Simulation results. The left panel shows how often the true best profile is discovered as we increase the threshold $\epsilon$. The right panel shows how often the true best profile is recovered in the highest posterior density (HPD) region. }
    \label{fig:reff-sim-2}
\end{figure}

The left panel of Figure \ref{fig:reff-sim-2} tells us how often the true highest expected outcome feature is correctly identified in RPS as a function of the threshold $\epsilon$.  Since $\epsilon$ is defined as in Equation \ref{eq:RPS}, its scale is dependent on the MAP for each method, and is not a fair representation across methods. Focusing on just $\epsilon$ masks the shape of the posterior. Therefore, we also present the highest posterior density (HPD) region. For a cutoff $0 \leq p \leq 1$, the HPD region includes the smallest region, centered around the mode, accounting for $p$ posterior probability. When the posterior is concentrated and the mode accounts for more than 90\% of the posterior mass, the HPD region is often the empty set for low cutoffs, as seen in the baselines. Additional insights are described in Figures \ref{fig:reff-sims-2-hpd} and \ref{fig:reff-sims-2-barchart} in Appendix \ref{appendix:additional-sims}.

For Bayesian Lasso, we set $\lambda = 10^{-5}$ and the Inverse-Gamma parameters to be $\tau_a = 10^{-1}, \tau_b = 10^{-1}$ where $\tau_a$ and $\tau_b$ are shape and scale respectively. We ran 3 chains for $4000$ iterations discarding the first $1000$ as burn-in and thinning every 2 samples.
For Spike-and-Slab Lasso, we set $\lambda_0 = 20$ and $\lambda_1 = 0.5$. We initialized $\theta = 0.3$ and used the hyperprior, $\text{Beta}(1, 1)$ to update $\theta$. We ran 4 chains for $3000$ iterations discarding the first $800$ as burn-in and thinning every 2 samples.
For PPMx, we used the cohesion function from the Dirichlet Process with $M = 1$. We used the auxiliary similarity function with Gaussian density with parameters $m_0 = 0, s_0^2 = 4, v = 0.5$. For details regarding these parameters, see \texttt{gaussian\_ppmx} of \citet{ppmSuite} and \citet{page2018calibrating}. We ran 1 chain for $1100$ iterations discarding the first $100$ as burn-in.
For frequentist Lasso, we used $\lambda = 5 \times 10^{-3}$ for regularization. When boostrapping, we sampled the data with replacement 1000 times. For bootstrap, the posterior is not clearly defined. We simply used the penalized loss, $L$, as a proxy for the posterior through $\Pr(\bbeta \mid \X, \y) \propto \exp\left\{-L\right\}$. It is important to note that this reflects only data sampling variability and not model variability unlike the Bayesian methods.

Figures \ref{fig:reff-sims-2-hpd} and \ref{fig:reff-sims-2-barchart} show additional results comparing and contrasting different methods. From Figure \ref{fig:reff-sims-2-hpd}, it is evident that that the Bayesian methods have a very concentrated posterior that is centered around the MAP. In other words, the MAP accounts for most of the posterior. This explains why the Bayesian methods are able to hit the ceiling 1.0 at much smaller $\epsilon$ values compared to RPS, which has a more flat posterior.

\begin{figure}[!tb]
    \begin{subfigure}[t]{0.4\textwidth}
        \centering
        \includegraphics[height=2in]{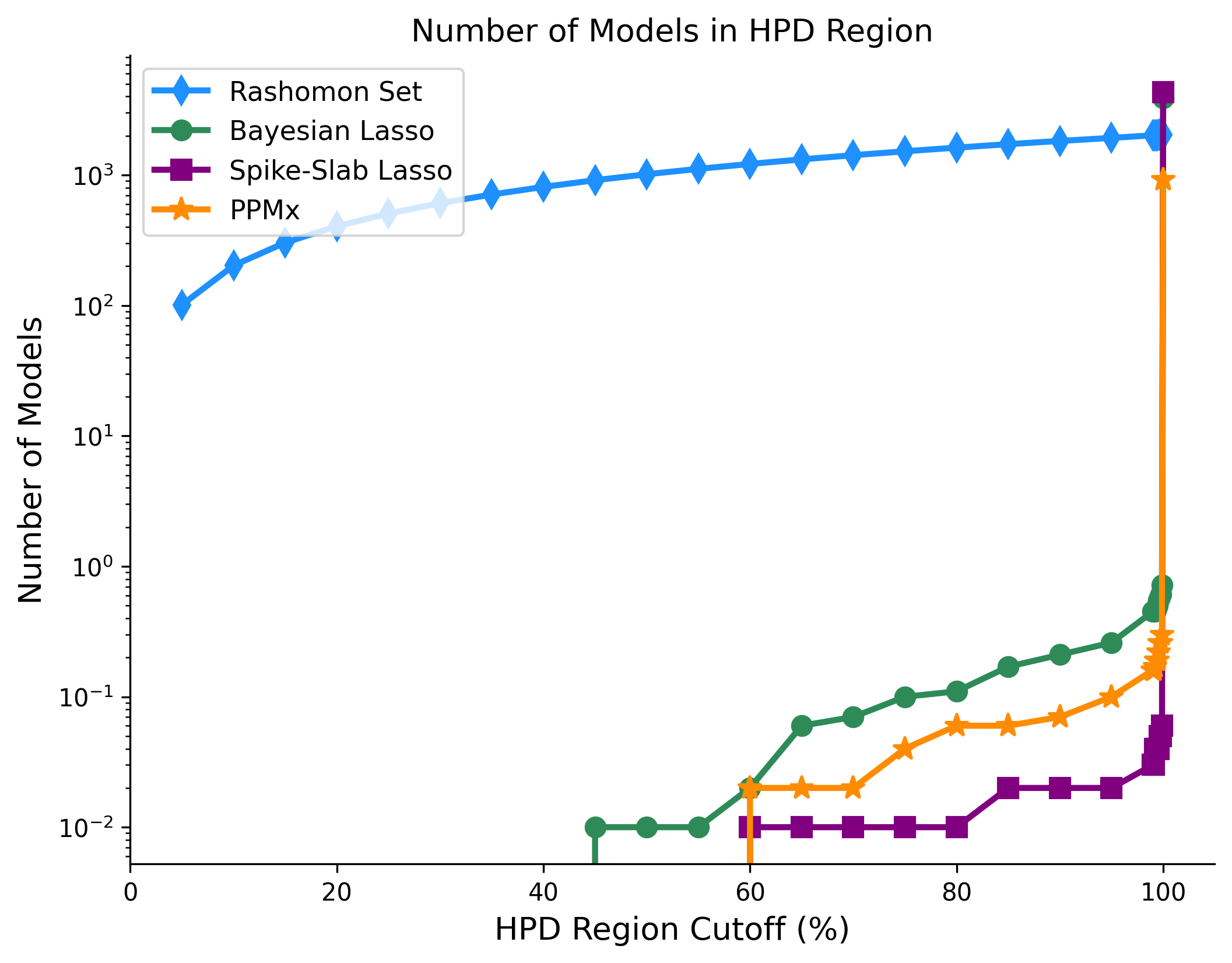}
    \end{subfigure}%
    ~ 
    \begin{subfigure}[t]{0.4\textwidth}
        \centering
        \includegraphics[height=2in]{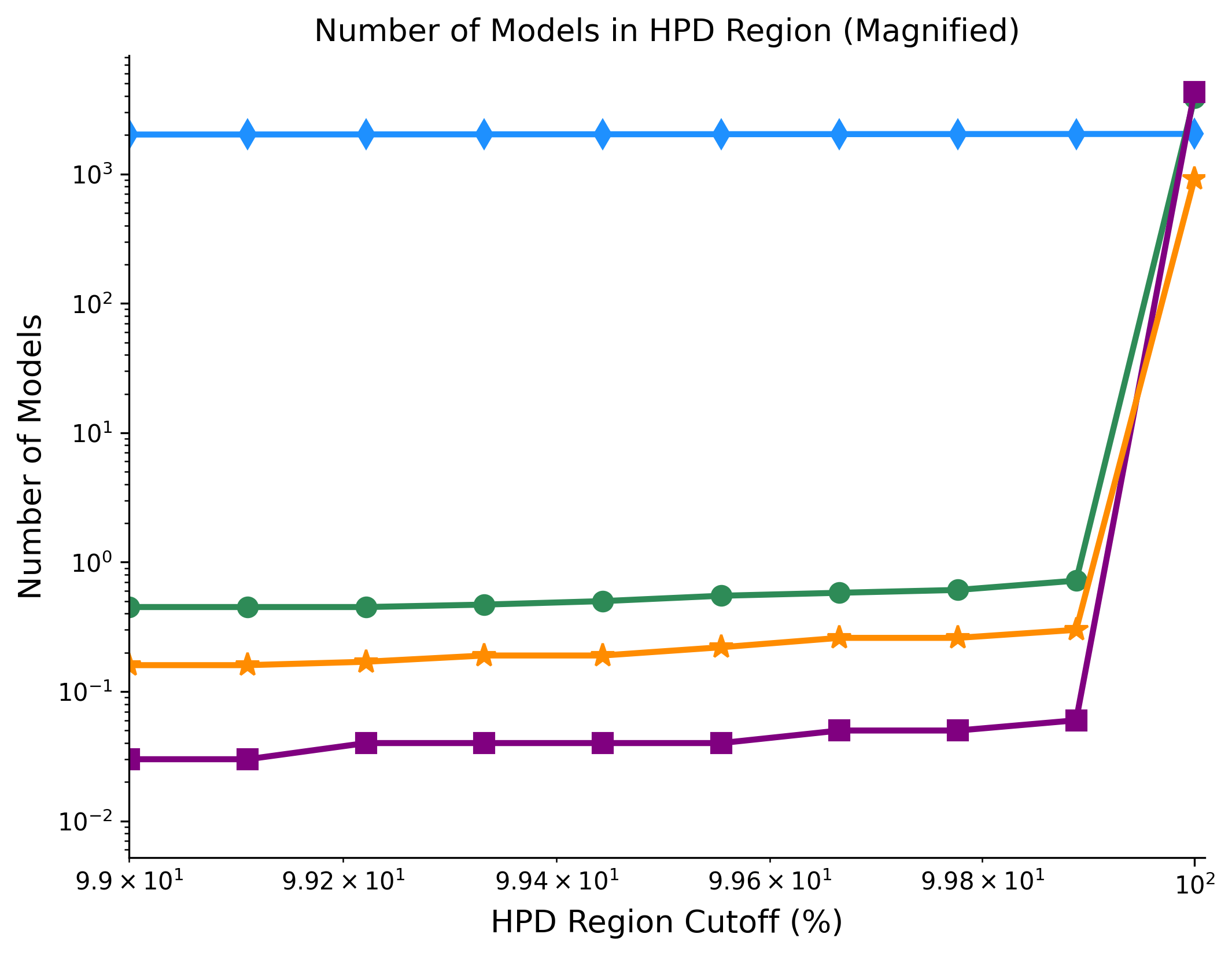}
    \end{subfigure}%
    \\
    \begin{subfigure}[t]{0.3\textwidth}
        \centering
        \includegraphics[height=2in]{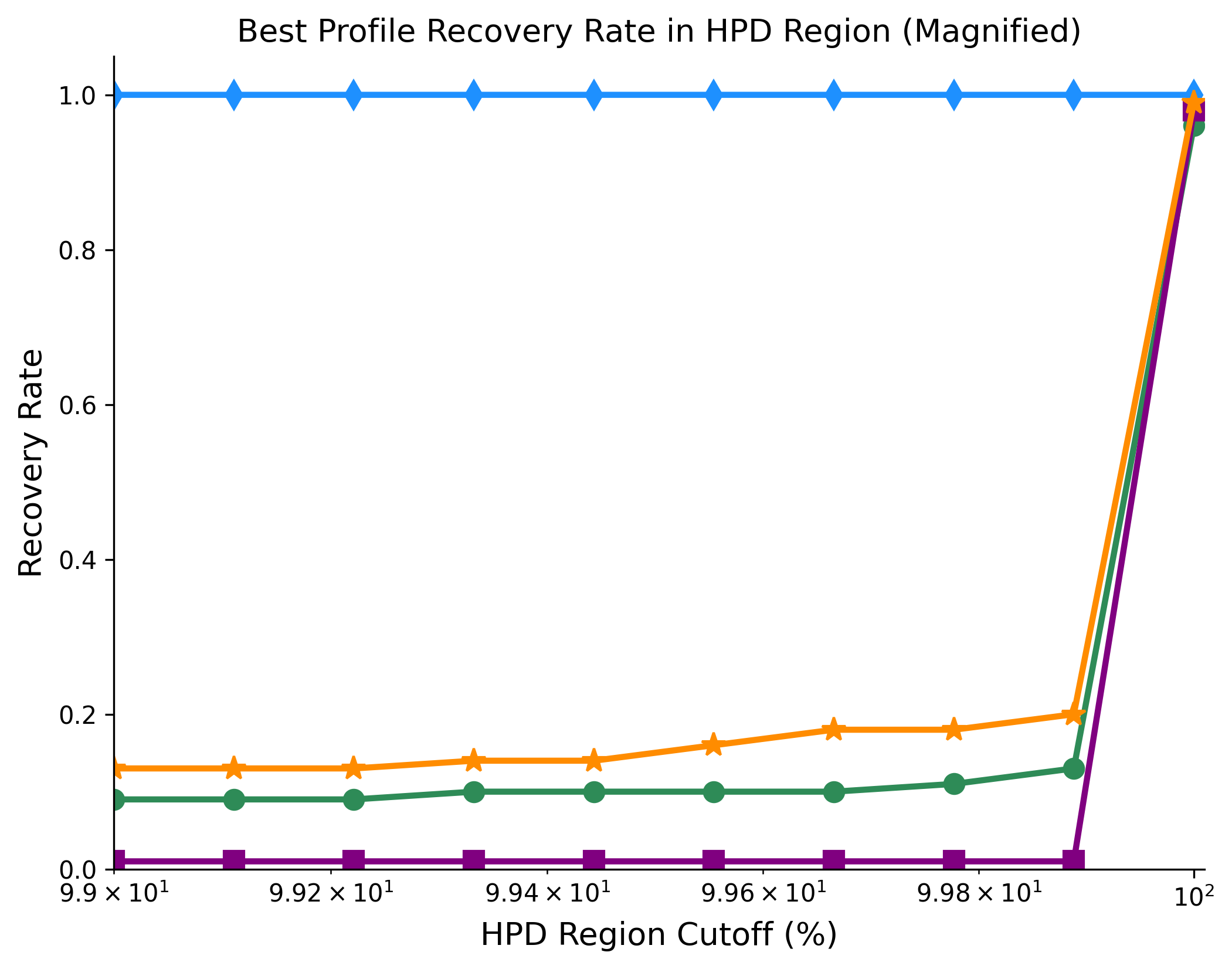}
    \end{subfigure}
    \caption{\small Simulation 1. The top two panels show the number of models in the highest posterior density (HPD) region for different Bayesian methods. The top right panel zooms in on the range [99\%, 100\%]. The bottom panel zooms in on the range [99\%, 100\%] in the left panel of Figure \ref{fig:reff-sim-2}.}
    \label{fig:reff-sims-2-hpd}
\end{figure}

\begin{figure}
    \centering
    \includegraphics[height=1.75in]{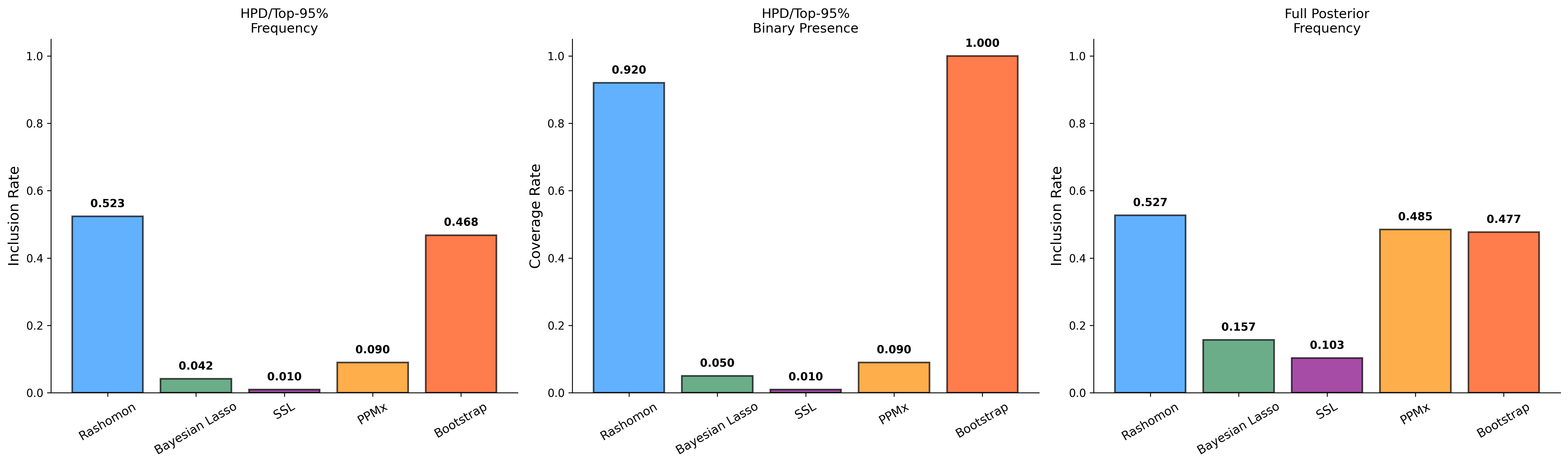}
    \caption{\small Simulation 1. The left bar plot shows how often there is a model in the 95\% HPD that correctly identifies the best profile. The middle bar plot shows how often there is at least one model in the 95\% HPD that correctly identifies the best profile. The right bar plot shows often there is a model in the full posterior (or the RPS) that correctly identifies the best profile.}
    \label{fig:reff-sims-2-barchart}
\end{figure}

\subsection{Runtime and approximation error.} 
\begin{figure}[!tb]
    \begin{subfigure}[t]{0.4\textwidth}
        \centering
        \includegraphics[height=2in]{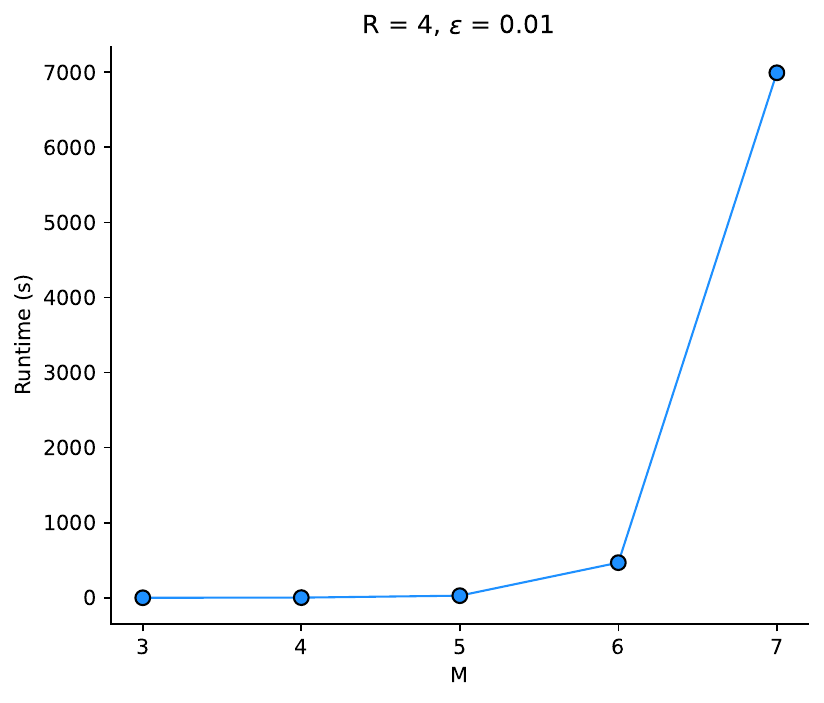}
    \end{subfigure}%
    ~ 
    \begin{subfigure}[t]{0.4\textwidth}
        \centering
        \includegraphics[height=2in]{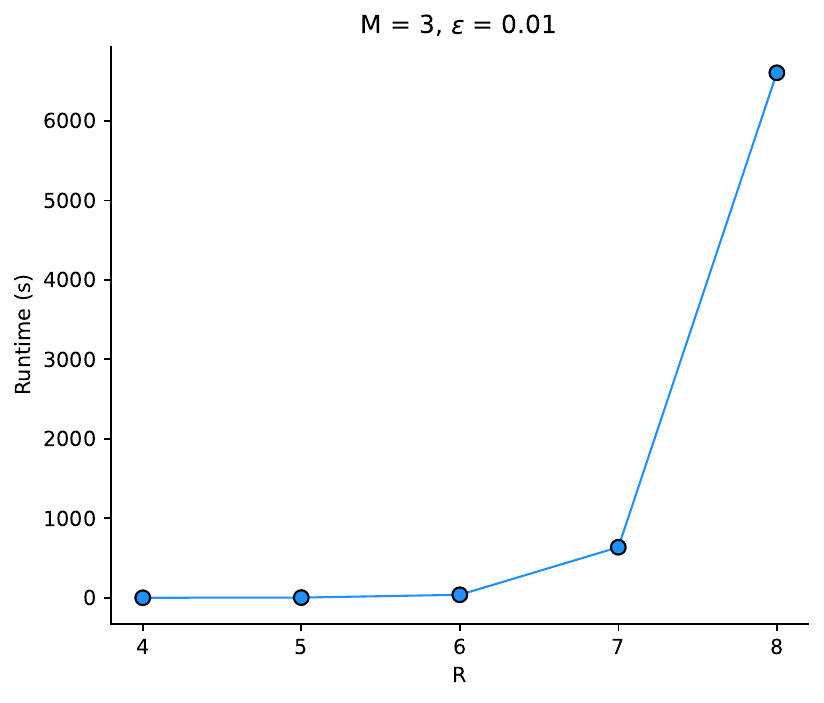}
    \end{subfigure}%
    \\
    \begin{subfigure}[t]{0.3\textwidth}
        \centering
        \includegraphics[height=2in]{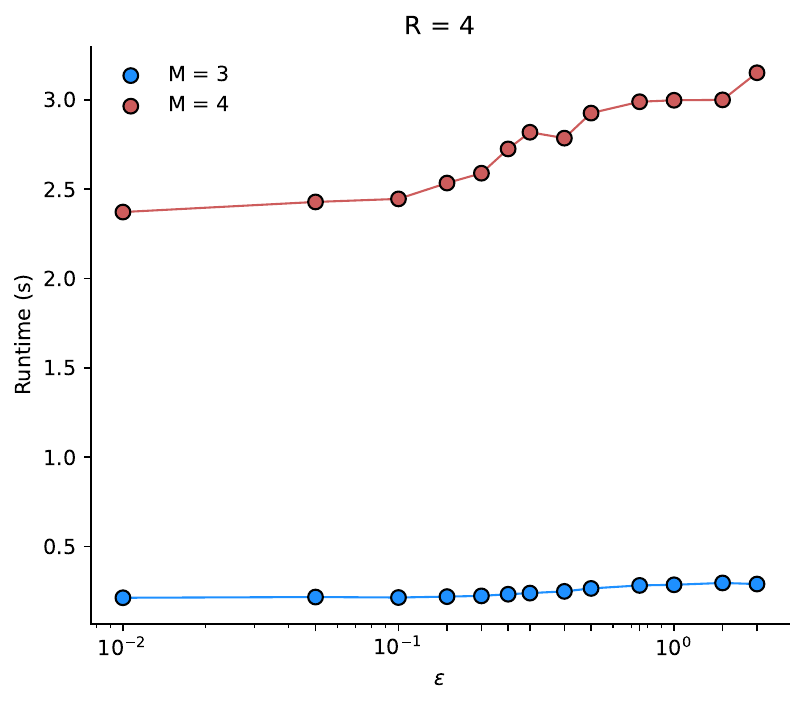}
    \end{subfigure}
    \caption{\small Runtime of enumerating RPS for a single profile.}
    \label{fig:runtimes}
\end{figure}

To illustrate the complexity of enumerating the RPS, we will only look at a profile with $M$ features where each feature $m$ takes on $R_m = R$ factor levels each at various $\epsilon$ thresholds. For each $(M, R)$ setting, we randomly generate a partition matrix with a Bernoulli(0.5) distribution. For each pool $\pi_i$ in this partition $\Pi$, we pick $\beta_i \sim \text{Uniform}(2, 6)$ and outcomes from $\mc{N}(\beta_i, 1)$. In every setting, we generate $n = 30$ samples per feature combination. The runtimes, averaged over 10 random seeds, are shown in \Cref{fig:runtimes}.

\begin{figure}[!tb]
    \begin{subfigure}[t]{0.4\textwidth}
        \centering
        \includegraphics[height=2in]{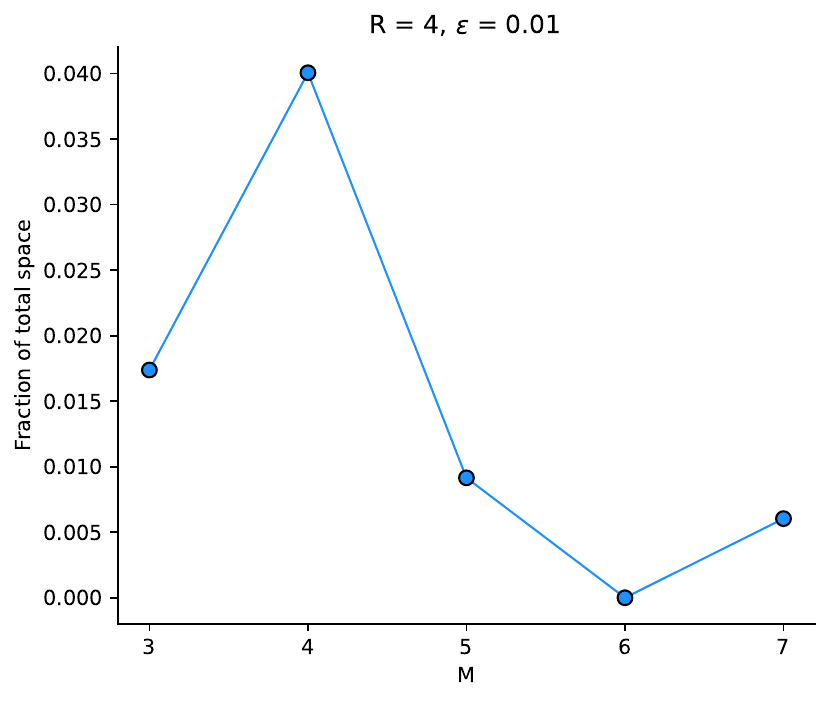}
    \end{subfigure}%
    ~ 
    \begin{subfigure}[t]{0.4\textwidth}
        \centering
        \includegraphics[height=2in]{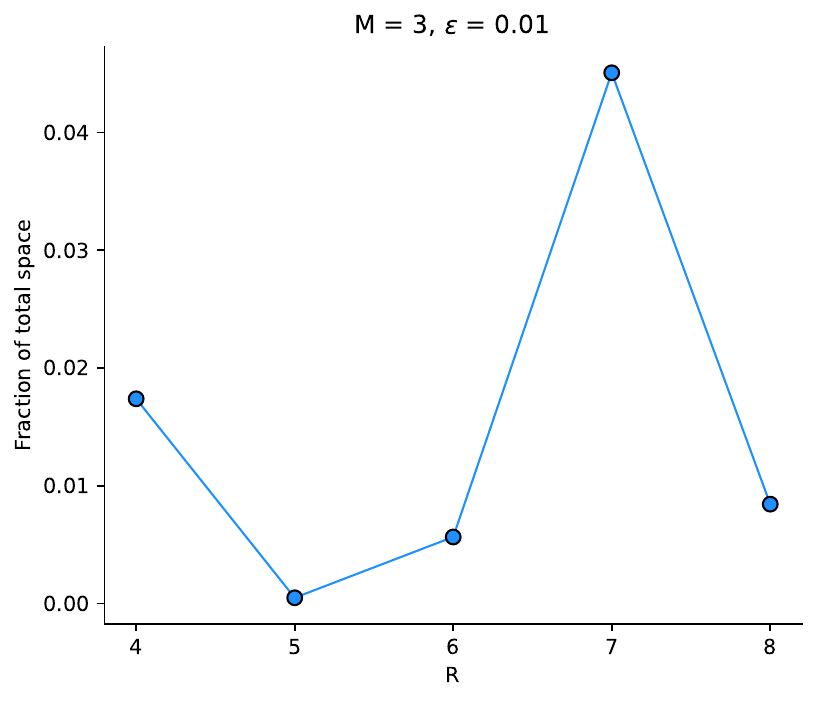}
    \end{subfigure}
    \caption{\small Fraction of the model space covered by the RPS.}
    \label{fig:frac-RPS-covered}
\end{figure}

As we can see, the runtime increases at an exponential rate in $M$ and $R$. On the other hand, for a fixed $\epsilon$, as $M$ and $R$ increase, the number of models in the RPS also increases even though the RPS covers roughly the same fraction of the entire model space (see \Cref{fig:frac-RPS-covered}). Our empirical analyses in Section \ref{section:real-data} took less than 3 hours on a personal computer despite having a larger number of features because we chose smaller Rashomon thresholds (thus fewer models in the RPS).\footnote{All simulations and real data analyses were performed on 2.3 GHz 8-Core Intel Core i9 processor with 16 GB memory.}

\begin{figure}[!tb]
    \begin{subfigure}[t]{0.4\textwidth}
        \centering
        \includegraphics[height=2in]{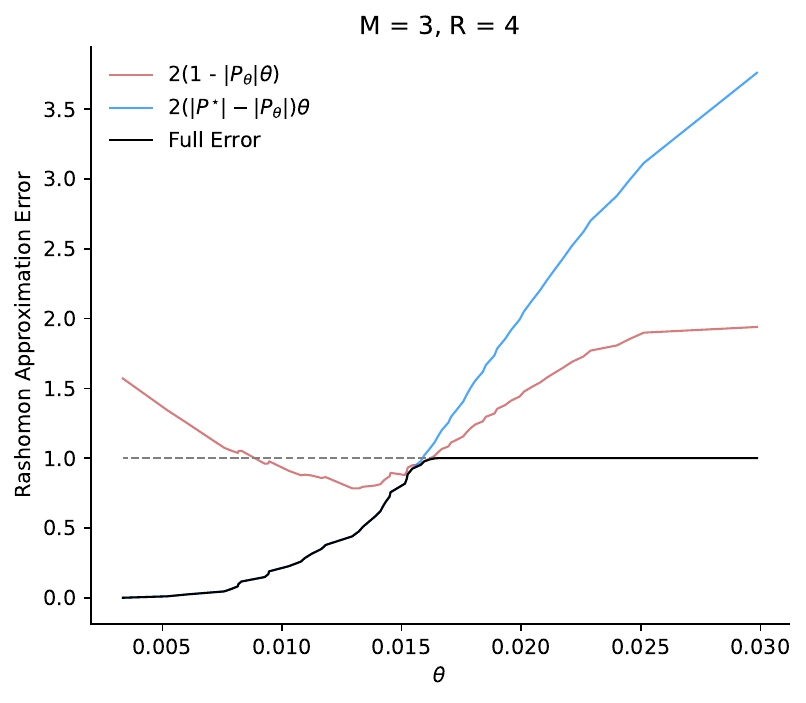}
    \end{subfigure}%
    ~ 
    \begin{subfigure}[t]{0.4\textwidth}
        \centering
        \includegraphics[height=2in]{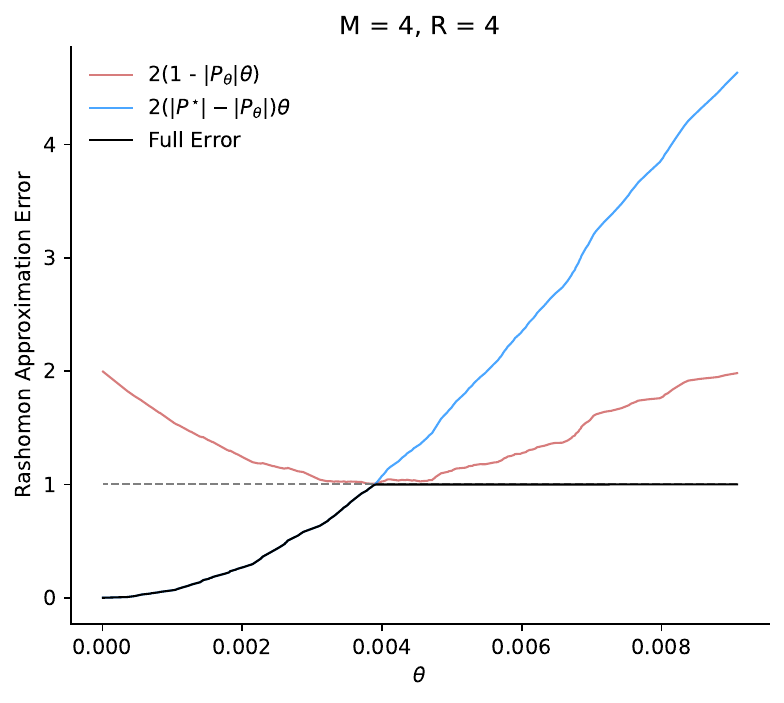}
    \end{subfigure}
    \caption{\small Rashomon approximation error as in \Cref{thm:marginal_approx_rashomon}.}
    \label{fig:sims-approx-error}
\end{figure}

In \Cref{fig:sims-approx-error}, we show the full Rashomon approximation error as in \Cref{thm:marginal_approx_rashomon} for two specific settings. We show both of the error curves illustrating when each case kicks in as we vary $\theta$.

\section{Generalizations}
\label{section:generalization}

Here, we consider two generalizations of the methods discussed so far. First, we consider a family of heterogeneous effects functions beyond just heterogeneity splits. For example, there might be some heterogeneity in \emph{slopes} (and slopes of slopes, and so on).  Second, we extend our method to pool on the space of the covariance between coefficients, rather than on the coefficients themselves. This means that coefficients no longer need to be exactly equal but, instead, related through a sparse covariance structure.

\subsection{Pooling higher order derivatives.}

We ask whether, given some feature combination $k = (k_1, \dots, k_M)$, the marginal effect of increasing, say, $k_1$ has a linear effect. That is, we can just as simply allow for outcomes as we increase the intensity up a given feature that is not just a step function, but one that checks if there is a linear relationship.\footnote{Extensions of this kind can be made to accommodate higher order derivatives and other bases as well, e.g., sinusoidal effects.} In this case, there is no ``large'' versus ``small'' effect and no natural pool in the space of levels. However, there is a natural low dimensional effect and even pools when considering the space of slopes. The result is a framework that captures extensions of Bayesian treed models (e.g.,~\cite{chipman2002bayesian}).

Before we proceed, we first generalize the notion of pools described in Definition \ref{def:pool}.

\begin{definition}[Generalization of pools]
\label{def:pool-general}
Given $M$ features taking on $R$ partially ordered values each and some function $g(k, \bbeta)$, a \textit{pool} $\pi$ is a set of feature combinations $k$ whose outcomes are given by $g(k, \bbeta_{\pi})$ where $\bbeta_{\pi}$ depends on $\pi$.
\end{definition}

It is easy to see that the original pool defined in Definition \ref{def:pool} is recovered by setting $g(k, \beta_{\pi}) = \beta_{\pi}$.

For instance, suppose we are interested in linear effects. Then the regression equation for pool $\pi$
\begin{align*}
    y = g(k, \bbeta_{\pi})  
    = \beta_{\pi, 0} + \sum_{m=1}^M \beta_{\pi, m} k_m
    = \beta_{\pi, 0} + \bbeta^{\intercal}_{\pi} k
\end{align*}
where $\bbeta_{\pi}$ is the linear slope within that pool. The estimated outcome for feature combination $k \in \pi$ is $\widehat{y} = f(k, \widehat{\bbeta}_{\pi} ; \Pi)$, where $\widehat{\bbeta}_{\pi}$ is estimated within each pool using some procedure like least squares.

For some partition $\Pi$, define the block vector $\bbeta = [\bbeta_{\pi_1}, \dots, \bbeta_{\pi_{\abs{\Pi}}}]$ where $\pi_i \in \Pi$. Then, the general outcome function for any feature combination $k$ can be written as
\begin{align*}
    y &= g(k, \bbeta; \Pi) = \sum_{\pi \in \Pi} \mathbb{I}\{k \in \pi\} g(k, \bbeta_{\pi}).
\end{align*}

The practitioner is free to choose any domain-specific parametric function. For example, $g$ could be a higher-order Taylor series-like expansion. Or, $g$ could even be sinusoidal because the practitioner believes the outcomes are (piece-wise) sinusoidal. Of course, the more complex the estimation procedure for $\bbeta$, the harder it is to enumerate the RPS.

Observe that the form of the posterior remains the same,
\begin{align*}
    \Pr(\Pi \mid \Z) &\propto \exp \left\{ -\eta\mathcal{L}(\Z) + \lambda H(\Pi) \right\} = \exp \left\{ - \eta\frac{1}{n} \sum_{i=1}^n (y_i - \widehat{y}_i)^2 + \lambda H(\Pi) \right\}.
\end{align*}
Therefore, the results in Section \ref{section:environment} still hold. We can freely choose any other non-negative loss function, $\mathcal{L}(\Z)$, and still use the same framework and algorithm to enumerate the RPS.

Further, the results in Section \ref{section:enumeration} are also valid when using an arbitrary parametric outcome function as discussed here. We summarize this in \Cref{thm:generalized-enumeration}.

\begin{theorem}
\label{thm:generalized-enumeration}
Suppose the outcome function is $g(k, \bbeta; \Pi)$ for feature combination $k$, admissible partition $\Pi$, and some unknown parameter $\bbeta$. Let us denote the estimated outcome for unit $i$ with feature combination $k$ by $\widehat{y}_i = g(k, \widehat{\bbeta}; \Pi)$ where $\widehat{\bbeta}$ is estimated from the data. If we use $\widehat{y}_i$ instead of $\widehat{\mu}_{\pi}$ in Equations \ref{eq:fixed-point-loss} and \ref{eq:equivalent-point-loss}, then
\begin{enumerate}[label=(\roman*)]
    \item Theorem \ref{thm:rashomon-fixed-bound} is still true,
    \item Theorem \ref{thm:rashomon-equivalent-bound} is still true, and
    \item Algorithm \ref{alg:r-aggregate} correctly enumerates the Rashomon partitions for outcome function $f$.
\end{enumerate}
\end{theorem}

\begin{proofof}[\Cref{thm:generalized-enumeration}]
The results follow directly from Theorems \ref{thm:rashomon-fixed-bound}, \ref{thm:rashomon-equivalent-bound}, and \ref{thm:enumeration-algorithm}.
\end{proofof}

\begin{figure}
\centering
\begin{subfigure}[t]{0.4\textwidth}
    \begin{centering}
     \begin{center}
     \begin{tikzpicture}[scale = 0.5] \def \n {5} \def \radius {2cm} \def \margin {8}
     \node[  minimum size=19pt] at (0,-3) (v1){$[y, 1, b]$};
     \node[ minimum size=19pt] at (-3,0) (v2){$[o, 1, b]$};
     \node[  minimum size=19pt] at (3,0) (v3){$[y, 2, b]$};
     \node[  minimum size=19pt] at (0,3) (v4){$[o, 2, b]$};
     \node[  minimum size=19pt] at (6,3) (v5){$[y, 3, b]$};
     \node[  minimum size=19pt] at (3,6) (v6){$[o, 3, b]$};
 
     \node[] at (1.5,-5) (pi1) {\small $\pi_1$};
     \node[] at (4.5,-2) (pi3) {\small $\pi_3$};
     \node[] at (7.5,1) (pi5) {\small $\pi_5$};
     \node[] at (-4.5,2) (pi7) {\small $\pi_7$};
     \node[] at (-1.5,5) (pi7) {\small $\pi_9$};
     \node[] at (2,8.2) (pi7) {\small $\pi_{11}$};
    
     \node[] at (11,3) (e) {\ };

      \draw[orange, line width=0.3mm] (0,-3) ellipse (1.75cm and 1.75cm);
      \draw[orange, line width=0.3mm] (-3,0) ellipse (1.75cm and 1.75cm);
      \draw[orange, line width=0.3mm] (0,3) ellipse (1.75cm and 1.75cm);
      \draw[orange, line width=0.3mm] (3,0) ellipse (1.75cm and 1.75cm);
      \draw[orange, line width=0.3mm] (6,3) ellipse (1.75cm and 1.75cm);
      \draw[orange, line width=0.3mm] (3,6) ellipse (1.75cm and 1.75cm);
     
     \draw[line width = 0.3mm, >=latex] (v1) to (v2);
     \draw[line width = 0.3mm,  >=latex] (v1) to (v3);
     \draw[line width = 0.3mm,  >=latex] (v3) to (v4);
     \draw[line width = 0.3mm,  >=latex] (v2) to (v4);
     \draw[line width = 0.3mm, >=latex] (v4) to (v6);
     \draw[line width = 0.3mm, >=latex] (v3) to (v5);
     \draw[line width = 0.3mm, >=latex] (v5) to (v6);
    \end{tikzpicture}
     \end{center} 
    \par
    \end{centering}
    \caption{Drug B dosages, $b \in \{1, 2\}$}
\end{subfigure}%
~
\begin{subfigure}[t]{0.4\textwidth}
    \begin{centering}
     \begin{center}
     \begin{tikzpicture}[scale = 0.5] \def \n {5} \def \radius {2cm} \def \margin {8}
     \node[  minimum size=19pt] at (0,-3) (v1){$[y, 1, b]$};
     \node[ minimum size=19pt] at (-3,0) (v2){$[o, 1, b]$};
     \node[  minimum size=19pt] at (3,0) (v3){$[y, 2, b]$};
     \node[  minimum size=19pt] at (0,3) (v4){$[o, 2, b]$};
     \node[  minimum size=19pt] at (6,3) (v5){$[y, 3, b]$};
     \node[  minimum size=19pt] at (3,6) (v6){$[o, 3, b]$};
 
     \node[] at (1.5,-5) (pi1) {\small $\pi_2$};
     \node[] at (4.5,-2) (pi3) {\small $\pi_4$};
     \node[] at (7.5,1) (pi5) {\small $\pi_6$};
     \node[] at (-4.5,2) (pi7) {\small $\pi_8$};
     \node[] at (-1.5,5) (pi7) {\small $\pi_{10}$};
     \node[] at (2,8.2) (pi7) {\small $\pi_{12}$};
     
     \node[] at (11,3) (e) {\ };

      \draw[orange, line width=0.3mm ] (0,-3) ellipse (1.75cm and 1.75cm);
      \draw[orange, line width=0.3mm ] (-3,0) ellipse (1.75cm and 1.75cm);
      \draw[orange, line width=0.3mm ] (0,3) ellipse (1.75cm and 1.75cm);
      \draw[orange, line width=0.3mm ] (3,0) ellipse (1.75cm and 1.75cm);
      \draw[orange, line width=0.3mm ] (6,3) ellipse (1.75cm and 1.75cm);
      \draw[orange, line width=0.3mm ] (3,6) ellipse (1.75cm and 1.75cm);
     
     \draw[line width = 0.3mm, >=latex] (v1) to (v2);
     \draw[line width = 0.3mm,  >=latex] (v1) to (v3);
     \draw[line width = 0.3mm,  >=latex] (v3) to (v4);
     \draw[line width = 0.3mm,  >=latex] (v2) to (v4);
     \draw[line width = 0.3mm, >=latex] (v4) to (v6);
     \draw[line width = 0.3mm, >=latex] (v3) to (v5);
     \draw[line width = 0.3mm, >=latex] (v5) to (v6);
    \end{tikzpicture}
     \end{center} 
    \par
    \end{centering}
    \caption{Drug B dosages, $b \in \{3, 4, 5\}$}
\end{subfigure}%

\caption{Hasse diagram for simulation with linear outcomes. $y$ is young and $o$ is old.}
\label{fig:hasse-linear-sim}
\end{figure}

To see the usefulness of the generalization, we motivate a simple example where we are interested in how a person's age affects their response to a treatment consisting of a combination of two drugs, A and B. Suppose that there are four possible dosages for drug A, $\{0, 1, 2, 3\}$, six possible dosages for drug B, $\{0, 1, 2, 3, 4, 5\}$, and people are classified as young aged or old aged where 0 indicates control. We assume that there is no treatment effect unless drug A and drug B are taken together. The partition matrix is
\begin{align*}
    \bSigma &= \begin{bmatrix}
        0 & - & - & - \\
        0 & 0 & - & - \\
        1 & 0 & 1 & 1
    \end{bmatrix}.
\end{align*}
We visualize the twelve pools in Figure \ref{fig:hasse-linear-sim} indicating heterogeneity in age and the dosages of drugs A and B.

Suppose that the treatment effects are piecewise linear (which generalizes the stepwise effects that we've assumed in previous simulations),
\begin{align*}
    \bbeta_{1} &= [0, -1, 0, 1] \\
    \bbeta_{2} &= [1.5, -4, 0, 1.5] \\
    \bbeta_{3} &= [0, -1, -0, 1] \\
    \bbeta_{4} &= [4.5, -4, 0, 0.5] \\
    \bbeta_{5} &= [4, -2, -1, 1] \\
    \bbeta_{6} &= [1, 1, 1, -1] \\
    \bbeta_{7} &= [-3, 2, -3, 1] \\
    \bbeta_{8} &= [0, 0, 0, 0] \\
    \bbeta_{9} &= [4, 2, -3, -1] \\
    \bbeta_{10} &= [0, 0, 0, 0] \\
    \bbeta_{11} &= [5, 2, -3, 0] \\
    \bbeta_{12} &= [5, -1, 0, -1],
\end{align*}
where the first coefficient is the intercept and the remaining elements are slopes on each feature. For feature profiles with zero treatment effect, we set the effect to be 0, a constant. For the feature profile where drugs A and B are administered together, a random error is drawn independently and identically from $\mathcal{N}(0, 1)$. We draw 10 measurements for each feature combination. We set $\lambda = 4 \times 10^{-3}$.

\begin{figure}[!tb]
    \centering
    \includegraphics[height=4in]{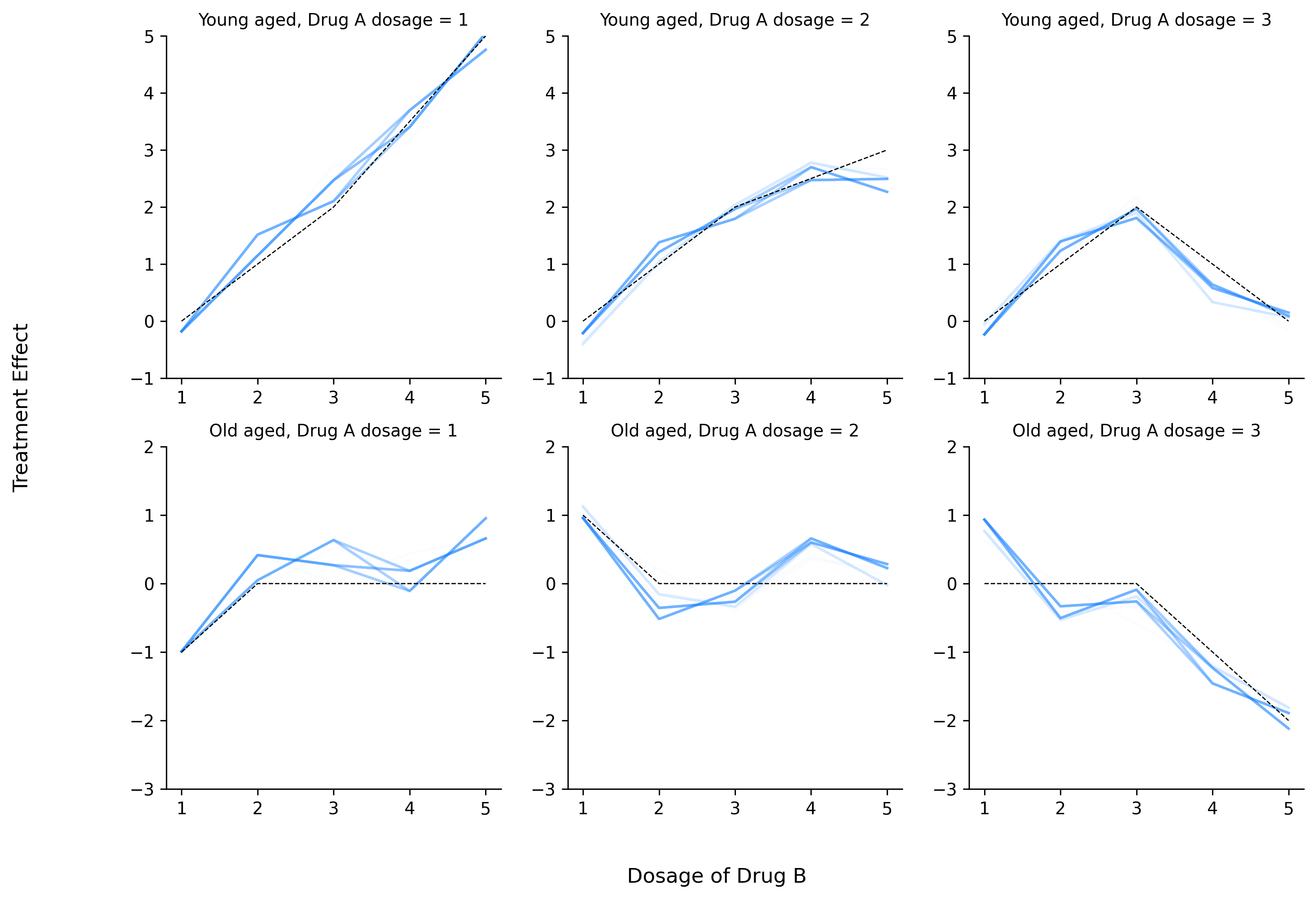}
    \caption{The black line corresponds to the true data-generating process and the blue lines correspond to effects estimated in each model in the Rashomon set. We estimate the outcome of each pool as a linear function of the features. The denser the blue line, the more often it appears in the Rashomon set.}
    \label{fig:slopes-sims}
\end{figure}

We illustrate the treatment effects for different combinations in black dashed lines in Figure \ref{fig:slopes-sims}. By choosing a linear function as the outcome for each pool, we can find the Rashomon set. In Figure \ref{fig:slopes-sims}, we show the estimated linear curves in 100 models present in the Rashomon set ($\epsilon \approx 5 \times 10^{-4}$) in blue. The denser the blue line, the more often it appears in the Rashomon set.

\subsection{Sparse correlation structure between coefficients.}
Next, we explore the space of potential (sparse) covariance matrices between the coefficients.  We now apply the Hasse structure to the elements of the variance-covariance matrix and pool on the space of covariances rather than the coefficients themselves. This generalization requires an additional distributional assumption on the coefficients.  Specifically, assume that
\begin{align*}
    \bbeta \mid \mu, \bLambda \sim \mathcal{N}(\mu, \bLambda)
\end{align*}
where $\mu$ is some mean matrix and $\bLambda$ is some covariance matrix. Then the posterior has the form
\begin{align*}
    \Pr(\bbeta, \bLambda, \Pi \mid \Z) \propto \Pr(\y \mid \bbeta, \bLambda, \D, \Pi) \cdot \Pr(\bbeta, \bLambda, \Pi)
\end{align*}

The likelihood component of the loss is
\begin{align*}
\Pr(\y \mid \bbeta, \bLambda, \D, \Pi) \propto \exp \left\{ - \frac{1}{N} (\y - \D \bbeta)^{\intercal} \bLambda (\y - \D\bbeta) \right\},
\end{align*}
where $N$ is the number of observed data points.

We do not have additional information about the covariance structure (though this could of course also be included in a prior) beyond the following three assumptions. First, we think that $\bLambda$ is dense i.e., $\bLambda$ is sparse in the number of uncorrelated outcomes. Second, we neither know nor want to know the correlation: it is an $\ell_0$ problem. Third, we assume independence across the mean and correlation conditional on the covariance pooling. That is, $\Pi$ is sufficient for the existence of dependence. Then
\begin{align*}
    \Pr(\bbeta, \bLambda, \Pi) &= \Pr(\bbeta \mid \bLambda, \Pi) \cdot \Pr(\bLambda \mid \Pi) \cdot \Pr(\Pi) 
\end{align*}

Suppose that we have a partition $\Pi = \{\pi_1, \dots, \pi_H\}$ where $H = \abs{\Pi}$ and $\Pi$ now is defined in the space of covariance matrices, so pooling means that the covariance values within a pool are non-zero. Then, consider the following procedure for drawing the covariance matrix, $\bLambda \in \mathbb{R}^{K \times K}$. For each pool $\pi_i \in \Pi$, draw $\bLambda_i \sim f_i$ independently where $f_i$ is some prior (for example, inverse Wishart). Then, $\bLambda = \text{diag}(\bLambda_i, \dots, \bLambda_H)$.
The number of non-zero elements of $\bLambda$ is given by $\norm{\bLambda}_0 = \sum_{i=1}^H h_i^2$. Therefore, we penalize the number of zero elements, $K^2 - \sum_{i=1}^H h_i^2$. Thus, the prior is
\begin{align*}
    \Pr(\Pi) &\propto \exp \left\{ -\lambda \left( K^2 - \sum_{i=1}^H h_i^2 \right) \right\}. 
\end{align*}

So our penalized loss function is just weighted mean-squared error penalized differently,
\begin{align}
    Q(\Pi; \Z) &= \eta\mathcal{L}(\Pi; \Z) + \lambda H(\Pi) = \frac{1}{n} \left( \y - \D \bbeta \right)^{\intercal} \bLambda \left( \y - \D\bbeta \right) + \lambda \left( K^2 - \sum_{i=1}^H h_i^2 \right). \label{eq:sparse-cov-loss}
\end{align}

 \begin{theorem}
\label{thm:covariance-enumeration}
Consider the same setup in Section \ref{section:enumeration} except the loss function is weighted mean squared error penalized by the number of zeros in the covariance matrix as in Equation \ref{eq:sparse-cov-loss}. Specifically, we hav
\begin{align}
    b(\bSigma, \mathcal{M}; \Z) &= \eta\frac{1}{n} \sum_{\pi \in \Pi_{\fix}} \sum_{k(i) \in \pi} \mathbb{I} \left\{ k(i) \in \pi_{\fix} \right\}  \widehat{\Lambda}_{k(i), k(i)}^2 (y_i - \widehat{\mu}_{\pi})^2  + \lambda H(\Pi, \mathcal{M}), \\
    b_{eq}(\bSigma, \mathcal{M}; \Z) &= \eta\frac{1}{n} \sum_{\pi \in \Pi_{\fix}} \sum_{k(i) \in \pi}  \mathbb{I} \left\{ k(i) \in \pi_{\fix}^c \right\} \widehat{\Lambda}_{k(i), k(i)}^2 (y_i - \widehat{\mu}_{\pi})^2.
\end{align}
where $\widehat{\Lambda}_{k, k}^2$ is the estimated variance of feature combination $k$.

Then
\begin{enumerate}[label=(\roman*)]
    \item Theorem \ref{thm:rashomon-fixed-bound} is still true,
    \item Theorem \ref{thm:rashomon-equivalent-bound} is still true, and
    \item Algorithm \ref{alg:r-aggregate} correctly enumerates the Rashomon partitions.
\end{enumerate}
\end{theorem}

\begin{proofof}[\Cref{thm:covariance-enumeration}]
The results follow directly from Theorems \ref{thm:rashomon-fixed-bound}, \ref{thm:rashomon-equivalent-bound}, and \ref{thm:enumeration-algorithm}.
\end{proofof}

\section{Further Details on Related Work.}
\label{section:related}
It is useful to contrast our method with several other (some recent) approaches to study heterogeneity. Specifically, we are interested in their application to settings with partial orderings (e.g., factorial structure and admissibility) which is easily interpretable.

We will focus on four main related approaches: (1) canonical Bayesian Hierarchical Models (BHM) \citep{rubin1981estimation,gelman2006prior,meager2019understanding}; (2) $\ell_1$ regularization of marginal effects to identify heterogeneity \citep{banerjee2021selecting}; (3) causal forests \citep{wager2018estimation}; and (4) machine learned proxies~\citep{chernozhukov2018generic}.  We intend this discussion to be a guide for practitioners considering implementing our proposed method or one of these state-of-the-art alternatives.  We discuss conceptually related work (e.g. Bayesian decision trees) in previous sections.   
an
Let us for the moment set aside the following immediate differences. Our focus on robustness, profiles, and enumerating the entire Rashomon Partition are all novel. Instead, it is useful to identify the philosophical differences across the various approaches and how they relate to us. 
Every approach, as we will note, effectively uses partitions $\Pi$ at some point to determine which data to pool or not. The specific techniques create distributions, possibly degenerate,  over these partitions, and these distributions are sampled from and marginalized to estimate treatment effects $\beta_k$. The interesting thing therefore is how one builds a distribution over $\Pi$.

\subsection{Bayesian Hierarchical Models.}
We now discuss how our work relates to a canonical representation of a Bayesian Hierarchical Model.  As discussed previously, our work is more similar to Bayesian Tree(d) models than to other methods for accounting for learning heterogeneity, such as Bayesian Model Averaging.  For context, however, we present how our approach compares to the canonical Bayesian approach.  The Bayesian perspective provides a compromise between complete and partial pooling. Partial pooling occurs by encouraging similarity in the values for parameters without requiring strict equality.  Using the notation from our model, for example, we could construct a model where 

\[\y \sim N( \D \bbeta,\sigma_y^2)\] and, for the sake of exposition, all $\beta$ are draw independently from  
\[\bbeta \sim N(\mu_{\beta}, \sigma^2_\beta).\]
Requiring that all values of $\bbeta$ come from the same distribution encourages sharing information across potential feature combinations and encourages the effects on heterogeneity to be similar (but not identical). ~\citet{meager2019understanding} uses this approach when comparing treatment effects across multiple domains.  In that paper, the goal is not to pool across potentially similar treatment conditions but instead to (partially) pool across geographic areas.  

As one example of the classical model,~\citet{meager2019understanding} has outcomes for household $i$ in study $k$ modeled as 

\[y_{ik}\sim N (\mu_k+\tau_k{\mathbf{T}_{ik}},\sigma^2_{yk}) \mbox{         }\forall\mbox{  } i,k\]

\begin{eqnarray*}
\begin{pmatrix}\mu_k\\
\tau_k
\end{pmatrix} & \sim & N\left[\left(\begin{array}{c}
\mu\\
\tau
\end{array}\right),\left(\begin{array}{ccc}
\sigma^2_\mu & \sigma_{\mu\tau} \\
\sigma_{\mu\tau} & \sigma^2_\tau 
\end{array}\right)\right] \mbox{  
      }\forall\mbox{  } i,k
\end{eqnarray*}

where $\tau_k$, $\mu_k$ are the overall mean and treatment effect at area $k$, respectively.  The vector ${\bf T_{ik}}$ is the treatment indicator for household $i$ in study $k$.

One way to measure the degree of pooling is the (partial) ``pooling factor'' metric defined in~\citet{gelman2006prior},
$\omega(\bbeta)={\sigma^2_y}/{\sigma^2_y+\sigma^2_\beta}$.  
The partial pooling metric quantifies how much the effect of treatment combinations varies compared to the overall heterogeneity in the outcomes. The partial pooling metric, the~\citet{meager2019understanding} context refers to the relative variation related to differences between studies compared to sampling variability.  

In contrast, we could think of our approach as using a prior on $\bbeta$ conditional on the partitions that potentially force some values of $\beta_k, \beta_k'$ to be equal.
In Appendix~\ref{appendix:generalized-bayes}, we show that the objective function we use in Equation~\ref{eq:objective_fcn} corresponds to a hierarchical model where we draw the $\bbeta$ vector as
\begin{align*}
    \bbeta \mid \Pi &\sim \mathcal{N}(\bmu_{\Pi}, \bLambda),
\end{align*}
where $\bmu_{\Pi}$ is structured such that $\mu_k = \mu_{k^{\prime}}$ for any $k, k^{\prime} \in \pi \in \Pi$. Then, given some feature combinations $\D$, we draw the outcomes as
\begin{align*}
    \y \mid \D, \bbeta &\sim \mathcal{N}(\D \bbeta, \bSigma).
\end{align*}
To understand the variation within the $\bbeta$ vector, we need to average across potential partitions, since some partitions will set $\beta_k=\beta_k'$ and others will not.  This amounts to replacing the $\sigma_\beta^2$ in the pooling factor with the variance of the distribution of $\Pr(\bbeta|Z)$.  

We could also conceptualize the above derivation in terms of equality on $\bbeta$ rather than the means $\mu_\Pi$.  If, for example, we replace $\bLambda$ with $\bLambda_\Pi$ where $Var(\mu_k, \mu_k')=0$ for any $k, k^{\prime} \in \pi \in \Pi$ (or equivalently, when $\mu_k = \mu_{k^{\prime}}$) then we enforce that $\beta_k=\beta_k'$. Of course, if we go the opposite direction and let the diagonal of $\bLambda_\Pi$ be unconstrained then there is essentially no sharing of information across feature combinations.

Finally, hierarchical models of this type are, of course, quite flexible and we could construct more complex models that capture features of our pooling approach.  Among those options would be to use the Bayesian version of penalized regression, such as the Bayesian Lasso, which would be philosophically related to the approach we describe in the next section.

\subsection{Lasso regularization.}\label{sec:lassolit} This is the approach taken in prior work by several of the authors of the present paper, in \cite{banerjee2021selecting}. There the setting was one in which the researcher faced a factorial experimental design: a crossed randomized controlled trial (RCT). The paper developed the Hasse structure described above and an approach that required transforming $\D$ into an equivalent form presented in Equation \eqref{eq:tva-regression} in Appendix \ref{appendix:environemnt}. Here every parameter $\alpha_k$ represents the marginal difference between $\beta_k$ and $\beta_{k'}$ where $\rho(k) = \rho(k')$ (they are the same profile) and $k$ exactly differs from $k'$ on one arm by one dose. The parameter vector $\balpha$ records the marginal effects. Notice the support of $\balpha$ therefore identifies $\Pi$ (since non-zero entries determine splits).

The first difficulty in applying this to general settings of heterogeneity is that $\ell_1$ regularization requires irrepresentability: that there is limited correlation between the regressors so that the support may be consistently recovered \citep{zhao2006model}. Unfortunately, the regression implied by the Hasse does not satisfy this so some pre-processing is required. \cite{banerjee2021selecting} apply the Puffer transformation of \cite{jia2015preconditioning} to retain irrepresentability and estimate the Lasso model. However, this is not free: the approach requires conditions on the minimum singular value of the design matrix. The authors leverage the structure of a crossed randomized controlled trial (which places considerable restriction on the design matrix) to argue that indeed these conditions are met. There is no guarantee and it is unlikely to be the case that these conditions are met for general factorial data of arbitrary covariates. So, tackling the much more general structure required moving away from regression (we use decision trees) and changing the regularization (we use $\ell_0$).

The second key observation is that the Bayesian lasso means that the $\ell_1$ penalty corresponds to priors $\Pr(\balpha)$ that are i.i.d. Laplace on every dimension $k$, which arises from first principles. That is
\begin{align*}
- \log \Pr(\balpha) &= \log \prod_k \Pr(\alpha_k) = \log \prod_k \exp(-\lambda \abs{\alpha_k}) = \lambda \sum_k \abs{\alpha_k}.
\end{align*}
Note that this is true whether one uses regular lasso, Puffer transformed lasso, spike-and-slab lasso, group lasso (up to the group level), and so on. No matter at whatever level the $\ell_1$ sum is being taken, it corresponds to independence at that level in the prior.

In practice what this means is that given two partitions $\Pi$ and $\Pi'$, which have the same number of pools and which have the same loss value, if one is more consistent with independent values of $\alpha_k$ than the other, it will receive a higher posterior. There are at least two problems.

The main philosophical problem is that there is no reason to place the meta-structure that the marginal differences between adjacent variants should have an i.i.d. distribution. In fact, one might think that the basic science or social science dictates \emph{exactly the opposite}. Independence means that a marginal increase in dosage of drug A, holding fixed B and C at some level, is thought to be \emph{independent} of increasing A holding fixed B and C at (potentially very similar) different levels. Similarly, the marginal value of receiving a slightly larger loan given that the recipient has 10 years of schooling and started 5 previous businesses is \emph{independent} of receiving a slightly larger loan if the recipient had 10 years of schooling and started 6 previous businesses. Independence is unreasonable in both examples. {In the case of general factorial data, even for very large samples, $\ell_1$ lacks theoretical guarantees. This is because as \cite{banerjee2021selecting} note, regressors are correlated, so naive LASSO over-selects the support. They apply the Puffer transformation of \cite{jia2015preconditioning} to shift bias into variance and recover support consistency. This works in crossed RCT settings because correlations are bounded. In general factorial designs, however, Puffer variance inflation can be arbitrarily large and destroy support recovery. These adversarial data distributions are not pathological: scientific structure \emph{often} implies positive correlation (“if there is response at low doses, there is more at higher doses”) or negative correlation (“effects plateau or decline”) among marginals.}

There is a second issue in that if an object of interest is $\Pi$, this approach provides no way forward. Regularization delivers posteriors over $\balpha$: $\Pr(\balpha \mid \y,\X)$. This implies a posterior over $S_{\balpha}$.  The map from $S_{\balpha}$ to $\Pi$ is deterministic, and is given by some $\phi(S_{\balpha}) = \Pi$, which means that
\[
\Pr(\Pi \mid \y,\X) = \int_{\balpha}  {\bf 1}\left\{\phi(S_{\balpha}) = \Pi \right\}  \cdot \Pr(\balpha \mid \y,\X)
\]
is the actual calculation of interest.

So the regularization approach requires the statistician to take all the marginal parameters to be i.i.d., and given this, integrate over possible coefficient vectors that are consistent with this specific aggregation. This makes calculating an RPS very difficult.

\subsection{Causal Random Forests.}
\label{appendix:crf}
We now compare our approach to Causal Random Forests (CRFs) introduced by ~\citet{wager2018estimation}.  CRFs construct regression trees over the space of potential combinations of covariates.  Trees partition the space of covariates into ``leaves.''  Unlike our setting, trees are heirarchical; the procedure to construct trees involves splitting the observed data in two based on $X_i$ being above or below a threshold.  They then partition recursively, dividing each subsequent group until the leaves contain very few observations.  This approach can also be thought of as finding nearest neighbors, where the number of neighbors is the number of observations in the leaf and using distance on the tree as the closeness metric.  CRFs construct a conditional average treatment effect at a pre-determined point $X=x$, $\tau(x)=\E[Y{(1)}-Y{(0)}|X=x]$ where $Y{(1)}$ is potential outcome for the treated and $Y{(0)}$ is the potential outcome for the control. 

Relating this back to our work, take $T$ to be a tree and $\pi\in \Pi(T)$ to be a leaf in the tree, which corresponds to a pool in our language.  Then, the estimated expected outcomes for each leaf is
\begin{align*}
    \widehat{\beta}_{\pi} &= \frac{1}{\abs{\{i:X_i \in \pi \}}}\sum_{\{i:X_i \in \pi \}} Y_i.
\end{align*}

 Further, taking $\tau_\pi$ to be the treatment effect of observations in pool (leaf) $\pi$ and $W_i$ as the treatment indicator, which we assume orthogonal to $X$ and $Y$, the estimated treatment effect for $\pi$ is 
 \begin{align*}
     \widehat{\tau}_{\pi} &= \frac{1}{\abs{\{i:W_i=1, X_i \in \pi \}}}\sum_{\{i:W_i=1,X_i \in \pi \}}\hspace{-20pt} Y_i \hspace{15pt}-\hspace{15pt} \frac{1}{\abs{\{i:W_i=0,X_i \in \pi \}}}\sum_{\{i:W_i=0,X_i \in \pi \}} \hspace{-20pt}Y_i.
 \end{align*}

To summarize, the approach for forming trees splits the observed covariate space into partitions, known as leaves.  Each leaf consists of a mix of people in treatment and control groups and, in fact, the specification of the tree depends on this balance across treatment and control groups since the algorithm requires that splitting be done in a way that preserves a minimum number of treatment and control in each leaf.  To compute a treatment effect conditional on a particular value of $X$, look at the difference in outcome between treated and control people in a given leaf.  Outcomes are not considered with constructing the tree (in contrast to our proposed approach) and treatment status is not used to split explicitly but does influence the construction of the tree through the sample size restriction.      

Despite being similar in that we both use geometric objects that partition the space of covariates, there are three fundamental differences between our approach and CRFs.  The first difference is geometric. CRFs use regression trees, whereas we use Hasse diagrams.  Regression trees are appealing in many settings because of their flexibility in representing complex, nonlinear, relationships between variables.  Regression trees, however, require imposing a hierarchy between variables that is not supported by the data.  This hierarchy is ``baked in'' to the structure of the trees and is evident from how we describe constructing trees in the previous paragraph.  The data, however, are not fully hierarchical and are instead partially ordered.  

This mismatch creates an identification issue.  Within education and within income, there are clear orderings.  There is, however, no heirarchy between education and income.  One tree may, therefore, split first based on income and then split on education conditional on income while another tree does the opposite.  In both cases, we can trace the trees to end up with the same estimated treatment effects for any group of covariates (as shown in~\citet{wager2018estimation}).  The trees themselves, however, arise from this arbitrary ordering and are, thus, not interpretable. Work such as \citet{benard2023variable} describe measures of variable importance in CRFs, but the problem of an arbitrarily imposed hierarchy is still present. Hasse diagrams, in contrast, are the natural geometry for partially ordered sets, alleviating this issue and allowing the researcher to interpret the pooling structure on the domain of the covariates directly.

The second difference is computational but has conceptual implications.  In both our approach and CRFs, we do not take the stucture of the partition as known.  Both approaches must, therefore, account for additional uncertainty in treatment effect estimates that arises from not knowing the partition.  In CRFs, bootstrap samples over the data propagate this uncertainty.  CRFs then aggregate over trees using Monte Carlo averaging over $b=1,..,B$ boostrap samples of the covariates and outcomes, $\{Z_1,...,Z_n\}$, 

\[ RF(\pi; Z_1,...,Z_n) \approx \frac{1}{B}\sum_{b=1}^B T(\pi; \xi_b; Z_{b1}^*,...,Z_{bs}^*),\]

where $\pi$ represents a pool or leaf specifying a combination of features and levels.  The $\xi_b$ term is an additional stochastic component.  The trees sampled as part of this process create a ``forest'' are, by definition, random draws given the data.  That is, given a different set of data, the distribution of likely trees would change.  They are also not guaranteed to be optimal or nearly optimal.  If the goal is to estimate average treatment effects, this approach represents a principled way to explore the space of trees. 
If the goal, however, is to identify potential models of heterogeneity, then sampling randomly is very unlikely to produce high quality trees.  With Rashomon partitions, by definition, we guarantee that all models in our set are of high posterior.

To this point, we have not discussed inference in CRFs.  A key contribution of~\cite{wager2018estimation} is forming so-called ``honest'' trees that account for issues that arise when using the same data to learn trees and then to make inference conditional on the group of trees. 
 In our work, we use a Bayesian framework to address this issue, which also has the advantage of being able to estimate functions of treatment effects (see \ref{thm:marginal_approx_rashomon}).  Future work, however, could consider Rashomon sets for honest regression trees. This work would build upon our own work as well as ~\citet{xin2022exploring} that introduces Rashomon sets for classification trees.  The algorithm for inference would begin with splitting as proposed by~\cite{wager2018estimation} to preserve honest inference, then construct Rashomon sets using the algorithm from~\citet{xin2022exploring}.  Since the space of trees is enormous, finding the ``best'' tree is impossible, which creates issues for finding the Rashomon set since it is used to define the reference partition.  Fortunately, a recent paper by~\cite{hu2019optimal} provides an algorithm.  While this approach would allow the CRF framework to find optimal trees, it does not address the identifiability issue that arises when using trees for data that are only partially ordered.  Similar work was explored in \citet{hahn2020bayesian}, who estimate heterogeneous treatment effects using a sum of Bayesian regression trees, which they refer to as the Bayesian causal forest. They decompose the outcome into a mean outcome and a treatment effect. Since they are only interested in the treatment effect, the mean outcome becomes a nuisance parameter. They impose a vague prior on the mean and a strong prior on the treatment effect. Otherwise, the tree estimation procedure is identical to Bayesian Additive Regression Trees \citep{chipman2010bart}.

Third, both our approach and CRFs impose regularization but do so in philosophically very different ways. 
We take the perspective that we do not know and cannot fully enumerate correlation structure in a high dimensional space. So we use the $\ell_0$ prior, which we show is the least informative prior in \Cref{thm:l0-minimax}. In other words, we regularize, and impose a prior, on the size of the partition. In doing so, we are trading off information on full distribution to robustly identify partitions. On the other hand, causal forests regularize on the number of observations in each leaf of the tree. Specifically, they require at least $k$ samples in each leaf. This choice is sensible because with insufficient data there is no information. At the same time, this is odd as the regularization depends directly on the data. Elaborating on this, we can write the posterior for some tree $T$ given data $\y, \X$ as
\begin{align*}
    \Pr(T \mid \y, \X) &\propto \Pr(\y \mid T, \X) \Pr( T \mid \X) \text{ where } \Pr(T \mid \X) \propto \exp \left\{ - \frac{\lambda}{\min_{\pi \in \Pi(T)} n_{\pi}(\X)} \right\},
\end{align*}
where $\Pi(T)$ is the set of pools (leaves) in $T$ and $n_{\pi}(\X)$ is the number of observations in $\X$ that belong to pool $\pi$.
This prior down-weights and discards partitions where for \emph{some} $\pi$ the observations are low. In that sense, the prior effectively assumes that in the background there is a kind of stratification -- that observations are sampled from some process such that all pools have enough observations, though of course \emph{the true partition is unknown}. This feels awkward as there is a relationship between the data collection process and the actual true partitioning wherein the user of the causal forest is assuming that they have effectively stratified data collection against the unknown partitions.

Together, these differences mean that the scope of our method is wider than CRFs.  While both methods can estimate heterogeneity in treatment effects and control for multiple testing, we also produce interpretable explanations of heterogeneity.  For the reasons outlined above, namely identification and sampling, it is not possible to extract information on the relationship between covariates from elements of the random forest.  We can, of course, test for any hypothesis about potential heterogneeity between arbitrary combinations of features, but CRFs require that we specify the hypothesis \emph{a priori}.  In our setting, however, finding the set of high posterior probability partitions gives a policymaker or researcher as set of potential models of heterogeneity and interaction between the covariates that can be used to design future policies or generate new research hypotheses.  On the other hand, our method assumes that the posterior has separated modes.  If the posterior distribution is very flat or has many (many) very similar modes, then the Rashomon set will be very large and our benefits in terms of interpretability will diminish.   

\subsection{Treatment heterogeneity via Machine Learning Proxies.}
\cite{chernozhukov2018generic} propose a general framework for using machine learning proxies to explore treatment effect heterogeneity.  They allow for estimation of multiple outcomes, including conditional average treatment effects and treatment effect heterogeneity between the most and least impacted groups.  Rather than search the space of covariates directly,~\citet{chernozhukov2018generic} uses a machine learning method to create a ``proxy'' for the heterogeneous effects.  This approach has the advantage that it can be applied in high dimensional settings.  A downside, however, is that the machine learning proxies are often uninterpretable in terms of the original covariates, making it necessary to post-process the treatment effect distributions to gain insights about particular covariates.

We now give a brief overview to unify notation but do not exhasutively cover all the estimators presented in~\citet{chernozhukov2018generic}.  Say that $s_0(Z)$ is the true conditional average treatment effect, $\E[Y(1)|Z]-\E[Y(0)|Z]$  Ascertaining the functional form of the relationship between the non-intervention covariates $X$ and the outcome $Y$, though, is complicated when $X$ is high dimensional. In response, ~\citet{chernozhukov2018generic} use a machine learning method (e.g. neural networks, random forests, etc) to construct a proxy for $s_0(Z)$ using an auxillary dataset.  In a heuristic sense, this proxy serves the role of a partition $\pi$, in that it aggregates across covariates to separate the data based on the treatment effect. This analogy is most direct when the machine learning model is a decision tree (which it need not be) since in that case leaves of the tree would correspond to partitions of the covariate space based on treatment effect. After computing the machine learning proxy,~\citet{chernozhukov2018generic} then project it back to the space of the observed outcomes.  It is then also possible to construct clusterings based on the proxies and related those clusterings back to the outcomes.  ~\citet{chernozhukov2018generic} differs from our approach in many of the same ways as the comparison with~\cite{wager2018estimation}, namely that we focus on identifying multiple explanations for heterogeneity and that we utlize the Hasse diagram as a geometric representation of partial ordering.  We also find that this structure is sufficient to explore models for heterogeneity on the space of the covariates without the need to use proxies.

\section{Appendix to Empirical Data Examples}
\label{appendix:real-data}

\subsection{Does price matter in charitable giving?}

\begin{figure}[!p]
    \centering
    \begin{subfigure}[t]{0.5\textwidth}
        \centering
        \includegraphics[height=2.5in]{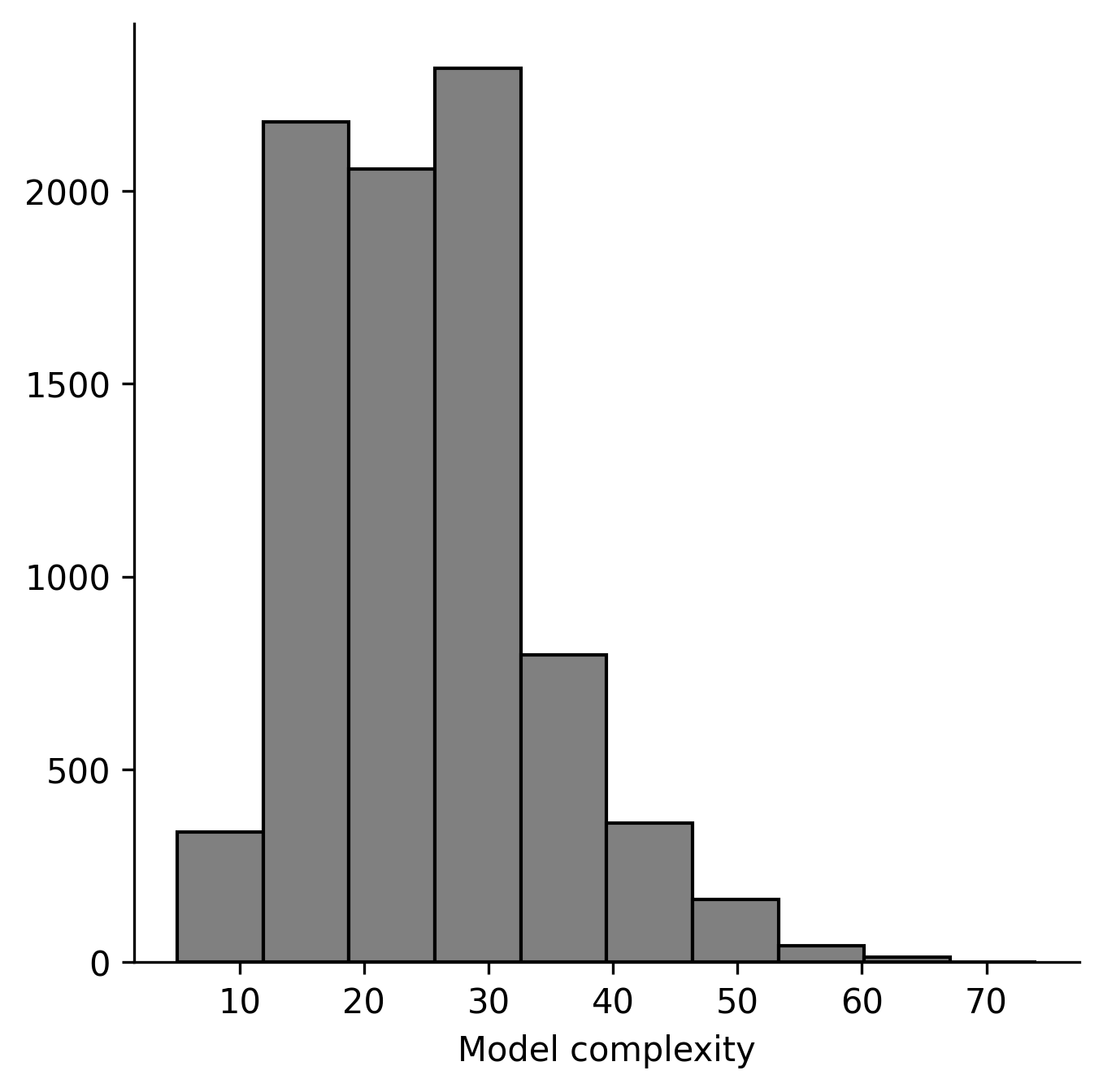}
    \end{subfigure}%
    ~ 
    \begin{subfigure}[t]{0.5\textwidth}
        \centering
        \includegraphics[height=2.5in]{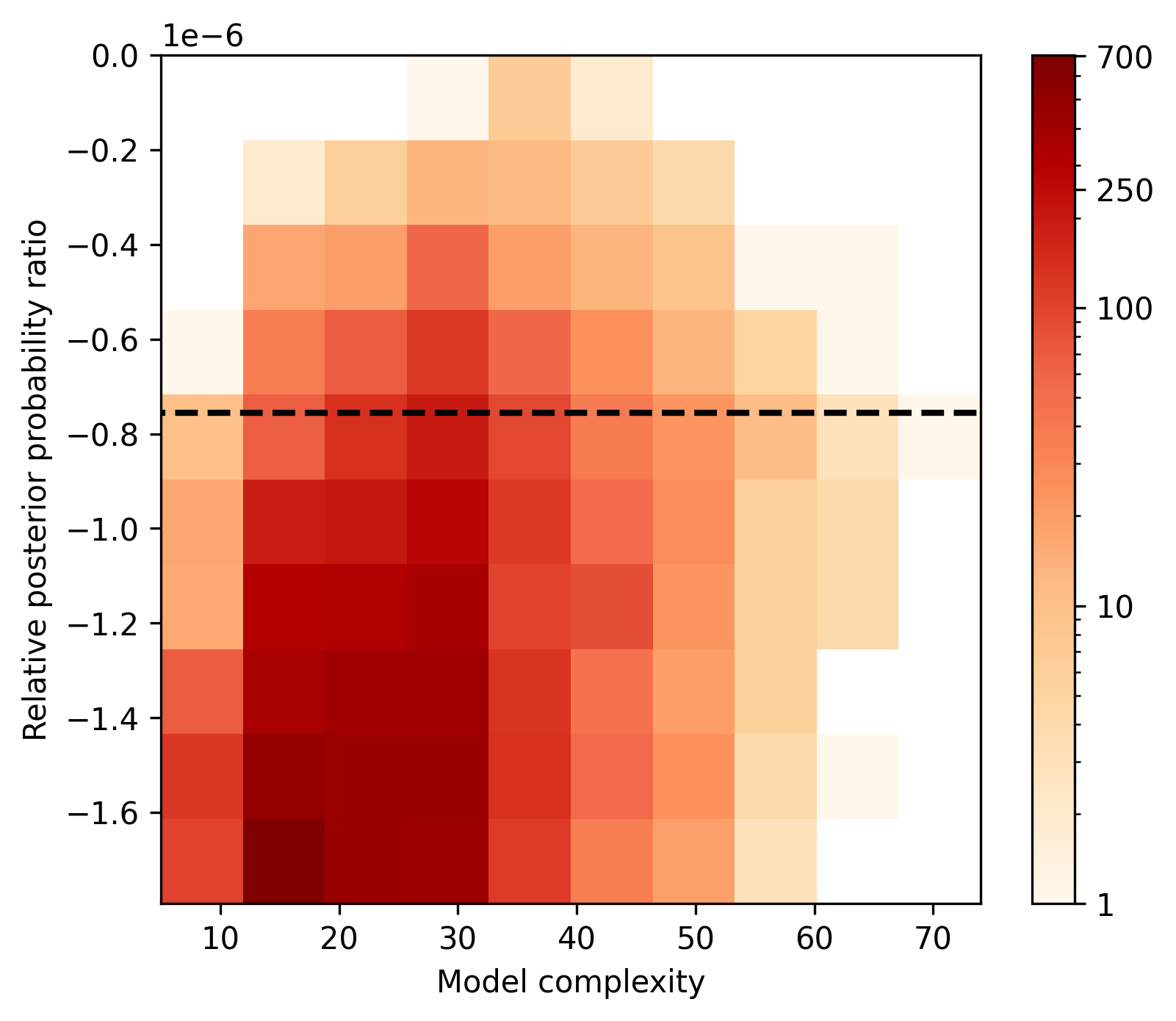}
    \end{subfigure}%
    \\
    \begin{subfigure}[t]{0.5\textwidth}
        \centering
        \includegraphics[height=2.5in]{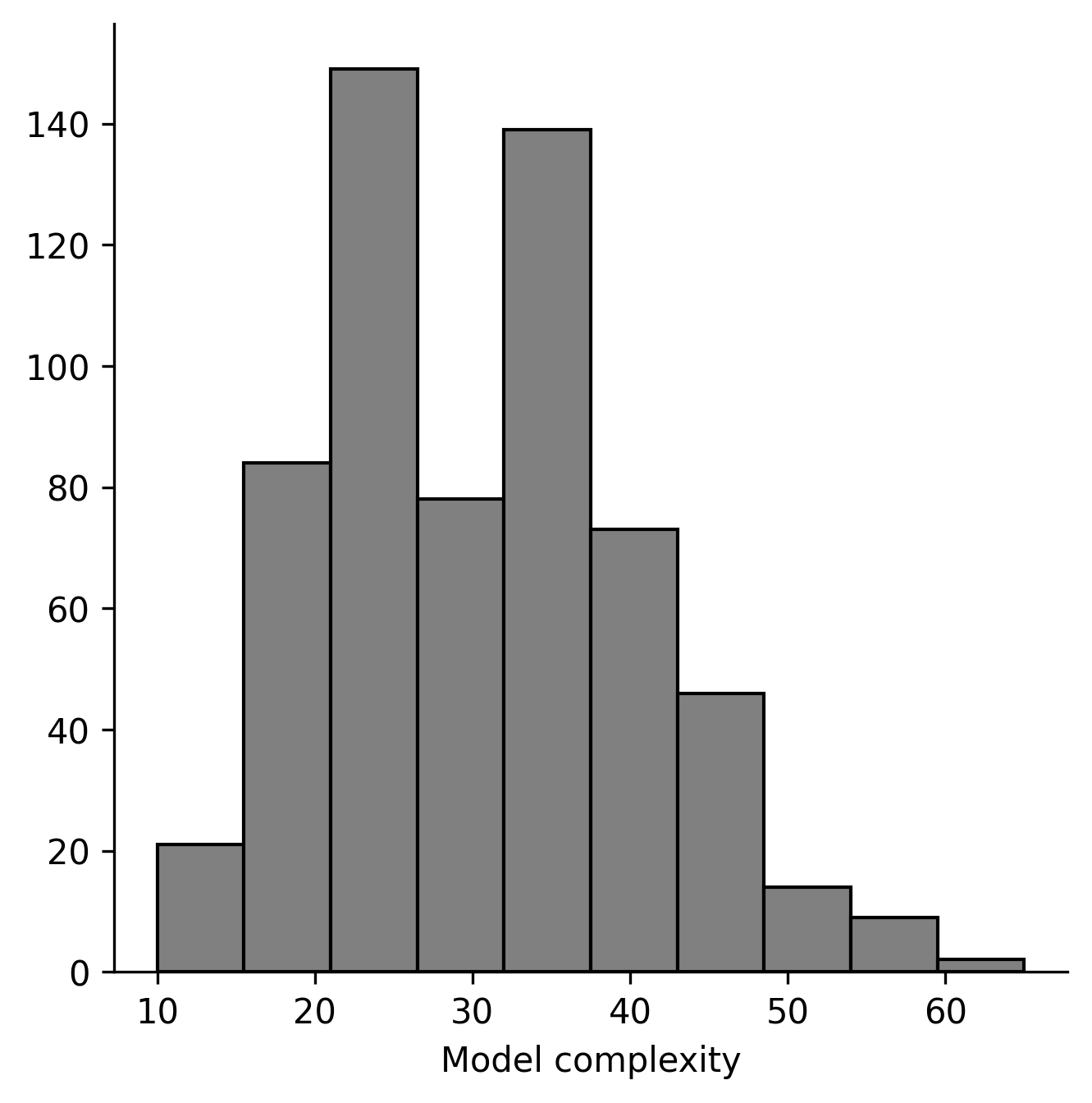}
    \end{subfigure}%
    ~ 
    \begin{subfigure}[t]{0.5\textwidth}
        \centering
        \includegraphics[height=2.5in]{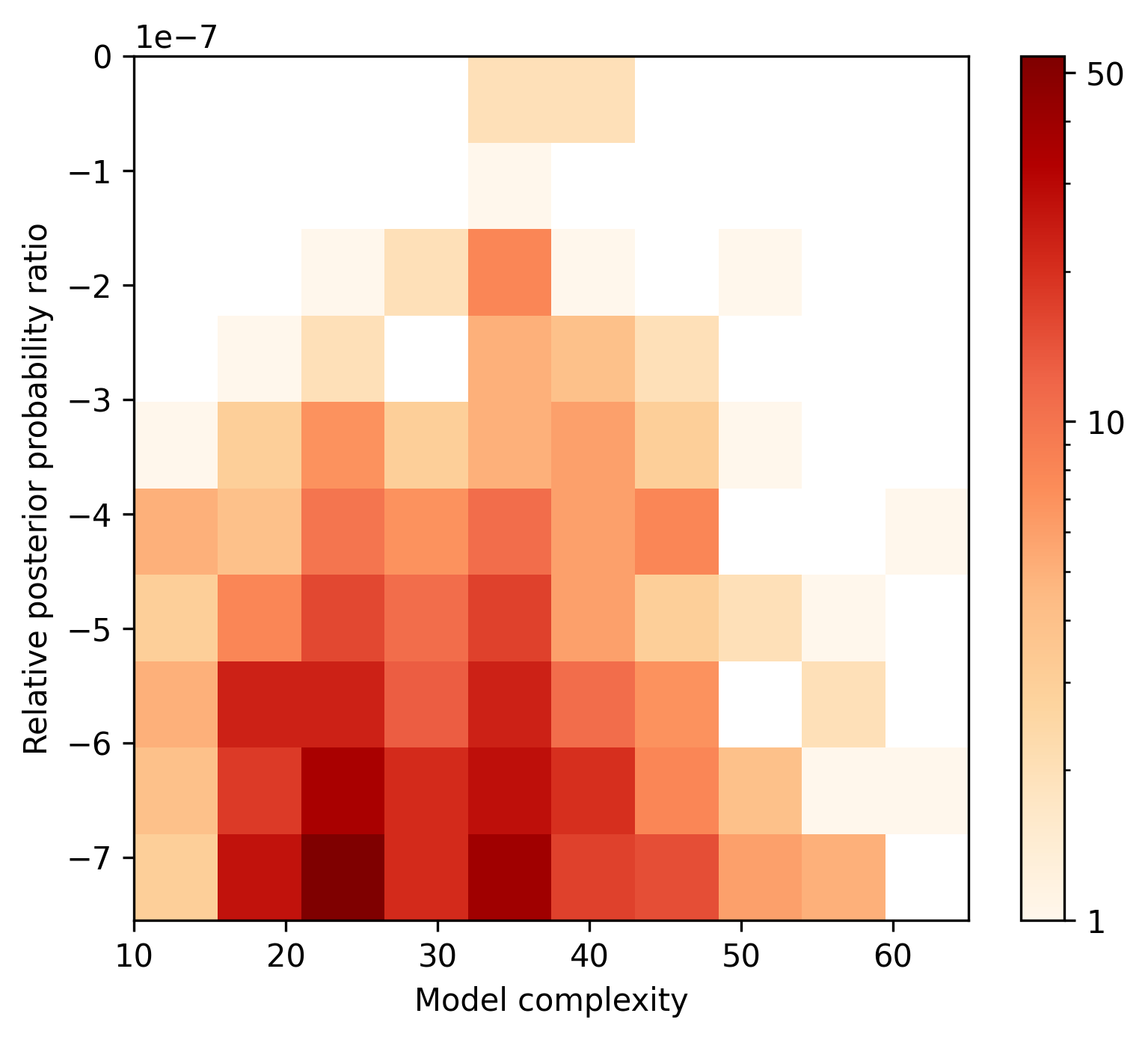}
    \end{subfigure}
    \caption{Visualizing the Rashomon set for \citet{karlan2007does} charitable donations dataset. The top two panels show the distribution of partition sizes and a 2D histogram of how partition sizes and relative posterior probabilities vary. The black dotted line in the 2D histogram shows our chosen Rashomon threshold. The bottom two panels show the same after pruning low-posterior models.}
    \label{fig:karlan-rset-dist-appendix}
\end{figure}

Figure \ref{fig:karlan-rset-dist-appendix} visualizes the Rashomon sets for the charitable giving datasets of \citet{karlan2007does} using the 2D histogram that is described in Figure \ref{fig:heatmap-example} in Appendix \ref{appendix:additional-sims}.

In Figure \ref{fig:karlan-het-dist-appendix}, we visualize the posterior densities of $t_{1, \mathbf{x}}$ and $t_{2, \mathbf{x}}$ described in Section \ref{section:real-data-karlan} over the RPS. We smoothed out the distribution using a Gaussian kernel density estimator. In Figure \ref{fig:karlan-rset}, we showed a quantized version of these plots. A single mode as in the last row of the top left panel (Unstated, 1.5x for \$2:\$1 Democrat) indicates homogeneity whereas multiple modes (such as \$25k, 1.5x for \$2:\$1 Republican) or a very flat distribution (such as \$100k, 1x for \$3:\$1 Republican) indicates heterogeneity.

\begin{figure}[!p]
    \centering
    \begin{subfigure}[t]{0.5\textwidth}
        \centering
        \includegraphics[height=3.5in]{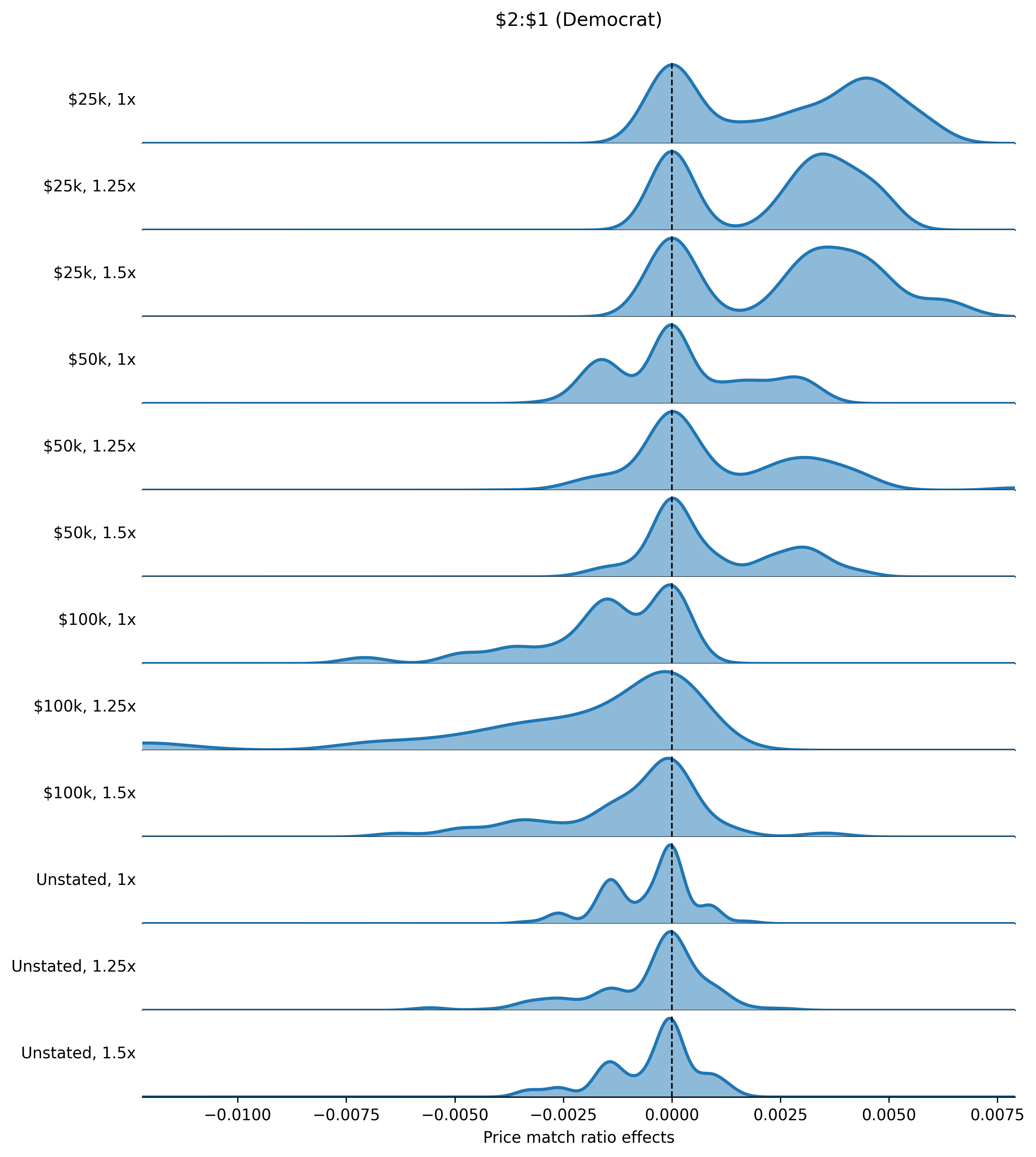}
    \end{subfigure}%
    ~ 
    \begin{subfigure}[t]{0.5\textwidth}
        \centering
        \includegraphics[height=3.5in]{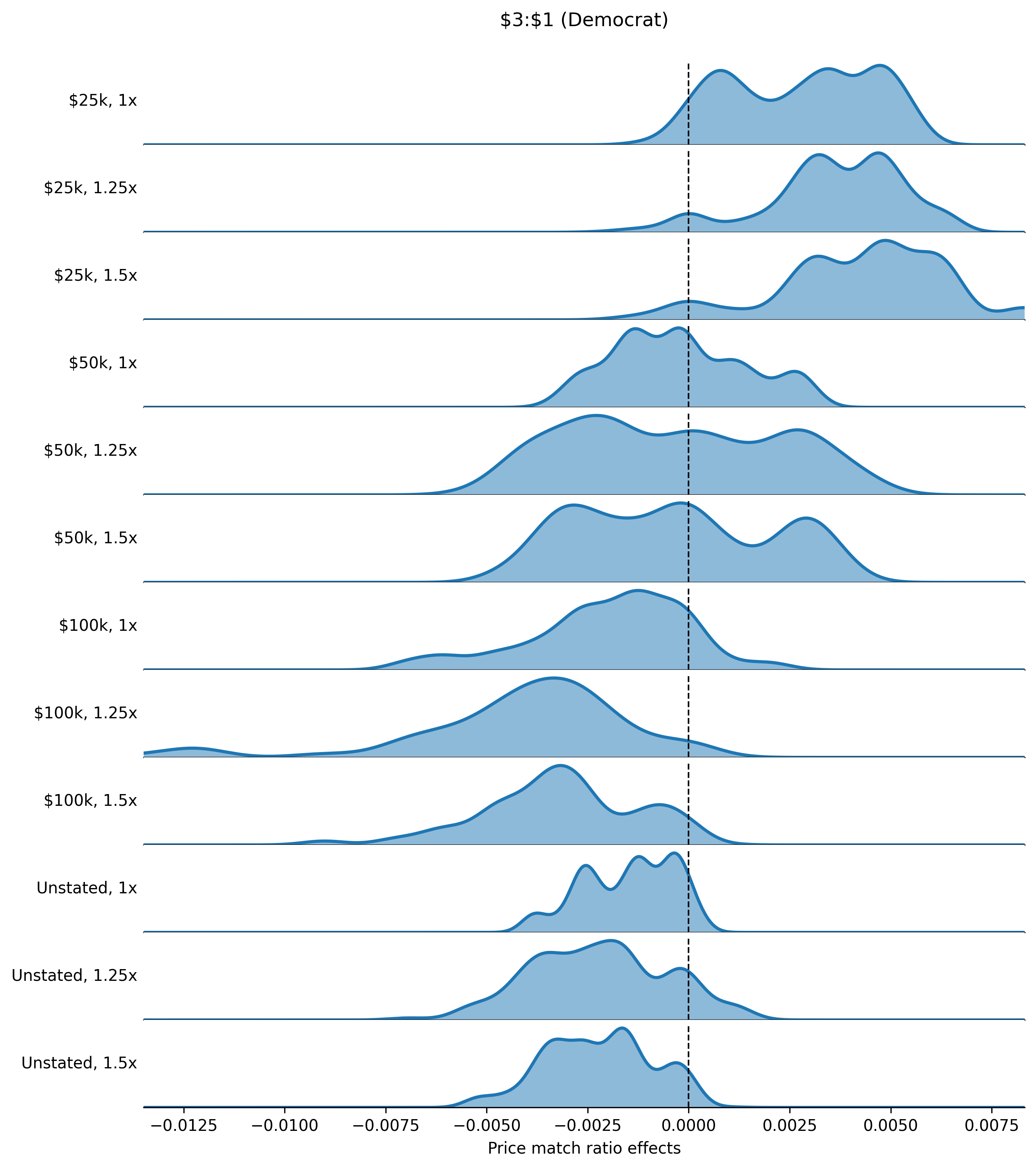}
    \end{subfigure}%
    \\
    \begin{subfigure}[t]{0.5\textwidth}
        \centering
        \includegraphics[height=3.5in]{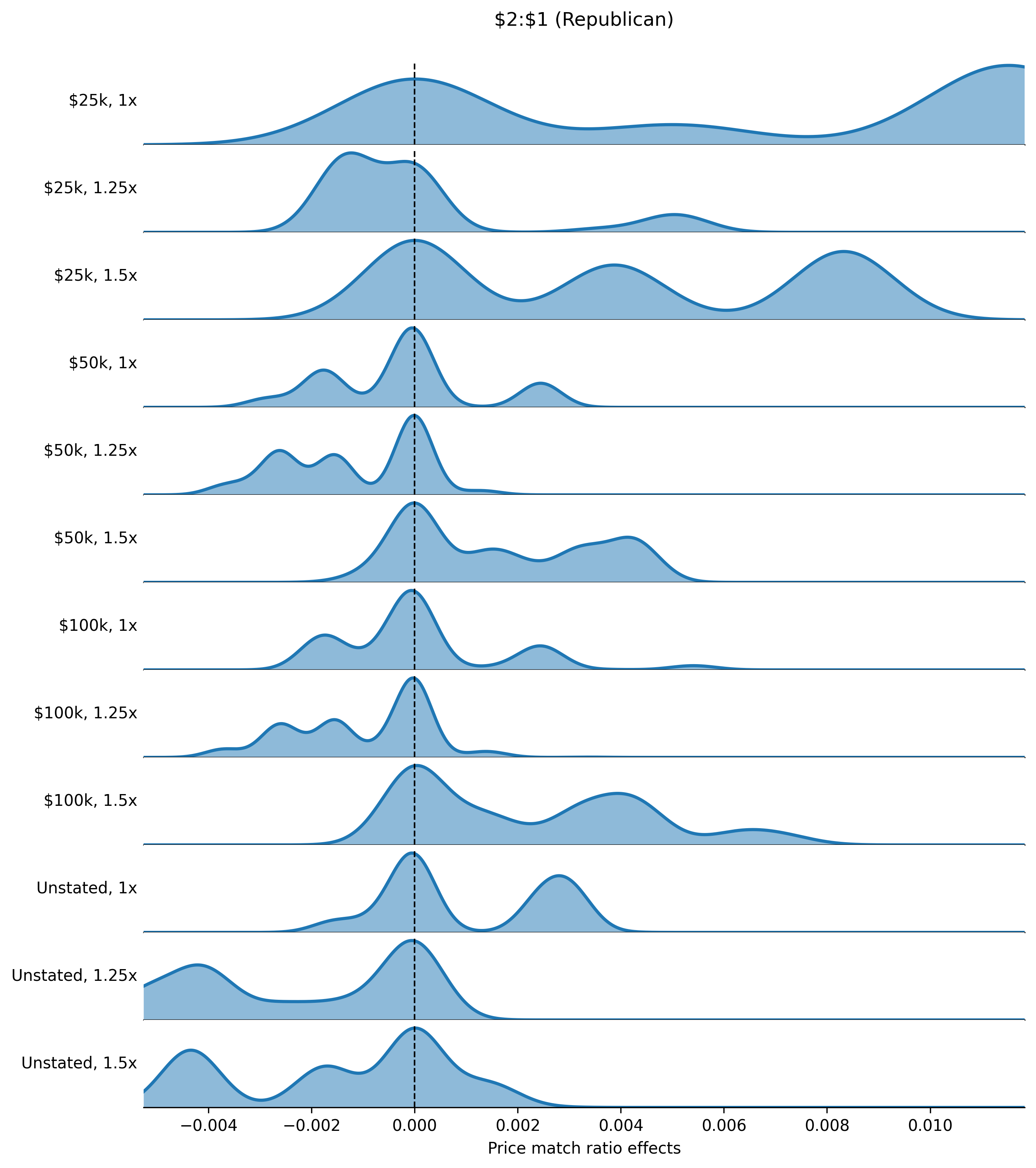}
    \end{subfigure}%
    ~ 
    \begin{subfigure}[t]{0.5\textwidth}
        \centering
        \includegraphics[height=3.5in]{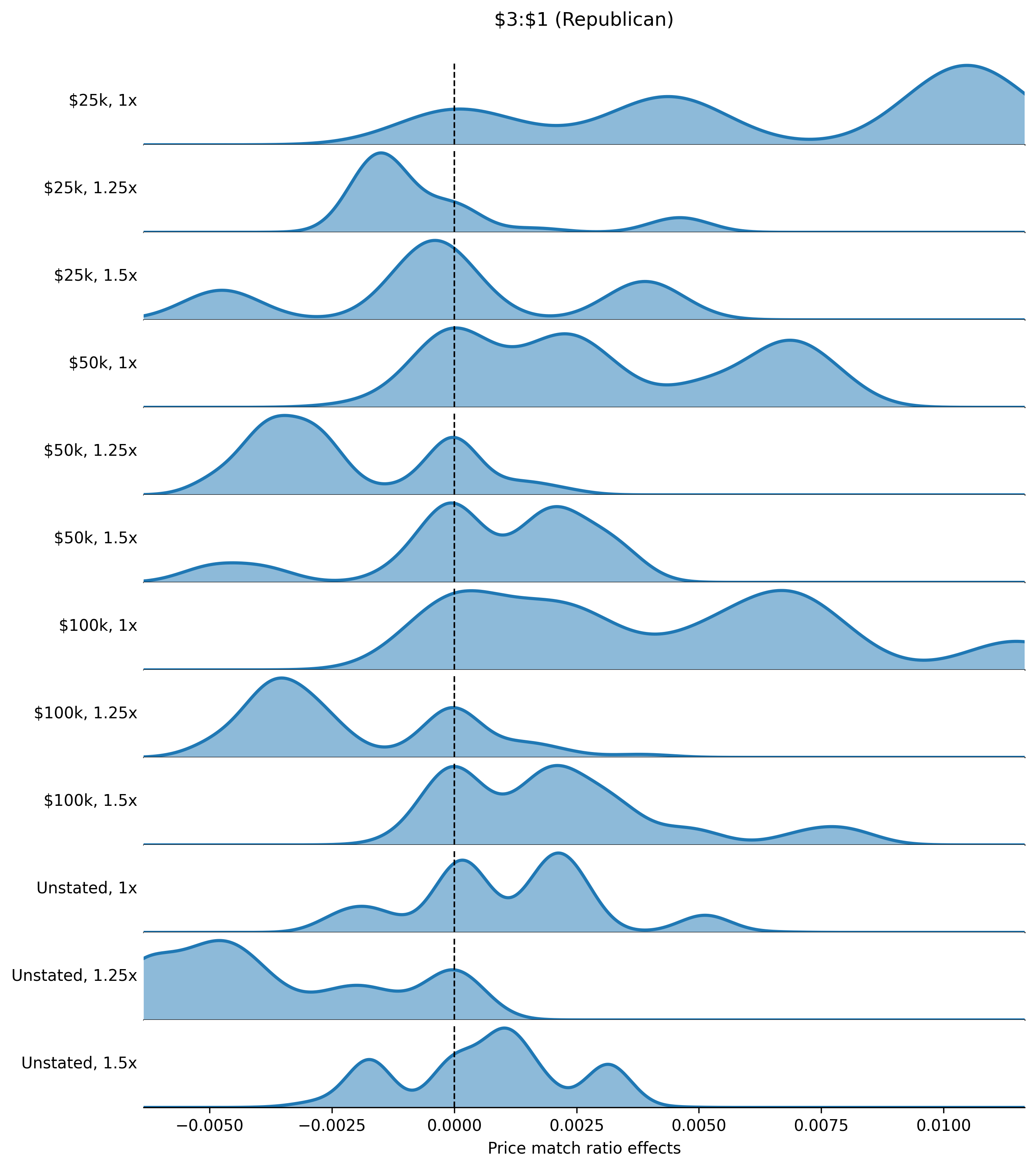}
    \end{subfigure}
    \caption{We visualize the posterior distribution of the functions $t_{1, \mathbf{x}}$ and $t_{2, \mathbf{x}}$ described in Section \ref{section:real-data-karlan} for \citet{karlan2007does} charitable donations dataset.}
    \label{fig:karlan-het-dist-appendix}
\end{figure}

\subsection{Telomere lengths}

\begin{figure}[!p]
    \centering
    \begin{subfigure}[t]{0.5\textwidth}
        \centering
        \includegraphics[height=2.5in]{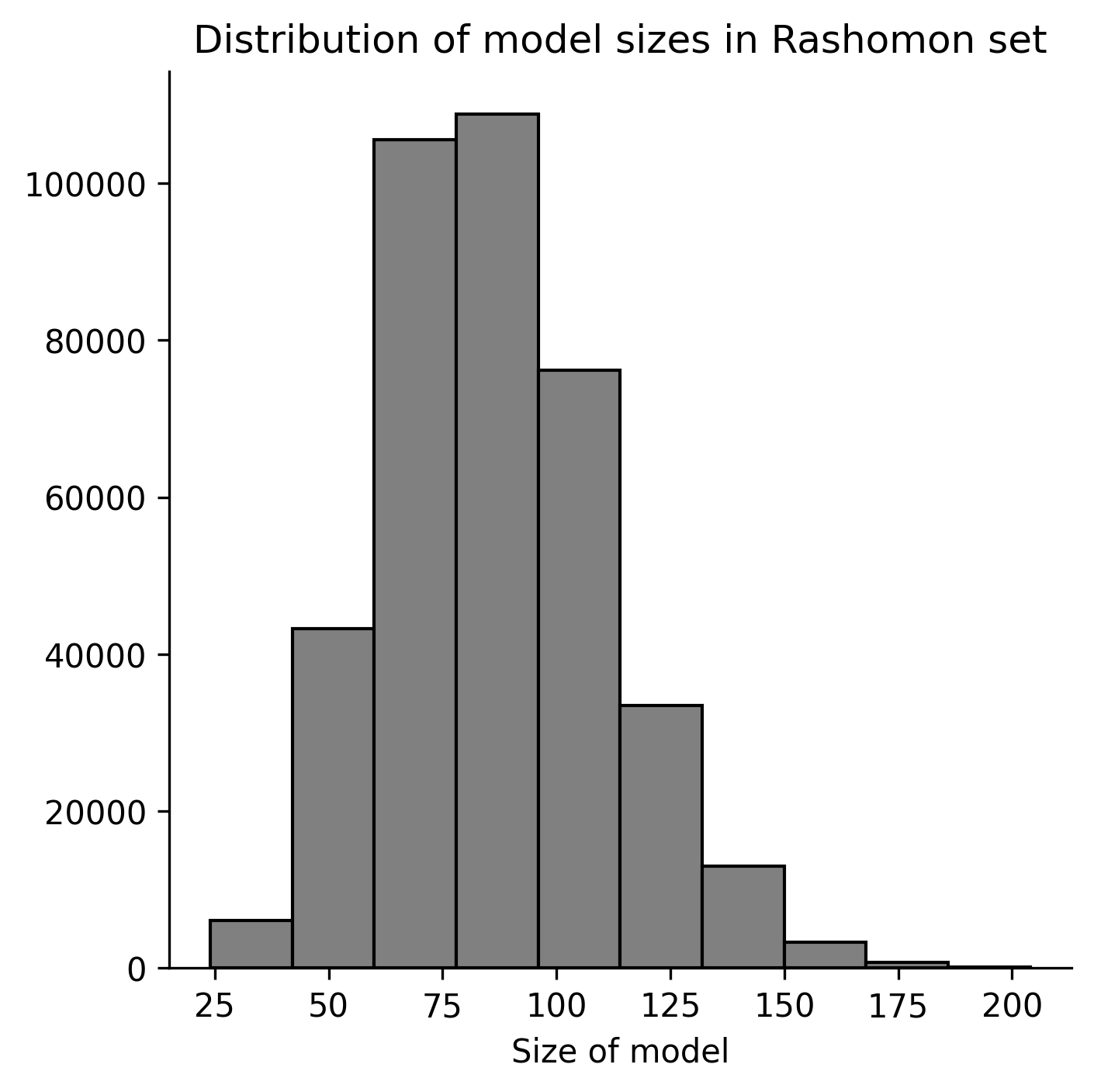}
    \end{subfigure}%
    ~ 
    \begin{subfigure}[t]{0.5\textwidth}
        \centering
        \includegraphics[height=2.5in]{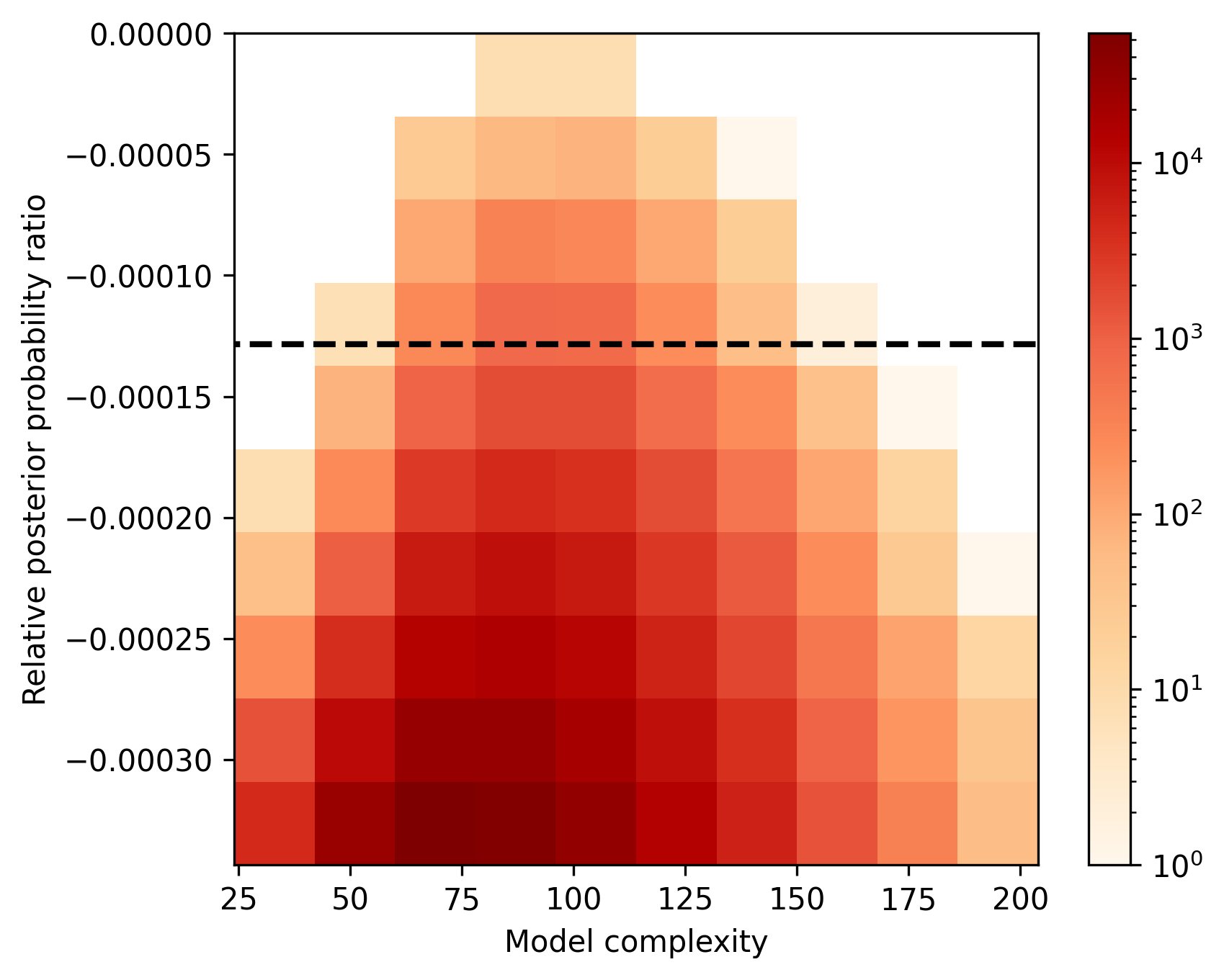}
    \end{subfigure} \\
    
    \begin{subfigure}[t]{0.5\textwidth}
        \centering
        \includegraphics[height=2.5in]{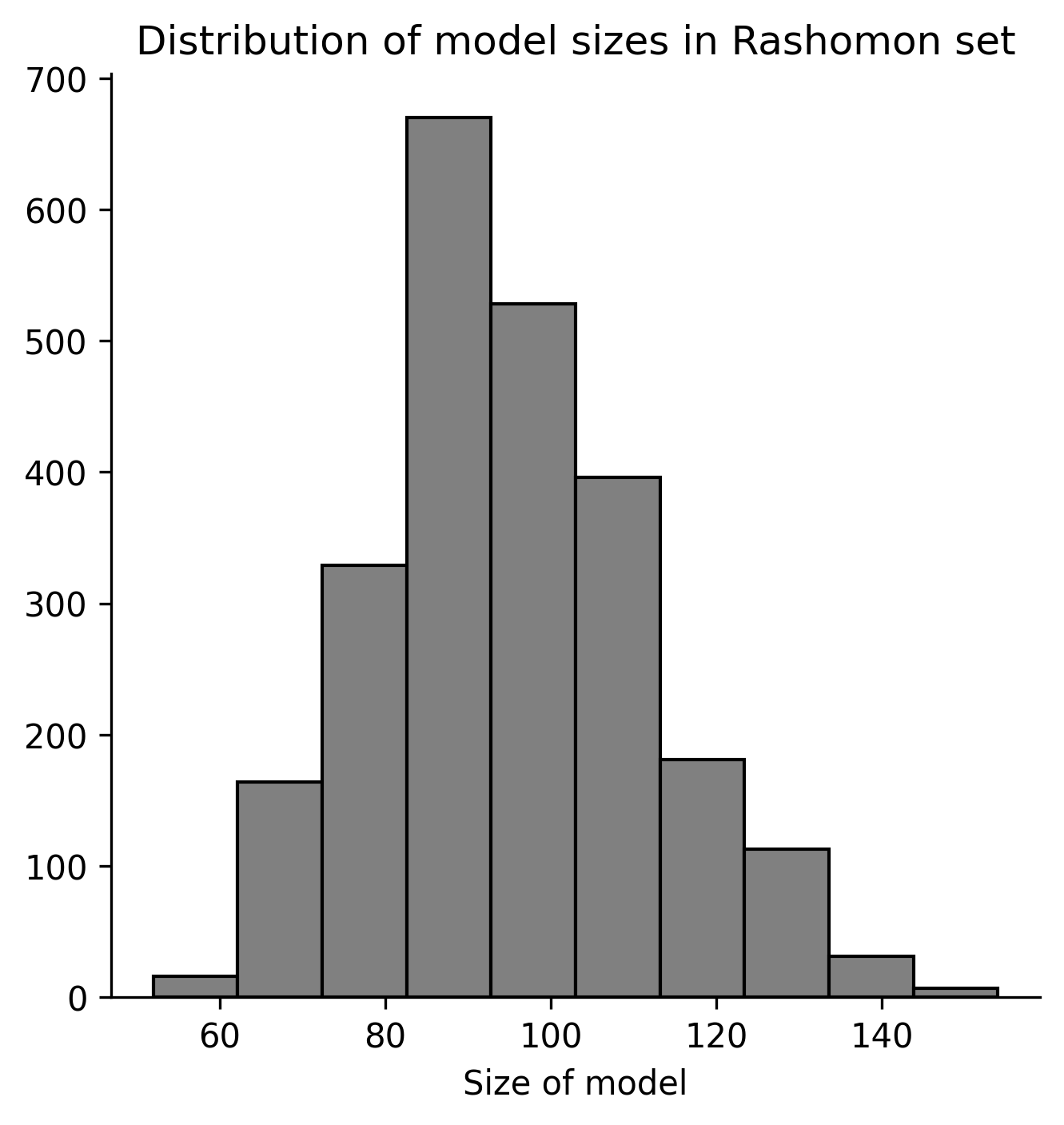}
    \end{subfigure}%
    ~ 
    \begin{subfigure}[t]{0.5\textwidth}
        \centering
        \includegraphics[height=2.5in]{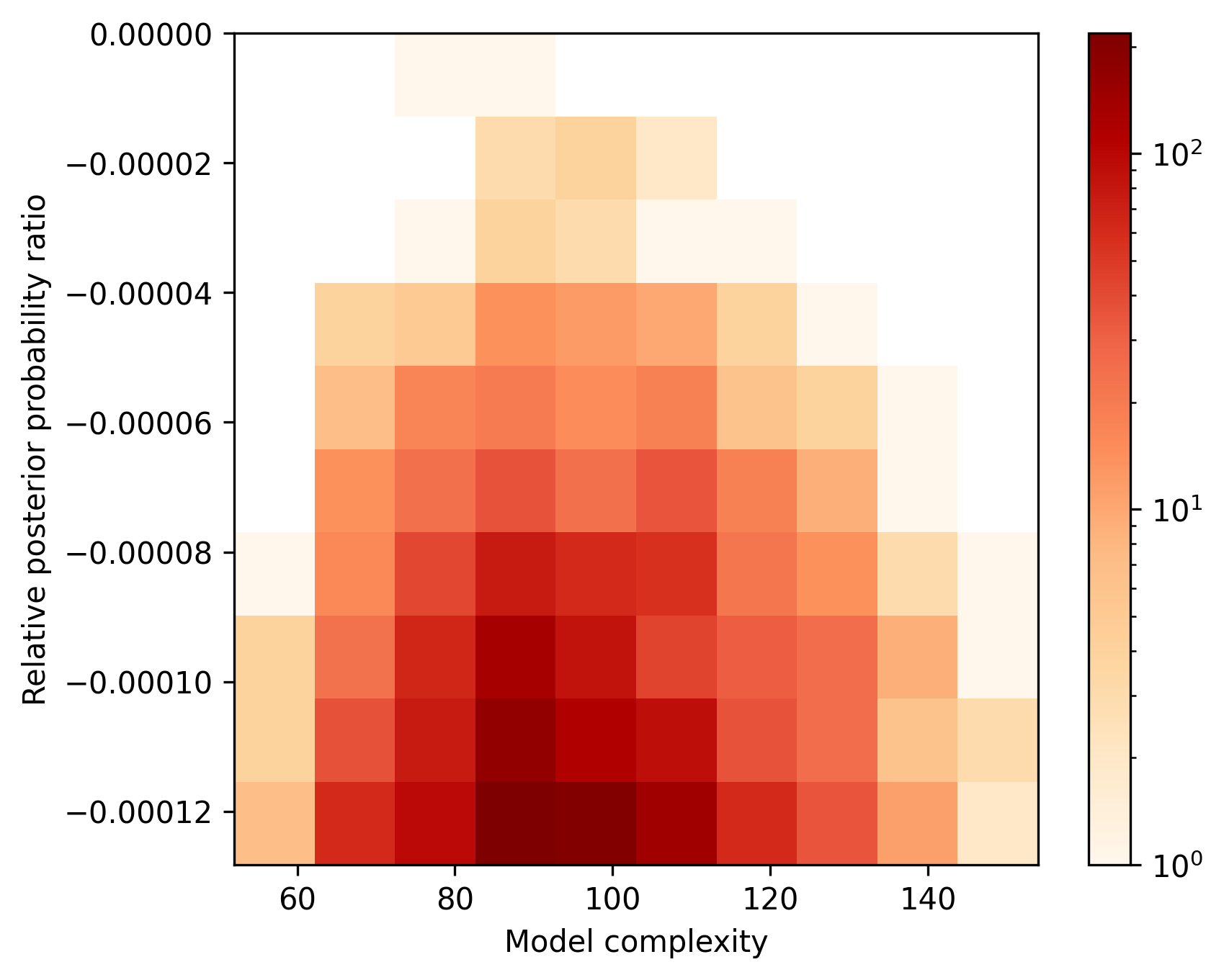}
    \end{subfigure}
    \caption{Visualizing the Rashomon set for NHANES telomeres dataset. The top two panels show the distribution of size of models and their relative posterior probability relative. The black dashed vertical and horizontal lines show the sparsity cutoff and Rashomon cutoff respectively. The bottom two panels show the same after pruning low-posterior models.}
    \label{fig:nhanes-rset-dist}
\end{figure}

\ref{fig:nhanes-rset-dist} visualizes the Rashomon sets for the NHANES telomere lengths using the 2D histogram that is described in Figure \ref{fig:heatmap-example} in Appendix \ref{appendix:additional-sims}. In Figure \ref{fig:nhanes-het-dist-appendix}, we visualize the posterior distributions, restricted to the RPS, of $t_{1, \mathbf{x}}$ for each race $r$ and covariates described in Section \ref{section:real-data-nhanes}.

\begin{figure}[!p]
    \centering
    \begin{subfigure}[t]{0.3\textwidth}
        \centering
        \includegraphics[height=1in]{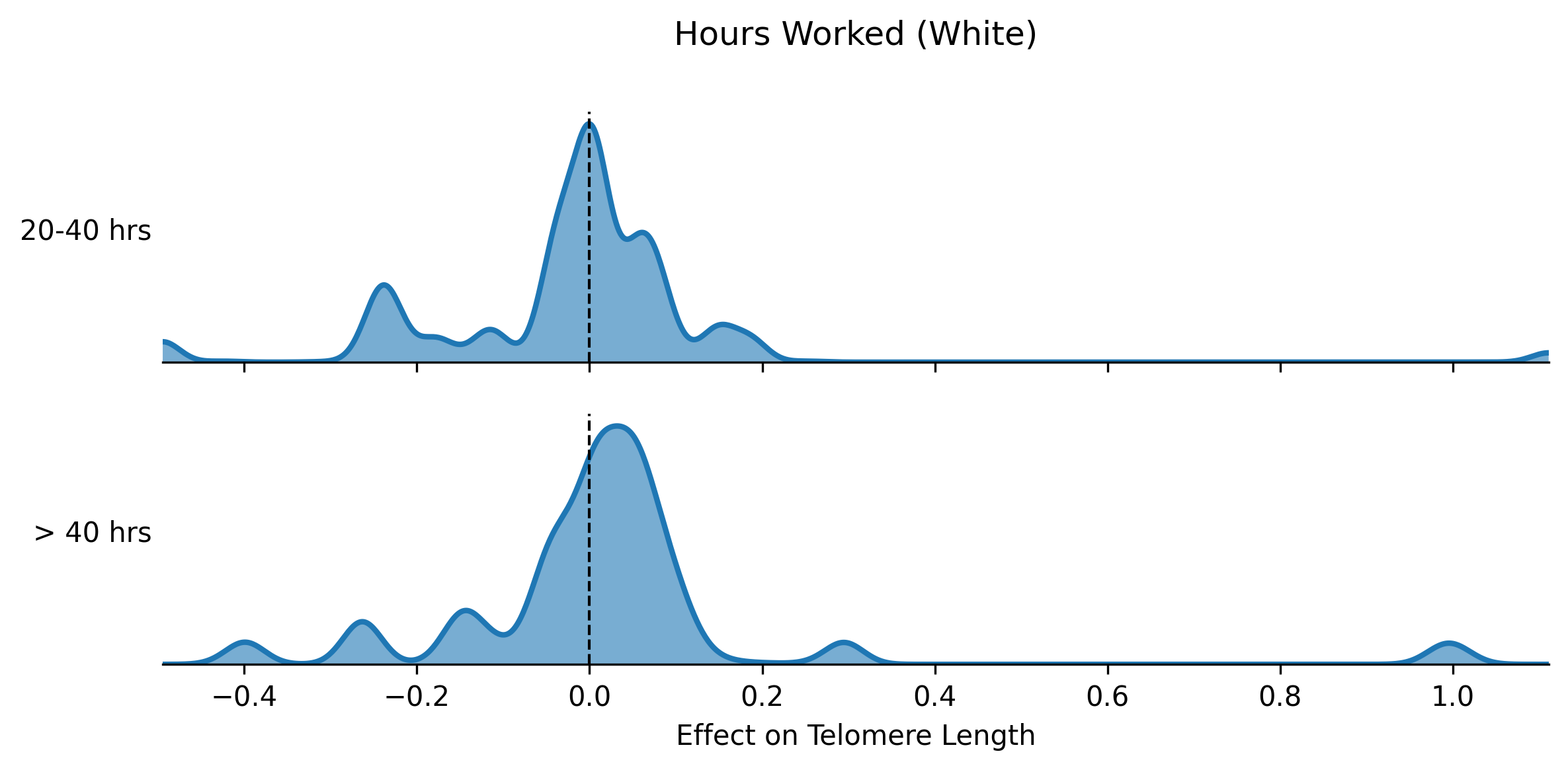}
    \end{subfigure}%
    ~ 
    \begin{subfigure}[t]{0.3\textwidth}
        \centering
        \includegraphics[height=1in]{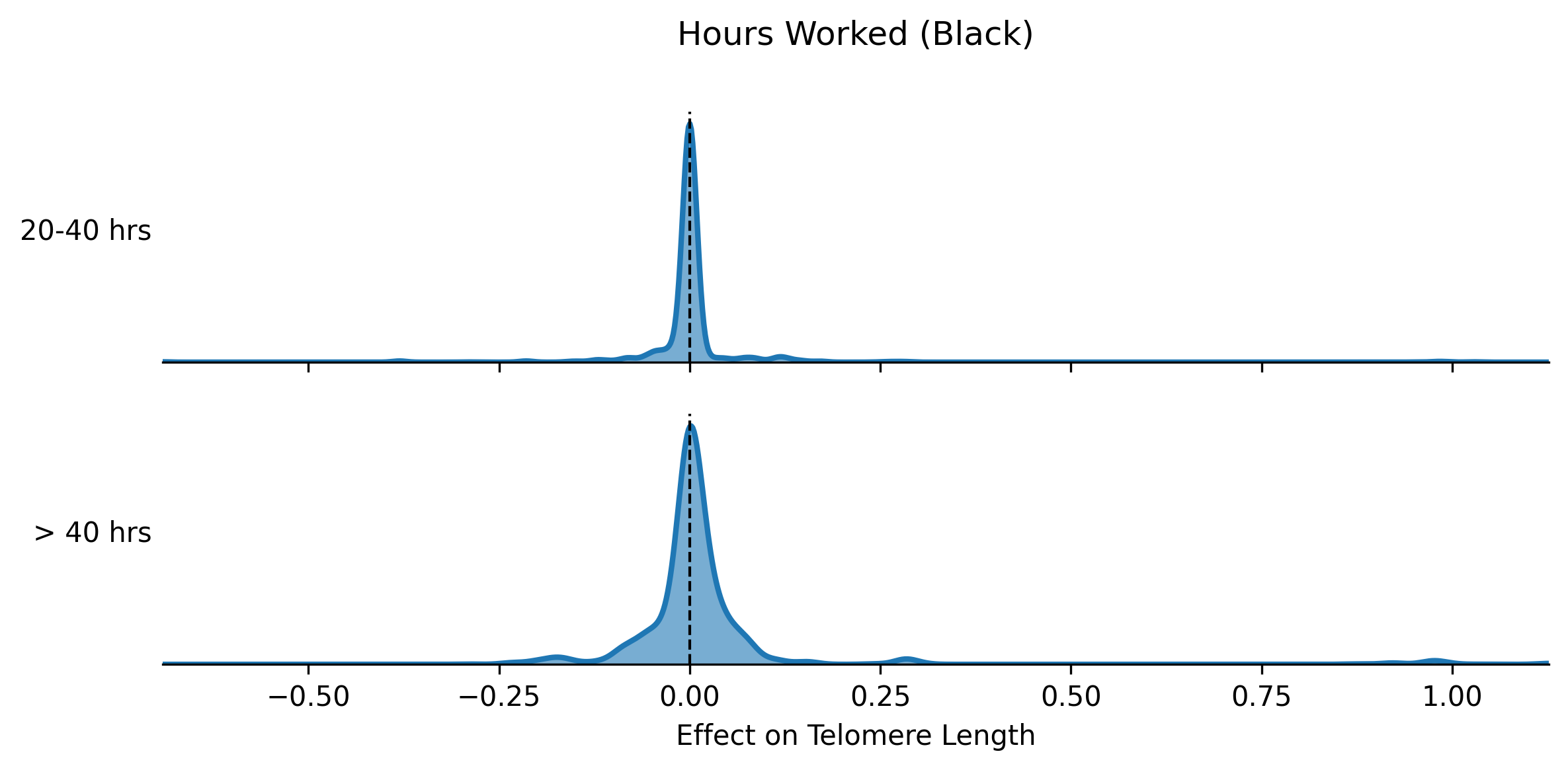}
    \end{subfigure}%
    ~
    \begin{subfigure}[t]{0.3\textwidth}
        \centering
        \includegraphics[height=1in]{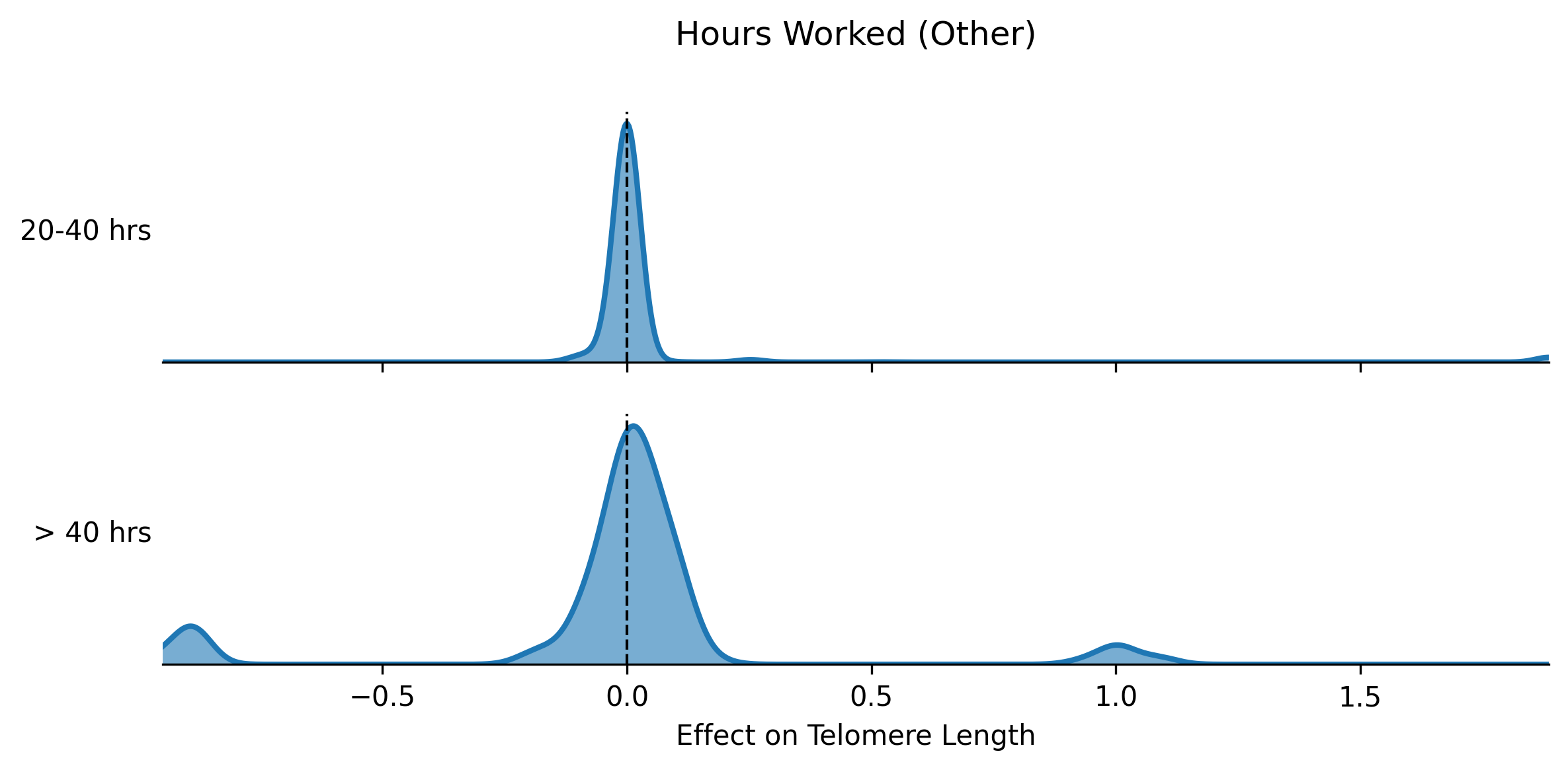}
    \end{subfigure}%
    \\ 
    \begin{subfigure}[t]{0.3\textwidth}
        \centering
        \includegraphics[height=1in]{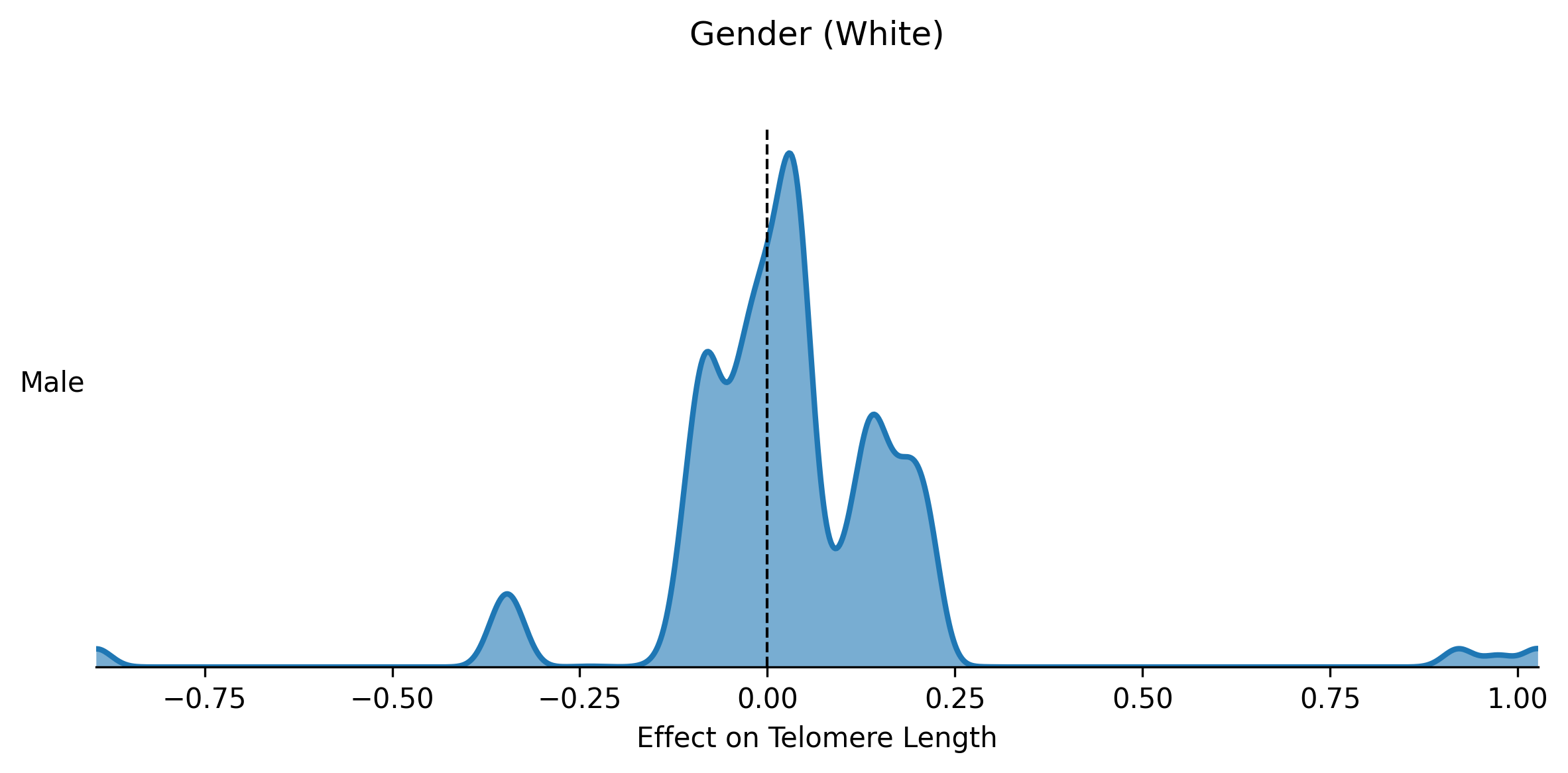}
    \end{subfigure}%
    ~ 
    \begin{subfigure}[t]{0.3\textwidth}
        \centering
        \includegraphics[height=1in]{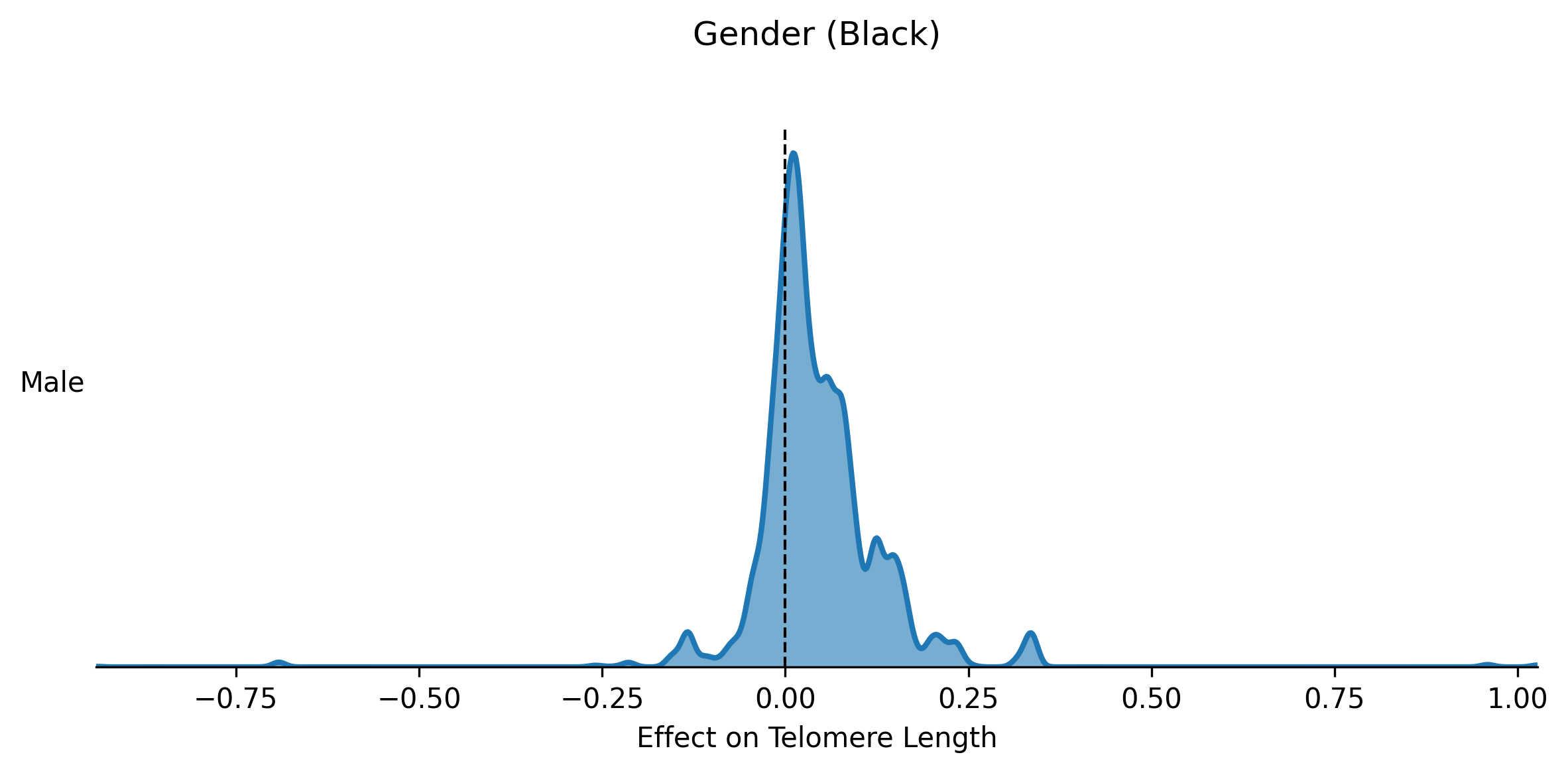}
    \end{subfigure}%
    ~
    \begin{subfigure}[t]{0.3\textwidth}
        \centering
        \includegraphics[height=1in]{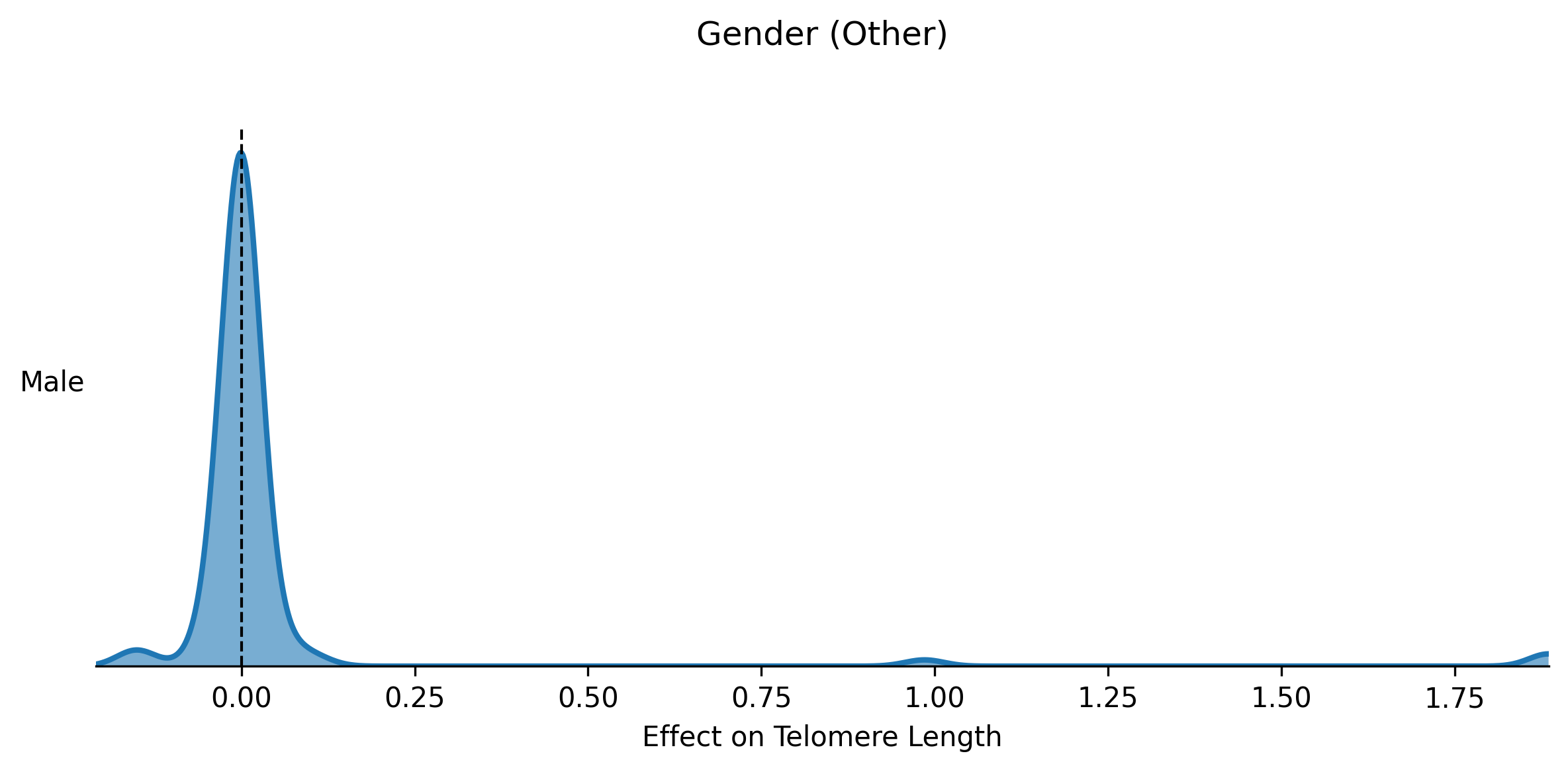}
    \end{subfigure}%
    \\
    \begin{subfigure}[t]{0.3\textwidth}
        \centering
        \includegraphics[height=1in]{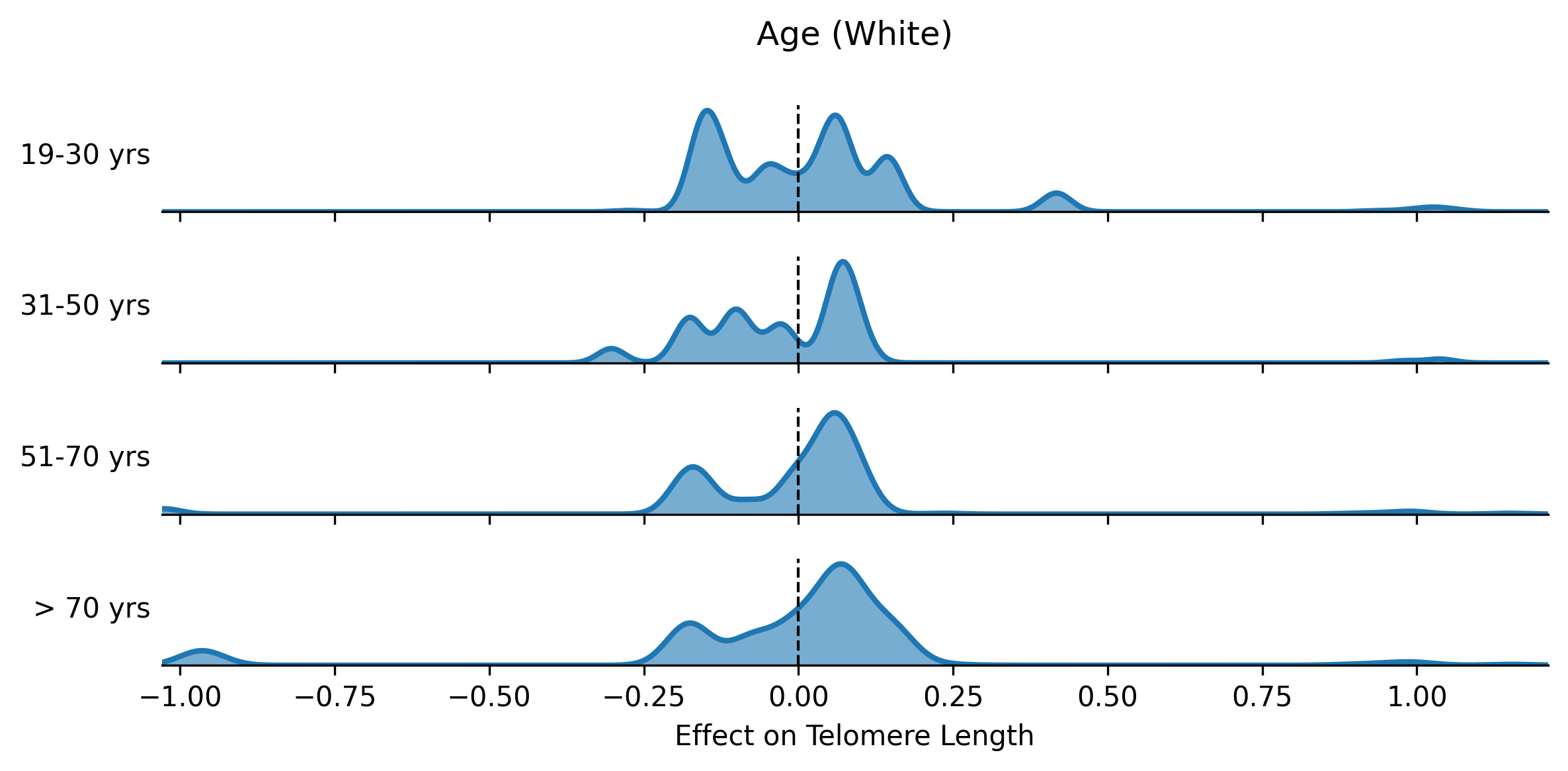}
    \end{subfigure}%
    ~ 
    \begin{subfigure}[t]{0.3\textwidth}
        \centering
        \includegraphics[height=1in]{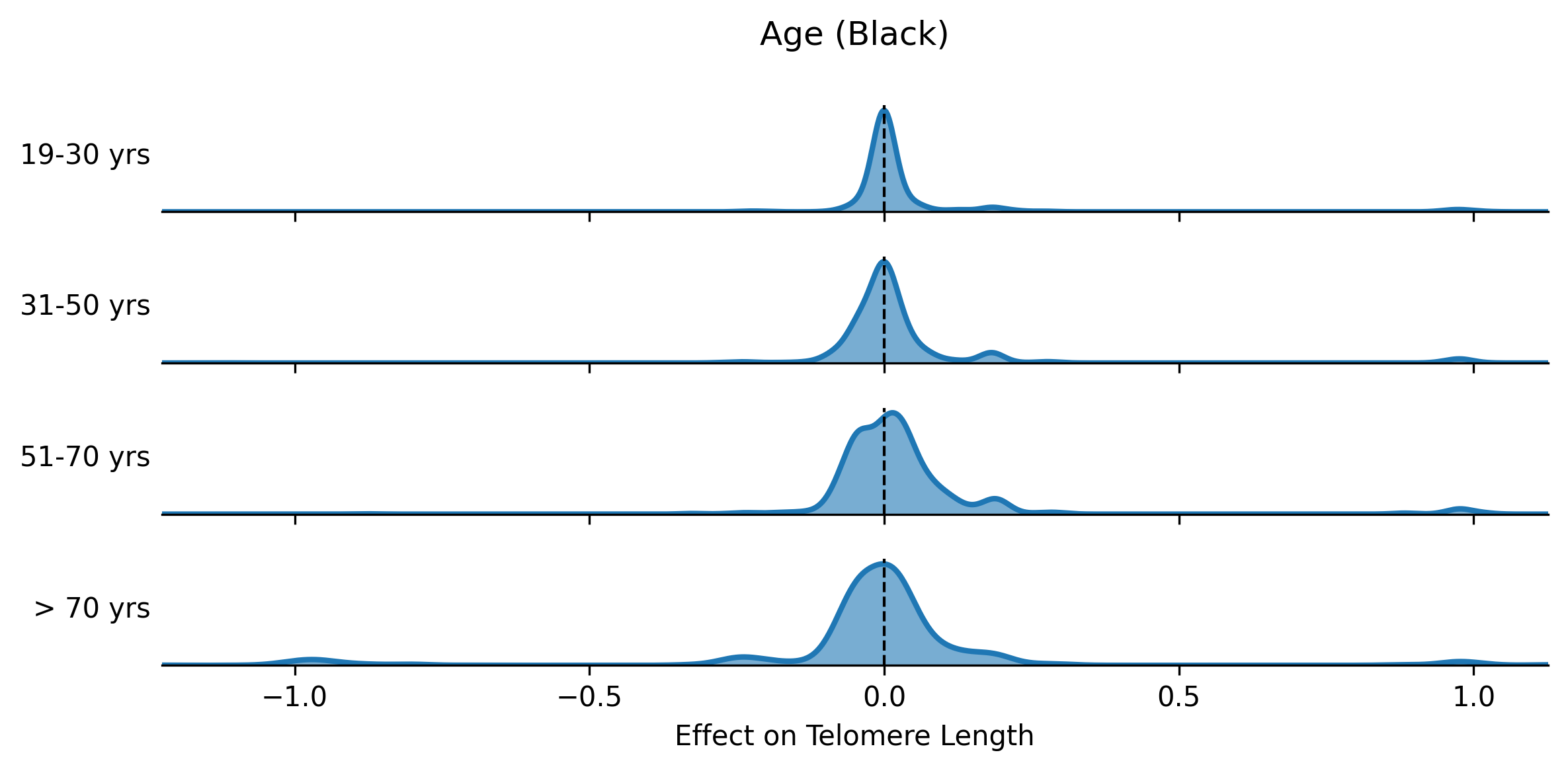}
    \end{subfigure}%
    ~
    \begin{subfigure}[t]{0.3\textwidth}
        \centering
        \includegraphics[height=1in]{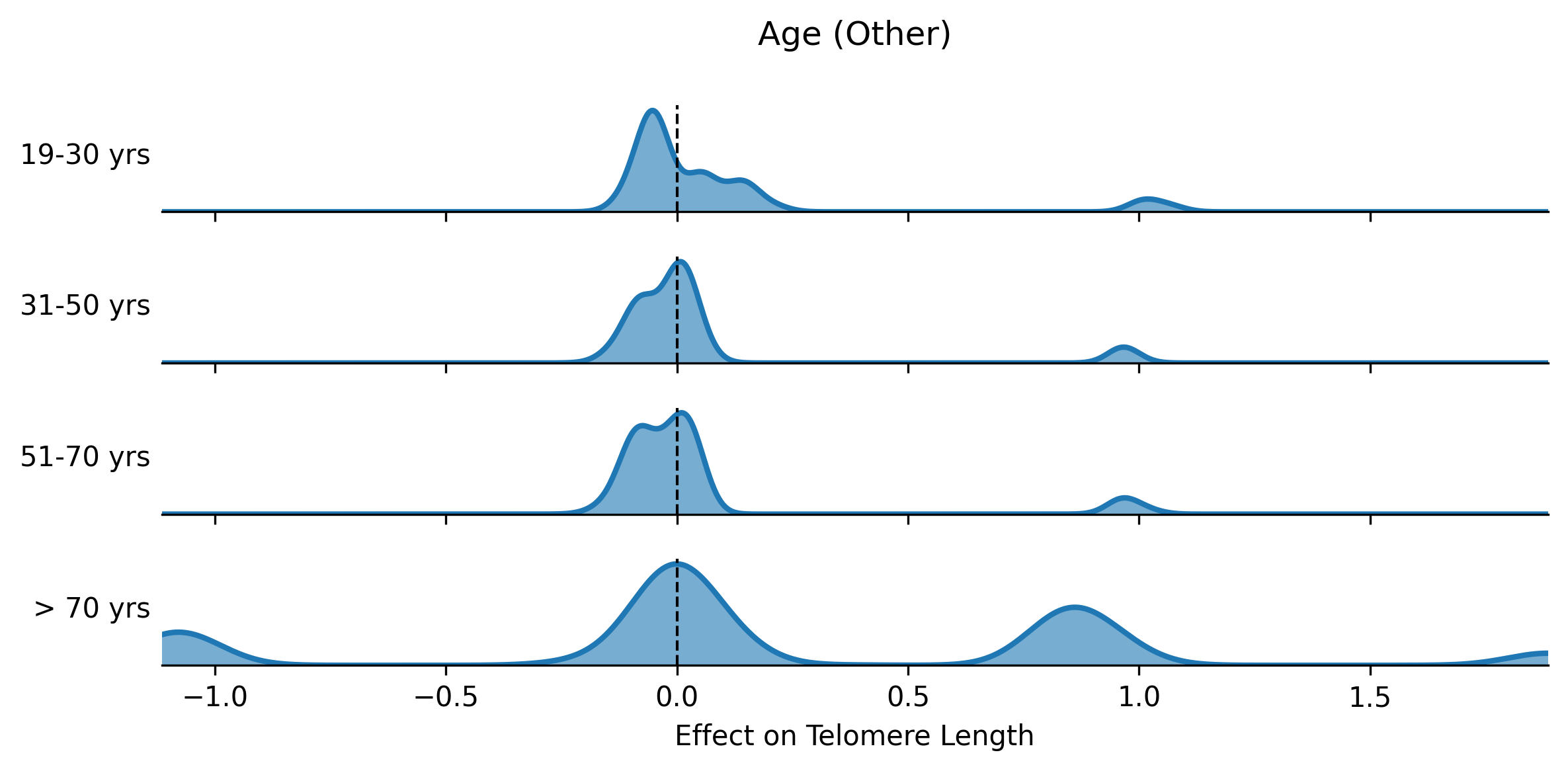}
    \end{subfigure}%
    \\ 
    \begin{subfigure}[t]{0.3\textwidth}
        \centering
        \includegraphics[height=1in]{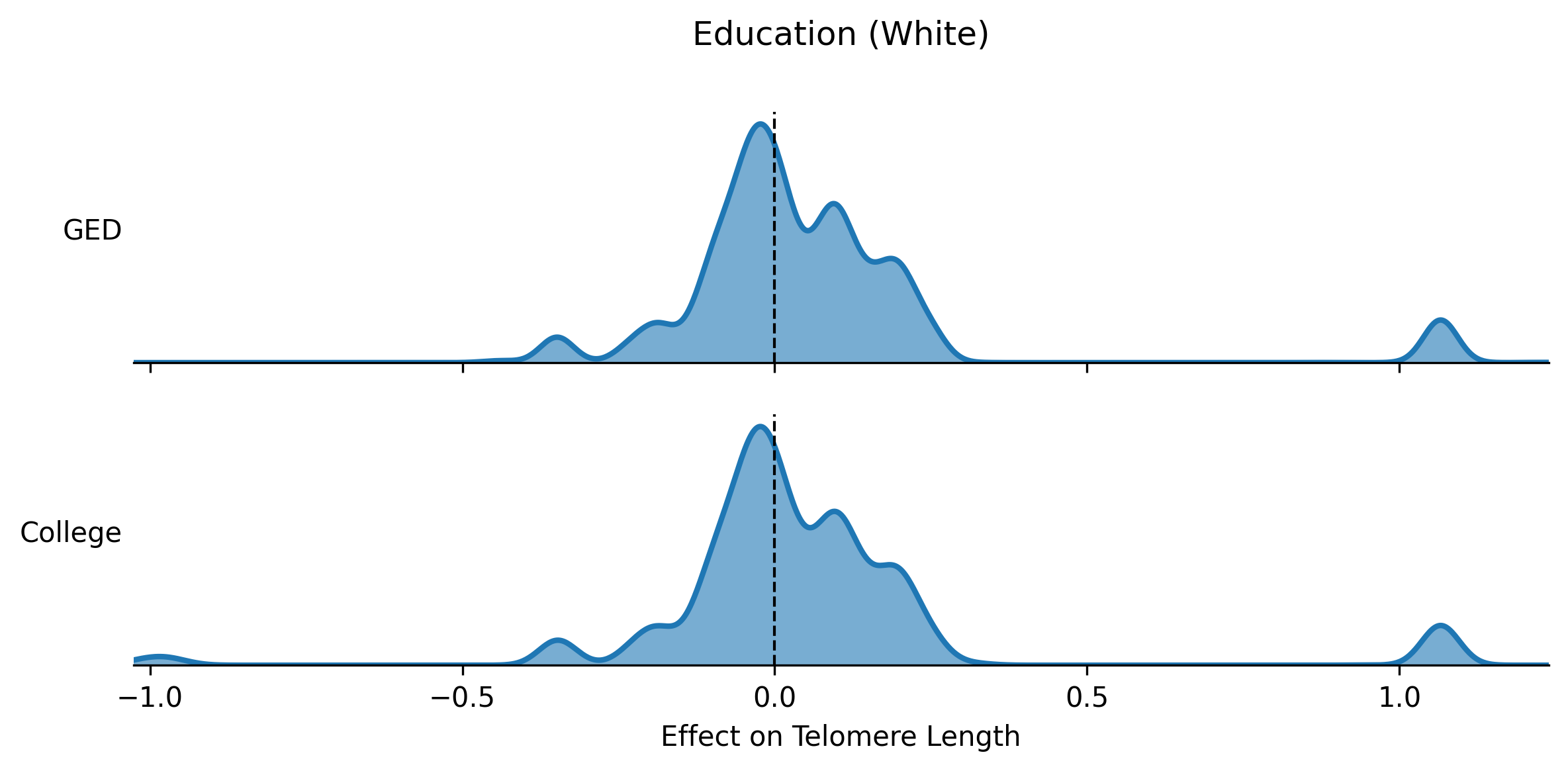}
    \end{subfigure}%
    ~ 
    \begin{subfigure}[t]{0.3\textwidth}
        \centering
        \includegraphics[height=1in]{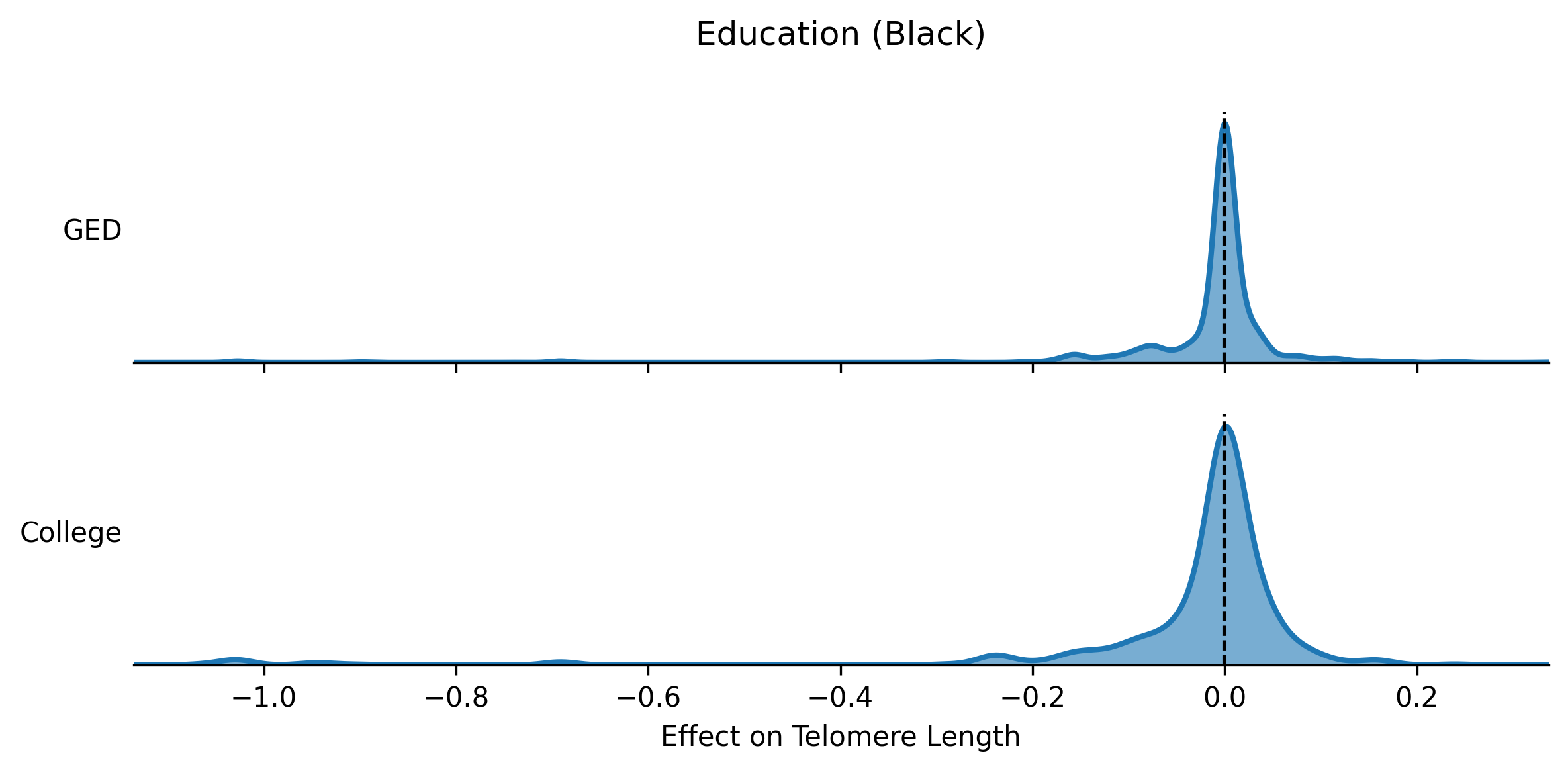}
    \end{subfigure}%
    ~
    \begin{subfigure}[t]{0.3\textwidth}
        \centering
        \includegraphics[height=1in]{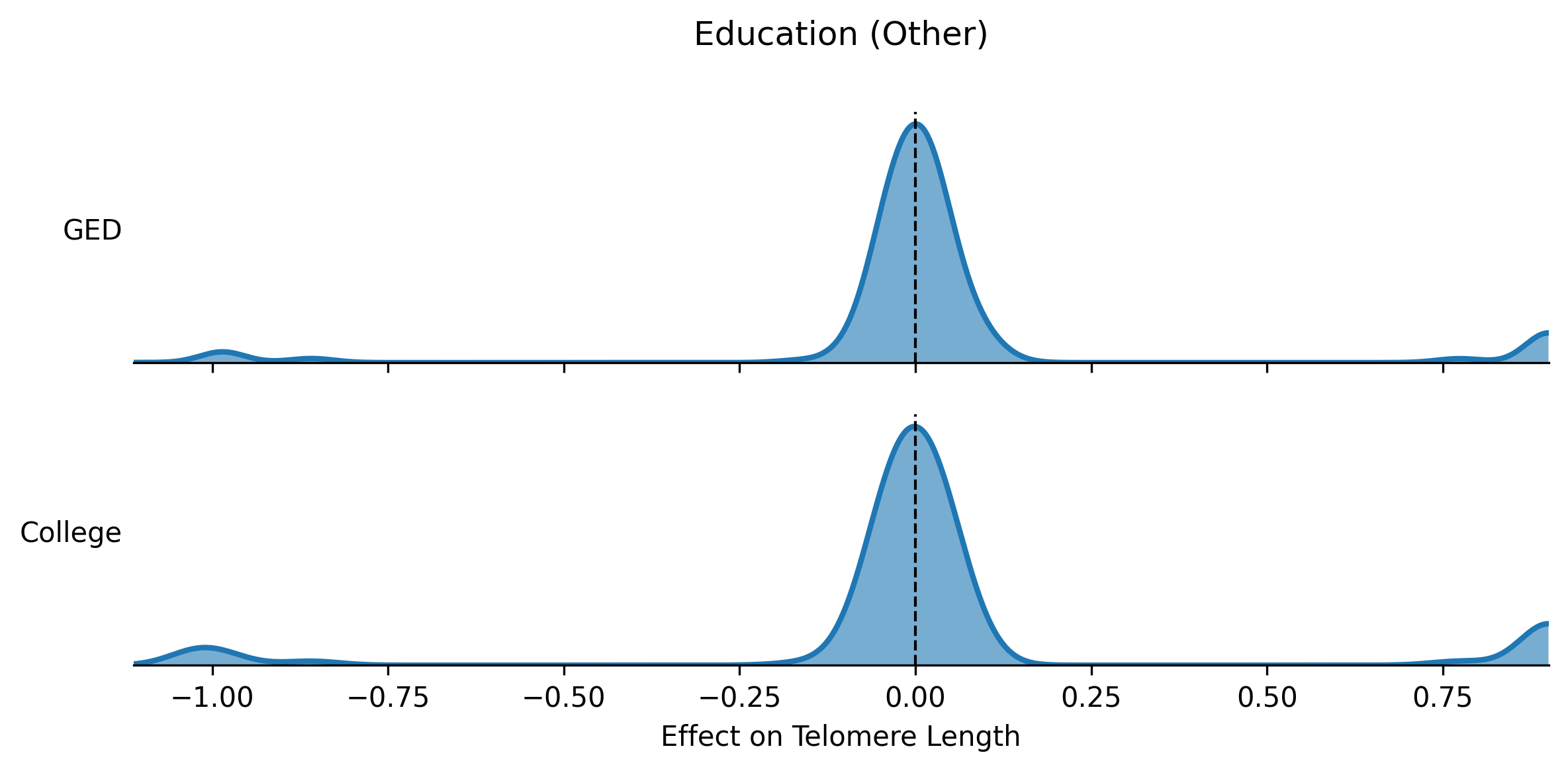}
    \end{subfigure}%
    \caption{We visualize the posterior distribution of the functions $t_{1, \mathbf{x}}$ for each race $r$ and covariates described in Section \ref{section:real-data-nhanes} for the NHANES telomere dataset.}
    \label{fig:nhanes-het-dist-appendix}
\end{figure}

\subsection{Heterogeneity in the impact of microcredit access.}

\begin{figure}[!tb]
    \centering
    \includegraphics[height=6.5in]{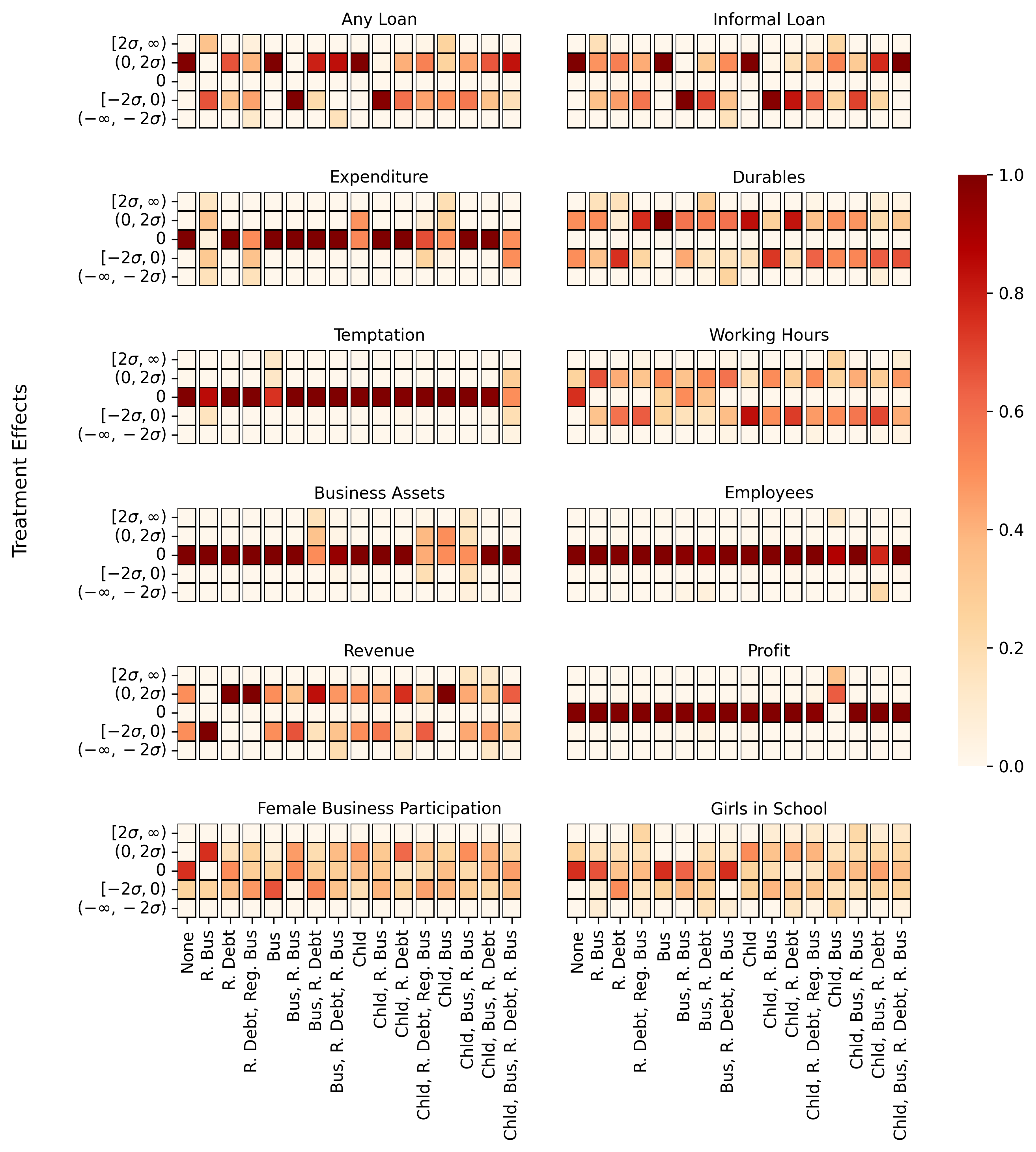}
    \caption{
    Here, we visualize the average number of models in the Rashomon set indicating a positive, zero, or negative effect. Each column corresponds to a different feature profile where the label denotes which features are active (i.e., do not take the lowest level). ``None'' means that all features are taking these lowest values. We also allow the gender of the household head and education status of the household head to take on any value in all of the sixteen feature profiles.
    }
    \label{fig:microcredit-mosiac-fx-full}
\end{figure}

\begin{figure}[!p]
    \centering
    \begin{subfigure}[t]{0.5\textwidth}
        \centering
        \includegraphics[height=3.5in]{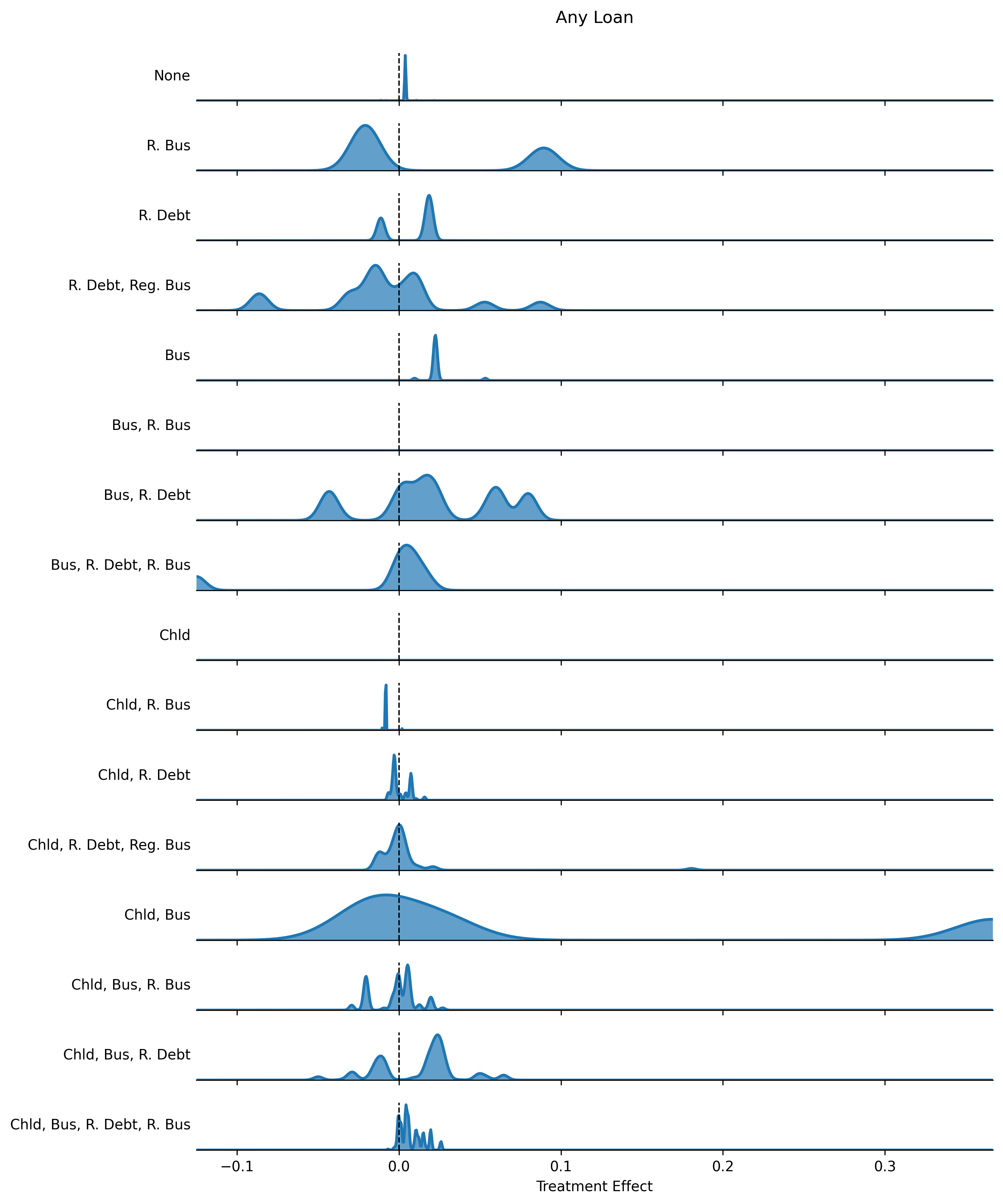}
    \end{subfigure}%
    ~ 
    \begin{subfigure}[t]{0.5\textwidth}
        \centering
        \includegraphics[height=3.5in]{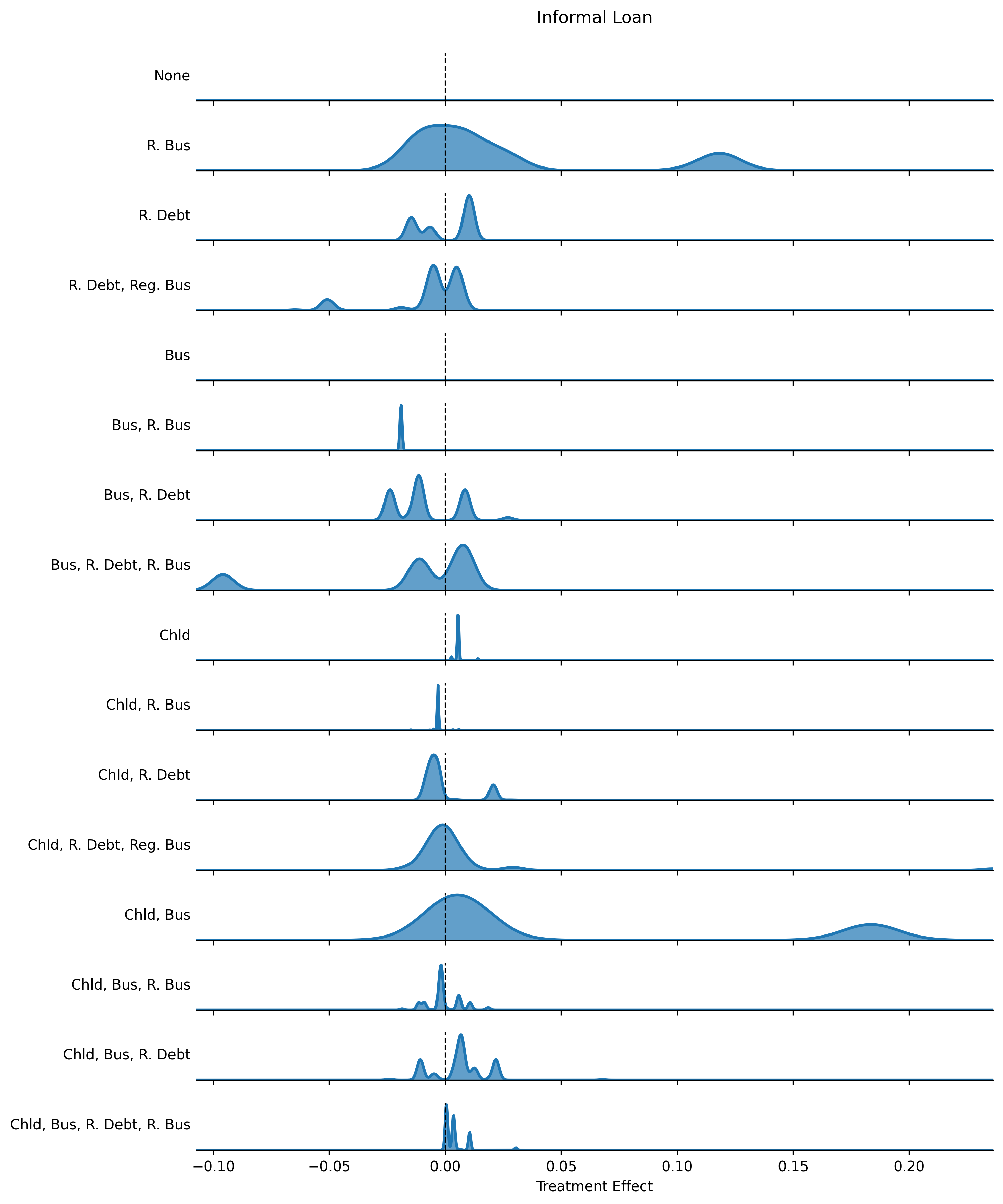}
    \end{subfigure}%
    \\
    \begin{subfigure}[t]{0.5\textwidth}
        \centering
        \includegraphics[height=3.5in]{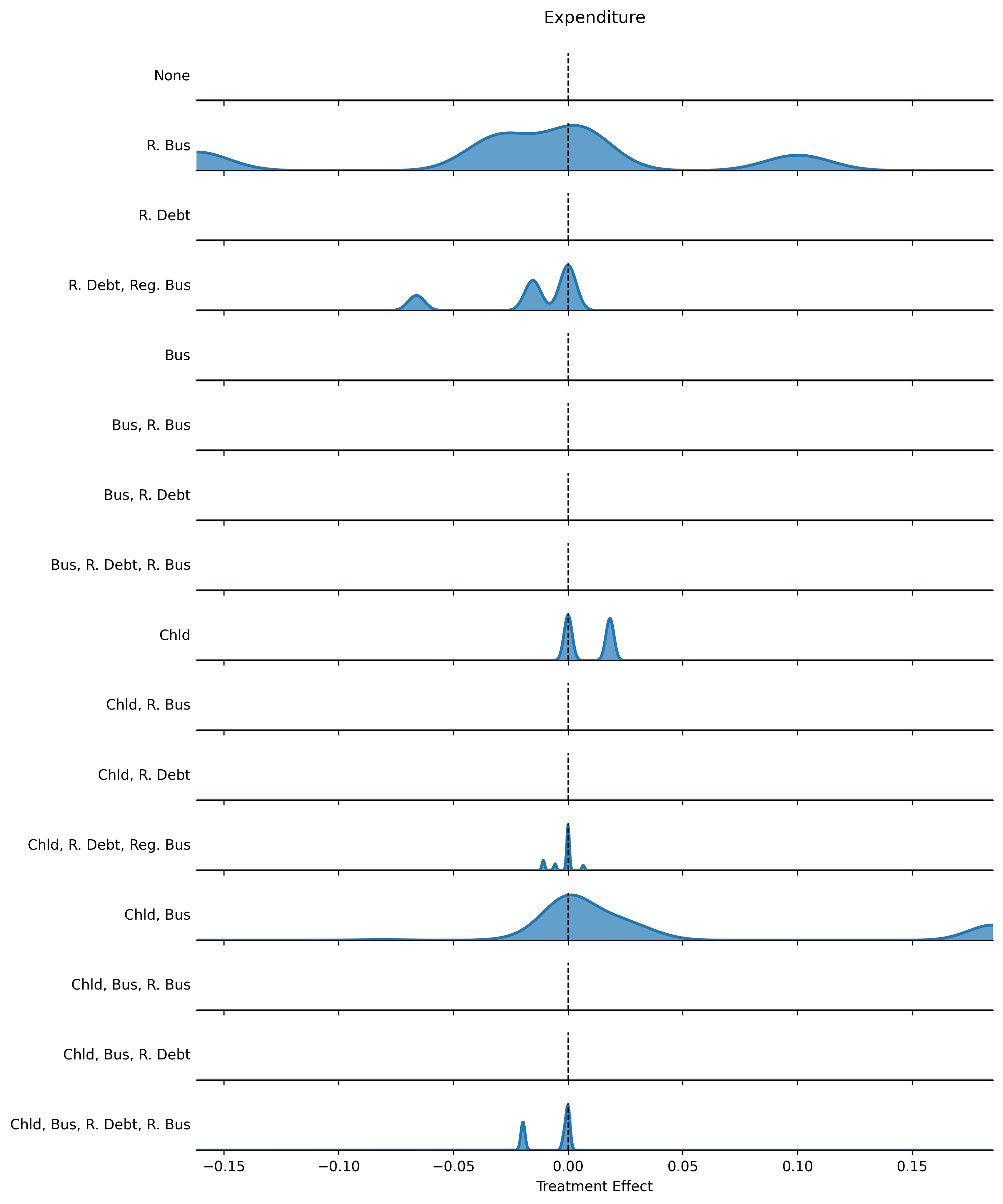}
    \end{subfigure}%
    ~ 
    \begin{subfigure}[t]{0.5\textwidth}
        \centering
        \includegraphics[height=3.5in]{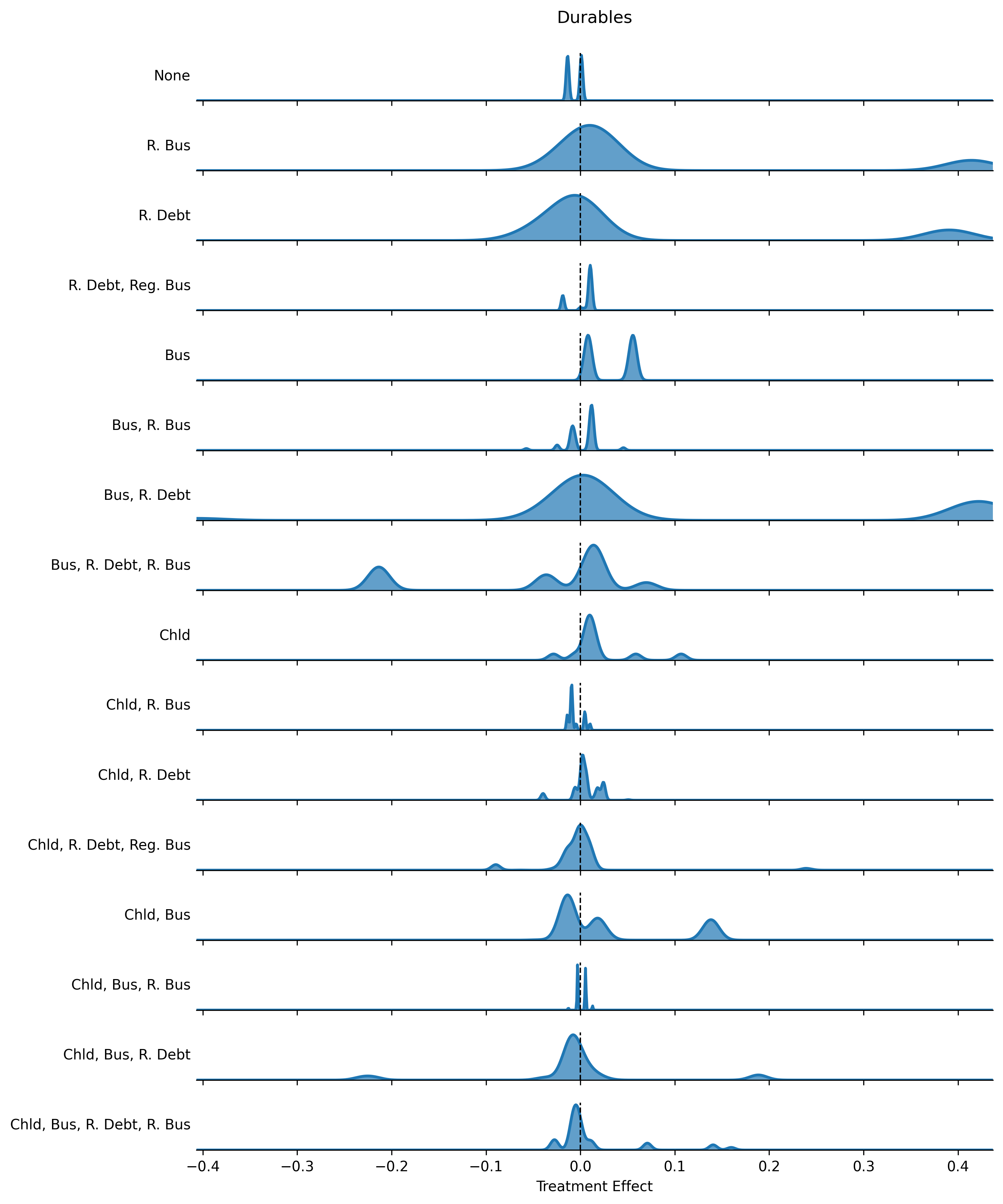}
    \end{subfigure}%
    \caption{The posterior distributions for the treatment effect on loan amounts and total and durable expenditures restricted to the RPS. See the corresponding quantized heatmaps in Figure \ref{fig:microcredit-mosiac-fx-full}.}
    \label{fig:microcredit-het-dist-appendix-loan}
\end{figure}

\begin{figure}[!p]
    \centering
    \begin{subfigure}[t]{0.5\textwidth}
        \centering
        \includegraphics[height=3.5in]{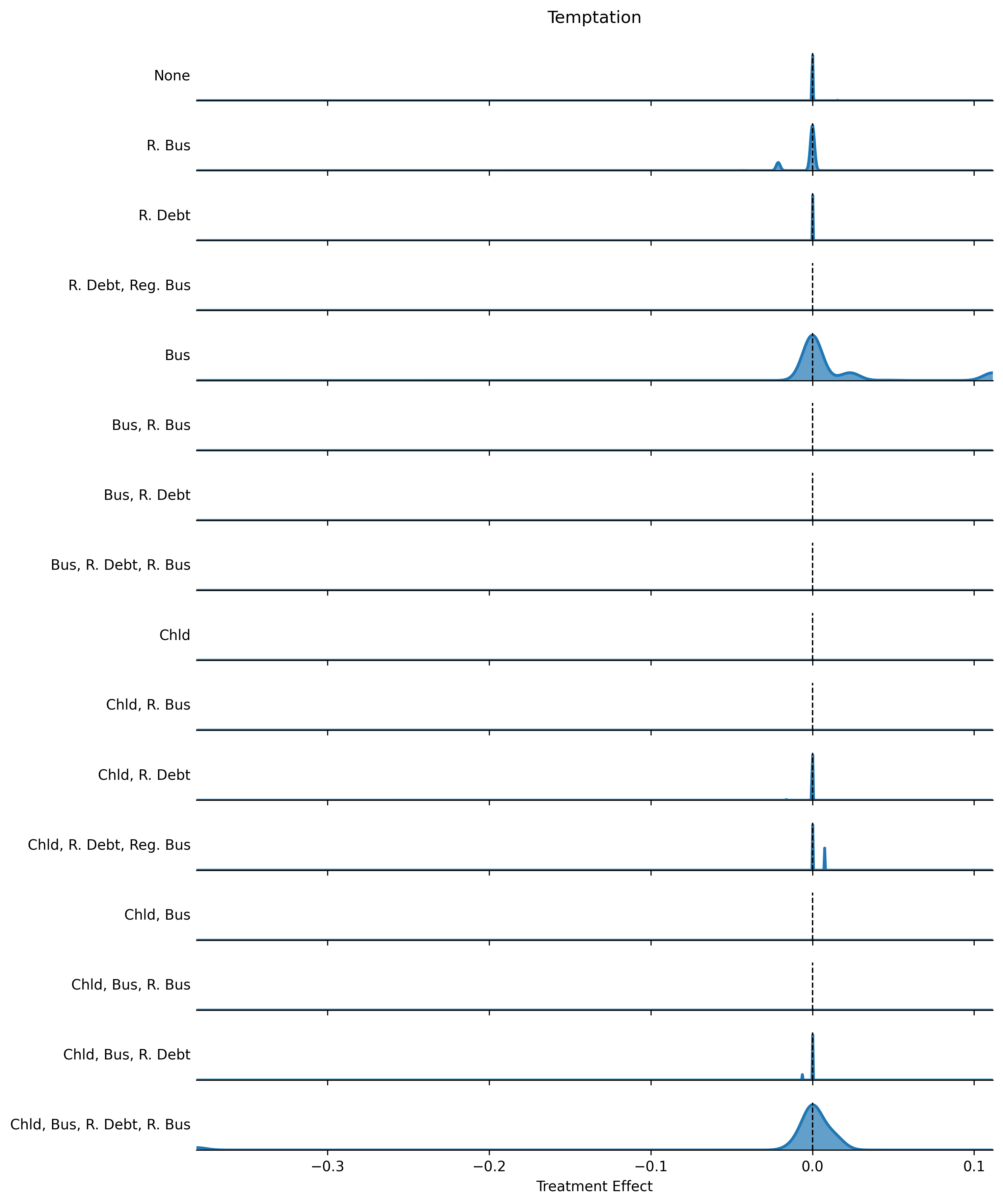}
    \end{subfigure}%
    ~ 
    \begin{subfigure}[t]{0.5\textwidth}
        \centering
        \includegraphics[height=3.5in]{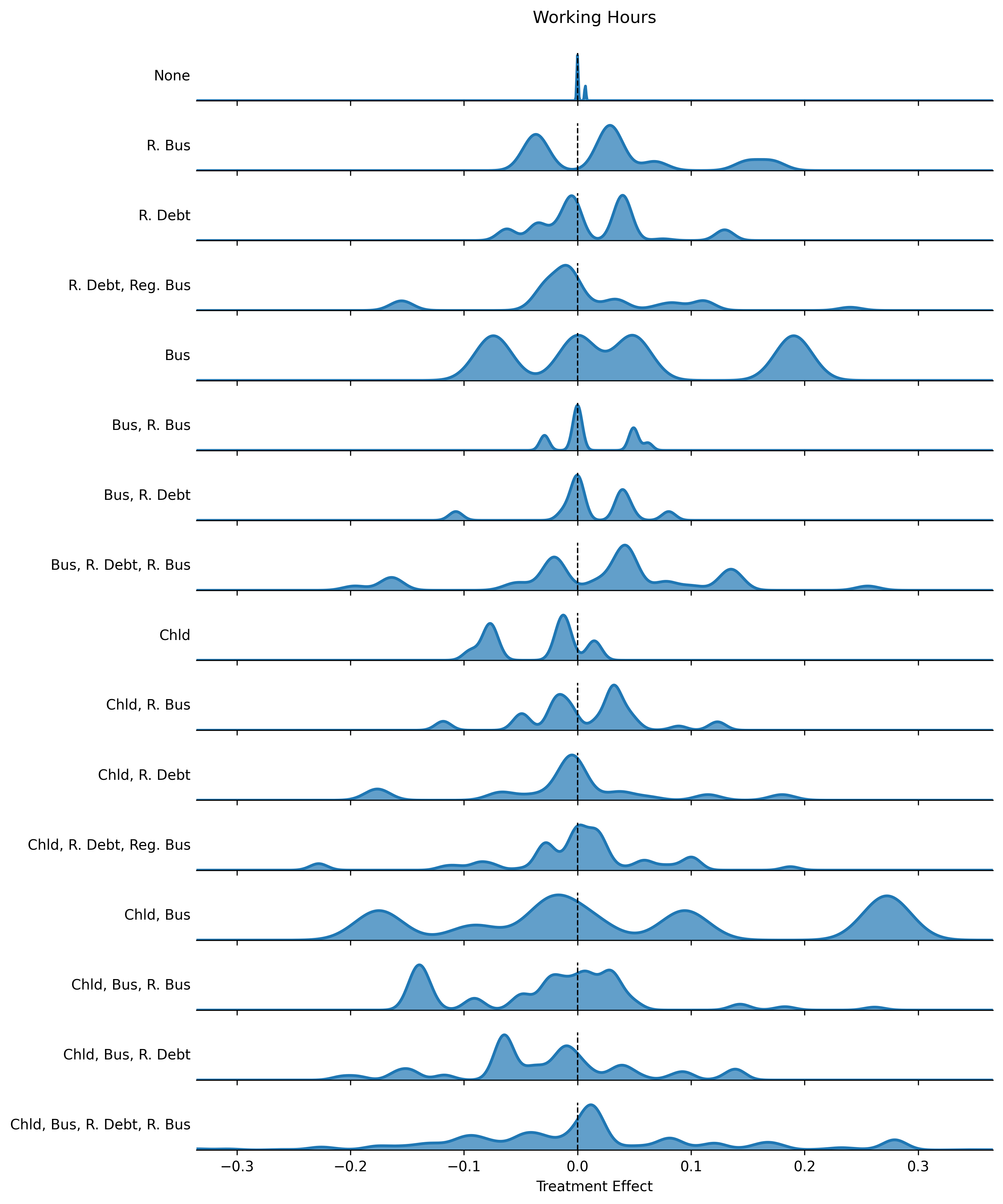}
    \end{subfigure}%
    \\
    \begin{subfigure}[t]{0.5\textwidth}
        \centering
        \includegraphics[height=3.5in]{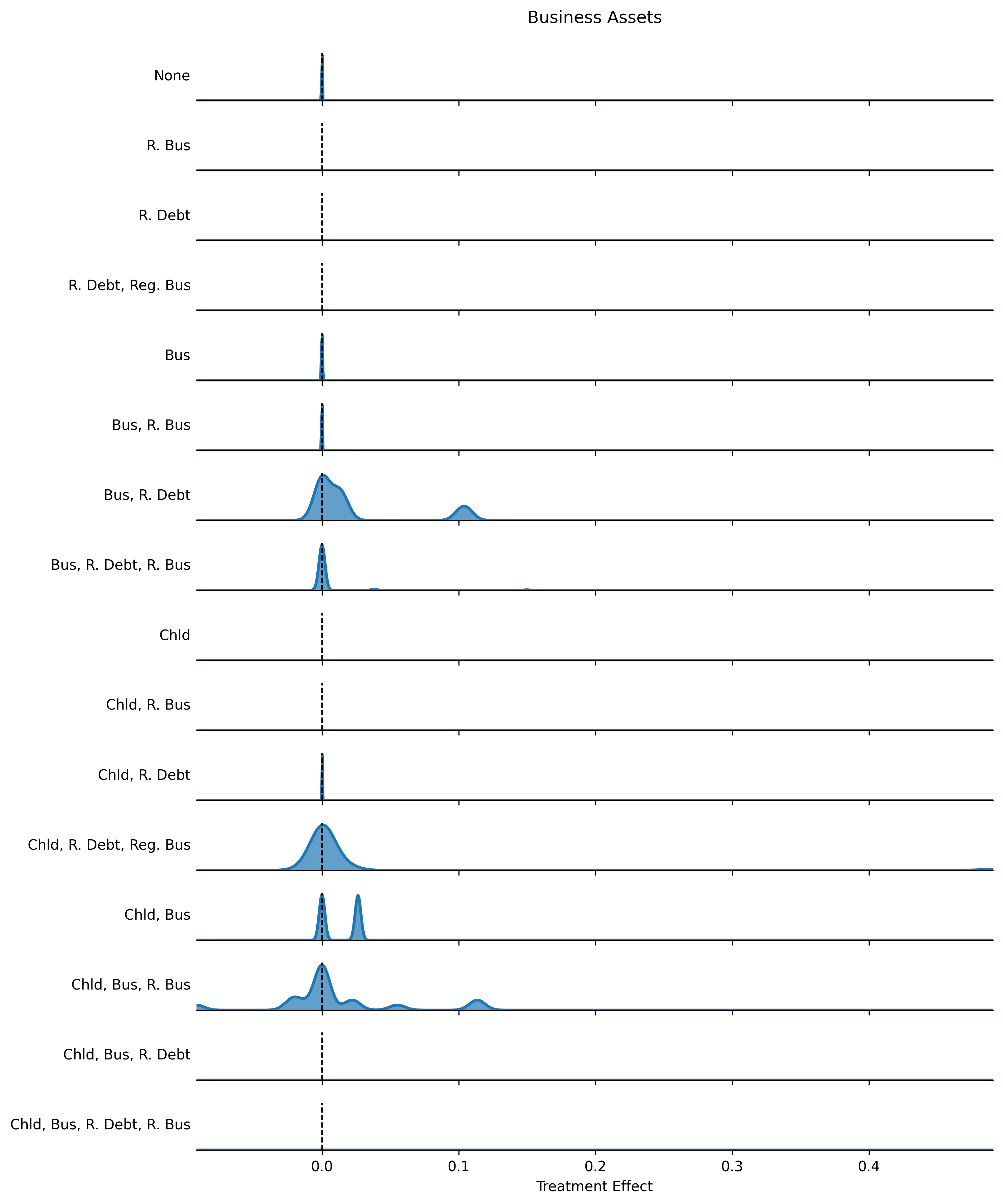}
    \end{subfigure}%
    ~ 
    \begin{subfigure}[t]{0.5\textwidth}
        \centering
        \includegraphics[height=3.5in]{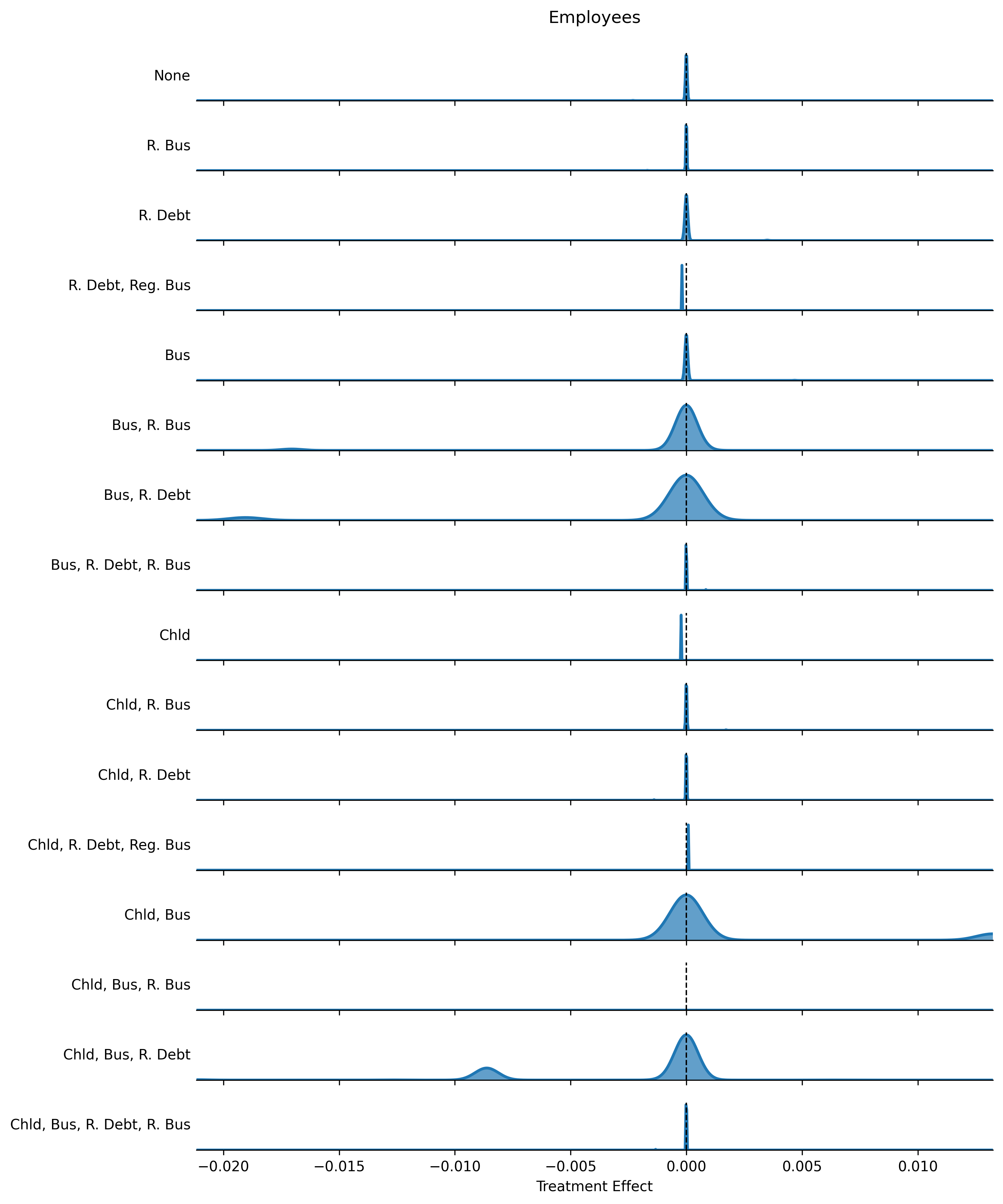}
    \end{subfigure}%
    \caption{The posterior distributions for the treatment effect on temptation expenses, working hours, business assets, and number of employees restricted to the RPS. See the corresponding quantized heatmaps in Figure \ref{fig:microcredit-mosiac-fx-full}.}
    \label{fig:microcredit-het-dist-appendix-temp}
\end{figure}

\begin{figure}[!p]
    \centering
    \begin{subfigure}[t]{0.5\textwidth}
        \centering
        \includegraphics[height=3.5in]{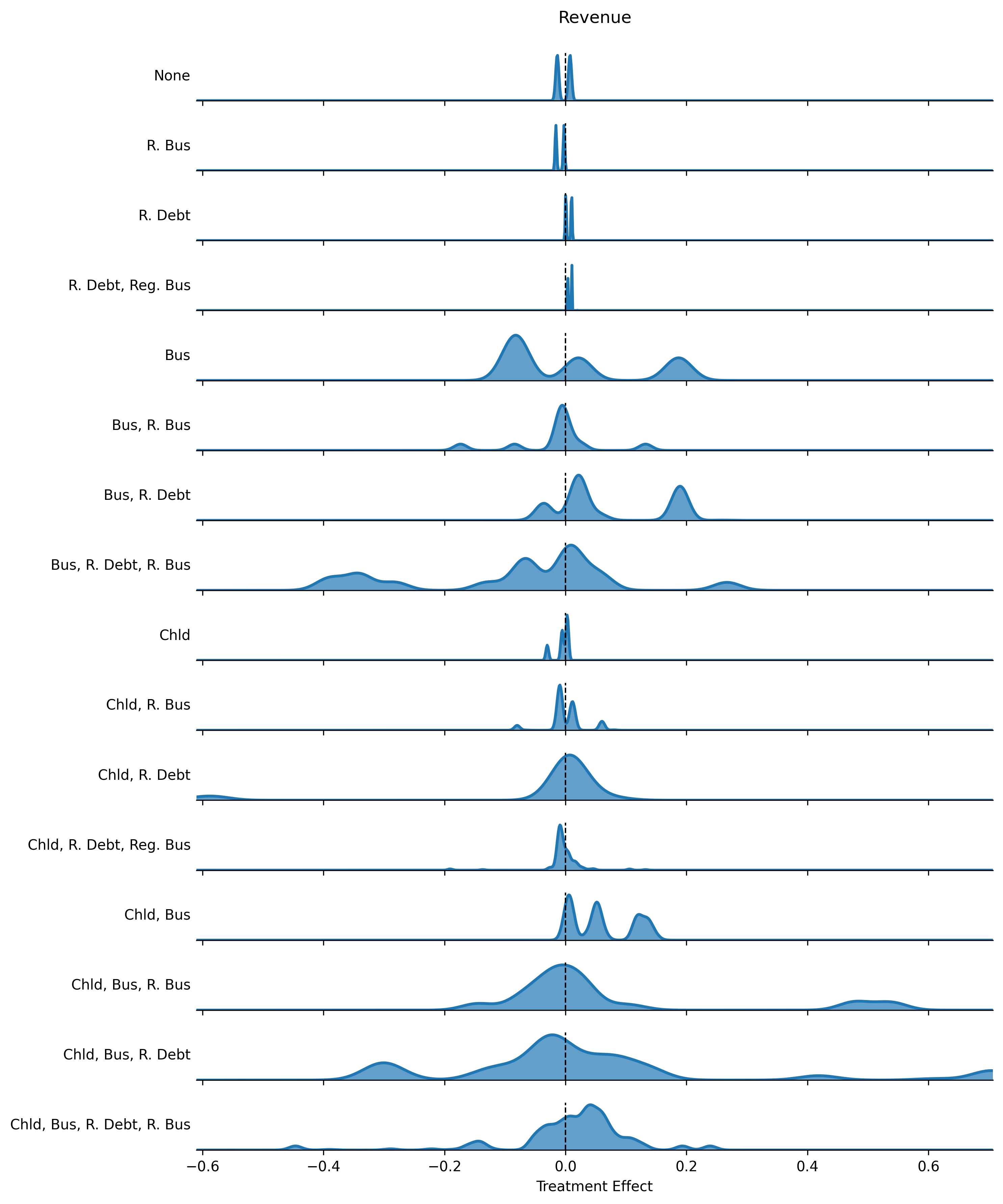}
    \end{subfigure}%
    ~ 
    \begin{subfigure}[t]{0.5\textwidth}
        \centering
        \includegraphics[height=3.5in]{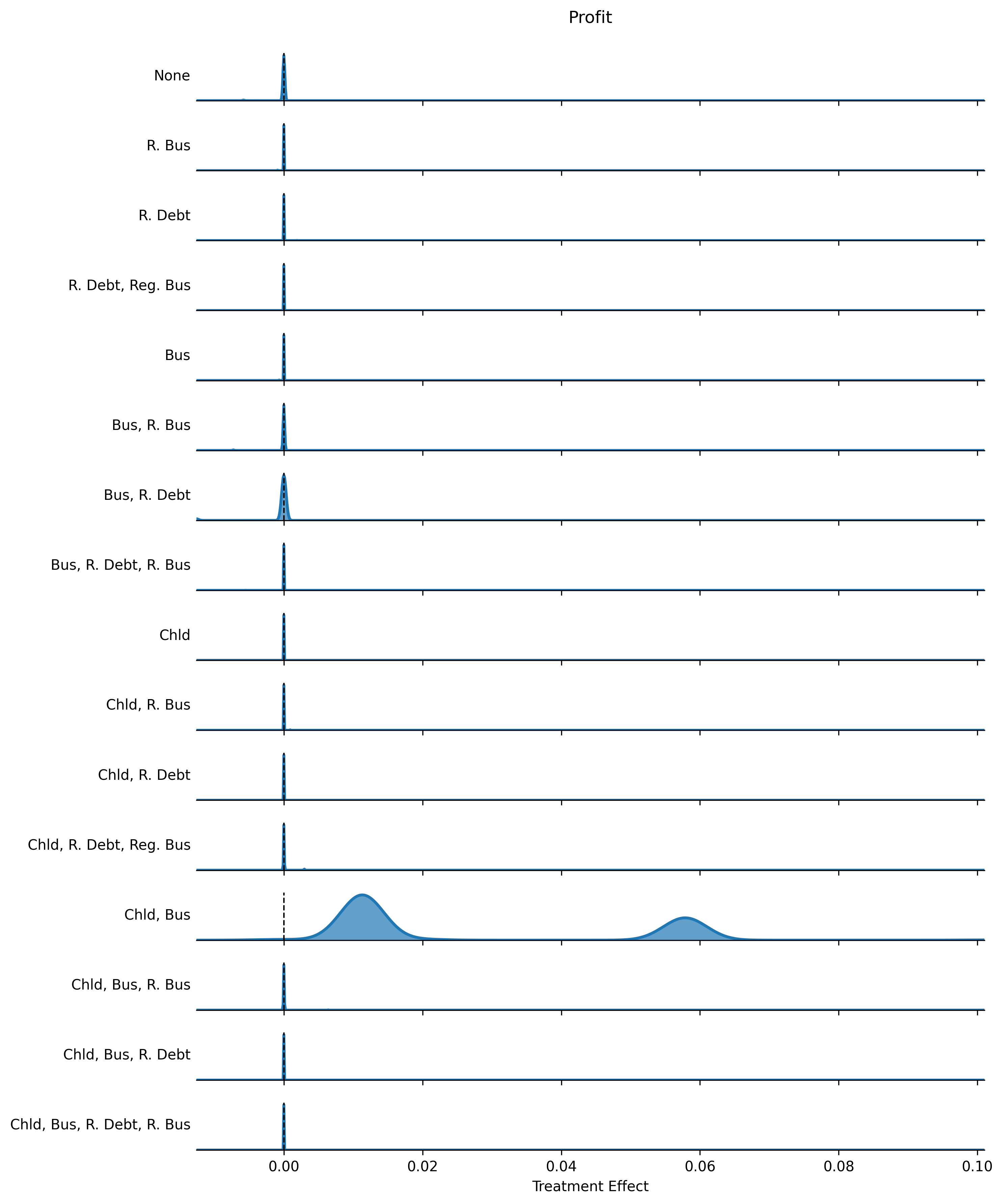}
    \end{subfigure}%
    \\
    \begin{subfigure}[t]{0.5\textwidth}
        \centering
        \includegraphics[height=3.5in]{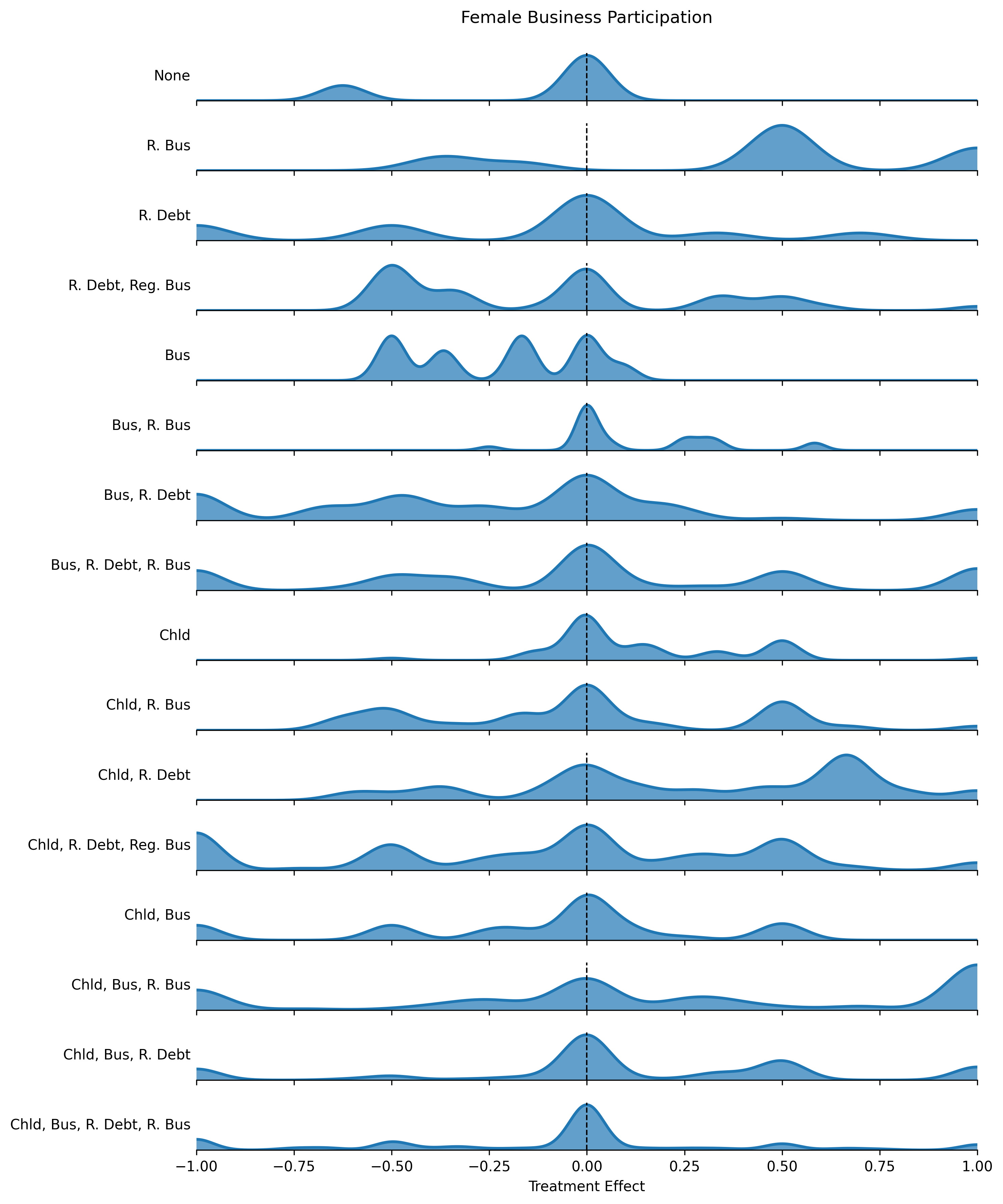}
    \end{subfigure}%
    ~ 
    \begin{subfigure}[t]{0.5\textwidth}
        \centering
        \includegraphics[height=3.5in]{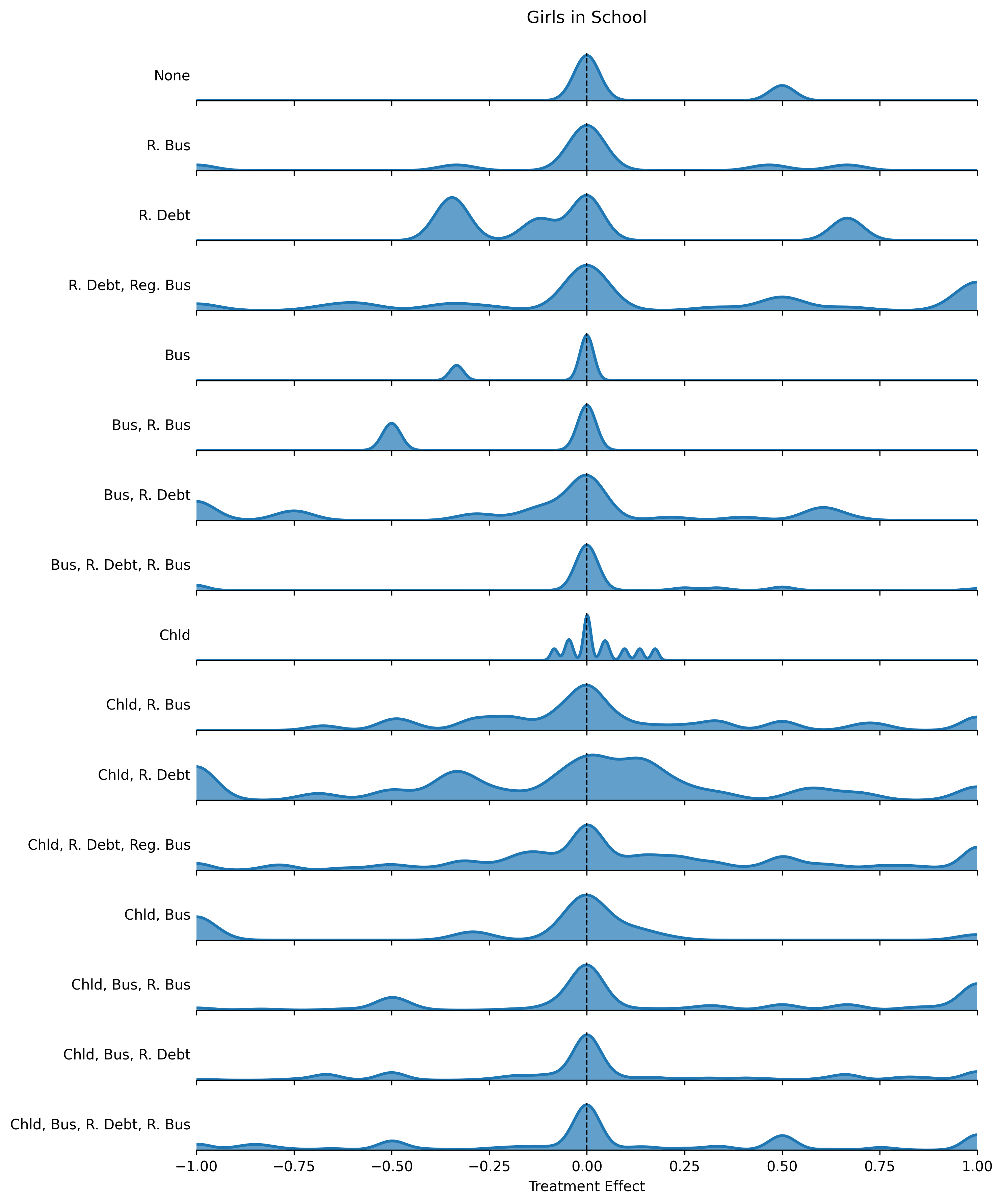}
    \end{subfigure}%
    \caption{The posterior distributions for the treatment effect on revenue, profit, female business participation, and education of girls restricted to the RPS. See the corresponding quantized heatmaps in Figure \ref{fig:microcredit-mosiac-fx-full}.}
    \label{fig:microcredit-het-dist-appendix-profit}
\end{figure}

For the microcredit data from \cite{banerjee2015miracle}, we present the results for all profiles in Figure \ref{fig:microcredit-mosiac-fx-full}. This includes the robust profiles we discussed in Figure \ref{fig:microcredit-mosiac-fx} as well as the non-robust ones.
We also show the posterior densities of these effects, restricted to the RPS, in Figures \ref{fig:microcredit-het-dist-appendix-loan}, \ref{fig:microcredit-het-dist-appendix-temp}, and \ref{fig:microcredit-het-dist-appendix-profit}.

\begin{figure}[!tb]
    \centering
    \includegraphics[height=6.5in]{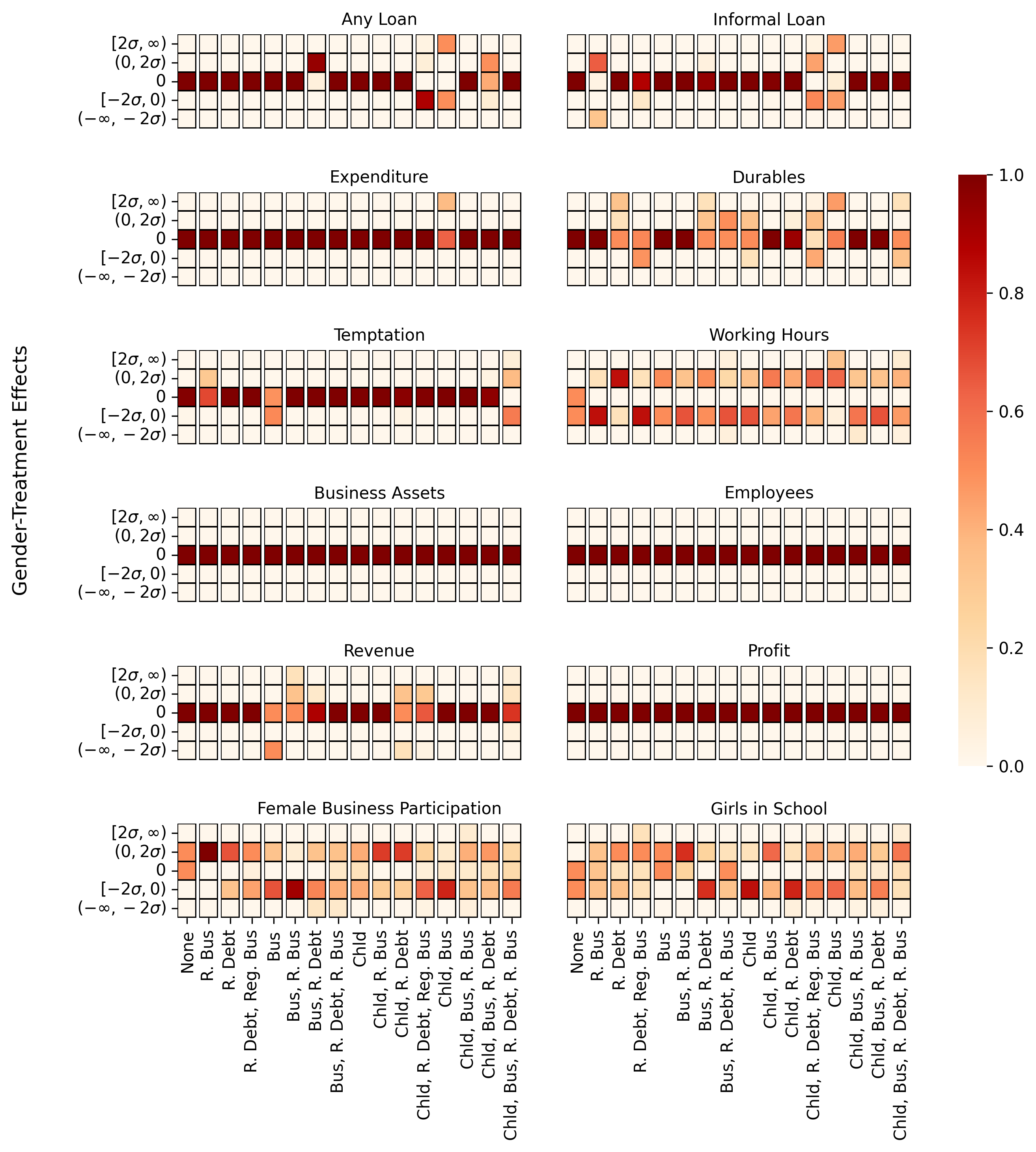}
    \caption{
    Here, we visualize the average number of models in the Rashomon set indicating a positive, zero, or negative effect. The axis labels should be read as in Figure \ref{fig:microcredit-mosiac-fx-full}.
    }
    \label{fig:microcredit-mosiac-fx-gender}
\end{figure}

\begin{figure}[!p]
    \centering
    \begin{subfigure}[t]{0.5\textwidth}
        \centering
        \includegraphics[height=3.5in]{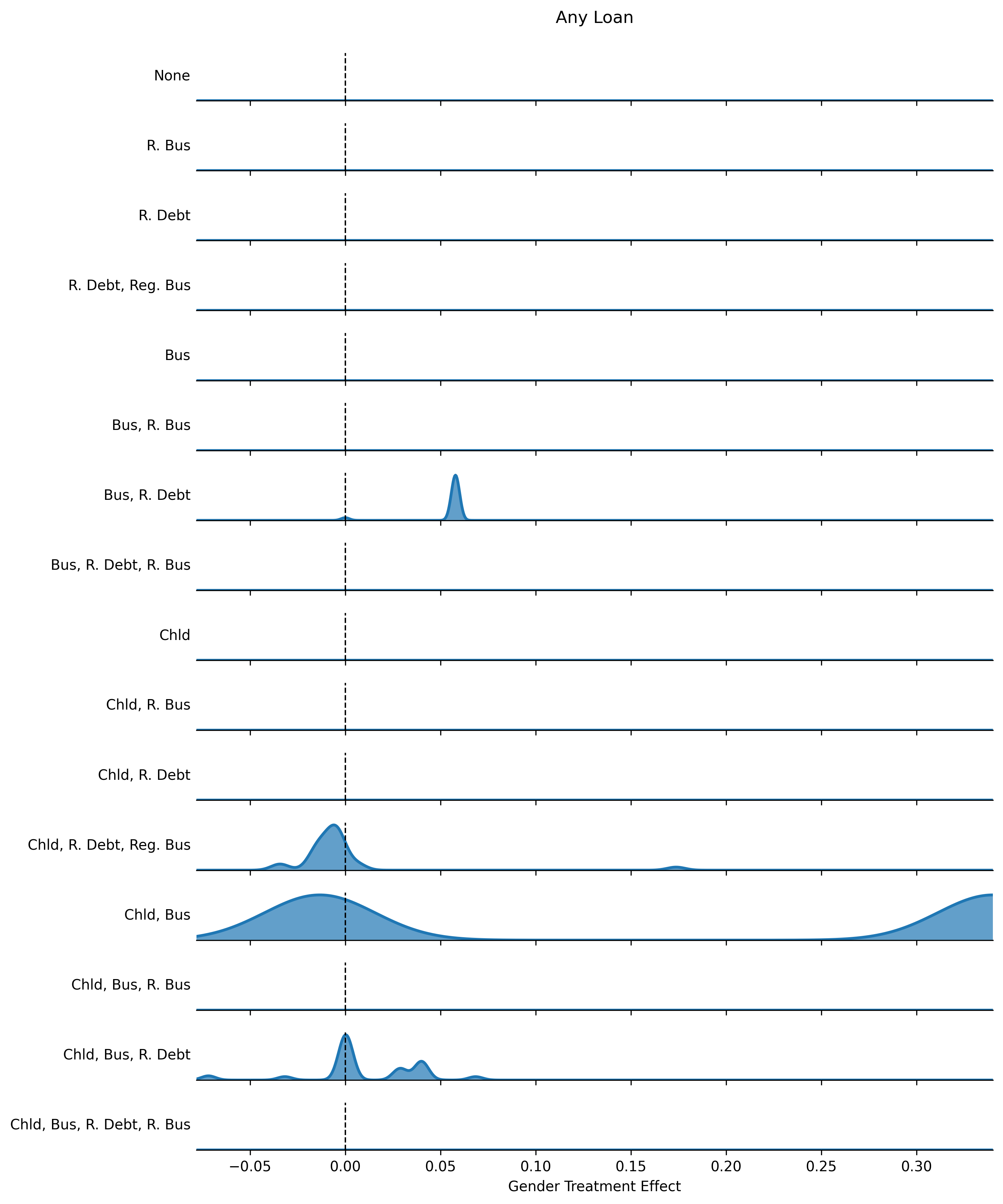}
    \end{subfigure}%
    ~ 
    \begin{subfigure}[t]{0.5\textwidth}
        \centering
        \includegraphics[height=3.5in]{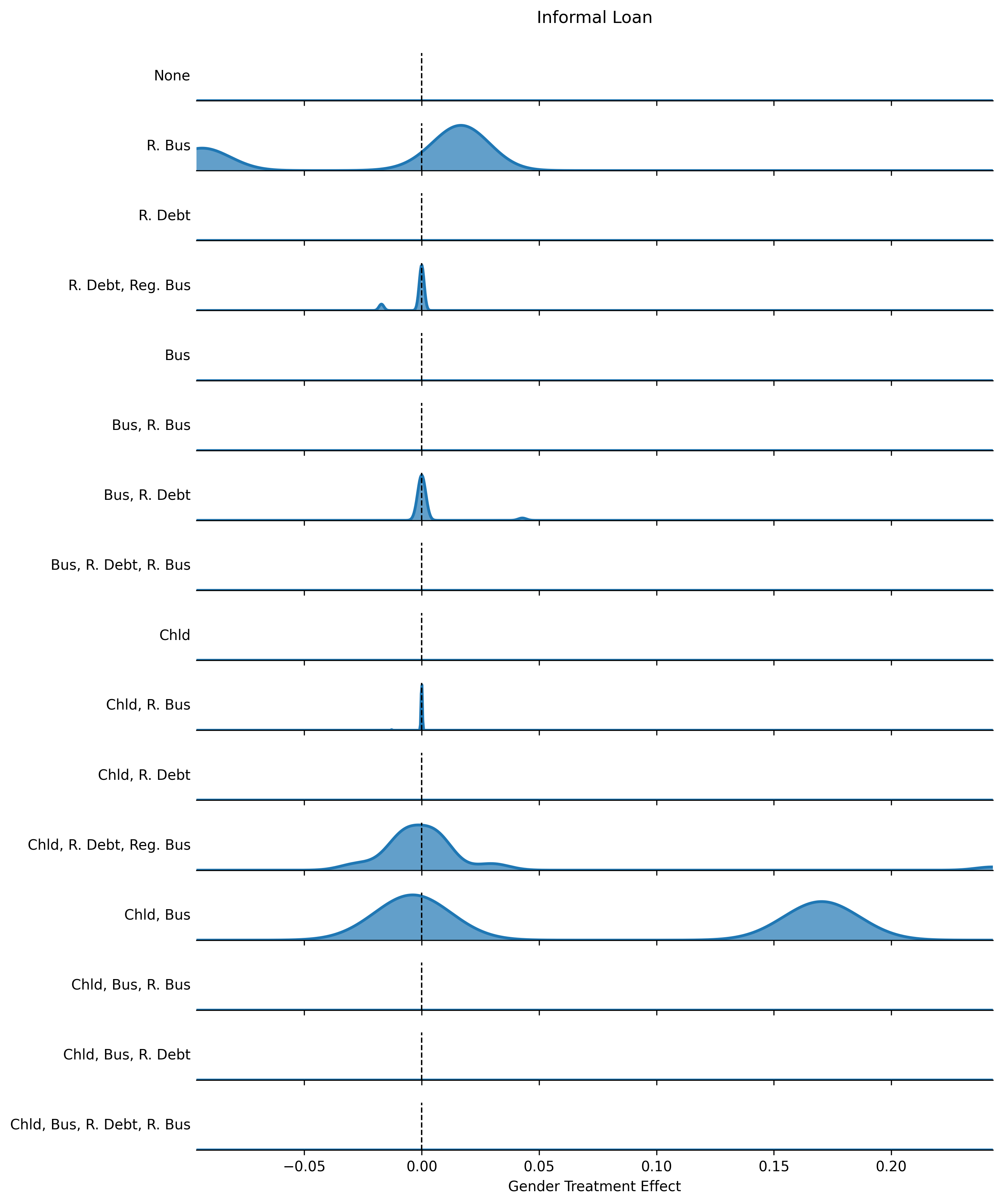}
    \end{subfigure}%
    \\
    \begin{subfigure}[t]{0.5\textwidth}
        \centering
        \includegraphics[height=3.5in]{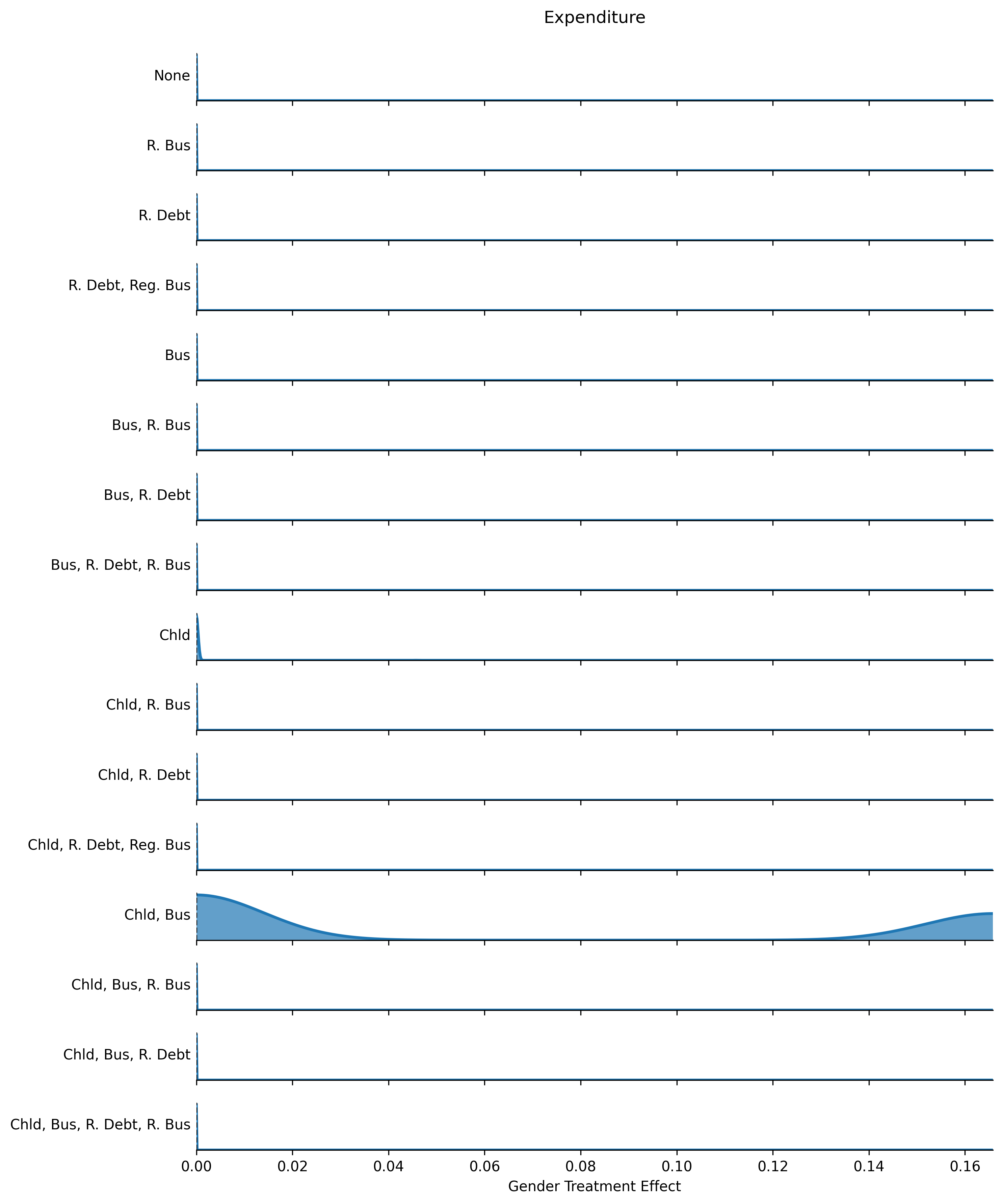}
    \end{subfigure}%
    ~ 
    \begin{subfigure}[t]{0.5\textwidth}
        \centering
        \includegraphics[height=3.5in]{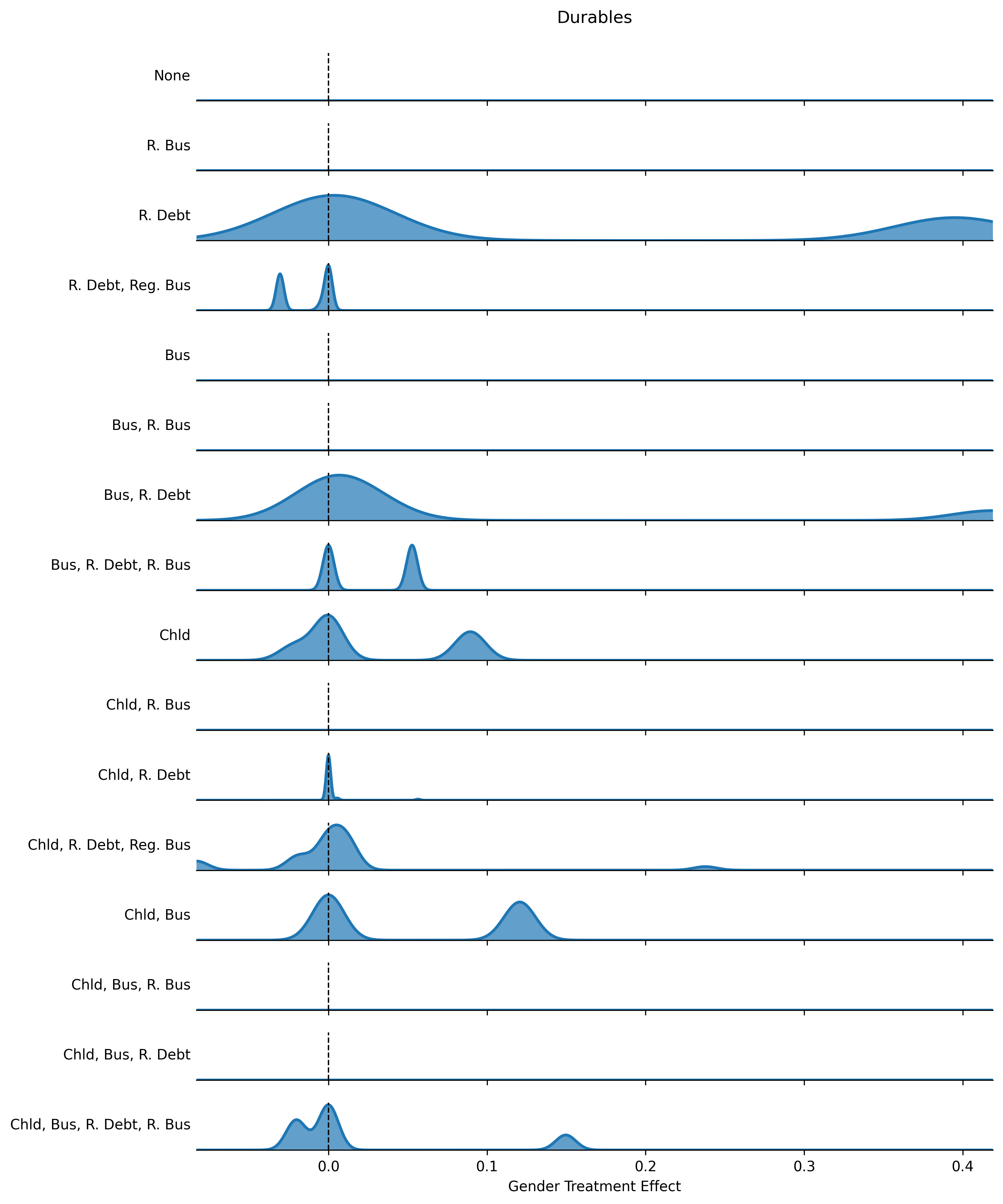}
    \end{subfigure}%
    \caption{The posterior distributions for the gender treatment effect on loan amounts and total and durable expenditures restricted to the RPS. See the corresponding quantized heatmaps in Figure \ref{fig:microcredit-mosiac-fx-gender}.}
    \label{fig:microcredit-het-dist-appendix-gender-loan}
\end{figure}

\begin{figure}[!p]
    \centering
    \begin{subfigure}[t]{0.5\textwidth}
        \centering
        \includegraphics[height=3.5in]{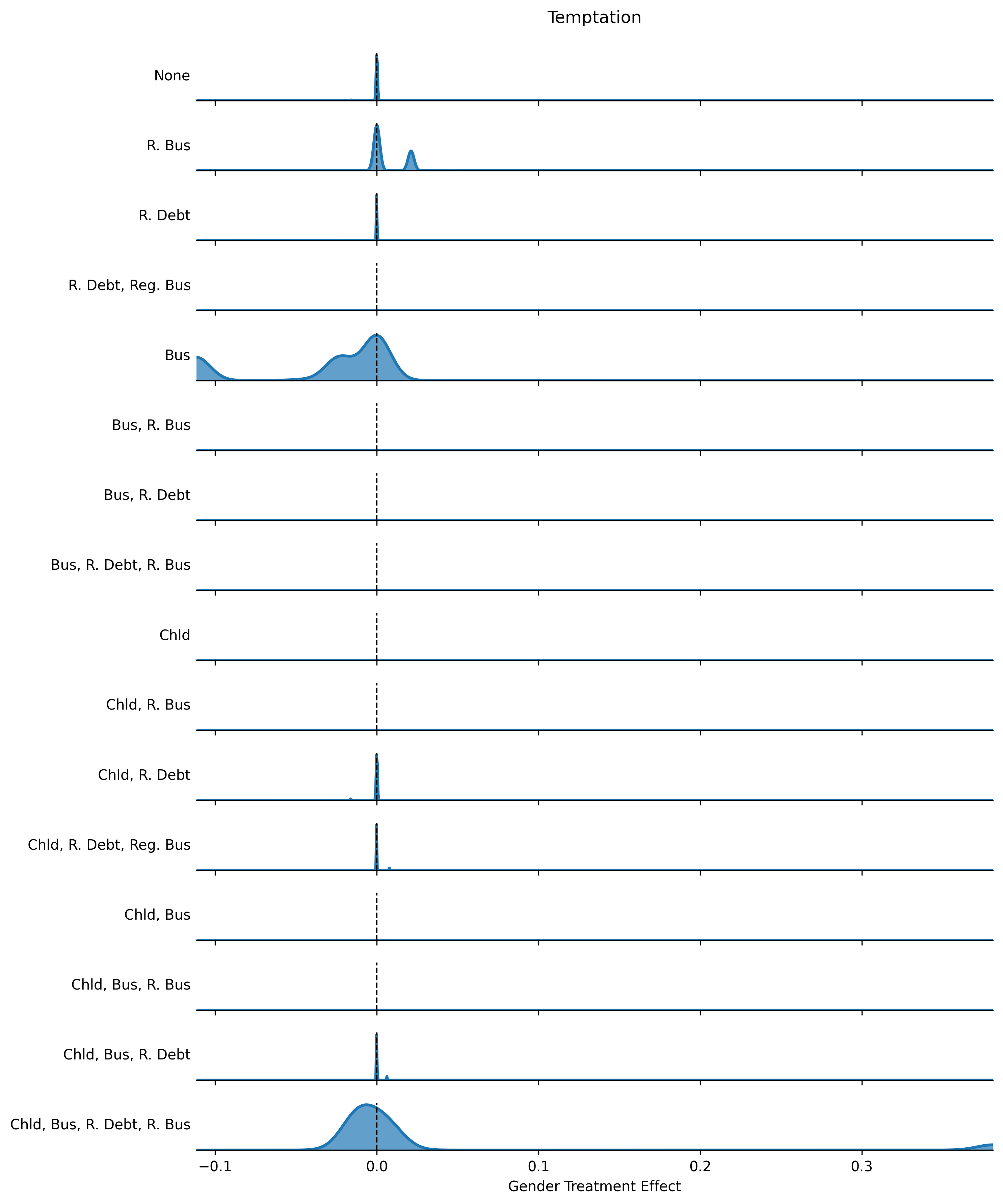}
    \end{subfigure}%
    ~ 
    \begin{subfigure}[t]{0.5\textwidth}
        \centering
        \includegraphics[height=3.5in]{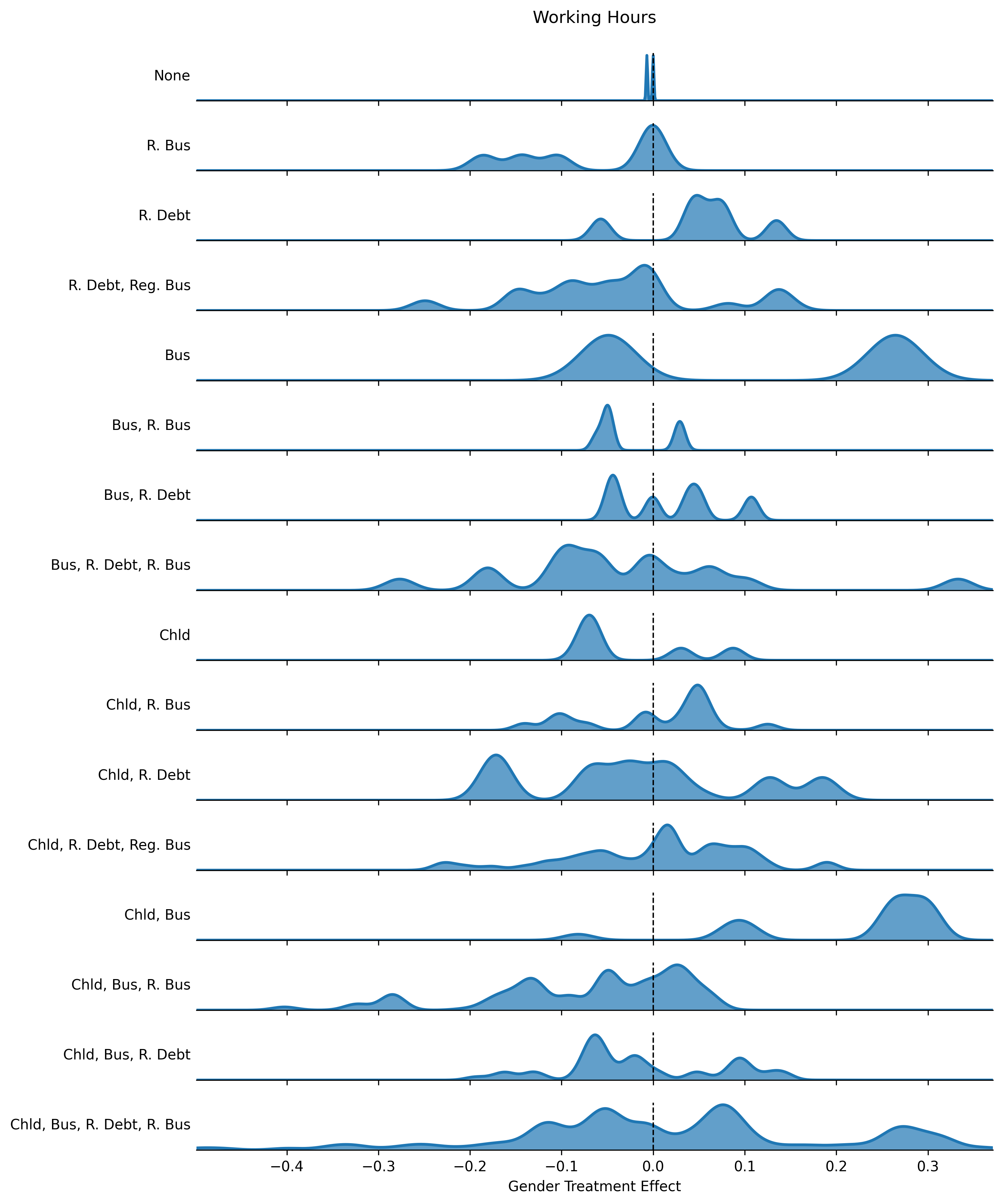}
    \end{subfigure}%
    \\
    \begin{subfigure}[t]{0.5\textwidth}
        \centering
        \includegraphics[height=3.5in]{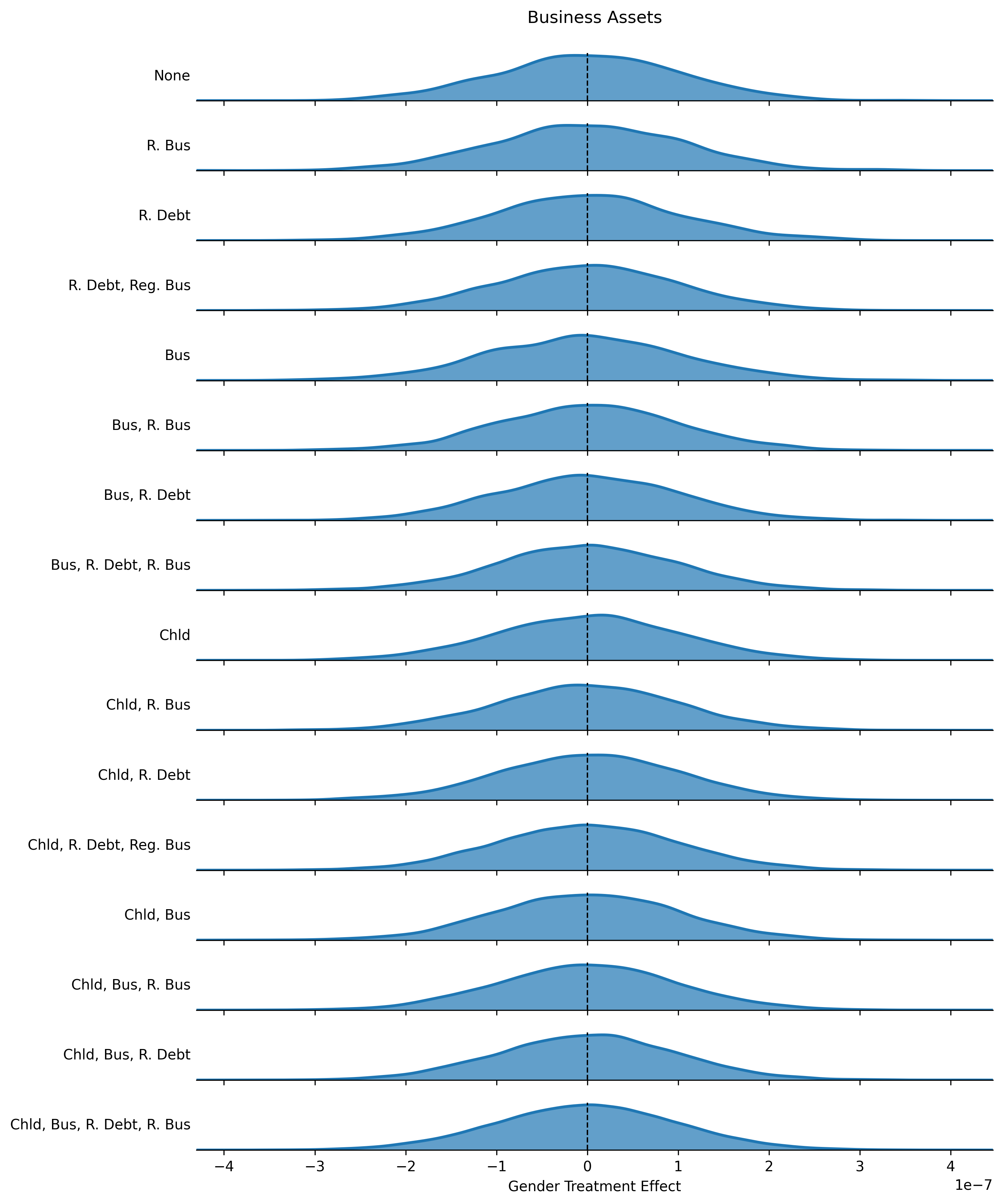}
    \end{subfigure}%
    ~ 
    \begin{subfigure}[t]{0.5\textwidth}
        \centering
        \includegraphics[height=3.5in]{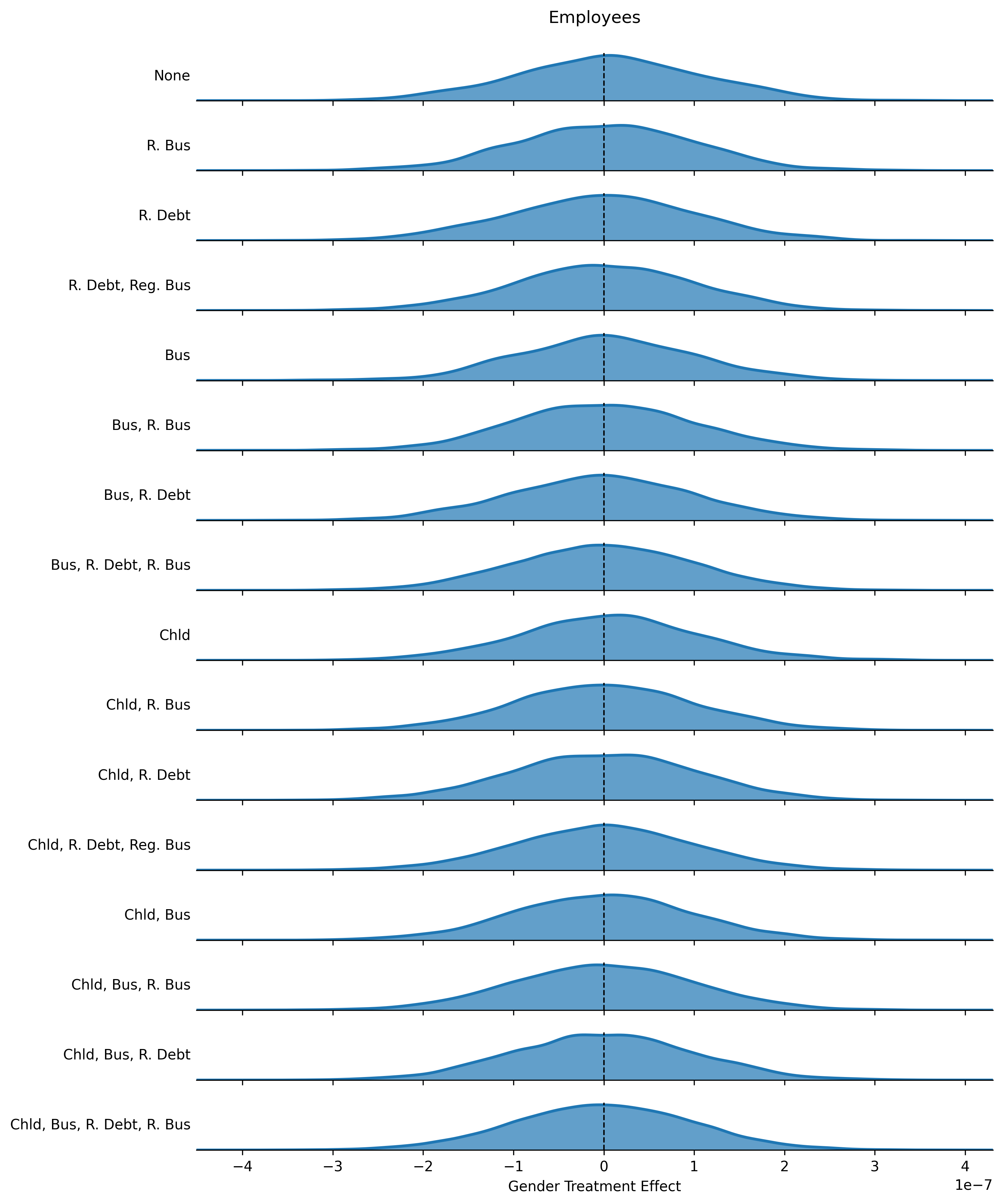}
    \end{subfigure}%
    \caption{The posterior distributions for the gender treatment effect on temptation expenses, working hours, business assets, and number of employees restricted to the RPS. See the corresponding quantized heatmaps in Figure \ref{fig:microcredit-mosiac-fx-gender}.}
    \label{fig:microcredit-het-dist-appendix-gender-temp}
\end{figure}

\begin{figure}[!p]
    \centering
    \begin{subfigure}[t]{0.5\textwidth}
        \centering
        \includegraphics[height=3.5in]{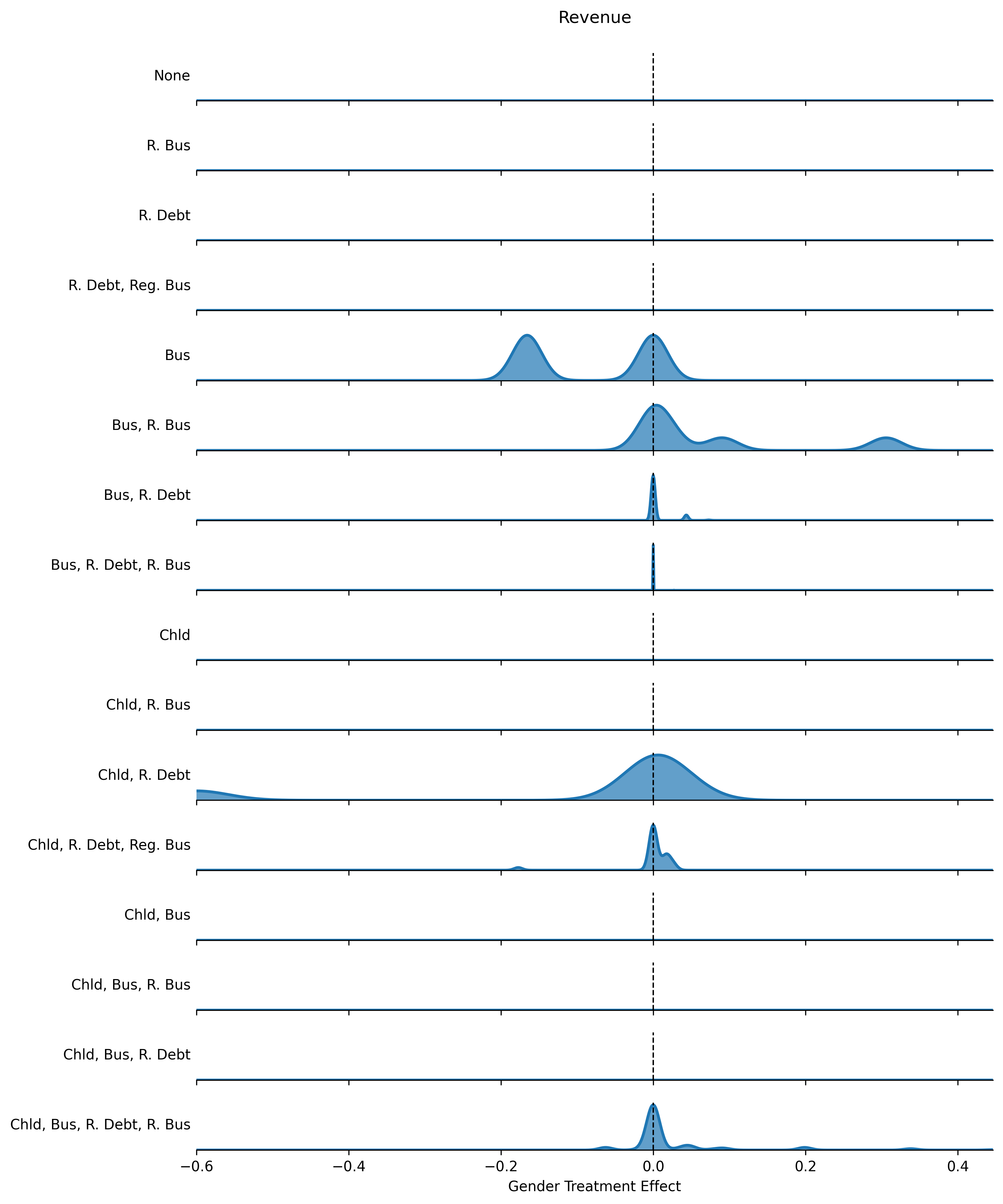}
    \end{subfigure}%
    ~ 
    \begin{subfigure}[t]{0.5\textwidth}
        \centering
        \includegraphics[height=3.5in]{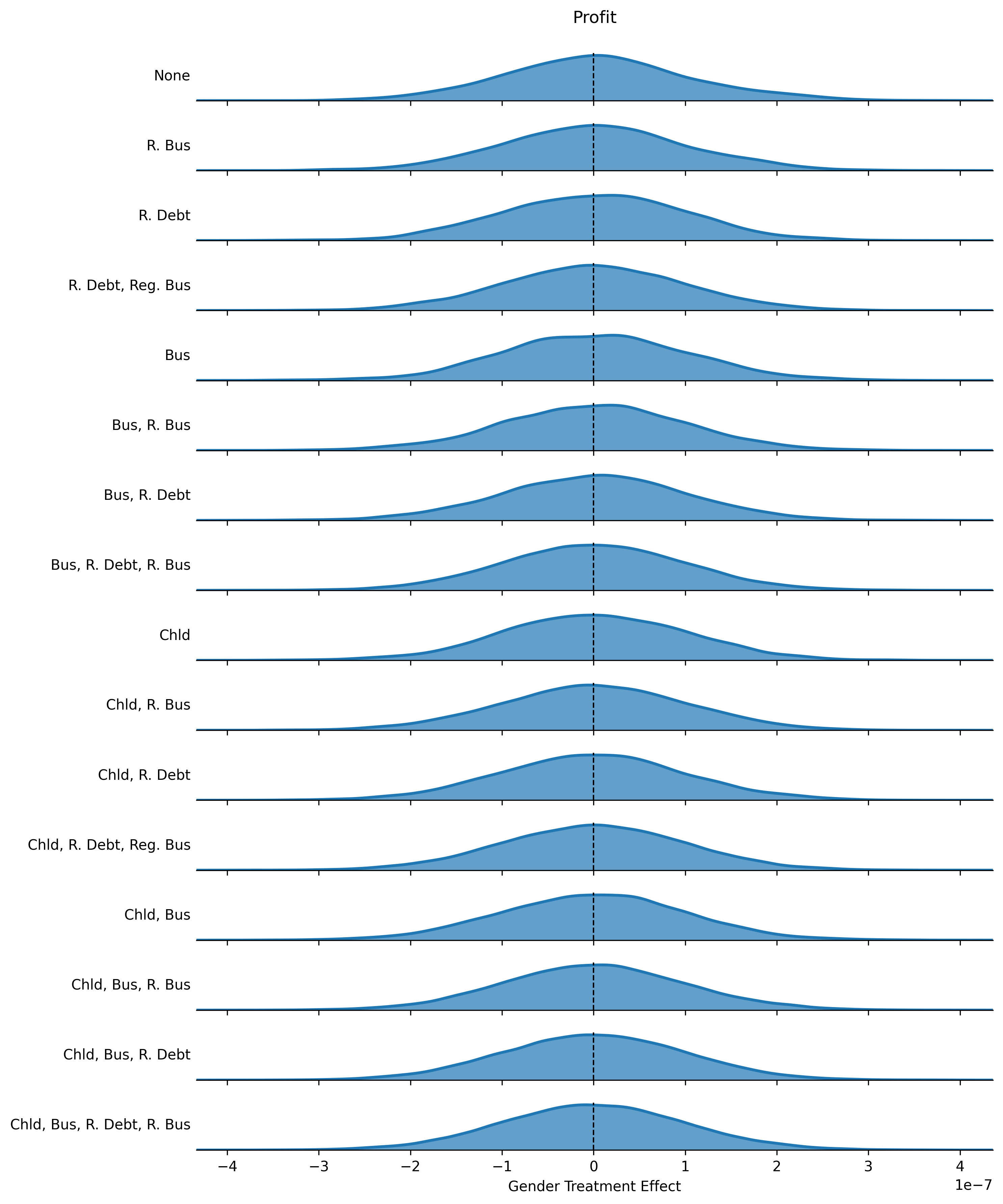}
    \end{subfigure}%
    \\
    \begin{subfigure}[t]{0.5\textwidth}
        \centering
        \includegraphics[height=3.5in]{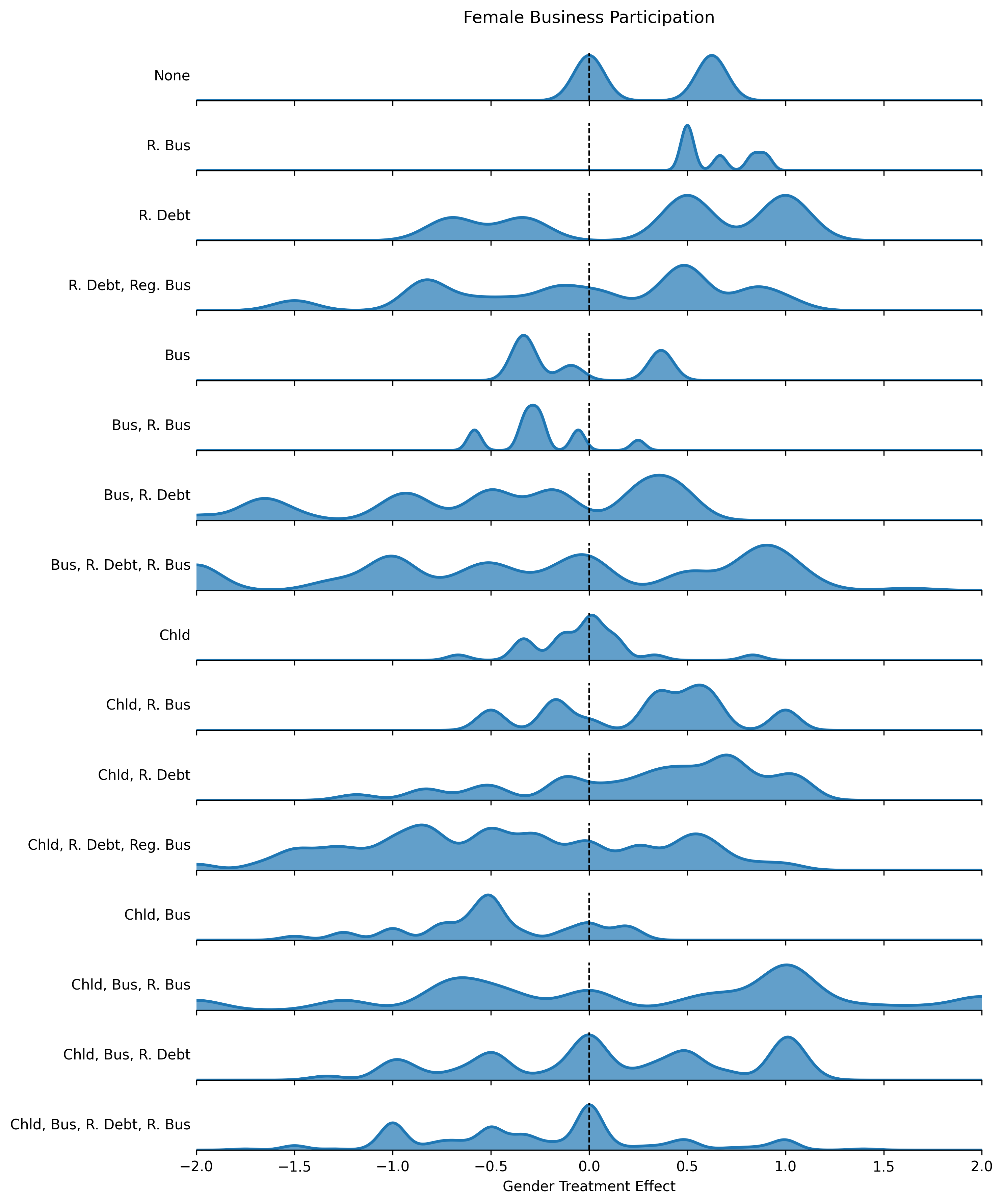}
    \end{subfigure}%
    ~ 
    \begin{subfigure}[t]{0.5\textwidth}
        \centering
        \includegraphics[height=3.5in]{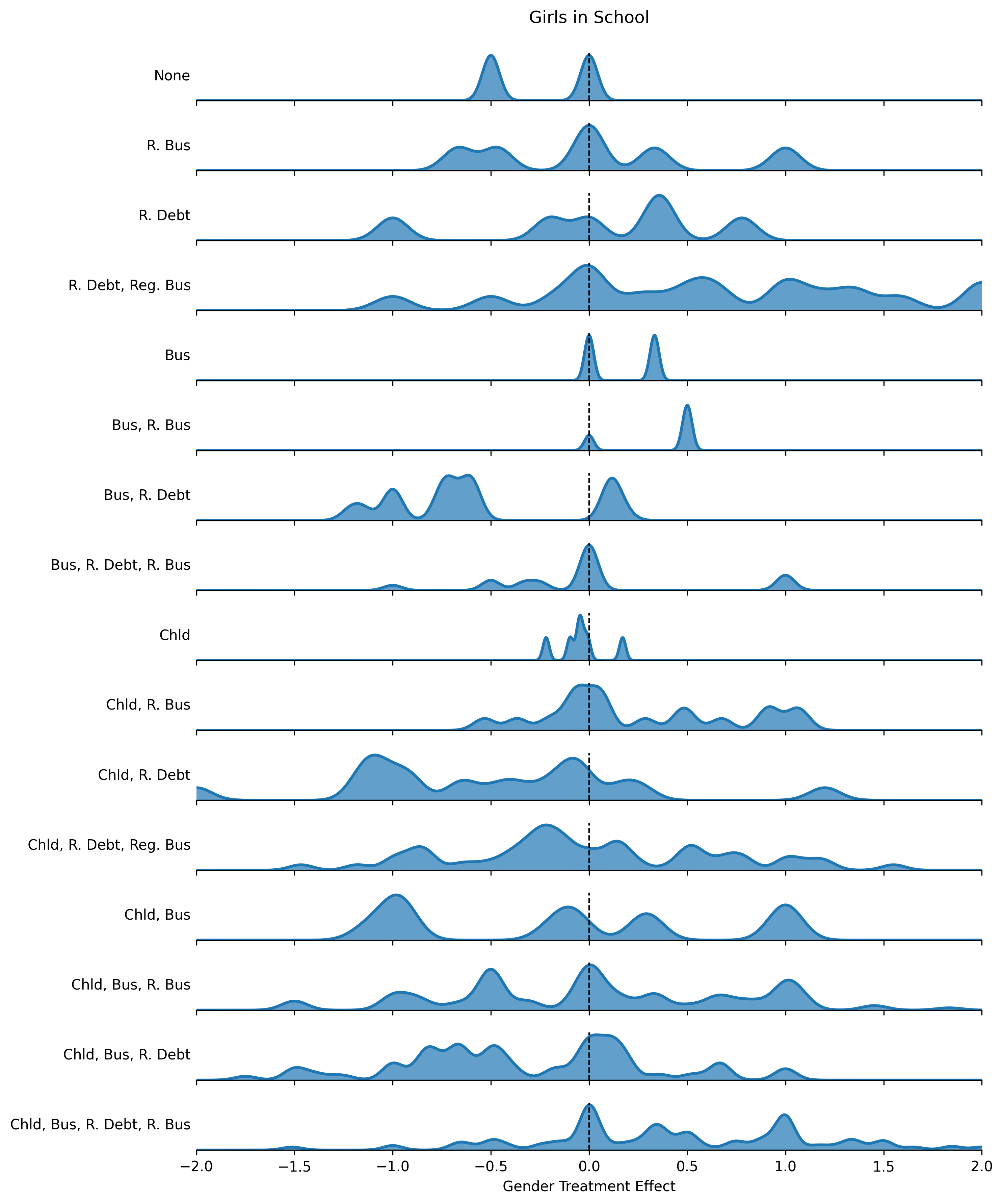}
    \end{subfigure}%
    \caption{The posterior distributions for the gender treatment effect on revenue, profit, female business participation, and education of girls restricted to the RPS. See the corresponding quantized heatmaps in Figure \ref{fig:microcredit-mosiac-fx-gender}.}
    \label{fig:microcredit-het-dist-appendix-gender-profit}
\end{figure}

Additionally, we look at the treatment effect heterogeneity across genders,
\begin{align*}
    \textrm{HTE}_{\mathbf{x}}( \Pi) &= \E \left[ \left\{Y_i(1, F, \mathbf{x}) - Y_i(0, F, \mathbf{x}) \right\} - \left\{ Y_i(1, M, \mathbf{x}) - Y_i(0, M, \mathbf{x})\right\} \mid \Pi \right],
\end{align*}
where $Y_i(\cdot, F, \cdot)$ is interpreted as the potential outcome of household $i$ were it headed by a woman, and $Y_i(\cdot, M, \cdot)$ is the potential outcome of household $i$ were it headed by a man. 
As before, we use the sample means $\widehat{y}(\cdot)$ to find $\widehat{\textrm{HTE}}_{\mathbf{x}}$ and $\textrm{sign}\{ \widehat{\textrm{HTE}}_{\mathbf{x}} \}$. Again, we repeat the same heterogeneity visualization exercise by calculating $c(\textrm{HTE}_{\mathbf{x}}, I)$ in Equation \ref{eq:RPS-confidence-counter} in Figure \ref{fig:microcredit-mosiac-fx-gender}. We also visualuze the posteriors, restricted to the RPS, in Figures \ref{fig:microcredit-het-dist-appendix-gender-loan}, \ref{fig:microcredit-het-dist-appendix-gender-temp}, and \ref{fig:microcredit-het-dist-appendix-gender-profit}. For most profiles, we see essentially no robust conclusions about gender heterogeneity in treatment effects. We highlight a few robust items below.

We see an increase in loans procured by households headed by women with past business experience when compared to households headed by men. When these households are already in debt with no previous experience, they tend to borrow less. We see no heterogeneity by gender in the amount of informal loans procured.

Households headed by women tend to consistently spend more. However, they spend more money on durable goods than households headed by men. We also see that, in the absence of past experience, there is a decline in expenditure on tempting goods compared to households headed by men. We also see a higher tendency for women to invest in business assets more than men.

We find that households headed by women with no past experience have a lower revenue than men. But this effect is reversed when the households do have previous business experience. However, there is no heterogeneity by gender in the profit or the number of employees. We also find that households headed by women tend to spend fewer hours working when they are in debt or when there is regional competition. But this makes a negligible difference in the profits.

We find that in households headed by women, there is less participation in the business by women if the household is in debt and there is competition from neighbors. We also find that fewer girls attend school in households headed by women with no previous experience than in households headed by men.

\end{document}